\documentclass[11pt,a4paper]{article}

\usepackage[pdftex]{graphicx}
\usepackage[integrals]{wasysym}
\usepackage{amsfonts}
\usepackage{ifsym}
\usepackage{pifont}
\usepackage{footmisc}
\usepackage[normalem]{ulem}
\usepackage{enumitem}
\usepackage{graphicx}
\usepackage{mathtools}
\usepackage{subcaption}

\usepackage{caption} %

\usepackage[section]{placeins} %

\usepackage{jheppub}

\usepackage{mwe} %

\usepackage{listings}
\usepackage[dvipsnames]{xcolor}
\usepackage{pifont}%

\graphicspath{{./images/}}

\usepackage{multirow}

\usepackage{ifsym}
\usepackage{pifont}
\usepackage{footmisc}
\usepackage[normalem]{ulem}
\usepackage{enumitem}
\usepackage{color}
\usepackage{graphicx}
\usepackage{dcolumn}
\usepackage{bm}
\usepackage{hyperref}
\usepackage[]{todonotes}

\newcommand{\figref}[1]{Figure~\ref{#1}}
\newcommand{\secref}[1]{Section~\ref{#1}}
\newcommand{\appref}[1]{Appendix~\ref{#1}}
\newcommand{\tabref}[1]{Table~\ref{#1}}

\makeatletter
\renewcommand{\todo}[2][]{%
    \@todo[caption={#2}, #1]{\begin{spacing}{0.5}#2\end{spacing}}%
} 
\makeatother 
\usepackage{setspace}

\newcommand{\ba}{\begin{aligned}} 
\newcommand{\ea}{\end{aligned}}

\def\be#1\ee{\begin{align}#1\end{align}}

\newcommand{\eps}{\varepsilon}

\renewcommand{\Re}{\mathop{\mathrm{Re}}}
\renewcommand{\Im}{\mathop{\mathrm{Im}}}

\newcommand{\Csphere}{{}^\bullet\kern-1.2pt C}
\newcommand{\Ctorus}{{}^\circ\kern-1.2pt C}

\newcommand{\nn}{\nonumber}

\newcommand{\COMMENT}[1]{}

\newcommand{\neqa}{\nonumber\end{eqnarray}}

\newcommand{\<}{{\langle}}
\renewcommand{\>}{{\rangle}}

\newcommand{\re}{\relax{\rm I\kern-.18em R}}

\def\su2{{SU(2)}}

\def\eps{{\epsilon}}

\def\[{\left[}
\def\]{\right]}

\def\l{\lambda}

\def\({\left(}
\def\){\right)}
\def\[{\left[}
\def\]{\right]}

\def\<{\langle}
\def\>{\rangle}

\def\i2{\frac{i}{2}}

\def\2F1{\,_2{\rm F}_1}

\newcolumntype{L}[1]{>{\raggedright\let\newline\\\arraybackslash\hspace{0pt}}m{#1}}
\newcolumntype{C}[1]{>{\centering\let\newline\\\arraybackslash\hspace{0pt}}m{#1}}
\newcolumntype{R}[1]{>{\raggedleft\let\newline\\\arraybackslash\hspace{0pt}}m{#1}}

\newcommand{\beq}{\begin{equation}}
\newcommand{\eeq}{\end{equation}}
\newcommand{\beqq}{\begin{equation*}}
\newcommand{\eeqq}{\end{equation*}}
\newcommand\beqa{\begin{eqnarray}}
\newcommand\eeqa{\end{eqnarray}}
\newcommand\beqaa{\begin{eqnarray*}}
\newcommand\eeqaa{\end{eqnarray*}}
\newcommand\bea{\begin{array}}
\newcommand\eea{\end{array}}

\usepackage{mmacells}

\begin{document}

\title{Scattering amplitudes from dispersive iterations of unitarity}

\author{Piotr Tourkine$^{a}$,  Alexander Zhiboedov$^b$}

\affiliation{$^a$Laboratoire d'Annecy-le-Vieux de Physique Th\'eorique (LAPTh), CNRS \& Universit\'e Savoie Mont-Blanc, 9 chemin de Bellevue, 74940, Annecy, France }
\affiliation{$^b$CERN, Theoretical Physics Department, CH-1211 Geneva 23, Switzerland}

\abstract{
We present and numerically implement a computational method to construct relativistic scattering amplitudes that obey
analyticity, crossing, elastic and inelastic unitarity  in three and four spacetime dimensions. 
The algorithm is based on the Mandelstam representation of the amplitude and iterations of unitarity. 
The input for the iterative procedure is given by the multi-particle double
spectral density, the S-wave inelasticity, and the value of the amplitude at the crossing-symmetric point. 
The output, obtained at the fixed point of the iteration of unitarity, is a nonperturbative scattering amplitude. The amplitudes we obtain exhibit interesting features, such as non-zero particle production, intricate high-energy and near the two-particle threshold behavior. 
Scattering amplitudes obtained by
initializing the iteration process with zero (or small) multi-particle input end up close to saturating the S-matrix
bounds derived by other methods. 
There is a version of the iterative algorithm that is directly related to Feynman diagrams: it effectively re-sums infinitely many two-particle reducible planar Feynman graphs in the $\phi^4$ theory, which remarkably produces a unitary nonperturbative 
scattering amplitude function. Finally, we discuss how the algorithm can be further refined by including multi-particle unitarity.

\vspace{1cm}
}

\begin{flushleft}
 \hfill \parbox[c]{40mm}{CERN-TH-2023-025}
\end{flushleft}
\maketitle

\section{Introduction and summary of results}

The S-matrix bootstrap arises from the delicate tension between the relativistic concept of causality and the quantum-mechanical notion of unitarity \cite{Eden:1966dnq,Sommer:1970mr,iagolnitzer2014scattering}. 
Causality manifests itself through analyticity of scattering amplitudes, as well as crossing symmetry which relates different parts of the complex scattering energy plane to different scattering processes. Unitarity relates different scattering processes to each other in a non-linear fashion.\footnote{The simplest example being the optical theorem which equates the imaginary part of the forward amplitude to the probability to scatter in an arbitrary final state.} %

At its infancy, the S-matrix bootstrap almost exclusively attempted to describe scattering of hadrons which turned out to be a hard problem, still unsolved today. Its multifarious modern incarnation pursues a  broader goal of exploring the space of relativistic, causal, quantum-mechanical theories (see e.g. the recent reviews \cite{Kruczenski:2022lot,deRham:2022hpx}). The full characterization of this space includes the tasks of finding its boundaries, and locating in it theories of physical interest, such as QCD. 
These tasks come with the general belief that if 
a theory resides on the boundary, it is conceivable that it can be ``solved''.  This phenomenon is familiar in the domain of the conformal bootstrap \cite{Poland:2018epd}, and $2d$ S-matrices \cite{Cordova:2019lot}. To this day, it remains an open question as to whether there exist solvable, physical theories whose S-matrices can be efficiently calculated nonperturbatively in spacetime dimension $d>2$. More generally, the S-matrix bootstrap techniques can be combined with inputs from experiment, perturbative calculations, lattice simulations, or other methods \cite{Henning:2022xlj}, to pinpoint a given physical theory.

Various approaches to the S-matrix bootstrap enforce various amounts of the basic constraints of analyticity, crossing and unitarity. The more analyticity, crossing and unitarity is put in, the more stringent the constraints on the space of amplitudes are. 
The novelty of the techniques that we develop and implement in this paper is that 
\begin{itemize}
    \item we enforce the constraints of two-particle (elastic and inelastic) unitarity fully;
    \item we put in certain constraints dictated by multi-particle unitarity, e.g. the support of the multi-particle double spectral density;
    \item we control the amplitudes up to arbitrarily high energies and for any scattering angles (or impact parameters).
\end{itemize}

We implement these constraints by solving unitarity iteratively, while keeping analyticity and crossing symmetry manifest at every step.
The basic idea of the iteration algorithm goes back to the early days of the S-matrix bootstrap and the works of Mandelstam \cite{Mandelstam:1958xc,Mandelstam:1959bc,Mandelstam:1963iyb}, Chew and Frautschi \cite{Chew:1961dci}, Ter-Martirosyan \cite{ter1961equations} and others.\footnote{In the lecture notes \cite{Mandelstam:1963iyb}, Mandelstam mentions in this context a preprint by Kenneth Wilson, but we did not manage to find it.} The idea was significantly sharpened and further developed by Atkinson in a series of papers
\cite{Atkinson:1968hza,Atkinson:1968exe,Atkinson:1969eh,Atkinson:1970pe},
and various follow-ups
\cite{Kupsch:1969sv,Atkinson:1972ppn,Atkinson:1976fg,Johnson:1976dt,Johnson:1976vs,Johnson:1977rx}, where the \emph{existence} of analytic, crossing-symmetric, unitary amplitudes, and the convergence of the iteration algorithm based on the Mandelstam representation, were rigorously proven using functional analysis methods. A pedagogical introduction to these ideas can be found in the lecture notes by Atkinson \cite{Atkinson:1970zza}.

To the best of our knowledge, this approach to the S-matrix bootstrap has never been implemented in $d>2$.\footnote{In $d>2$, we could only find one attempt to numerically implement a simplified unitarity-inspired set of equations in $d=4$ by Boguta \cite{Boguta:1974bm}. In $d=2$ the algorithm was successfully implemented in our previous work \cite{Tourkine:2021fqh}, as a warm-up for the present paper.} In other words, while the \emph{equations} have been written long time ago, no \emph{solutions} have been produced. This is the problem we address in the present paper using numerical methods.

The precise goal of this paper is to apply these techniques to the simplest setup: $2 \to 2$ scattering amplitude of identical scalar particles in three and four spacetime dimensions. %
The amplitudes functions\footnote{We call them amplitude functions to emphasize  that we do not discuss an algorithm to generate the S-matrix beyond the $2 \to 2$ amplitude. In the rest of the paper, to avoid cluttering we will simply call them amplitudes.} that we construct here
satisfy \emph{maximal analyticity, crossing, elastic and inelastic unitarity}. They exhibit many interesting features, such as clearly identifiable particle production (consistent with the Aks theorem \cite{Aks:1965qga}),\footnote{The amplitudes that we obtain have much  larger inelasticity compared to the very weak asymptotic bound obtained in \cite{Martin:2017ndt}.} the detailed expected analytic structure coming from the Landau curves, and a nontrivial fixed angle and Regge behavior~\cite{Gribov:1961fm}. 
They are specified by a certain multi-particle production data which is the subject of our choice, and which serves as an input for the algorithm. %
To describe a given physical theory, such as QCD, one then needs to develop a model for the multi-particle production data. We describe all these notions in detail below. 

\begin{figure}[htb!]
  \centering
  \includegraphics[scale=1.2]{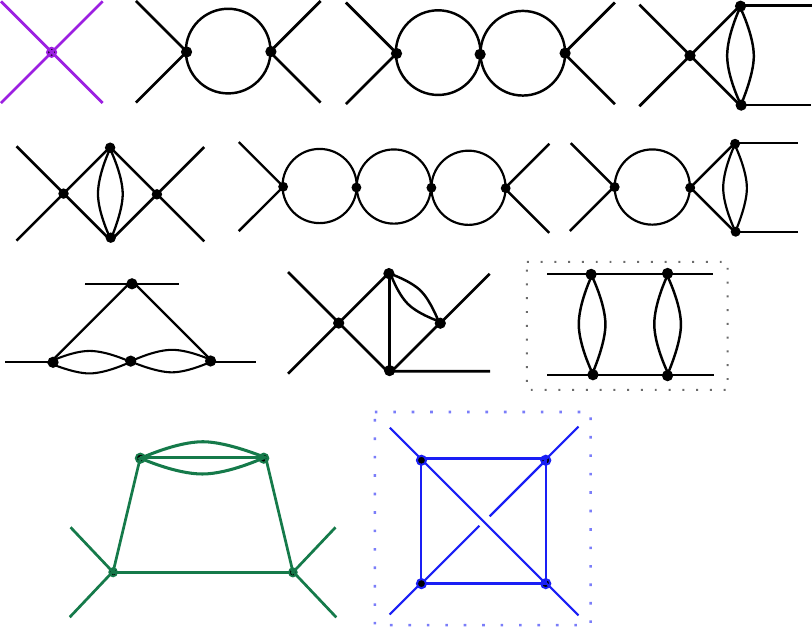}
  \caption{%
  Diagrams that generate $2\to 2$ scattering amplitude in $\phi^4$ theory up to three loops. Diagrams inside a dashed frame are those for which the double spectral density is non-zero. In our iteration scheme the tree-level contact diagram (magenta) has the meaning of the fixed value of the amplitude at the crossing-symmetric point $\textcolor{violet}{\lambda} \equiv T({4 m^2 \over 3}, {4 m^2 \over 3})$. The black diagrams are then generated by iterations of two-particle unitarity and crossing-symmetric Mandelstam representation. The ``frog'' diagram (green) describes multi-particle production in the S-wave, which serves as an input for the algorithm $\textcolor{OliveGreen}{\eta_{\text{MP}}(s)}$. The ``open envelope'' diagram (blue) is the first one that contributes to the multi-particle double spectral density $\textcolor{blue}{\rho_{\text{MP}}(s,t)}$ which we also take as an input. In this way the iterative algorithm is specified by the triad $\Big(\textcolor{violet}{\lambda}, \textcolor{OliveGreen}{\eta_{\text{MP}}(s)}, \textcolor{blue}{\rho_{\text{MP}}(s,t)}\Big)$. Given multi-particle data, the nonperturbative amplitude is effectively generated by iterations of two-particle unitarity gluing and dispersive crossing-symmetrization via the Mandelstam representation. Importantly, imposing multi-particle unitarity would lead to extra unitarity relations capturing both the frog and the open envelope diagrams.}
  \label{fig:diagramspic}
\end{figure}

Before delving into the details of these techniques, let us mention that an equivalent, maybe more accessible way to state what our algorithm achieves is a \emph{crossing-symmetric unitarization} of scattering amplitudes via a re-summation of a class of two-particle reducible planar graphs, whose vertices are made of the quartic coupling, and an additional effective vertex which accounts for multi-particle physics, see \figref{fig:diagramspic}.\footnote{See e.g. \cite{Oller:2020guq} for a review of  standard approaches to unitarization.} 
In general, these graphs are not Feynman diagrams, but signify the discontinuity structure of the amplitude in the spirit of generalized unitarity \cite{Bern:1994cg,Bern:1994zx,Bern:2011qt}, see also \cite{Correia:2021etg}. For a specific, restricted choice of the input parameters to be described below, our algorithm literally re-sums a class of Feynman diagrams, allowing to make direct connection with perturbation theory at low energies.

We now explain our set-up. Firstly, we work with a gapped theory with a single stable particle of mass $m$.
We parameterize the two-to-two scattering amplitude $T(s,t)$ using the Mandelstam representation \cite{Mandelstam:1959bc}
\begin{equation}
  \label{eq:mandelstamRep1}
  \begin{aligned}
T(s,t) &= \lambda + B(s,t) + B(s,u) + B(t,u) ,  \\
B(s,t) &={1 \over 2} \int_{4 m^2}^\infty {d s' \over \pi} {\rho(s') \over s'-s_0} \Big( {s-s_0 \over s'-s} + {t-s_0 \over s'-t} \Big) \\
&+ \int_{4 m^2}^\infty {d s' d t' \over \pi^2}
{(s-s_0)(t-t_0) \rho(s',t') \over (s'-s)(t'-t)(s'-s_0)(t'-t_0)} \ ,
\end{aligned}
\end{equation}
where $\rho(s)$ is called the single spectral density,
$\rho(s,t) = \rho(t,s)$ is called the double spectral density,\footnote{We sometimes interchangeably use the term single/double spectral \emph{function}.} and we set $s_0,t_0,u_0$ to the crossing-symmetric value,
$s_0 = t_0 = u_0 = {4 m^2 \over 3}$.
The subtraction constant $\lambda$ has dimensionality $m^{4-d}$ and we will always measure it in the units of external particle mass $m$.
The representation \eqref{eq:mandelstamRep1} makes
\emph{analyticity} and \emph{crossing symmetry} 
\be
\label{eq:crossing}
T(s,t) = T(t,s) = T(u,t), ~~~ s+t+u=4m^2 ,
\ee
manifest.

Let us briefly comment on the assumptions that go into the Mandelstam representation. Firstly, the analyticity domain of the amplitude given 
by the Mandelstam representation, the so-called \emph{maximal analyticity} in which the only singularities of the amplitude are those required by unitarity, i.e. the amplitude is analytic in $(s,t,u) \in \mathbb{C} \setminus \{\text{unitarity cuts}\} $. Maximal analyticity goes beyond the axiomatic analyticity domain proven from the axioms of QFT \cite{Bros:1964iho,Sommer:1970mr}. 
It is however a natural domain of analyticity for the scattering of lightest particles, where it is believed that no ``anomalous thresholds'' arise, see \cite{Correia:2021etg,Hannesdottir:2022bmo,Correia:2022dcu} for  recent discussions.
Maximal analyticity plays a central role in our analysis since we use it to analytically continue unitarity relations, as reviewed for example in~\cite{Correia:2020xtr,Correia:2021etg}.
Secondly, the Mandelstam representation with $N$ subtractions assumes that the amplitude is polynomially bounded, $|T(s,t)| < |s|^N$ for any $t$ on the \emph{physical sheet}. For our purposes, the physical sheet can be defined by analytic continuation from the kinematical region $0<s,t,u<4m^2$ to the complex plane without crossing any unitarity cuts.\footnote{One might also wonder, why instead of the unproven Mandelstam representation not to iterate crossing-symmetric dispersion relations, see e.g. \cite{Sinha:2020win}? The reason is that in the latter case the required analytic properties of the amplitude are not manifest.}

The nontrivial task therefore is to impose \emph{unitarity} on \eqref{eq:mandelstamRep1}. To discuss unitarity, it is
convenient to introduce the partial wave expansion  of the amplitude. In $d=4$ we have\footnote{Here we present the formula in $d=4$ not to clutter the notations. In the main text and appendix \ref{app:definitions} we present $d$-dimensional formulas.}
\be
T(s,t) &=16 \pi \sum_{J=0, J - \text{even}}^{\infty} (2J+1) f_J(s)  P_J \Big( 1+{2 t \over s - 4 m^2} \Big), \\
S_J(s) &\equiv 1 + i {\sqrt{s - 4 m^2} \over \sqrt{s} } f_J(s) , \ee
where $P_J(z)$ are the usual Legendre polynomials. 

It is convenient to distinguish the following unitarity conditions:
\paragraph{ Elastic unitarity.} 
For $s$ between the elastic and inelastic thresholds, we have:
\be
\label{eq:elastic}
| S_J(s) | = 1, ~~~   4 m^2 \leq s < s_{\text{MP}},  %
\ee
where $s_{\text{MP}}$ is the minimal energy required for multi-particle production. This condition exists only in the gapped theories, where production of two and three or more particles are separated by energy.
Even though it is most simply formulated for positive integer $J$, 
elastic unitarity also holds for complex $J$ with ${\rm Re} J > J_0$, where $\lim_{|t| \to \infty} |T(s,t)| \lesssim |t|^{J_0}$ for $s_{\text{MP}} > s \geq 4 m^2$, 
see e.g. \cite{Gribov:2003nw} and \cite{Correia:2020xtr}.

\emph{The Mandelstam equation}, first derived in \cite{Mandelstam:1958xc}, is an expression of elastic unitarity at complex $J$ and it takes the form\footnote{Loosely speaking, one obtains it by taking another discontinuity of the usual $\Im T\sim \int |T|^2$ elastic unitarity equation.}
\be
\label{eq:mandelstam}
 \rho(s,t) &= {(s - 4 m^2)^{1 \over2} \over  (4\pi)^{2} \sqrt{s} }\int\limits_{z_1}^{\infty} d \eta'  \int\limits_{z_1}^{\infty} d \eta'' \theta(z- \eta_+) %
 {T_t(s+i \eps, t(\eta') ) T_t(s- i \eps, t(\eta'') ) \over \sqrt{(z - \eta_-)(z - \eta_+)}} , 4 m^2 \leq s <  s_{\text{MP}}
\ee
where $\eta',\eta''$ are cosines of the complexified  scattering angles that parameterize the two-particle unitarity cut,
$ \eta_{\pm} = \eta' \eta'' \pm \sqrt{\eta'^2 -1} \sqrt{\eta''^2 - 1} $, $t(\eta) = {s - 4 m^2 \over 2} (\eta - 1)$, $z_1=1+{8m^2 \over s-4m^2}$, and $T_t \equiv {T(s,t+i \eps) - T(s, t-i \eps) \over 2 i}$ is the $t$-channel discontinuity of the amplitude. Via the Froissart-Gribov formula, see e.g. \cite{Gribov:2003nw,Correia:2020xtr} and \appref{app:definitions},
\eqref{eq:mandelstam} guarantees elastic unitarity for spins ${\rm Re}[J]>J_0$, where $J_0$, called the \emph{Regge intercept}, is mentioned above. For the amplitudes discussed in the present paper $J_0=0$.

\paragraph{Inelastic unitarity.}
\be
\label{eq:inelastic}
| S_J(s) | \leq 1, ~~~ s \geq s_{\text{MP}} , \ee
expresses the fact that the total probability of the initial state of
two-particle with angular momentum $J$ to go to the two-particle final
state is less or equal than one. %
This constraint can be restated as a positive semi-definiteness
condition and was successfully implemented first in \cite{Paulos:2017fhb}, and then in many follow-ups.

The condition \eqref{eq:inelastic} implies at large $J$ unitarity for scattering at fixed impact parameter~$b$
\be
\label{eq:impactunitarity}
| S_{{\sqrt{s-4m^2} b \over 2}}(s) | \leq 1, ~~~ s \gg m^2,
\ee
which we also check.\footnote{Here, we recall that ${\sqrt{s-4m^2} \over 2}$ is the spatial momentum, so $J=\sqrt{s-4m^2} b/2$ is just the standard definition of the angular momentum.}
\paragraph{Multi-particle unitarity.}%
The conditions above are the simplest consequences of unitarity of the S-matrix $\hat S^\dagger \hat S = \hat 1$. Naively they exhaust constraints that can be imposed purely at the level of the $2 \to 2$ scattering amplitude. 
While it is true if we use unitarity for physical kinematics only, it is not true for the analytic continuation of unitarity away from the physical region. 
Such an analytic continuation is particularly natural for amplitudes that satisfy maximal analyticity, such as the ones considered in the present paper, or which are common in the literature starting from \cite{Paulos:2017fhb}. In this case multi-particle unitarity manifests itself at the level of the two-to-two scattering amplitude in various ways.

First of all, in this paper, we assume the asymptotic Hilbert space to be the Fock space of multi-particle states made of a single stable particle. This puts natural constraints on the analytic structure of the amplitude. For example, this implies the existence of a set of branch points when $s,t,u = n^2 m^2$ (with $n\in\mathbb{N},\,n\geq2$), known as \emph{normal thresholds}.
Another, maybe slightly less familiar fact, is that when combined with maximal (or extended) analyticity this structure also implies that the two-to-two amplitude exhibits a complicated but computable structure of the 
\emph{Landau curves}, which are the thresholds for the double spectral density $\rho(s,t)$, see e.g. \cite{KORS,Mizera:2021icv,Correia:2021etg}. The double discontinuity of the amplitude that develops along these  Landau curves can be sometimes further expressed in terms of the two-to-two scattering amplitude. The simplest example of this type is the Mandelstam equation described above that expresses the double spectral density above the leading Landau curve in terms the square of the discontinuity of the amplitude \eqref{eq:mandelstam}, and follows from analytic continuation of elastic unitarity. An example of a condition which follows from the analytic continuation of multi-particle unitarity is the extended regime of validity of the Mandelstam equation below the leading multi-particle Landau curve identified in \cite{Correia:2021etg}.  More generally, the double spectral density $\rho(s,t)$ that is developed across the leading multi-particle Landau curve can be again expressed in terms of the $2\to 2$ amplitude leading to \emph{extra} unitarity constraints that go beyond \eqref{eq:elastic} and \eqref{eq:inelastic}. We do not implement these type of constraints in the present paper and leave this task for future work. These extra constraints are nonperturbative analogues of the familiar generalized unitarity relations that we have already mentioned above. 

The three unitarity conditions above are nonperturbative.
In contrast, in perturbation theory they are only satisfied order by order in the expansion parameter. However,  perturbation theory computations guarantee that \emph{multi-particle unitarity} is satisfied order by order in the coupling as well. This goes well beyond the current nonperturbative S-matrix analysis which focuses almost exclusively on the two-to-two scattering amplitude.

Let us now explain how we solve unitarity by means of an iterative procedure.

\subsection{Basic idea of the iteration algorithm}

The amplitude \eqref{eq:mandelstamRep1} is fully specified by the triad $\Big(\lambda, \rho(s), \rho(s,t) \Big)$. We keep $\lambda$ fixed at all times,  whereas $\rho(s)$ and $\rho(s,t)$ are subject to iterations using unitarity. 

Let us start with the double spectral density $\rho(s,t)$. Given an amplitude that satisfies elastic unitarity, we explain later that it can always be written as
\begin{align}
  \label{eq:iterationdoublespectral}
  \rho(s,t) &= \rho_{\text{el}}(s,t) + \rho_{\text{el}}(t,s) +   \rho_{\text{MP}}(s,t) , ~~~ \rho_{\text{MP}}(s,t) = \rho_{\text{MP}}(t,s),
\end{align}
where $\rho_{\text{el}}(s,t)$ satisfies the Mandelstam equation \eqref{eq:mandelstam}, and $\rho_{\text{MP}}(s,t)$ has nontrivial support only in the multi-particle region $s,t\geq s_{\text{MP}}$. In fact, multi-particle unitarity implies that it has support in a smaller region $s,t\geq s_{\text{MP,LC}}(t)$
given by the position of the leading multi-particle Landau curve, see e.g. \cite{KORS,Correia:2021etg}. In the iteration algorithm considered in this paper we take $\rho_{\text{MP}}(s,t)$ to be \emph{given} and \emph{fixed}, and $\rho_{\text{el}}(s,t)$ is computed iteratively using the Mandelstam equation \eqref{eq:mandelstam} that guarantees elastic unitarity of partial waves with ${\rm Re}[J] >0$. Intuitively, $\rho_{\text{MP}}(s,t)$ controls physics which is ``multi-particle in all channels.'' In the language of Feynman diagrams, it is captured by the two-particle irreducible graphs, see \cite{Correia:2021etg} for a detailed explanation.

Because the Mandelstam equation involves only the double-discontinuity, we also  need to enforce a separate unitarity condition to make sure that elastic unitarity is satisfied for the $J=0$ partial wave. Without loss of generality, we can write it as follows
\be
\label{eq:partialwavezero}
1 - |S_{0}(s)|^2 = \eta_{\text{MP}}(s)   ,
\ee
where $0 \leq \eta_{\text{MP}}(s) \leq 1$ characterizes particle production in the $J=0$ partial wave. We use \eqref{eq:partialwavezero} to compute $\rho(s)$. For $\eta_{\text{MP}}(s)$ we consider two schemes: a) it is \emph{given} and \emph{fixed}; b) it is expressed in terms of $(\lambda,\rho_{\text{MP}}(s,t))$. We describe both approaches in the text.

In the most general case, the iteration process is therefore \emph{initialized} by the triad $(\lambda, \eta_{\text{MP}}(s),\rho_{\text{MP}}(s,t))$ which are kept fixed, and $\rho(s)$ and $\rho(s,t)$ are then computed iteratively using 
\eqref{eq:iterationdoublespectral} and \eqref{eq:partialwavezero}.

Loosely speaking, we can say that we are reconstructing \emph{scattering} from \emph{production}, specified by $\eta_{\text{MP}}(s),\rho_{\text{MP}}(s,t)$, and \emph{subtraction}, specified by $\lambda$.\footnote{As we explain below, one role of subtraction in our dispersion relations is to fix the value of the amplitude at the crossing-symmetric point to $\lambda$.} At the fixed point of the iteration process the scattering amplitude $T(s,t)$ automatically satisfies:
\begin{itemize}
    \item maximal analyticity
    \item crossing
    \item elastic unitarity
\end{itemize}
We then separately check that the amplitude also satisfies
\begin{itemize}
    \item inelastic unitarity 
\end{itemize}
This last condition is not guaranteed by the algorithm, but for the class of amplitudes considered in this paper we will find that inelastic unitarity is always satisfied.

\subsection{Summary of results and plan of the paper}

In this paper, we describe a numerical implementation of the iteration algorithm presented above and report on a class of nonperturbative amplitudes in $d=3$ and $d=4$ which we produced through this  procedure. Specifically, we have constructed three different classes of amplitudes depending on the input $(\lambda, \eta_{\text{MP}}(s),\rho_{\text{MP}}(s,t))$:
\begin{itemize}
    \item \textit{toy-model amplitudes}, for which the double discontinuity is zero and scattering in S-wave is purely elastic. These serve a pedagogical purpose and also speed up the full algorithm thanks to a ``hot-start'' procedure explained below.
    \item \emph{two-particle quasi-elastic amplitudes} (2QE)  correspond to the input $(\lambda,0,0)$. This algorithm is characterized by the fact that the $J=0$ partial wave is purely elastic at all energies. From the point of view of our methods, these are truly minimal amplitudes driven solely by their value at the crossing-symmetric point.
    \item \emph{two-particle reducible amplitudes} (2PR) correspond to the input $(\lambda,\eqref{eq:analyticityinspinJ0},0)$, where we made an extra assumption that the imaginary part of partial waves is analytic in spin all the way to $J=0$. This fixes $\eta_{\text{MP}}(s)$ in terms of the double spectral density, see \eqref{eq:analyticityinspinJ0}, which in turn is fixed by $\lambda$. We expect that this scheme is equivalent to re-summation of fully two-particle reducible planar Feynman diagrams computed in the renormalization scheme $\lambda = T({4 m^2 \over 3},{4 m^2 \over 3})$.
    \item amplitudes with $\rho_{\text{MP}}(s,t) \neq 0$. All known physical theories are of this type, and we only initiate the exploration of this space in the present paper with a specific example using the ``acnode graph'' \cite{eden1961acnodes}.
\end{itemize}
After we have constructed the amplitudes we can explore their physical properties. These include particle production, near-threshold behavior, the Regge limit, fixed-angle high energy scattering, elastic and inelastic cross sections. We find that they exhibit non-zero but small particle production, therefore we can call them \emph{quasi-elastic amplitudes}.\footnote{Let us emphasize that, while the third class of amplitudes we construct have an inelastic input, we still refer to it as ``quasi-elastic'', because the overall inelasticity, as measured in the partial waves, is \text{small} and vanishes at high energies.} We can also place various amplitudes in the space of allowed couplings explored recently in~\cite{Chen:2022nym,EliasMiro:2022xaa}. All amplitudes constructed in the present paper live very close to the boundary of the allowed region. The most ``extremal'' amplitudes are the 2QE amplitudes. 

Both 2QE and 2PR amplitudes have $\rho_{\text{MP}}(s,t)=0$. In the old S-matrix bootstrap literature, the condition $\rho_{\text{MP}}(s,t)=0$ was called ``strip approximation'' and it was applied to pion scattering, see e.g. \cite{Weinberg:1996kw}. Based on our results, we expect this approximation might work reasonably well for scalar, $\phi^4$-like theories, but not for confining gauge theories for which $\rho_{\text{MP}}(s,t) \neq 0$ should be very important.

The plan of the paper is as follows:
\begin{itemize}
\item In section \ref{sec:algorithm} we introduce the iteration algorithm in detail. We also discuss its numerical implementation. The key ingredient in our work is the Mandelstam representation of the amplitude that allows us to combine unitarity and crossing in a elegant way.
\item In section \ref{sec:toymodel}, we introduce a simplified version of the iteration algorithm where we construct quasi-elastic amplitudes with \emph{zero double discontinuity}, $\rho(s,t)=0$ and elastic S-wave scattering $\eta_{\text{MP}}(s)=1-|S_0(s)|^2=0$, in spacetime dimensions $3 \leq d \leq 4$. This section is mostly self-contained and can provide an accessible, less technical way to delve into the formalism.

\item In sections \ref{sec:QE3d} and \ref{sec:QE4d} we numerically construct examples of what we call \emph{quasi-elastic} amplitudes. These are amplitudes which exhibit non-zero, but small particle production. We construct both 2QE and 2PR amplitudes in $d=3$ and $d=4$
in a finite range of $\lambda$ around $0$. We explicitly check that the amplitudes satisfy both elastic and inelastic unitarity. We analyze their behavior at high energies both at fixed angles and in the Regge limit.
\item In section \ref{sec:acnode} we derive the double discontinuity of the acnode graph in four dimensions. Following the work of Gribov and Dyatlov \cite{gribov1962contribution}, we analytically continue the three-particle phase space integral from negative to positive $t$. We then proceed to use the acnode double spectral density as a proxy for $\rho_{\text{MP}}(s,t) \neq 0$, and we construct the corresponding amplitude functions.
\item In section \ref{sec:lowenergy} we look at some low energy observables, to make comparison with the parameter space analyzed recently in \cite{Chen:2022nym,EliasMiro:2022xaa}. We find that the quasi-elastic amplitudes constructed in this paper all lie close to the boundary of the allowed region. We find that the 2QE amplitude is the most ``extremal.''
\item In section \ref{sec:mandelstamQG} we consider the question whether the Mandelstam representation holds in gravity. In this context, we revisit the old argument by Cerulus and Martin \cite{Cerulus:1964cjb}, which connects the Regge limit, the fixed angle scattering and the polynomial boundedness required for the validity of the Mandelstam representation. 
\item In section \ref{sec:conclusions} we conclude and collect thoughts on open problems and future directions.
\end{itemize}

\section{Algorithm}
\label{sec:algorithm}

In this section we first review the key concepts and equations that serve as our starting point. We then present details of the numerical implementation of the iteration procedure.

\subsection{Basic concepts}

\paragraph{Subtracted Mandelstam representation.}

We parameterize our amplitude using the double-dispersive
representation called Mandelstam
representation~\cite{Mandelstam:1959bc}, introduced in~\eqref{eq:mandelstamRep1}. This rewriting of the Mandelstam
representation involves one \textit{subtraction}. Subtractions are used to write down dispersion relations  when the arc of the Cauchy contour at infinity does not vanish. For
instance, let $f(z)$ be a function analytic in the complex
plane apart from a branch cut on the real axis starting from $z=z_1$ and extending to infinity. If $f(z)$ goes to a constant $a$ at infinity, the usual
dispersion integral that would give
$f(z) = \frac{1}{2i \pi}\int_{z_1}^\infty\frac{\text{Disc} f(w)}{z-w}dw$ cannot
be written because the arc of the Cauchy contour around infinity does not vanish. A dispersion relation can only be written for the
\textit{subtracted} function $f(z)-a$. If the function has a polynomial growth $f(z)\sim z^n$, $n+1$ subtractions need to be
used. The same result can be achieved by dividing instead of subtracting,
and obtaining dispersion relations for $\frac{f(z)}{z^{n+1}}$, in terms of $n+1$ undetermined constants. 

The Mandelstam representation is a double dispersive representation: we consider the dispersion relation for the discontinuity of the function and then we plug it into the usual dispersion relation. Similarly, double dispersion relations can be written. 

The subtractions constants are unknowns, which are not determined by the dispersion relation. For us, eq.~\eqref{eq:mandelstamRep1} has one unknown, $\lambda$, the value of the amplitude at the crossing symmetric point,
\begin{equation}
  \label{eq:lambda-def}
  \lambda \equiv T \Big( {4 m^2 \over 3}, {4 m^2 \over 3} \Big).
\end{equation}
which is usually taken to represent the quartic coupling of the amplitude. We can turn this to our advantage, as it allows us to input the nonperturbative value of the coupling for any amplitude, thus we can explore in a controlled manner the space of couplings.

Furthermore, non-zero $\lambda$ acts as \textit{a source} for the iterations, and creates amplitudes which are necessarily non-trivial. 

Note finally that \eqref{eq:mandelstamRep1} allows to describe a class of functions that go at worst to a constant at infinity, but we do not put in the value of this constant. The behavior at infinity, $T(s,t)\to0$ or $T(s,t)\to\mathrm{const}\neq 0$ is dynamically generated by the algorithm.

\paragraph{Mandelstam equation.}

The Mandelstam equation is the analytic continuation in $t$ of unitarity in the elastic strip in $s$. It relates the double discontinuity of the amplitude to the square of the single discontinuity of the amplitude. Equivalently, the Mandelstam equation expresses elastic unitarity for complex spins $J$. Schematically, it is obtained by taking a second discontinuity in the standard $\Im T \propto |T|^2$ formulation of elastic unitarity. It is convenient to introduce the following notation
\be
{\cal T}(s,z) \equiv T(s, t(z) ), ~~~ t(z) = - {s-4m^2 \over 2} (1-z) \ . 
\ee

\begin{figure}[t]
  \centering
  \includegraphics[scale=1.]{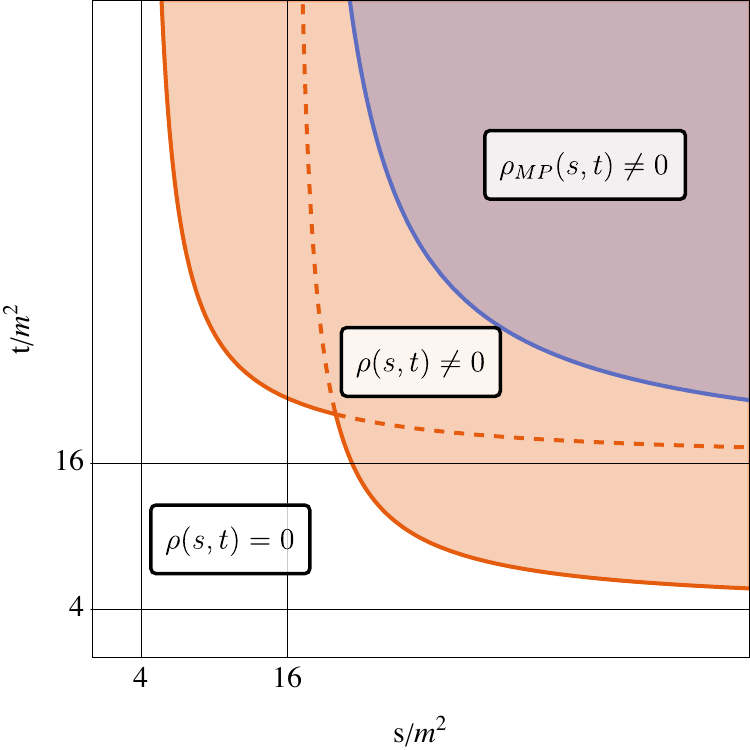}
  \caption{Domain of support of the double spectral density $\rho(s,t)$ in a theory with $\mathbb{Z}_2$ symmetry ($\phi^4$-type interaction) with $s_{\text{MP}}=16m^2$. The Landau curves which separate the region $\rho(s,t)=0$ from $\rho(s,t) \neq 0$ are given in \eqref{eq:LCleading}.
  }
  \label{fig:rho-LC-def}
\end{figure}

\begin{figure}[t]
  \centering
  \includegraphics[scale=1.4]{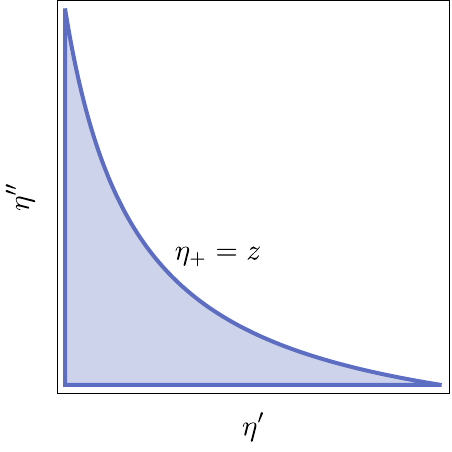}
  \caption{Typical integration domain of the Mandelstam equation \eqref{eq:mandelstam-eqn}. The integration range for the complexified angles is $\infty > \eta',\eta'' \geq 1+{4 m^2 \over s-4m^2}$ and $\eta_+ \leq z$, where recall that $\eta_+ = \eta' \eta'' + \sqrt{(\eta'^2-1)(\eta''^2-1)}$.}
  \label{fig:hyperbola-mandelstam}
\end{figure}

For a full modern account of the Mandelstam equation and elastic unitarity, see~\cite{Correia:2020xtr}. For $s$ in  the elastic strip, and $4m^2<t<\infty$, we have
\begin{equation}
\label{eq:mandelstam-eqn}
  \rho(s,t) = {(s - 4 m^2)^{d-3\over2} \over 4\pi^2 (4\pi)^{d-2} \sqrt{s} }\int\limits_{z_1}^{\infty} d \eta'  \int\limits_{z_1}^{\infty} d \eta''\, {\cal T}_t^{(+)}(s, \eta') {\cal T}_t^{(-)}(s, \eta'')\, \tilde K_d(z, \eta', \eta'') 
\end{equation}
where $\eta',\eta''$ are cosines of the complexified scattering angles, ${\cal T}_t$ is the $t$ channel discontinuity of $\cal T$ and $z_1=1+{4 m^2 \over s-4m^2}$. The upper index $\pm$ specifies the direction of real $s$ from the complex plane, namely $\pm i \epsilon$. The double discontinuity of the amplitude is only non-zero
above the so-called \textit{Landau curves} and vanishes below them:
\begin{equation}
  \label{eq:LC-def}
s\leq \sigma(t):  ~~~\rho(s,t)=0 
\end{equation}
where $\sigma(t)$ is a function given by the union of the leading Landau curves in two channels, see solid-orange curve in \figref{fig:rho-LC-def},
\begin{equation}
\label{eq:LCleading}
  \sigma(t) = \min \left(\frac{4t}{t-16 m^2},\frac{16t}{t-4 m^2}\right)
\end{equation}
The kernel $\tilde K_d$ is of purely kinematic origin and is given by, in $d>3$ dimensions:
\begin{equation}
  \label{eq:kernel-mandelstam}
  \tilde K_d(z, \eta', \eta'')= {4\pi^{d+1\over2} \over \Gamma({d-3\over 2})}  \Theta( z - \eta_+){(z^2 - 1)^{4-d \over 2}  \over (z-\eta_-)^{5-d\over2}(z-\eta_+)^{5-d\over2}} .
\end{equation}
where $\eta_\pm=\eta' \eta'' \pm \sqrt{(\eta'^2-1)(\eta''^2-1)}$. In $d>3$, the kernel reduces the double integral $\int\limits_{z_1}^{\infty} d \eta'  \int\limits_{z_1}^{\infty} d \eta''$ to a domain contained below
a hyperbola-shaped curve define by $\eta',\eta''>z_1$ and $\eta_+<z$, see \figref{fig:hyperbola-mandelstam}.
In $d=3$ we get
\begin{equation}
  \label{eq:kernel-3d}
  \tilde K_3(z, \eta', \eta'')= 4
  \pi^2\delta(z-\eta_+){\sqrt{z^2 - 1}\over \eta_+ - \eta_-} \ ,
\end{equation}
where the delta-function localizes the 2d integral to a one-dimensional segment.

In terms of $\rho(s)$ and $\rho(s,t)$, the $t$-channel discontinuity $T_t(s,t)$ is given by
\begin{align}
\label{eq:PMdef}
T_t (s,t)&= \rho(t)+\int_{4 m^2}^\infty {d s' \over \pi} {(s-s_0)
           \rho(s',t) \over (s'-s)(s'-s_0)}+\int_{4 m^2}^\infty {d u'
           \over \pi} {(u-u_0) \rho(u',t) \over (u'-u)(u'-u_0)} .
\end{align}

Equation~\eqref{eq:mandelstam-eqn} is defined for $4m^2 \leq s<s_{MP}$. Let
us now define a function $\rho_{el}(s,t)$ by the RHS of
eq.~\eqref{eq:mandelstam-eqn} \textit{for all values of
  $s \geq 4m^2$}. Considering a purely multi-particle function
$\rho_{MP}(s,t)$ supported in the quadrant $s,t \geq s_{MP}$, we have now
defined with precision the model of the amplitude written in \eqref{eq:iterationdoublespectral}, which is the model that we
use. For us, as we said, $\rho_{MP}(s,t)$ will be treated as
\textit{an input}, and we will typically define it above some
realistic multi-particle Landau curves, following the analysis of~\cite{Correia:2021etg}. 

It is important to note that no assumption was made in writing~\eqref{eq:iterationdoublespectral}: any double spectral density of an amplitude that satisfies elastic unitarity can be written in the form \eqref{eq:iterationdoublespectral}. The key aspect of this formula is the fact that adding the double spectral density needed to satisfy elastic unitarity in the $t$-channel, namely $\rho_{\text{el}}(t,s)$, restores crossing symmetry and at the same time does not spoil elastic unitarity in the $s$-channel. This happens due to the fact that $\rho_{\text{el}}(t,s)$ has only support for $s\geq s_{\text{MP}}$. In this way we can have both unitarity and crossing! %

\paragraph{$J=0$ unitarity.}

The Mandelstam equation is a dispersive version of elastic unitarity.
Indeed, by integrating both sides of \eqref{eq:mandelstam-eqn} against the Legendre Q-function
and using the Froissart-Gribov formula for partial waves, see \appref{app:definitions} and \cite{Correia:2020xtr},
one recovers $|S_J|=1$ for complex $J$ as long as ${\rm Re}[J]>J_0$, where $J_0$ is the Regge
intercept of the amplitude. The last condition is required for validity of the Froissart-Gribov
representation of partial waves. In this paper, we consider amplitudes with $J_0=0$. Therefore,
the Mandelstam equation does not guarantee elastic unitarity of $J=0$ partial wave and we need
an extra unitarity constraint to impose it.
It was described in the introduction and is much
simpler to state than the Mandelstam equation. We simply enforce the
$J=0$ partial wave unitarity equation \eqref{eq:partialwavezero} as follows:
\begin{equation}
  \label{eq:spinzeropartialwave}
  2 {\rm Im} f_0(s) = {(s - 4 m^2)^{{d-3 \over 2}} \over \sqrt{s}} |
  f_0(s)|^2 + {\sqrt{s} \over (s - 4 m^2)^{{d-3 \over 2}} }
  \eta_{\text{MP}}(s) ,
\end{equation}
where $\eta_{\text{MP}}(s) \neq 0$ only in the multi-particle region
$s \geq s_{\text{MP}}$. Now, to be practically implemented in
terms of the data of our amplitude, and in particular $\rho(s)$ and
$\lambda$ (but also $\rho(s,t)$), we need to perform the partial wave
projection to $J=0$ of the Mandelstam representation
eq.~\eqref{eq:mandelstamRep1}. We display the lengthy expression of the
result of the projection in~\eqref{eq:spinzeroAgenerald} in
general dimension $d$.

To unpack \eqref{eq:spinzeropartialwave} a bit more, and prepare it
for the iterations explained below, we keep $|f_0(s)|^2$ in the
right-hand side implicit, but spell out the left-hand side using using \eqref{eq:spinzeroAgenerald} in terms of $\rho(s)$, $\rho(s,t)$. We thus obtain the following intermediate step equation, in spacetime dimension $d$:
\begin{multline}
  \label{eq:single-disc-unitarity}
  \rho(s) + 2 \int_{4 m^2}^\infty {d t' \over \pi} \rho(s,t')
  \Big( {1 \over t_0 - t'} + {1 \over t'}
  \ _2 F_1 (1, {d-2 \over 2}, d-2, {s-4m^2 \over t'}) \Big)=\\
  4 (16 \pi)^{{d-3 \over 2}} \Gamma({d-1 \over 2})
  \left( {(s-4 m^2)^{1/2} \over \sqrt{s}} |  f_0(s) |^2 +
    {\sqrt{s} \over (s - 4 m^2)^{{d-3 \over 2}} } \eta_{\text{MP}}(s) \right) \,.
\end{multline}
Here, just like in \eqref{eq:spinzeroAgenerald}, in integer
dimensions $d=3,4,5,\dots$, the hypergeometric function simplifies to
functions with logarithmic or square roots singularities, depending on whether $d$ is even or odd,
respectively.\footnote{The ${ }_2 F_1$ function simply comes from performing the partial
wave projection of the $t$- and $u$-channel cuts in the Mandelstam representation.} Note finally that
although $\lambda$ is not visible in this is equation, it is explicitly
entering $f_0(s)$ in the RHS, see \eqref{eq:spinzeroAgenerald}.

\subsection{Iteration algorithm.} 
\label{sec:iteration-algorithm}

We are now well equipped to explain the iteration algorithm. As we already mentioned above, the basic idea behind it goes back to the S-matrix program pursued in the 60's \cite{Mandelstam:1959bc,Mandelstam:1958xc,Mandelstam:1963iyb,ter1961equations,Atkinson:1968hza,Atkinson:1968exe,Atkinson:1969eh,Atkinson:1970pe}. 

The idea is to \textbf{input fixed values} of the triad
$(\lambda, \eta_{\text{MP}}(s), \rho_{\text{MP}}(s,t))$, and construct the amplitude $T(s,t)$ that satisfies analyticity, crossing and
elastic unitarity, that is to say \textbf{ obtain the
  functions} $\rho(s)$ and $\rho(s,t)$ which satisfy the equations
\eqref{eq:mandelstam-eqn} and \eqref{eq:single-disc-unitarity}. The algorithm does not guarantee inelastic unitarity \eqref{eq:inelastic} as an output. It has to be checked separately on the fixed point of the iteration algorithm. In this sense, inelastic unitarity constrains \textit{a posteriori} the space of inputs $(\lambda, \eta_{\text{MP}}(s), \rho_{\text{MP}}(s,t))$.

The unitarity equations can be formally rewritten in terms of a functional
$\Phi$, whose elements are already defined above, in the following form:
\begin{equation}
  \label{eq:Phi-fixed-point}
( \rho(s) , \rho(s,t))= \Phi_{\text{MP},\lambda}\left[ \rho(s),\rho(s,t)\right]
\end{equation}
where we specified the input triad elements as an index of $\Phi$ to
insist on the fact that these are external, fixed parameters.

We solve these equations by iterating the map $\Phi$: initialising on
a starting point $\rho^{(0)}(s),\rho^{(0)}(s,t)$, we define a sequence
of functions  $\rho^{(n)}(s),\rho^{(n)}(s,t)$, $n=1,2,\dots$ by
\begin{equation}
  \label{eq:Phi-fixed-point2}
(\rho^{(n+1)}(s,t),\rho^{(n+1)}(s))= \Phi_{\text{MP},\lambda} \left[\rho^{(n)}(s),\rho^{(n)}(s,t)\right]
\end{equation}
If the map converges, the fixed point,
$(\rho^{(\infty)}(s,t), \rho^{(\infty)}(s) )=\Phi_{\text{MP},\lambda}[\rho^{(\infty)}(s),\rho^{(\infty)}(s,t)]$
gives by definition a scattering amplitude which satisfies all of the equations above,
therefore satisfies analyticity, crossing and
elastic unitarity (with inelastic unitarity  checked  separately). 

\paragraph{Iteration steps.}
The iteration algorithm proceeds in the following steps, in order:
\begin{enumerate}
    \item Single discontinuity iteration, from \eqref{eq:single-disc-unitarity},
      \begin{align}
        \label{eq:iterationrho0}
        \rho^{(n+1)}(s) &=4 (16 \pi)^{{d-3 \over 2}} \Gamma({d-1 \over 2})  \left( {(s-4 m^2)^{1/2} \over \sqrt{s}} |  f_0^{(n)}(s) |^2 + {\sqrt{s} \over (s - 4 m^2)^{{d-3 \over 2}} } \eta_{\text{MP}}(s) \right) \nn \\
                        &- 2 \int_{4 m^2}^\infty {d t' \over \pi} \rho^{(n)}(s,t')  \Big( {1 \over t_0 - t'} + {1 \over t'} \ _2 F_1 (1, {d-2 \over 2}, d-2, {s-4m^2 \over t'}) \Big)  ,
      \end{align}
where $f_0^{(n)}(s)$ is obtained from formula \eqref{eq:spinzeroAgenerald}, adding $(n)$ iteration counters exponents to the spectral $\rho$ functions. 
\item Double discontinuity iteration
\be
\label{eq:atkinson0}
 \rho_{\text{el}}^{(n+1)}(s,t) &= \text{M}_d [T_t^{(n)},T_t^{(n)}] (s,t) ,  \\
  \rho^{(n+1)}(s,t) &= \rho_{\text{el}}^{(n+1)}(s,t) + \rho_{\text{el}}^{(n+1)}(t,s) + \rho_{\text{MP}}(s,t) .
  \label{eq:doublespectraliter}
\ee
where $ \text{M}_d[\cdot, \cdot]$ is the quadratic functional of the
Mandelstam equation, which is exactly the RHS of
eq.~\eqref{eq:mandelstam-eqn}, with $T_t^{\pm}$ replace by
$T_t^{\pm,(n)}$, so we do not reproduce the expression here.

We simply rewrite the $t$-channel single disc, to clarify that we do not iterate any function with that step, but simply reconstruct the single disc. from the double disc.:
\begin{equation}
\label{eq:t-channel-SD-iteration}
T_t^{(n)} (s,t)= \rho^{(n)}(t)+\int_{4 m^2}^\infty {d s' \over \pi} {(s-s_0) \rho^{(n)}(s',t) \over (s'-s)(s'-s_0)}+\int_{4 m^2}^\infty {d u' \over \pi} {(u-u_0) \rho^{(n)}(u',t) \over (u'-u)(u'-u_0)} .
\end{equation}

\end{enumerate}

In practice, for the initialization step, we set
$\rho^{(0)}(s) = \rho^{(0)}(s,t)=T_t^{(0)} (s,t)=0$. The condition $T^{(0)}(s,t) = \lambda$ is automatically generated by the subtracted representation we use, and it enters $|f_0|^2$ in the RHS of \eqref{eq:iterationrho0} at step 1.

Let us now imagine a situation where this iterative process converges and we reach the fixed point. The physical content of
\eqref{eq:iterationrho0} at the fixed point is then nothing but
unitarity for the $J=0$ partial wave
\eqref{eq:spinzeropartialwave}. Elastic unitarity for partial waves
with $J>0$ is satisfied at the fixed point due to
\eqref{eq:atkinson0}. Crossing symmetry is manifest due to
\eqref{eq:doublespectraliter}, thanks to adding
$\rho_{\text{el}}^{(n+1)}(t,s)$ which restores crossing, and at the same time does not
spoil elastic unitarity since it has only support for
$s \geq s_{\text{MP}}$.

Lastly, inelastic unitarity for $J>0$ partial waves
\eqref{eq:inelastic} is not guaranteed to hold at the fixed point of
the iteration. It presents a nontrivial constraint on the triad
$(\lambda, \eta_{\text{MP}}(s), \rho_{\text{MP}}(s,t))$ which we check below. 

\subsection{Numerical implementation}
\label{sec:num-impl}

Let us now describe how we numerically implemented this procedure. %

\subsubsection{Discretization}

The numerical implementation is rather simple in essence. We discretize the functions $\rho(s),\,\rho(s,t),\,T_t(s,t)$ on grids and define linear interpolants as the functions we iterate. To perform an iteration, we evaluate the updated functions of step $n+1$ on the grid points, and obtain the iterated interpolants in this way.

For the discretization, we need three grids:
\begin{itemize}
    \item a one-dimensional grid for $\rho(s)$, that spans $s\geq 4m^2$,
    \item a two-dimensional grid for $T_t(s,t)$ that spans the whole quadrant $s,t\geq 4m^2$ (the single discontinuity of the amplitude is non zero after the first threshold, this is standard),
    \item a two-dimensional grid for $\rho(s,t)$ that spans the portion of the quadrant $s,t\geq 4m^2$ which is above the Landau curves.
\end{itemize} 

We start by mapping these domains to $[0,1]$ and the unit square $[0,1]^2$ via the following change of variables: \begin{equation}
  \label{eq:xydef}
(s,t)\to  (x,y)=\left(\frac{4m^2}{s},\frac{4m^2}{t}\right)
\end{equation}
By doing so we effectively introduce a UV and IR cutoffs related to the sampling of the functions close to $0$ and $1$, which we discuss below. We provide more details on the explicit construction of these grids in appendix~\ref{app:grids}.

\begin{figure}[t]
  \centering
  \includegraphics[scale=1.]{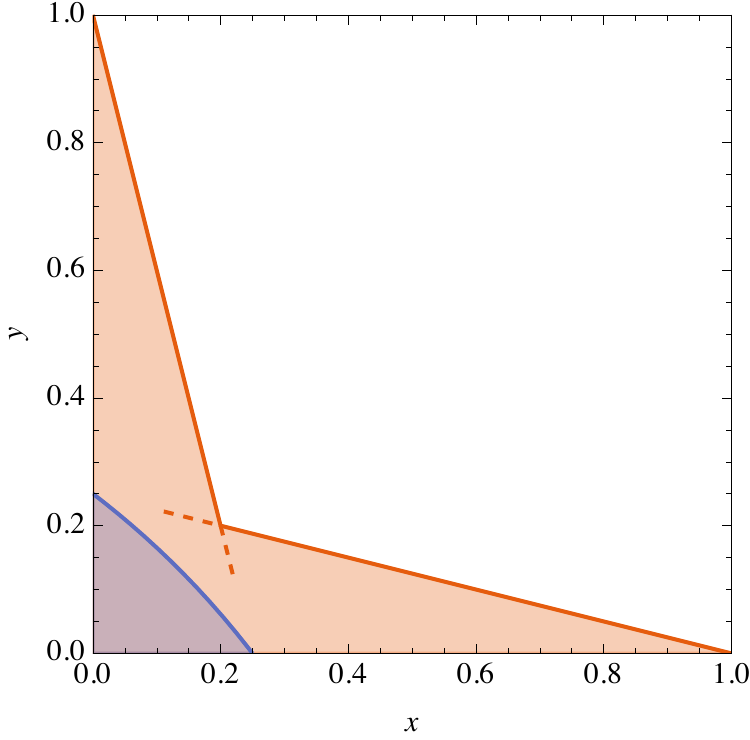}
  \caption{The region of the nontrivial support of the double spectral density in the $x,y$ variables. It is given in terms of the location of the leading Landau curves \eqref{eq:LC-def}.}
  \label{fig:plxy}
\end{figure}

\paragraph{Grid for $\rho(s,t)$.}
As was said above, $\rho(s,t)$ is non-zero only above the union of the two Landau curves $(\frac{16s}{s-4},\frac{4s}{s-16})$. In the $x,y$ variables, these curves become lines, defined by
\begin{equation}
  \label{eq:LCxy}
  \Big( \frac{1-x}{4},1-\frac x 4 \Big)\,,
\end{equation}
and, as a function of $x,y$, the double spectral function
$\rho(x,y)$ has support below these lines, as is depicted in fig.~\figref{fig:plxy}.

The various high energy limits of the amplitude are now described as follows: the Regge limits correspond to the segments $x=0$, $y=0$. The fixed (imaginary)\footnote{Recall that we are in the unphysical kinematics quadrant $s,t\geq0$.} angle region corresponds to $x,y\to0$ with $x/y$ fixed. The threshold region, which matters in $d=3$ but is trivial in $d=4$, is mapped to $x\to1$ and $y\to1$. The grids have to sample accurately these various regions and for this reason, we choose grids with exponentially many points up to some cut-off. 

After experimenting a lot with various types of grids, we settled for a default ``fine'' $(s,t)$-grid with 11445 points, with cutoff at $x=10^{-12}$, which provided a good interplay between the need to describe inelastic effects to very high energies while keeping manageable computation times. We give more detail on this grid in appendix.

\begin{figure}[htb!]
  \centering
  \includegraphics{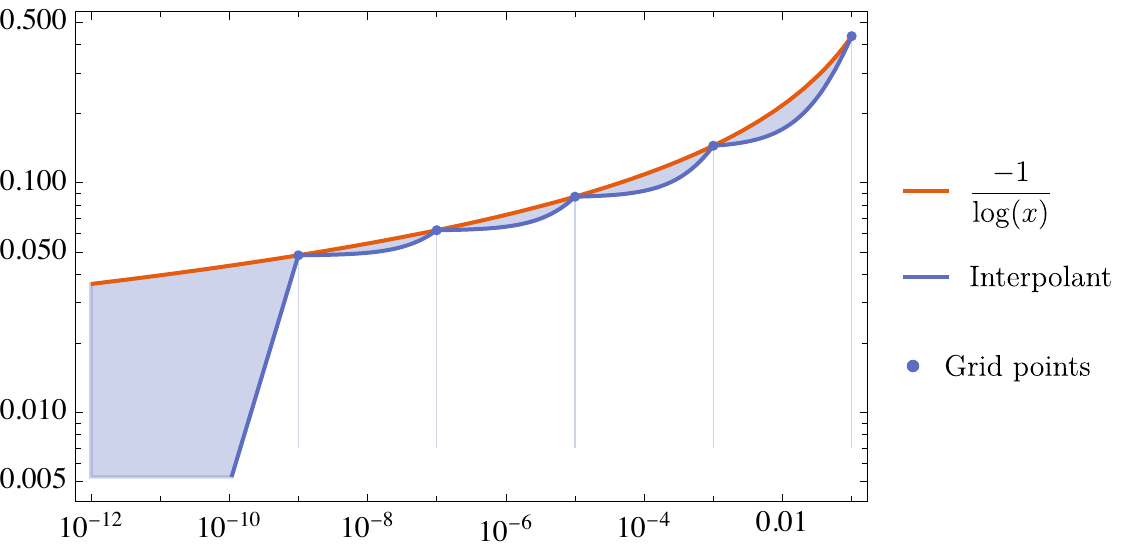}
  \caption{Linear interpolation of a slowly decaying function, $-1/\log(x)$ in this example. The first grid point is at $10^{-9}$, and then the interpolating function sharply connects to zero. 
  The dispersive integrals used in the iteration process have $1/s$ kernels which suppresses the integration at the cutoff scale and renders this effect innocuous, if the cutoff is high enough. Note also that the garland-shape of the piecewise-linear interpolating function is due to the log-log scale.}
  \label{fig:interp-ex}
\end{figure}

\paragraph{A note about the cutoff.} For the class of amplitudes considered in this paper, we always have $\rho(x),\rho(x,y)\underset{x,y\to0}{\rightarrow}0$ at high energies. The cutoff mentioned above is to be understood in the following sense. Above the cut-off, even though we use linear interpolation, the functions are dynamically generated by the iteration process. Between the cut-off and zero, we continue to use linear interpolation in $x,y$, which means, in the $s,t$ variables, that we force a $1/s$ or $1/t$ decay of the functions. An illustration of this phenomenon is given in \figref{fig:interp-ex}.

\begin{figure}
    \centering
    \includegraphics[scale=1.4]{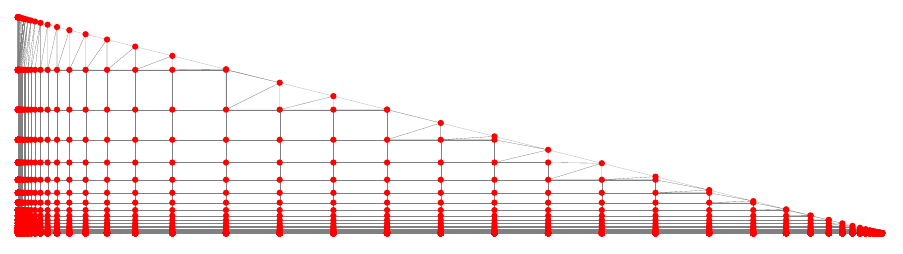}
    \caption{Fine mesh for the double spectral density (11445 points). Red dots are points $(x_i,y_i)$ on which
      the functions is evaluated, while the interpolation is linear on
      each facet.}
    \label{fig:mesh}
\end{figure}

One last important detail about the 2d grid is that when we compute $\rho_{\text{el}}^{(n)}(x_i,y_j)$ in step 2 of the iteration, it is only non-vanishing below the Landau curve $y=(1-x)/4$, thus we only need to discretize the wedge between $y=0$ and $y=(1-x)/4$, for $0<x<1$. The full double-discontinuity is obtained after crossing-symmetrizing, $\rho(x,y)=\rho_{\text{el}}(x,y)+\rho_{\text{el}}(y,x)+\rho_{\text{MP}}(x,y)$ and is correctly defined below the union of both Landau curves, as it should.

The fine grid with which we generated most of the data is represented in \figref{fig:mesh}. Later in section \ref{sec:acnode} we comment on the influence of density of the grids on the precision of our result.  This grid has essentially has a subleading influence on the significant digits of the low-energy observables described in section~\ref{sec:lowenergy}. However, this grid is essential to produce high energy quantities, such that inelasticity in the partial waves, accurately.

In general, all of the errors we face in this paper are errors coming from the ``trapezoidal rule'', familiar from approximating one-dimensional integrals with piecewise linear functions.

\paragraph{Grid for $T_t(s,t)$}
The single discontinuity $T_t$ is defined on the whole
quadrant $[0,1]^2$. Therefore, we need another grid to sample this domain. For the latter we simply chose a uniformly-spaced grid of points at $(x_i,y_j) = (i/N,j/N)$ with $i,j=0,1,\dots,N$. In practice, this appeared sufficient. We used a regular square grid with $72\times72=5184$ points. This grid has essentially no influence low-energy observables nor high energies. This comes from the fact that $T_t(s,t)$ is dominated by $\rho(s)$, which has its own grid.

\paragraph{Grid for $\rho(s)$.}
Finally, the $t$-independent simple spectral function  $\rho(s)$ is
discretized on a 1-dimensional grid which also samples
logarithmically the extremities $x\to0,1$ so as to have more
information on threshold and Regge behavior. We also took a cut-off
$10^{-12}$ in order to have consistency between our grids and 91 points in total. The explicit definition of the grid is given by the following sequence of points:
\begin{equation}
    x_0=0,~x_1=10^{-12},~\dots,~x_N=1 \ . 
\end{equation}
In section~\ref{sec:lowenergy}, we produce some estimation of the convergence of low-energy observables as we densify this grid, and find it to be the major source of error. Again, this error appears to be of the simple trapezoidal-rule type, as it scales with the square of the inverse of the number of points, $1/N_{\text{grid-points}}^2$.

\subsubsection{Numerical integrals.}
Once we have defined the interpolating functions, we can run the algorithm. At step $(n+1)$, we compute numerically all the right-hand sides in the equations defined in section~\ref{sec:iteration-algorithm}. These numerical integrals need to be handled with care. One important point relates to all the dispersion integrals. The real part of these is computed using the standard principal value prescription (the imaginary part is just a delta function and poses no problem). These principal values can be improved in a standard way, using the following identity
\begin{equation}
    \text{P.V.}\int_0^1 \frac{f(x')}{x'-x}dx' = \int_0^1 \frac{f(x')-f(x)}{x'-x}dx' +f(x)\log {1-x \over x} \,.
\end{equation}
 and variations thereof when the singular point lies at the extremity and for two-dimensional integrals.

\begin{figure}
    \centering
    \includegraphics[scale=0.7]{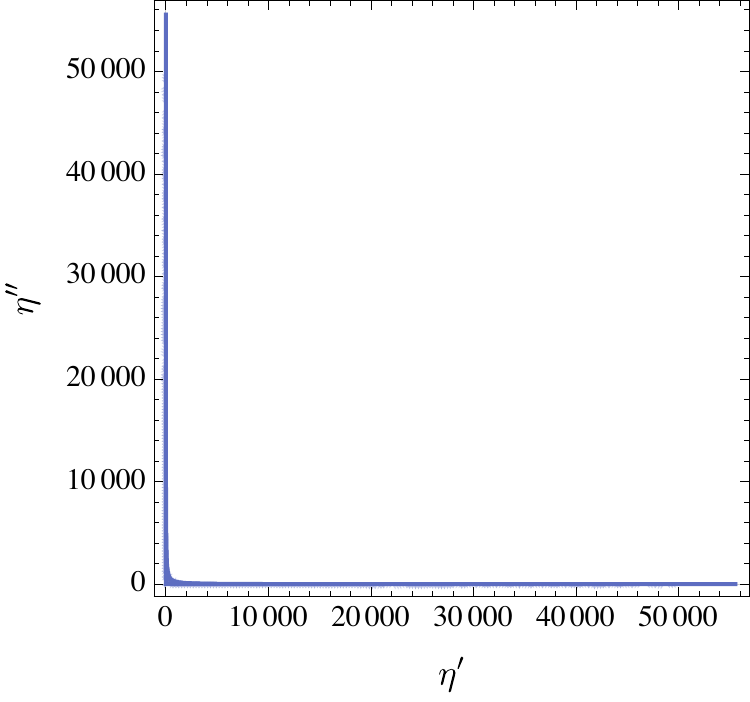}
    \caption{An example of the integration domain for the Mandelstam equation \eqref{eq:mandelstam-eqn} in the Regge limit for $s=5 m^2,t=10^6 m^2$.}
    \label{fig:plhyperskew}
\end{figure}

Another source of numerical intricacies come from the deformation of the integration domain in the RHS of the Mandelstam equation in the Regge limit. In a very deep Regge kinematics, the two-dimensional hyperbola-shaped domain of integration of the Mandelstam integral becomes very skewed and requires care in the numerical evaluation, see \figref{fig:plhyperskew}. For these integrals, we have found that we obtained reliable results using the standard \texttt{GlobalAdaptive} method of Mathematica, supplied with various \texttt{PrecisionGoal}, which we would in general keep to 5 to keep good runtimes and control on the numerics. We did not give much importance to the estimated error of the numerical integrations, because we have a precise and independent way to check our results. which is to measure violation of the imposed unitarity condition, and we have found that it is satisfied to a very good precision, e.g. $1-|S_0|\sim 10^{-10}$ for the 2QE amplitudes.

We plan to release the code in the near future\footnote{Definitely before the earth reaches a warming of $+1^{\circ}$, which is unfortunately bound to happen soon.}, once we have optimized it fully. In the meantime, we make available with the arXiv submission a notebook with  a few of the most representative examples of our amplitudes.

\subsubsection{Convergence}

In this program we can discuss two types of convergence. The first one refers to the convergence of the fixed-point iteration with respect to the number of iterations, whereas the second one is the convergence of the discretized solution to the continuum limit solution  as we take the grid size and cut-offs to zero. %

\paragraph{Convergence of the discretized algorithm.}

Fixed-points can be attractive or repulsive. Our iterations converge in some range of the input parameter space, which depends on the size of the inputs $(\lambda,\eta_{\text{MP}}(s),\rho_\text{MP}(s,t) )$. To first approximation, we find that the iteration algorithm converges when the input is small enough. Physically, it corresponds to interactions which are not too strong. 
In $d=3$, we found convergence for $-3\pi\leq\lambda\leq0$, and in $d=4$ for $0\leq \lambda\leq 5\pi$. Convergence ranges are recapitulated in table~\ref{tab:conv}, for the toy-model of section~\ref{sec:toymodel} and the full algorithm of sections~\ref{sec:QE3d},~\ref{sec:QE4d}.

\begin{table}[]
\centering
\begin{tabular}{l|l|l|l|}
\cline{2-4}
                                           &  Model       & $\lambda_{\text{min}}$ & $\lambda_{\text{max}}$ \\ \hline
\multicolumn{1}{|c|}{\multirow{2}{*}{$d=3$}} & toy-model    & $-5\pi$         & 0               \\ \cline{2-4} 
\multicolumn{1}{|c|}{}                     & back-reacted & $-3\pi$         & 0               \\ \hline
\multicolumn{1}{|l|}{\multirow{2}{*}{$d=4$}} & toy-model    & 0               & $25\pi$          \\ \cline{2-4} 
\multicolumn{1}{|l|}{}                     & back-reacted  & 0               & $5 \pi$        \\ \hline
\end{tabular}
\caption{Approximate range of convergence of the algorithm.  ``Back-reacted'' refers to the amplitudes with non-zero double spectral density constructed in ~\secref{sec:QE3d} and ~\secref{sec:QE4d}. }
\label{tab:conv}
\end{table}

Let us next talk about the speed of convergence. A sequence $(f)_{n,n\geq0}$ is said to have rate of convergence $q$ if there exists a non-zero constant $\mu$ such that 
\be
\label{eq:convergencerate}
\lim_{n\to\infty}\frac{||f_{n+1}-f_*||}{||f_n-f_*||^q} = \mu. 
\ee
where we stay generic about the norm and the space in which $f_n$ lives. If $q=1$, the convergence is called \textit{linear}, this is the case for fixed-point iterations in general, and what we observe here, see \figref{fig:lin-conv-hotstart}. Gradient methods such as Newton-Raphson have \emph{quadratic} convergence, which we observed explicitly in $d=2$ \cite{Tourkine:2021fqh}, where we implemented both methods. As can be seen from \eqref{eq:convergencerate}, \emph{linear convergence} implies that the approach to the fixed point is \emph{exponentially fast}.

\paragraph{Convergence in the continuum limit.}

Let us next discuss convergence of the algorithm to the full, smooth, solution to the equations, as the grid size goes to zero. 

First of all, let us notice that no continuum limit proofs for the existence of solutions to the equations described in the present paper exist up-to-date. In \cite{Atkinson:1969eh}, Atkinson considered a $d=4$ amplitude given by the Mandelstam representation with one subtraction, but when imposing unitarity he used a modified set of equations which effectively introduces a UV cutoff that trivializes the Regge limit. We describe this phenomenon in a simplified setting in \appref{app:atkinsonproofd34}. When trying to remove this cutoff, the parameter space in which convergence of the algorithm can be proven shrinks to zero size, unless one restricts to $3<d<4$. Note finally that in the original paper, Atkinson works with no subtractions and in this case no such cut-off is needed, \cite{Atkinson:1968hza,Atkinson:1968exe}, but this setup does not allow to describe $\phi^4$-like interactions, which do not decay in the Regge limit in $d=4$ and in all kinematics in $d=3$.

As a consequence, the only way for us to describe non-discretized solutions in $d=3$ and $d=4$ is empirical: we generate solutions on different grids of increasingly smaller minimum spacing and observe convergence to a smooth solution.

Next, when taking the continuum limit, we observed two qualitatively different behaviors: either the solutions were simply becoming more and more smooth and valid in a larger range of energies, or we were loosing the solutions. We observed that the continuum limit is smooth as we are removing all cutoffs for $\lambda<0$ in $d=3$, and $\lambda>0$ in $d=4$. For the opposite signs of couplings, positive in $d=3$ and negative in $d=4$, we observed that the algorithm would converge on a finite grid with small number of points, and start to diverge as we took the cutoff to infinity and grid spacing to zero. In $d=3$, the continuum limit solution in the class of functions that we considered requires $\lambda <0$ via the simple dispersive argument, see \eqref{eq:sumrule3dlambda}. In $d=4$, we believe that the divergence of the algorithm is related to the existence of the Landau pole, which causes the theory become strongly coupled as we send the UV cutoff to infinity. For us, it has manifested itself through the fact that the amplitude becoming bigger in the UV eventually caused a divergence of the algorithm.

In conclusion, we find that we can define a continuous limit while maintaining the contracting property of the algorithm.

\subsubsection{Computation times.} 

To close this section, let us just comment briefly on the computation times. We chose our typical grids for $\rho(s,t)$ 11,455 points and $\rho(s)$ with 100 points, respectively, such that computation times were manageable. As a  the data plotted in this paper could be re-ran on a modern laptop in less a week. Furthermore, the codes are easily parallelized, since the iteration proceeds by evaluating integrals of the same functions $\rho^{(n)}(s),\rho^{(n)}(s,t)$ at all the grid points $(s_i,t_j)$ and are completely independent of each other. To produce a typical dataset, with 10 different couplings, and 10 iterations (with hotstart to achieve good precision), can by done in $O(10h)$ with 50 CPUs, using mathematica's native lightweight grid-computing environment.
Given how easy it is to parallelize this code, which essentially performs thousands of independent integrals at each step, we expect a huge speed-up if it were adapted to run on a GPU.

\subsection{Fixing $\eta_{\text{MP}}$: analyticity in spin up to $J=0$ and Feynman graphs}
In the discussion above we considered the S-wave inelasticity $\eta_{\text{MP}}(s)$ and $\rho_{\text{MP}}(s,t)$ to be independent inputs. However, there is a simple situation,
where the two are not independent: consider an amplitude for which discontinuity $T_s(s,t)$ vanishes in the Regge limit $t \to \infty$ for any $s$. In this case we can use
the Froissart-Gribov formula for the imaginary part of the partial wave 
\be
\label{eq:FGimf}
{\rm Im} f_J(s) = {2 {\cal N}_d \over \pi} \int_{z_1}^\infty d z (z^2 - 1)^{{d-4 \over 2}} Q_J^{(d)}(z) \rho(s,t(z)) . 
\ee
all the way to $J=0$.
Recall that we can always decompose the double spectral density as follows
\be
\rho(s,t) &= \rho_{\text{el}}(s,t) + \rho_{\text{el}}(t,s) + \rho_{\text{MP}}(s,t) .
\ee
Plugging this decomposition into the Froissart-Gribov formula \eqref{eq:FGimf}, we get the desired relationship between $\eta_{\text{MP}}(s)$ and $\rho_{\text{MP}}(s,t)$:
\begin{equation}
\begin{aligned}
&{\text{Analyticity in spin up to } J=0:} \\
&\eta_{\text{MP}}(s) = {(s-4m^2)^{{d-3 \over 2}} \over \sqrt{s}} {4 {\cal N}_d \over \pi} \int_{z_1}^\infty d z (z^2 - 1)^{{d-4 \over 2}} Q_0^{(d)}(z) \left( \rho_{\text{el}}(t(z),s) +  \rho_{\text{MP}}(s,t(z))   \right) \ ,
\label{eq:analyticityinspinJ0}
\end{aligned}
\end{equation}
where we also used the fact that $\rho_{\text{el}}(s,t)$ satisfies the Mandelstam equation and therefore is responsible for elastic scattering. In the numerical iterations, we use the single spectral function $\rho(s)$, so to implement \eqref{eq:analyticityinspinJ0} it is convenient to rewrite it as follows
\be
\label{eq:2prdefinition}
\rho_{\text{MP}}(s) = {4 \Gamma({d-1 \over 2}) \over \pi^{3/2} \Gamma({d-2 \over 2})}  \int_{z_1}^\infty d z (z^2 - 1)^{{d-4 \over 2}} Q_0^{(d)}(z) \left( \rho_{\text{el}}(t(z),s) +  \rho_{\text{MP}}(s,t(z))   \right)  ,
\ee
where $\rho_{\text{MP}}(s)$ is the multi-particle single spectral density.

We call the algorithm in which we fix the S-wave particle production using \eqref{eq:2prdefinition} \emph{two-particle reducible} (2PR).
The reason is that it has simple relationship to Feynman diagrams in $\phi^4$ theory, which we discuss now. Let us start by setting $\rho_{\text{MP}}(s,t)=0$: this corresponds to throwing away all the graphs which are not two-particle irreducible, e.g. the open envelope graph in \figref{fig:diagramspic}. We can then generate the amplitude by iterations which is graphically summarized in figure \ref{fig:atkinson-graphical}. We also set $\eta_{\text{MP}}(s)$ to be given by \eqref{eq:analyticityinspinJ0}. We expect that this procedure corresponds to re-summing \emph{all the two-particle reducible Feynman graphs generated by two-particle unitarity and crossing}. In \figref{fig:diagramspic} these are graphs depicted in black.
This expectation is based on the assumption that analyticity in spin of inelasticity up to $J=0$ is a true property of this subset of Feynman graphs, something that we verified explicitly up to three loops.

In the full theory we do not expect \eqref{eq:analyticityinspinJ0} to hold. The reason is that we get diagrams similar to the frog diagram, see \figref{fig:diagramspic}. On one hand, this diagrams contributes to $\eta_{\text{MP}}(s)$. On the other hand, it does not have any double spectral density. Interestingly, the frog diagram is just a first in a series of planar melonic diagrams of this type.

Such diagrams have been recently studied in the context of the SYK model, see e.g. \cite{Maldacena:2016hyu,Sarosi:2017ykf}, and it would be very interesting to explore them in the context of multi-particle scattering. Similarly, understanding unitarity structure of these graphs nonperturbatively is an interesting problem.

\subsection{Graph interpretation and summability.}

After having exposed this whole procedure, the reader might legitimately ask: why would this iteration procedure converge? After all, iterating unitarity is very similar in essence to the standard perturbative computations, and perturbation theory is known to produce an asymptotic series due to a factorial growth of the total number of Feynman graphs. 

\begin{figure}
  \centering
  \includegraphics[scale=0.9]{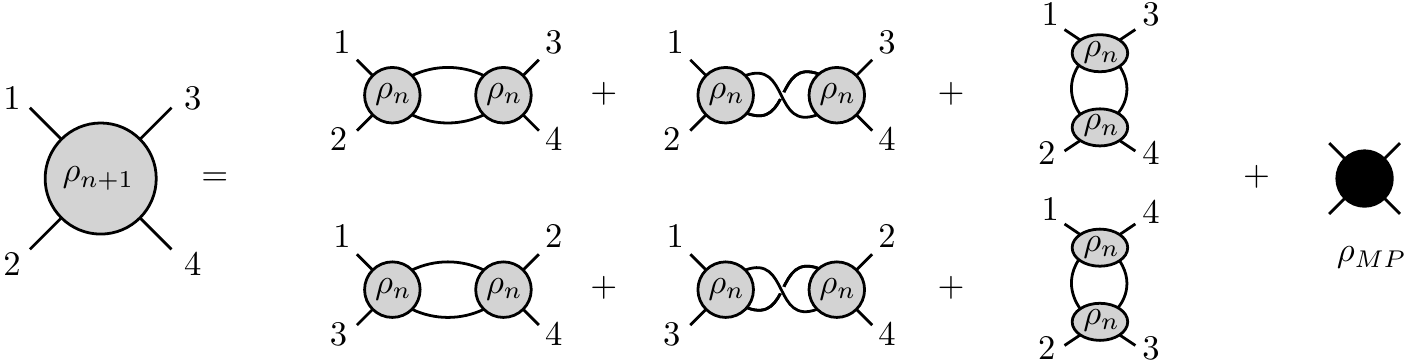}
  \caption{Graphical representation of the iteration process for the double spectral density. The first three graphs in the RHS are generated by the Mandelstam double integral and the Mandelstam equation. Graphs in the second row are generated through the addition of the crossed term $\rho_{\text{el}}(t,s)$ in the iteration process for $\rho(s,t)$. Finally, the last term specifies the addition of production multi-particle term $\rho_{\text{MP}}(s,t)$ at every iteration.}
  \label{fig:atkinson-graphical}
\end{figure}

Two simple arguments indicate that our iteration scheme should be convergent. The iteration can be schematically presented as a simple graphical recursion relation of \figref{fig:atkinson-graphical}. The first element in favor of convergence is that this recursion only generates \textit{planar} graphs, and the planar sector is known to be convergent in some cases, see e.g. \cite{Brezin:1977sv}).
Secondly, a more precise counting can be made. The number of terms $N_{n+1}$ at the $n+1$-st step can be seen to be roughly given in terms of the  number of terms $N_{n}$ at the $n$-th step by
\begin{equation}
  \label{eq:number-graphs}
  N_{n+1}= 6 (N_n)^2+1 .
\end{equation}

Compared to Feynman graphs at step $n$, we generate graphs with
$N_{n-1}^2$ loops, as opposed to $N_{n-1}+1$ loops which we would have
in perturbation theory by adding a loop. Therefore, the relation
between $n$ and the number of loops $L$ is not linear, but rather exponential:
\begin{equation}
    L \simeq 2^n\,.
\end{equation}
From \eqref{eq:number-graphs}, we can estimate that
$N_n\simeq e^{\log c_0 L} \simeq c_0^L$ where $c_0$ depends on the initialization. This does not exactly tell us the
 number of $L$-loop graphs
 yet it motivates that the growth should be a power-law, rather than a factorial which implies a finite radius of convergence of the algorithm. A slightly more refined expectation is that we have a series of the type $\sum_{L} (\lambda_p c_0)^L$, where $\lambda_p$ is the strength of scattering at energy scale $p$. For the series to converge therefore we expect that the theory should stay weakly coupled at all scales. All the amplitudes that we construct in this paper are indeed of this type.\footnote{As a side remark, let us notice that our split of Feynman diagrams into the multi-particle graphs which serve as a building block for loops of two-particle reducible graphs is reminiscent of the ``loops of loops'' and ``hard" \text{vs.} ``easy'' graphs in the discussion in \cite{Arkani-Hamed:2021iya}.}

A closely related problem was solved in \cite{sundberg1998rate}, and later in \cite{Zinn-Justin:2003ecd} using matrix model techniques. The generating function for the number of two-particle fully two-particle reducible planar graphs was found to be given by the following generating function
\be
\label{eq:ZJZ}
\Gamma(g) = {1 \over 2} (1-g- \sqrt{1-6 g + g^2}) = g + 2 g^2 + 6 g^3 + 22 g^4 + \dots .
\ee
Here the coefficients describe the number of different graphs at order $g^n$. For example, $2 g^2$ stands for horizontal plus vertical one-loop bubble; $6 g^3$ comes from horizontal and vertical two-loop bubbles as well as four different orientations of the bubble plus triangle diagram, etc. From this expression we immediately see that the number of graphs indeed grows like a power, and the radius of convergence of the small $g$ expansion is finite $g < g_*$, and is controlled by the smallest root of $\sqrt{1-6 g + g^2}$
\be
\label{eq:convergenceradius}
g_* = 3 - 2 \sqrt{2} \simeq 0.171573 \ .
\ee
Given that the actual perturbation theory expansion parameter involves extra numerical factor, e.g. ${\lambda \over 16 \pi^2}$ in $d=4$, we expect that the radius of convergence of our algorithm to be bigger than given by \eqref{eq:convergenceradius}. This is indeed what we observed in practice.

It is also interesting to list the number of non-automorphic graphs both two-particle fully reducible and total number \cite{MiguelCor}, see \tabref{table:graphs}.
\begin{table*}[htb!]
\begin{center}
\begin{tabular}{|c | c | c | c| c | c | c | c| c | c | c |} 
 \hline
 \# of vertices & 1 & 2 & 3 & 4 & 5 & 6 & 7 & 8 & 9 & 10 \\ 
 \hline
  All graphs & 1 & 1 & 2 & 8 & 26 & 124 & 627 & 3794 & 25306 & 188746 \\
 \hline
 2PR graphs & 1 & 1 & 2 & 6 & 19 & 74 & 310 & 1451 & 7130 & 35398\\ 
 \hline
  Planar graphs & 1 & 1 & 2 & 7 & 23 & 95 & 411 & 2005 & 10214 & 54873 \\
 \hline
\end{tabular}
\end{center}
\caption{\small The number of non-automorphic graphs contributing to $2\to 2$ scattering in the $\phi^4$ theory as a function of number of vertices. We distinguish two-particle fully reducible graphs which are automatically generated by the iteration algorithm considered in the present paper and all graphs. Given that the number of 2PR graphs grows like a power, we see that most of the graphs (which add up to the factorial growth) contribute to the multi-particle input. Similarly, if we take a given multi-particle graph and add it to the iteration algorithm, it will generate the number of extra graphs that grow like a power. We thank Miguel Correia for producing the counting presented in this table.}
\label{table:graphs}
\end{table*}
As expected the number of full graphs grows much faster very quickly only a small fraction of diagrams comes from two-particle iterations, whereas most of the graphs contribute to $\eta_{\text{MP}}(s)$ and $\rho_{\text{MP}}(s,t)$. When one accounts for automorphisms $(1,1,2,6,...)$ from \tabref{table:graphs} becomes $(1,2,6,22,...)$ from \eqref{eq:ZJZ}.
Generalizing the estimate above for the number of graphs,  from the two-particle reducible to $n$-particle reducible, we still get a power-law behavior (though with a larger power). The factorial growth thus originates from the fact that $n$ is unbounded from above.

A few comments are in order. Although our summation involves only planar graphs and fewer terms than expected, we still obtain a fully nonperturbative amplitude in the end, which is generated by the inputs $(\lambda, \eta_{\text{MP}}(s),\rho_{\text{MP}}(s,t) )$.  These inputs parameterize the space of amplitude functions: navigating the space of inputs formally maps to navigating the space of outputs, i.e. the space of amplitudes. In particular, the actual physical amplitudes within this space.  Practically, this construction allows us to formally ignore the problem of re-summation of all Feynman graphs (which should result in $\eta_{\text{MP}}(s)$ and $\rho_{\text{MP}}(s,t)$), and yet be able to describe nonperturbative physical amplitudes.

\section{Toy-model amplitudes: zero double discontinuity}
\label{sec:toymodel}

Having described the full set of equations and the algorithm in detail, we now turn to a simplified version of the problem. 
The amplitudes we consider are obtained by discarding the double spectral function of the Mandelstam representation, and assume the form:
\begin{equation}
  \label{eq:mandelstamone}
T(s,t) = \lambda + \int_{4m^2}^\infty {d s' \over \pi } {\rho(s') \over s' - {4 m^2 \over 3}} \left( {s-{4 \over 3} m^2 \over s'-s}+  {t - {4 m^2 \over 3} \over s' - t}+  {u - {4 m^2 \over 3} \over s' - u}\right) \ . %
\end{equation}
Thanks to the subtraction, we still have that
\begin{equation}
  \label{eq:couplingdef}
T \left( {4  m^2\over 3},{4 m^2 \over 3} \right) =\lambda \,.
\end{equation}
This representation is manifestly crossing symmetric and has maximal analyticity.
However, given zero double spectral density, unitarity can only be imposed in the $J=0$ sector.\footnote{Later we study the full solution, with nonzero double spectral density, which restores unitarity for all spins. This will be the subject of the following sections. } It takes the form
\begin{equation}
    |S_0(s)|^2 = 1-\eta_{\text{MP}}(s), 
\end{equation}
where $1 \geq \eta_{\text{MP}}(s)\geq0$  for $s\geq s_{\text{MP}}$.

However, in this section, we describe in detail the case $\eta_{\text{MP}}(s)=0$, so  we impose that the scattering in the S-wave is elastic at all energies:\footnote{Notice that existence of such
  S-matrices in $d>2$ is consistent with the Aks theorem \cite{Aks:1965qga}, which only
  requires particle production for partial waves with $J \geq 2$.}
\begin{equation}
  \label{eq:unitarity0spin}
| S_0 (s) | = 1 , ~~~ s>4 m^2.
\end{equation}
We further recall that in this paper we work with the assumption that we have no bound states below the two-particle threshold. Below, we refer to the amplitudes that satisfy \eqref{eq:unitarity0spin} and \eqref{eq:mandelstamone} as \emph{toy-model amplitudes}.

This toy-model serves three main purposes. Firstly, it is a simpler pedagogical setup to learn the general techniques of this paper. Secondly, we actually use this algorithm in the first step of generating our solutions, to reach a first fixed-point which is close to that of the solution to the full set of equations. We refer to this procedure as a \textit{hotstart}.
Lastly, these amplitudes are interesting in their own right, and they provide a sort of analogue of the CDD two-dimensional integrable S-matrices, see e.g. \cite{Paulos:2016but} and \appref{app:CDDs}.

Note finally that for the problem discussed in this section, all integrals are one-dimensional. This implies that all the methods, including the gradient-based methods developed in our former work \cite{Tourkine:2021fqh} can be applied to implement a more efficient iteration algorithm. The use of gradient-guiding would be expected to increase the space of parameters in which the iteration converges. 
The reduction of numerical integrations to tensorial operations also speeds up significantly the process. This would consequently render  hotstart procedure  described here very efficient. We did not attempt this here and leave this improvement for future work. 

Let us now begin the detailed study of this toy-model. We split our discussion according to the number of spacetime dimensions. The situation in $d=2$ is reviewed in \appref{app:CDDs}. We analyze in detail the new cases $d=3$ and $d=4$, which turn out to be quite subtle, because of threshold and Regge behavior, respectively, and the iteration algorithm requires extra care to converge in practice.

Note finally that, as we commented on before, the situation is the simpler in $3<d<4$, and for this case we can even present an argument about the existence of the toy-model amplitudes directly in the continuum limit, along the lines of the original work by Atkinson \cite{Atkinson:1970zza}, see appendix~\ref{app:atkinsonproofd34}. We do not attempt to solve the problem in $d>4$.

\subsection{$d=3$}
\label{sec:3dnp}
In three dimensions, the near two-particle threshold region requires extra care. We start the discussion by explaining the equations to be solved iteratively, derive the expected two-particle threshold behavior, and then present the results of our numerical implementation. The section is meant to be, to a large extent, self-contained.

\subsubsection{The algorithm}

Our starting point are amplitude functions that admit once-subtracted
Mandelstam representation with zero double
discontinuity~\eqref{eq:mandelstamone}. We start with discussing
S-wave unitarity \eqref{eq:unitarity0spin}, which is the sole
unitarity condition to be satisfied. In \secref{sec:algorithm}, we
referred to \appref{app:definitions} for the derivation of
eq.~\ref{eq:single-disc-unitarity}. Here, the absence of
double-discontinuity makes this derivation much less cumbersome, and
to keep the discussion self-contained, let us derive the analogue of
\eqref{eq:single-disc-unitarity} in a fully explicit manner (with
explicit $f_0$).
To start, we expand the amplitude in three-dimensional partial waves\footnote{We follow the conventions of  \cite{Correia:2020xtr} which are review in reviewed in \appref{app:definitions}.}
\be
T(s,t) &= 16 \sum_{J=0}^\infty {1 \over 1+\delta_{J,0}}f_J(s) \cos (J \theta), ~~~ t = -{s-4 m^2 \over 2} (1-\cos \theta), \\
f_J(s) &= {1 \over 16 \pi} \int_0^{2 \pi} d \theta \cos (J \theta) T \Big(s,-{s-4 m^2 \over 2} (1-\cos \theta) \Big) . 
\label{eq:3dpartialwave}
\ee
The unitarity equation that we are solving, $|S_0(s)|^2=1$, takes the form
\be
\label{eq:unitarity3d}
2 {\rm Im}f_0(s) = {1 \over \sqrt{s}} | f_0(s) |^2 , ~~~ s \geq 4 m^2 ,
\ee
where we used that, in $d=3$, %
\begin{equation}
  \label{eq:Sj-fj-3d}
  S_J(s) = 1 + {i \over \sqrt{s}} f_J(s)\,.
\end{equation}

We are interested in finding an amplitude, or rather its discontinuity $\rho(s)$,  that solves \eqref{eq:unitarity3d} iteratively. Using \eqref{eq:3dpartialwave}, we can compute the spin zero partial wave in terms of the spectral density $\rho(s)$. We plug \eqref{eq:mandelstamone} into \eqref{eq:3dpartialwave} to get
\begin{equation}
  \begin{aligned}
\label{eq:spin0partiald3}
f_0(s) &= {1 \over 8} \left( \lambda + \int_{4m^2}^\infty {d s' \over \pi } {\rho(s') \over s' - {4 m^2 \over 3}} \left( {s-{4 \over 3} m^2 \over s'-s} +K_0^{(d=3)}(s',s) \right) \right) , \\
K_0^{(d=3)}(s',s) &= 2 \left( {s' - {4 m^2 \over 3} \over \sqrt{s'} \sqrt{s+s'-4 m^2}} - 1  \right). 
  \end{aligned}
\end{equation}
From this equation, it follows readily that 
\begin{align}
    \Im f_0(s) &= {1 \over 8} \rho(s)\,,\\
    \Re f_0 &= {1 \over 8} \left( \lambda +  \text{P.V.} \int_{4m^2}^\infty {d s' \over \pi } {\rho(s') \over s' - {4 m^2 \over 3}} \left( {s-{4 \over 3} m^2 \over s'-s} +K_0^{(d=3)}(s',s) \right) \right)\,.
\end{align}
In this way we obtain a closed equation for $\rho(s)$ given by
\begin{equation}
\label{eq:unitartiy3diter}
  \rho(s) ={1 \over 16 \sqrt{s}}\left[ \Big(\rho(s) \Big)^2 + \left( \lambda + \text{P.V.}\int_{4m^2}^\infty {d s' \over \pi } {\rho(s') \over s' - {4 m^2 \over 3}}  \left( {s-{4 \over 3} m^2 \over s'-s} +K_0^{(d=3)}(s',s) \right) \right)^2  \right]
\end{equation}
We start the iteration process by setting $\rho^{(0)}(s)=0$ and at step $n+1$, we define $\rho_{n+1}(s)$ as a function of $\rho_n(s)$ using the RHS of \eqref{eq:unitartiy3diter}, as was explained in \secref{sec:algorithm}.
One can check by inspection that the first two iterations reproduce exactly the two-loop computation of the amplitude in the ${\lambda \over 4!} \phi^4$ theory in a renormalization scheme defined by \eqref{eq:couplingdef}. At step 0, we have
\begin{equation}
T^{(0)}(s,t) = \lambda . 
\end{equation}
Plugging this into the unitarity relation \eqref{eq:unitartiy3diter}, we get
\begin{equation}
\label{eq:firstiteration3d}
\rho^{(1)}(s) = {\lambda^2 \over 16 \sqrt{s}} .
\end{equation}
Plugging this into the Mandelstam representation, we get $T^{(1)}(s,t)$, from which the two-loop $\rho^{(2)}(s) $ can be computed using unitarity, see \appref{sec:d3phi4twoloops} for details. Starting from three loops, or equivalently at order $\lambda^4$, we would have to introduce the double spectral density in order to produce further actual $\phi^4$ theory graphs, as well as $\eta_{\text{MP}}(s)$. This will be the subject of next sections. 

Let us note the following features of the iteration process. First, we see that after the first iteration the spectral density $\rho^{(1)}(s) $ goes to a constant at the two-particle threshold $s = 4m^2$. Upon performing the dispersive integral, this leads to the appearance of logarithms $\int_{4 m^2} {d s' \over s'-s} \sim \log (s-4m^2)$, and therefore we expect perturbation theory to break down close to the two-particle threshold due to large logarithms $\left(\lambda \log (s-4m^2) \right)^k$. Second, in the Regge limit, we find that $\rho^{(1)}(s) \sim s^{- 1/2}$, and by analyzing the iteration equations it is easy to convince oneself that this behavior is preserved, and should hence be the behavior of the solution.\footnote{We will see later that in the back-reacted solution, this Regge behavior, a fixed Regge pole, is not compatible with unitarity and indeed gets dressed by logarithmic corrections.}

The nonperturbative behavior of the amplitude close to the two-particle threshold is actually known and it takes the following form \cite{Chadan:1998qm,Bros:1998tt}
\begin{equation}
\label{eq:univ-thresh-3d}
\rho(s) = {4 \over \sqrt{s}} \left[ \left(b_0(\lambda) + O({s \over 4 m^2}-1) + {1 \over 2 \pi \sqrt{s}}\log ({s \over 4 m^2}-1) \right)^2 + {1 \over 4 s} \right]^{-1} \,.
\end{equation}
The solutions we found matched very will this behavior, and we could observe empirically that $b_0(\lambda) \sim {1 \over \lambda}$ for small $\lambda$.\footnote{The same behavior was also argued to hold in the full $\phi^4$ theory in \cite{Chadan:1998qm,Bros:1998tt}. } The form of the partial wave is dictated by unitarity near the threshold, see formula (5.6) in \cite{Correia:2020xtr}. In this way the nonperturbative spectral density at the fixed point actually vanishes, albeit at a slow rate, at the two-particle threshold:
\be
\label{eq:nonpthreshold3d}
\text{Threshold}:~~~\lim_{s \to 4m^2} \rho(s) \simeq {32 \pi^2 m \over \log^2 ({s \over 4 m^2}-1)} \to 0 .
\ee
This is in contrast with the result of the first iteration \eqref{eq:firstiteration3d}. Numerically, we observe that this tension creates a divergence in the iterations if we try to solve the discretized system on a grid which goes too far in the IR (close to the two-particle threshold).

To resolve this tension, we used a set of successive grids, that extend further and further near the threshold. In this way, we find that the iteration algorithm converges and smoothly matches to the expected asymptotic behavior \eqref{eq:nonpthreshold3d}. We describe this in detail below.

Let us also notice, that given \eqref{eq:nonpthreshold3d}, unitarity \eqref{eq:unitarity3d} implies that ${\rm Re} f_0(4m^2)=0$. This condition leads to the following sum rule
\be
\label{eq:sumrule3dlambda}
\lambda + \int_{4m^2}^\infty {d s' \over \pi } {\rho(s') \over s'} {32 m^4 \over 3 (s'-4 m^2) (s'-{4 m^2 \over 3}) } = 0.
\ee
This sum rule is an interesting example of a situation where the ``coupling constant'' $\lambda$ is dispersive, even though the amplitude $T(s,t)$ goes to a constant (different from $\lambda$) at infinity.
On the other hand, for the spectral density $\rho(s)$ in the Regge limit we get
\be
\label{eq:nonpertregge3d}
\text{Regge}:~~~\lim_{s \to \infty} \rho(s) \sim s^{-1/2} \to 0 .
\ee

Let us also comment briefly on the sign of $\lambda$. The standard $\phi^4$-theory with a repulsive potential bounded from below corresponds in our conventions to $\lambda \leq 0$. This is the case for which we find our iterations to converge.
Positive $\lambda>0$, on the other hand, are excluded by the sum rule \eqref{eq:sumrule3dlambda} and the fact that unitarity implies that $\rho(s)\geq 0$ in the toy-model. Consistently, we find that our algorithm diverges in this case as we try to remove the IR cutoff.

\subsubsection{Numerical implementation: hotstart near the two-particle threshold}

We now turn to the description of the numerical implementation of this algorithm, and present some explicit results. We use the setup described in sec.~\ref{sec:num-impl}, and interpolate the single-spectral function $\rho(x)$ on grids $x_0=0,\dots,x_N=1$.

Then we perform iterations of unitarity on this interpolant. We evaluate the value of the spectral density at the grid point point $\rho_{n+1}(x_i)$ at $(n+1)$-th iteration using the piece-wise linear $\rho_n(x)$ plugged into the unitarity equation \eqref{eq:unitartiy3diter} after performing the change of variable $s'\to4m^2/x'$.
We also impose the following boundary conditions for the spectral density $\rho(s)$ which are conserved by the iteration process
\be
\rho(0)=\rho(1)=0.
\ee
In the limit of grid spacing becoming small, we observe that the function describes a smooth curve, and we naturally expect the function to approach the true continuous solution. 

In practice, the simple algorithm described above requires care because of the threshold behavior, as was explained in the previous subsection. The unitarity equation \eqref{eq:unitartiy3diter} fixes the behavior close to $x=0$ and $x=1$ to be of particular form. For the example, close to the two-particle threshold $x=1$, we expect the universal behavior \eqref{eq:nonpthreshold3d} and not a simple linear behavior. Moreover, as discussed above, initializing the iteration process with $\rho_0(x)=0$ leads to large logarithms close to the threshold, which destabilize the iteration process. 

Luckily, there is an efficient way to circumvent the difficulty above and find the desired solution  in an arbitrary large range of energies using a version of the hotstart described above.

In the present case, we use the hotstart in the following way. The iteration process is not defined on one fixed grid, but on a family of grids, with increasing near two-particle threshold density of points and higher-energy cut-off. We initialize the iteration process of a finer grid using the fixed point solution of the previous, rougher, grid. In this way, and proceeding with sufficient caution, we are able to build functions which can be extended arbitrarily close to the threshold, and probe arbitrarily far the slow logarithmic decay when $x\to1$.

This procedure provides a guidance in the fixed-point iteration procedure. The fixed-point on a fine grid is not reachable by a standard iteration because of the large logarithms that destabilize the iterations. However, the guidance provided by the hotstart at each improvement of the grid induces a little ``kick'' in the right direction which allows to push to arbitrary high energies the problem and can converge to a solution up to arbitrarily high cutoffs. We then find that the solution converges to the expected universal behavior  \eqref{eq:nonpthreshold3d}.

\begin{figure}[!htb]
    \centering
    \includegraphics[scale=1.1]{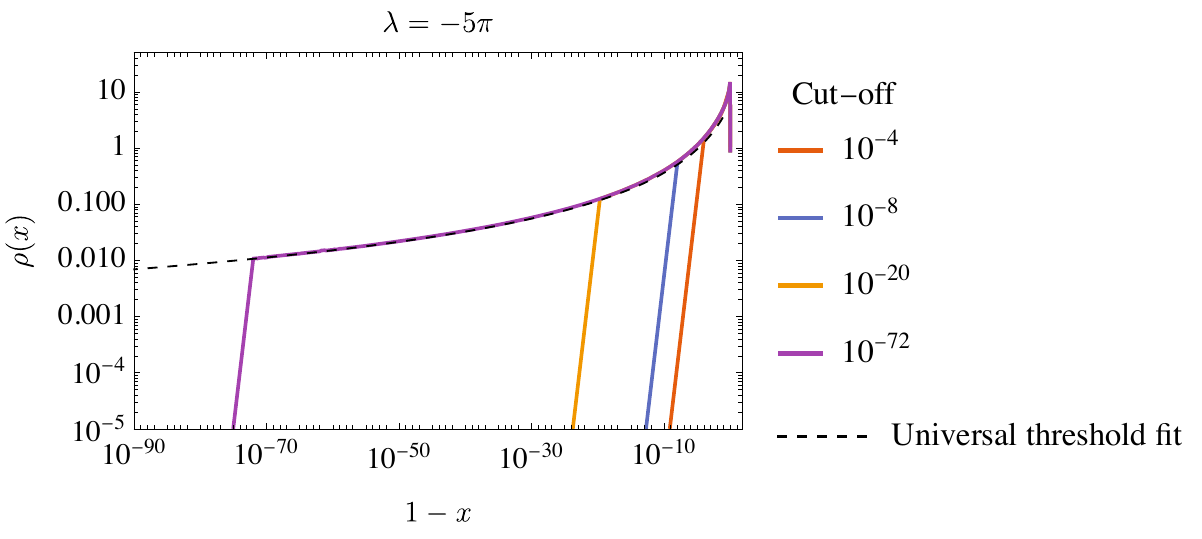}
    \caption{A typical example of the hotstart iterations. Each solid line is the interpolating function $\rho_n(x)$ at the final stage of its iteration ($n=8$ in this example). Without these intermediate steps, the algorithm would not converge. The list of $x_1$'s (closest point to 1) is listed on the right. The gray line is the threshold behavior given in \eqref{eq:univ-thresh-3d} with $b_0(\lambda)=8/\lambda$ determined experimentally.}
    \label{fig:hotstart3d}
\end{figure}

Let us illustrate the procedure described above with an example. Let us say we take $\lambda=- 5 \pi$, we find a minimum grid first point $x_{N-1}$ ($x_N=1$). For concreteness we can take $x_{N-1} = 1-10^{-4}$. We run the iteration algorithm \textit{starting from} $\rho_0(x)=0$ and we find that it converges after $n \sim 10$ iterations. Then, we push the cut-off higher and set $x_{N'-1} = 10^{-8}$. On this new grid, with possibly more elements than the first ($N'\geq N$) the iteration would not have converged starting from $\rho_0=0$, but it does so if we set $\rho_0(x)$ to the converged value of the previous round of iterations. In this way we can empirically push the cut-off  to values arbitrarily close to 1.\footnote{For instance, we could easily reach cut-offs of order $1-10^{-100}$. A small technical details is that we encountered a bug with mathematica's interpolation routines which would stop working at precision of order \texttt{\$MachinePrecision} and so we had to interpolate $\rho(1-x)$ near $0$ instead of $\rho(x)$ near $1$.} An example of this procedure is provided in figure \ref{fig:hotstart3d}. Note that, as was said above, if we start directly on a cut-off which is too close to the threshold, for instance $10^{-16}$, the algorithm does not converge. Note also that since the Regge limit is trivial, no problem is found near $x\to0$ and we do not need any caution there.

\begin{figure}[!htb]
    \centering
    \includegraphics[scale=1.2]{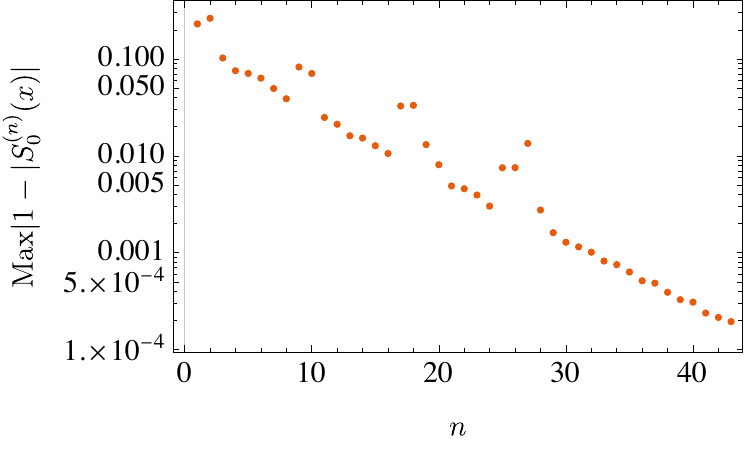}
    \caption{Convergence of the algorithm with hotstart, represented as the convergence of $|S_0(x)|$ to 1. Every 8 steps (the total number of iteration per given grid size), a small glitch is generated by the hotstart but is soon erased (the last step has 20 iterations). Overall, the process induces nice linear convergence, as expected in a fixed-point iteration (the definition of linear convergence was given around eq.~\ref{eq:convergencerate}).}
    \label{fig:lin-conv-hotstart}
\end{figure}

\begin{figure}[!htb]
    \centering
    \includegraphics[scale=0.9]{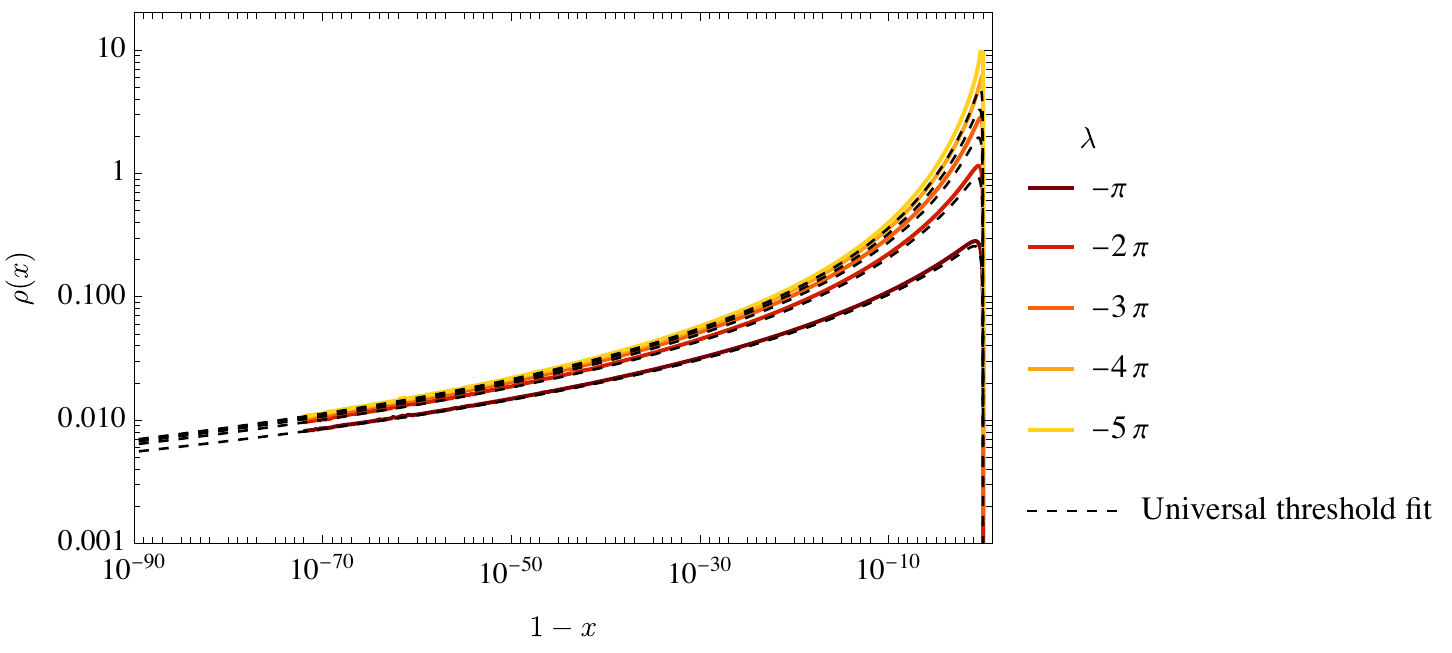}
    \caption{Endpoint of the iteration for various couplings. Solid/colors = numerics, dashed = fit with universal solution with $b_0 = 8/\lambda$. This demonstrates agreement with universal threshold behavior in the near-threshold limit.}
    \label{fig:resultsofiterations3dtoy}
\end{figure}

\begin{figure}
    \centering
    \includegraphics{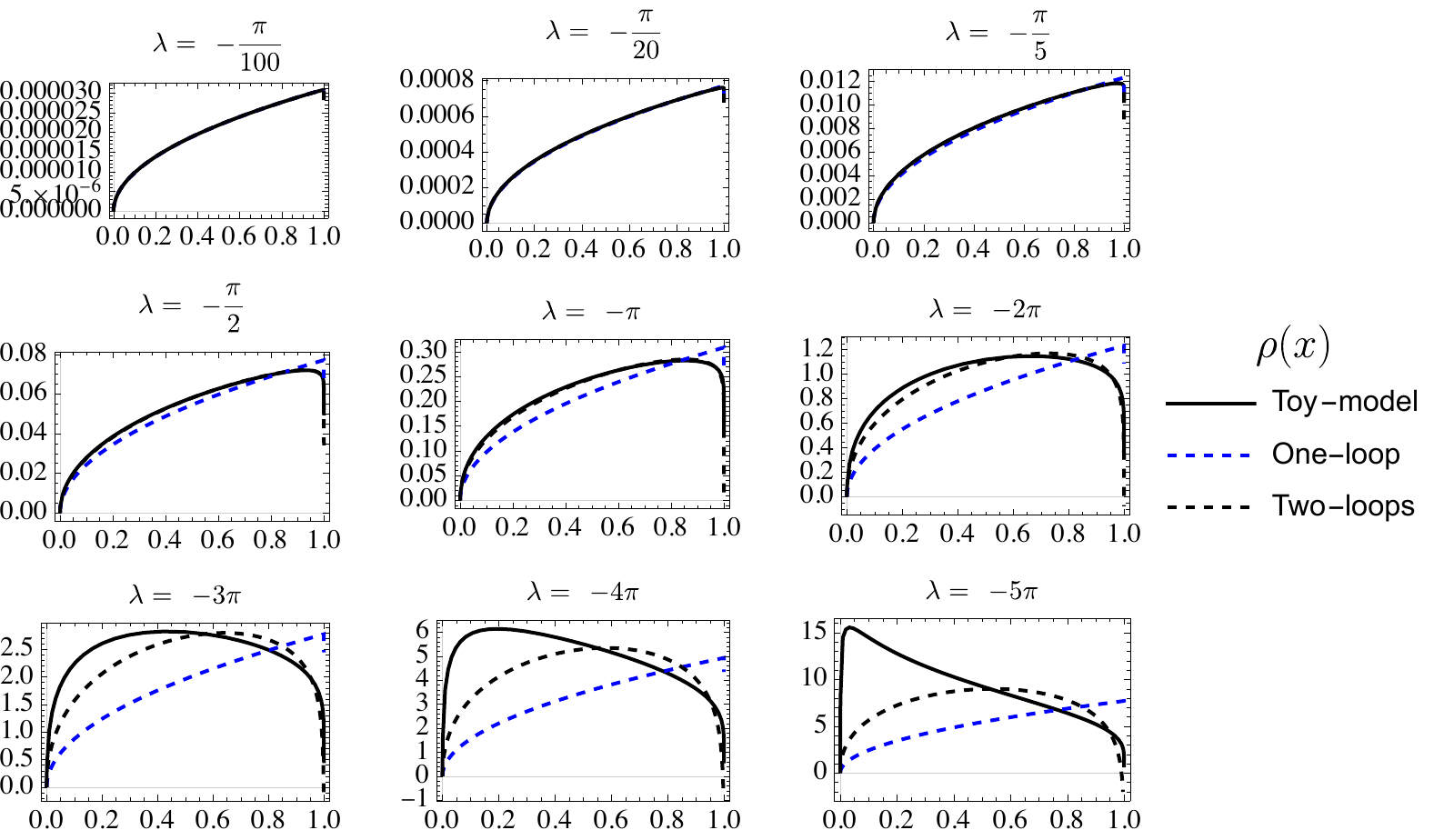}
    \caption{$\rho(x)$ vs perturbation theory for toy-model in $d=3$. We observe that at small $\lambda$ the matching between $\rho(s)$ and the perturbative results derived in \appref{app:perturbativephi4} is good for all $x$. At higher $\lambda$'s the difference becomes significant.}
    \label{fig:3dtoymodeldifferentcouplings}
\end{figure}

\begin{figure}[!htb]
    \centering
    \includegraphics[scale=1.1]{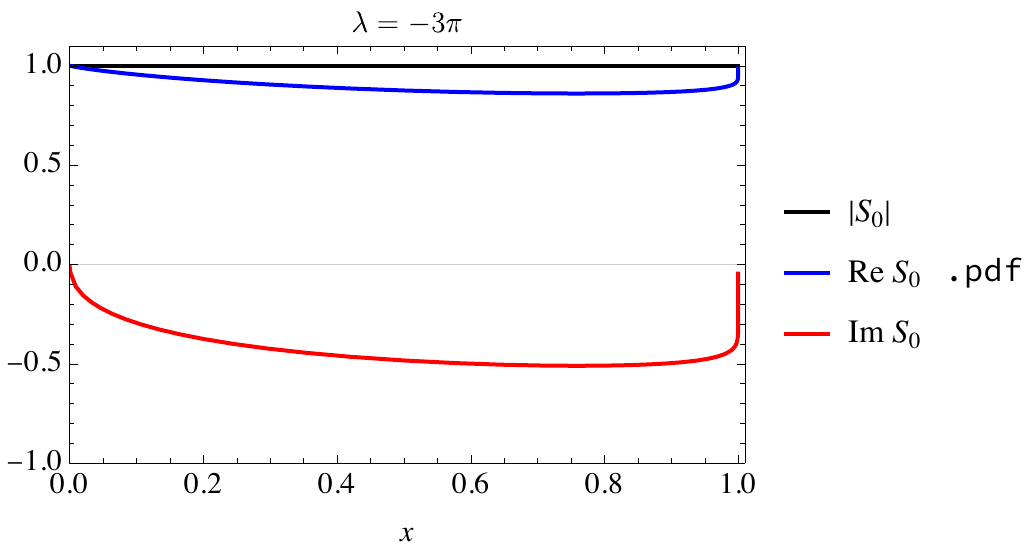}
    \caption{We plot the spin zero partial wave $S_0$ as a function of energy. The scattering becomes trivial, $S_0=1$, both at the two-particle threshold, $x=1$, and at high energies, $x=0$.}
    \label{fig:plS03dtoy}
\end{figure}

\begin{figure}[!htb]
    \centering
    \includegraphics[scale=1.1]{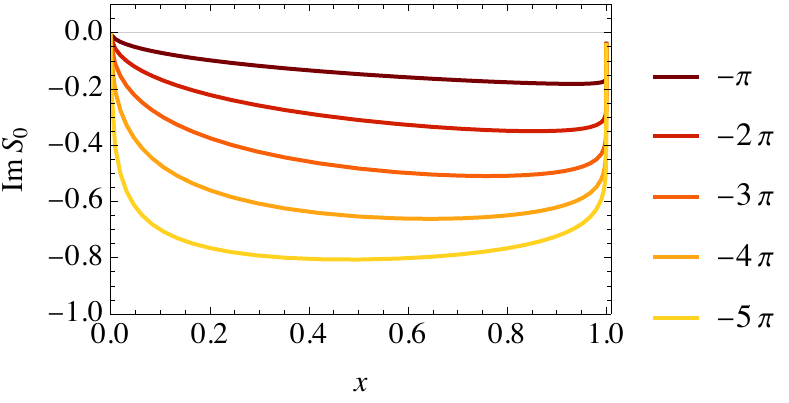}
    \includegraphics[scale=1.1]{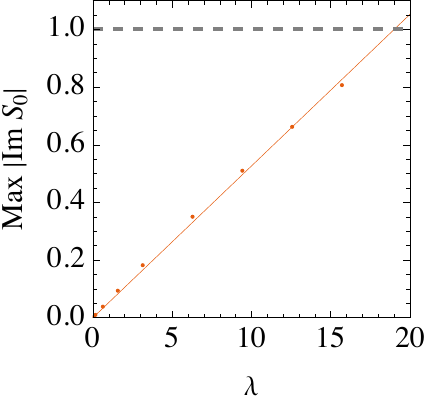}
    \caption{{\bf Left panel}: ${\rm Im} S_0$ as a function of energy for various $\lambda$. {\bf Right panel}: maximum of ${\rm Im} S_0$ of the fixed-point solution as a function of $\lambda$. For $\lambda\simeq6\pi$, the solution would appear to exceed the unitarity bound, which is consistent with the fact that in our method $\lambda \lesssim 6\pi$ is the convergence bound. Note further that the bound is reached inside the elastic band ${1 \over 4} \leq x \leq 1$, so  inelasticity could only help indirectly. It would be interesting to understand the mechanism by which amplitudes with larger couplings $\lambda>6\pi$ can be generated, as in \cite{Chen:2022nym}.}
    \label{fig:maxims03d}
\end{figure}

\begin{figure}[!htb]
    \centering
    \includegraphics[scale=1.]{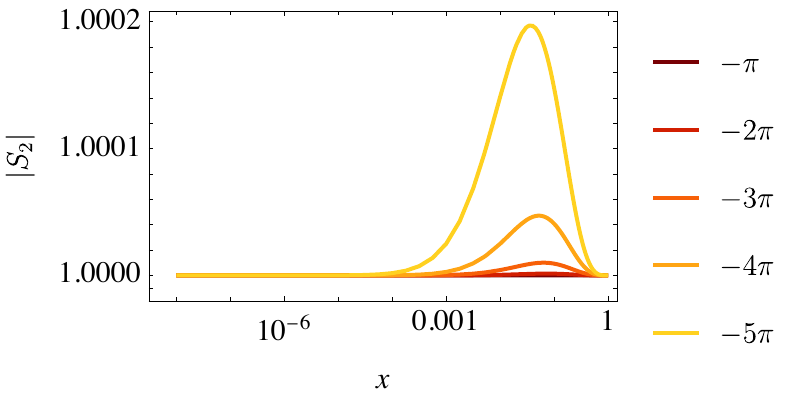}
    \caption{$S_2$ of the toy-model amplitudes in $d=3$. Unitarity violation is manifest ($|S_2|>1$), but is not imposed in the model so it is not surprising. This is cured in the full algorithm later when the double spectral density is added, see e.g. fig.\ref{fig:3piSJ}. 
    } 
    \label{fig:unit-viol-S2-3d}
\end{figure}

\begin{figure}[!htb]
    \centering
    \includegraphics{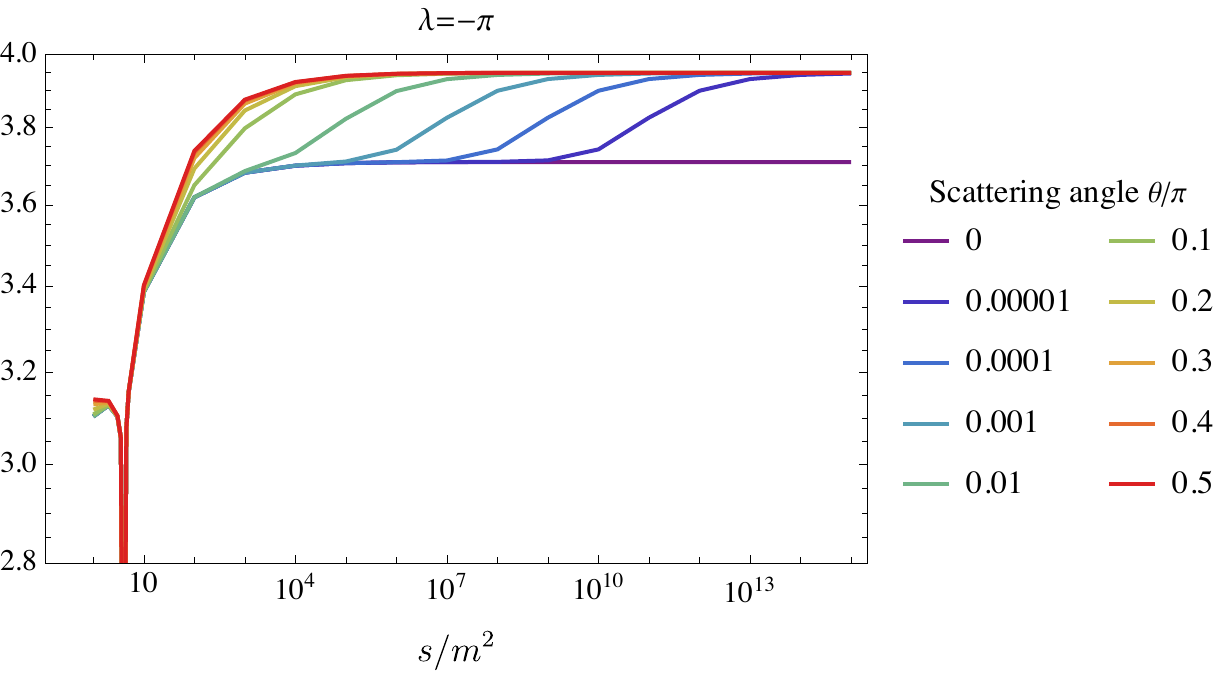}
    \caption{Here we plot the absolute value of the toy model amplitude, $|T(s,t)|$, at several different fixed-angles as a function of energy ${s \over m^2}$ in $d=3$. This plot exhibits that the amplitude has two asymptotic constant regimes: Regge, and fixed-angles, which are constants. The plot also shows that, as a finite angle is slowly turned on from the strict Regge limit, the amplitude leaves Regge to join fixed angles at energies given by $  s=(\theta/\pi)^{-2}$ (or impact parameters $b \sim {1 \over m}$). These two asymptotic constants persist in the back-reacted model, and the amplitudes are indistinguishable by eye, so the reader is invited to think of this plot as the representation of the fully back-reacted quasi-amplitudes themselves.}
    \label{fig:amp-toy-model-3d}
\end{figure}

We present various results regarding the amplitudes we obtain in the list of figures that follows:
\begin{itemize}
\item \figref{fig:hotstart3d} illustrates the hotstart procedure described above,
\item in \figref{fig:lin-conv-hotstart} we show the linear convergence of the algorithm on a typical process,
\item in \figref{fig:resultsofiterations3dtoy} we plot the resulting spectral densities $\rho(x)$ for various values of the coupling $\lambda$. In \figref{fig:3dtoymodeldifferentcouplings} we compare our results with the perturbative calculation.
\item in \figref{fig:plS03dtoy} we plot the spin zero partial wave $S_0$ as a function of energy $x$ for $\lambda = - 3 \pi$.
\item in \figref{fig:maxims03d} we plot the maximum value of ${\rm Im} S_0$ as a function of $\lambda$. A linear interpolation suggests that our algorithm would violate unitarity at $\lambda \sim - 6 \pi$.
\item in \figref{fig:unit-viol-S2-3d} we plot the absolute of the spin $J=2$ partial wave and observe that it violates inelastic unitarity \eqref{eq:inelastic}. This is not surprising given that we set the double spectral density to zero.
\end{itemize}
In summary, we were able to construct the desired amplitudes in $d=3$ for $-5 \pi \lesssim \lambda \leq 0$. These amplitudes exhibit the universal near two-particle threshold behavior \eqref{eq:nonpthreshold3d} and the Regge limit \eqref{eq:nonpertregge3d}. They also show interesting behavior for fixed angle scattering, see \figref{fig:amp-toy-model-3d}.

\FloatBarrier

\subsection{$d=4$}

Next we repeat the same exercise in $d=4$. The region that requires care is now the Regge limit, $x\to0$.

\subsubsection{The algorithm}

The dispersive representation \eqref{eq:mandelstamone} stays intact, however the unitarity equation changes in an important manner. 
We start again with the partial wave expansion, which now reads
\be
T(s,t) &=16 \pi \sum_{J=0, J - \text{even}}^{\infty} (2J+1) f_J(s)  P_J \Big( 1+{2 t \over s - 4 m^2} \Big), \\
f_J (s) &= {1 \over 32 \pi} \int_{-1}^1 d z P_J(z) T(s,t(z)), 
\label{eq:4dpartialwave}
\ee
where $P_J(z)$ are the ordinary Legendre polynomials.
The unitarity equation $|S_0(s)|^2=1$ can be written as
\be
\label{eq:unitarity4d}
2 {\rm Im}f_0(s) = {\sqrt{s-4m^2} \over \sqrt{s}} | f_0(s) |^2 , ~~~ s \geq 4 m^2 ,
\ee
where we used that $S_0(s) = 1 + i {\sqrt{s-4m^2} \over \sqrt{s}} f_0(s)$ in $d=4$.

Our goal is again to solve \eqref{eq:unitarity4d} iteratively. For this purpose it is convenient to compute the spin zero partial wave in terms of the spectral density $\rho(s)$. We therefore plug \eqref{eq:mandelstamone} into \eqref{eq:4dpartialwave} to get
\be
f_0(s) &= {1 \over 16 \pi} \left( \lambda + \int_{4m^2}^\infty {d s' \over \pi } {\rho(s') \over s' - {4 m^2 \over 3}} \left( {s-{4 \over 3} m^2 \over s'-s} +K_0^{(d=4)}(s',s) \right) \right) , \\
K_0^{(d=4)}(s',s) &= 2 \left( {s' - {4 m^2 \over 3} \over s-4 m^2} \log {s'+s-4m^2 \over s'} - 1  \right). 
\ee

We start the iteration by setting $\rho^{(0)}(s)=0$. The first two iterations are then nothing but a two-loop computation in the ${\lambda \over 4!} \phi^4$ theory
. Indeed, we have
\be
T^{(0)}(s,t) = \lambda . 
\ee
Plugging this into the unitarity relation we get
\be
\rho^{(1)}(s) ={\lambda^2 \over 32 \pi} \sqrt{1-{4m^2 \over s}} ,
\ee
where we used that $\rho(s) = 16 \pi {\rm Im}f_0(s)$. This is exactly the single discontinuity of the one-loop bubble. In  \appref{app:perturbativephi4}, we explain the next order of the calculation.

The location of the ``trivial'' and ``nonperturbative'' regimes in $d=4$ is exchanged compared to $d=3$. More precisely, we see that the near-threshold behavior of $\rho^{(1)}(s) \sim \sqrt{s-4m^2}$ is preserved under the iterations of unitarity. On the other hand, the Regge limit receives large logarithmic corrections under iterations because $\int^\infty {ds' \over s'} {1 \over s'-s} \sim \log s$.

Thus, the nonperturbative spectral density at the fixed point has a square-root singularity at the threshold
\be
\text{Threshold}:~~~\lim_{s \to 4m^2} \rho(s) \sim \sqrt{s-4m^2} \to 0.
\ee
In the Regge limit the amplitude exhibits the following behavior which is non-obvious from the perturbation theory point of view 
\be
\label{eq:regged4sumrule}
\text{Regge}:~~~\rho(s) \underset{s \to \infty}{\sim} {32 \pi^3 \over 9 \log^2 s} \to 0. 
\ee

Let us now present the nonperturbative solution of the Regge limit for this toy model. It turns out that the expansion is naturally organized in terms of $y = \log s$ and it takes the following form (see appendix \ref{sec:analyticregge} for the derivation)
\be
\rho(e^y) &= {32 \pi^3 \over 9} {1 \over y^2} \left( 1 - {8 \over 3} {\log y \over y} + {8 +{\pi^2 \over 3} - {32 \over 9} \log y +{16 \over 3} (\log y)^2 \over y^2} + ... \right) \nn \\
&+{b_{\infty}(\lambda) \over y^3} \left( 1 + {4 \over 3 y} - {4 \log y \over y^4} + ... \right) \nn \\
&+ {27 \over 128 \pi^3} {b_{\infty}^2(\lambda) \over y^4} + ... \ , 
\label{eq:ansatz-rhos-4d}
\ee
where asymptotically, around $s=\infty$, $b_{\infty}(\lambda) $ is a free parameter. Empirically, we observed that $b_\infty(\lambda) \sim {1 \over \lambda}$ at small coupling. We comment more on this below, when we discuss the numerics.

It is also very interesting to understand the Regge limit of the full amplitude in this case. Given that $\rho(s) \to 0$ at large $s$, we can rewrite \eqref{eq:mandelstamone} as
\be
\label{eq:mandelstamzero}
T(s,t) = \left( \lambda - 3 \int_{4m^2}^\infty {d s' \over \pi } {\rho(s') \over s' - {4 m^2 \over 3}} \right)  + \int_{4m^2}^\infty {d s' \over \pi } \rho(s') \left( {1 \over s'-s}+  {1 \over s' - t}+ {1 \over s' - u}\right). 
\ee
Moreover, consistency with unitarity, or ${\rm Re} f_0(\infty)=0$, in the Regge limit leads to  the following sum rule
\be
 \lambda - 3 \int_{4m^2}^\infty {d s' \over \pi } {\rho(s') \over s' - {4 m^2 \over 3}} = 0,
 \label{eq:sum-rule-rhos}
\ee
which, together with \eqref{eq:regged4sumrule}, guarantees that ${\rm Re} f_0(s) \sim {1 / \log s}$ at large energies. 
Note that \eqref{eq:sum-rule-rhos} immediately implies that $\lambda \geq 0$ thanks to unitarity $\rho(s) \geq 0$. Therefore, interactions in this model are attractive. It is interesting to notice that $\lambda$ is dispersive (expressible through the discontinuity of the amplitude), despite the fact that in the Regge limit $T(s,t) \sim \text{const}$.

Let us next discuss the sign of $\lambda$ and the behavior of the amplitude at fixed angles ${\cal T}(s, \theta) \equiv T \Big(s, -{s-4 m^2 \over 2}(1-\cos \theta) \Big)$. In $d=4$, $\lambda <0$ corresponds to theory which has the Landau pole in the UV. From the point of view of the unitarity iterations, it corresponds to the fact that $\rho(s)$ grows in the UV, which eventually destabilizes the convergence of iterations. On the other hand, for $\lambda >0$, which from the Lagrangian point of view corresponds to the potential which is unbounded from below, the theory is asymptotically free. This is the sign for which our algorithm converges and the Regge limit is the one observed above. Let us recall that the running of the coupling in the $\phi^4$ theory is given by
\be
\label{eq:running}
\lambda_p = {\lambda \over 1 + {3 \lambda \over 16 \pi^2} \log {p \over m}} \ ,
\ee
where we adopted our sign convention for $\lambda$ which differs from the one in the standard textbooks (see e.g. Peskin and Schroeder \cite{Peskin:1995ev}, Eq. (12.82)). Below we will see that \eqref{eq:running} correctly captures the qualitative behavior of the fixed angle scattering amplitude as a function of energy ${\cal T}(s, \theta)$. In particular ${\cal T}(s, \theta) \sim {32 \pi^2 \over 3} {1 \over \log {s \over m^2}}$ at high energies. The amplitude we obtain also exhibits the nontrivial Regge limit as $\theta\to0$, and transitions between Regge and fixed-angle at ${s \over m^2} \sim 1/\theta^2$. Finally, let us discuss the behavior at $\theta=0$. In this case we find that ${\cal T}(s, 0) \to {\rm const} \sim \lambda$ when $s \to \infty$.

\subsubsection{Numerical implementation: ``hotstart'' in the Regge limit}

\begin{figure}
    \centering
    \includegraphics{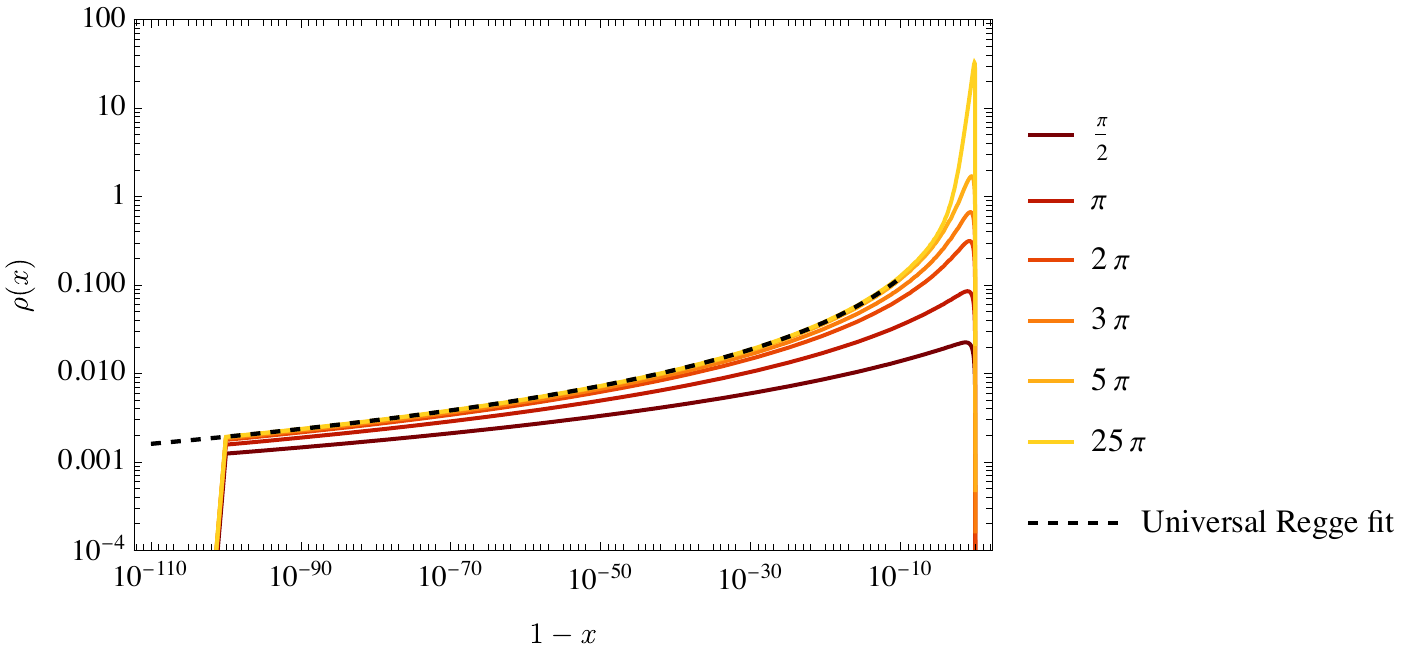}
    \caption{$\rho(s)$ for various couplings in $d=4$ (color on the right). The Regge limit ansatz is given in \eqref{eq:ansatz-rhos-4d} (dashed line). The only place where our one-parameter enters the ansatz is in front of the $\log(\log(s))$ term, which makes coupling dependence impossible to detect  at these scales. For $\lambda\gtrsim2\pi$, all the dotted curves are essentially on top of each other.}
    \label{fig:rhos-all}
\end{figure}

The numerical implementation of the iterations in $d=4$ goes through the same basic steps as in three dimensions
except that 
the two-particle threshold and Regge regions get interchanged. In terms of difficulty the threshold region is now trivial, with a square root fall-off, and the Regge region is non-trivial, with $1/(\log s)^2$ behavior leading order behavior, with fixed coefficient, given in \eqref{eq:ansatz-rhos-4d}. We show the results of our fixed-point iteration as a function of the coupling $\lambda$ in fig.~\ref{fig:rhos-all}.

In $d=3$, it was easy to work out the $\lambda$-dependence of the asymptotic behavior. In $d=4$, the question is more subtle, see the discussion around \eqref{eq:ansatz-rhos-4d}. In particular, a curious phenomenon arises, when solving asymptotically unitarity with a series expansion in $1/(\log s)^n$. The term $1/(\log s)^3$ is in the kernel of the RHS of unitarity equation, therefore is not canceled and would appear to produce unitarity violation. We have been able to find one correction that allows to generate a $1/(\log s)^3$ term which can then be fixed to remove the $1/(\log s)^3$ term in the LHS, which adds a $\log(\log(s))$ term in the numerator, $1/(\log s)^3\to(a+b\log(\log(s)))/(\log s)^3$. 
This means that the coupling-dependence of our solution is governed by the ${\log(\log(s)) \over (\log s)^3}$ term, and therefore we cannot test it unless we go to super-exponential grids, which is not feasible for our numerics. Nevertheless, assuming that $b_\infty\sim 1/\lambda$, as in 3d, a consistent picture emerges. At large lambda, all curves are on top of each other. At smaller lambda, curves start to differ, but the contribution of the ${\log(\log(s)) \over (\log s)^3}$ term is simply not detectable before very high scales.

\begin{figure}[!htb]
    \centering
    \includegraphics[scale=1]{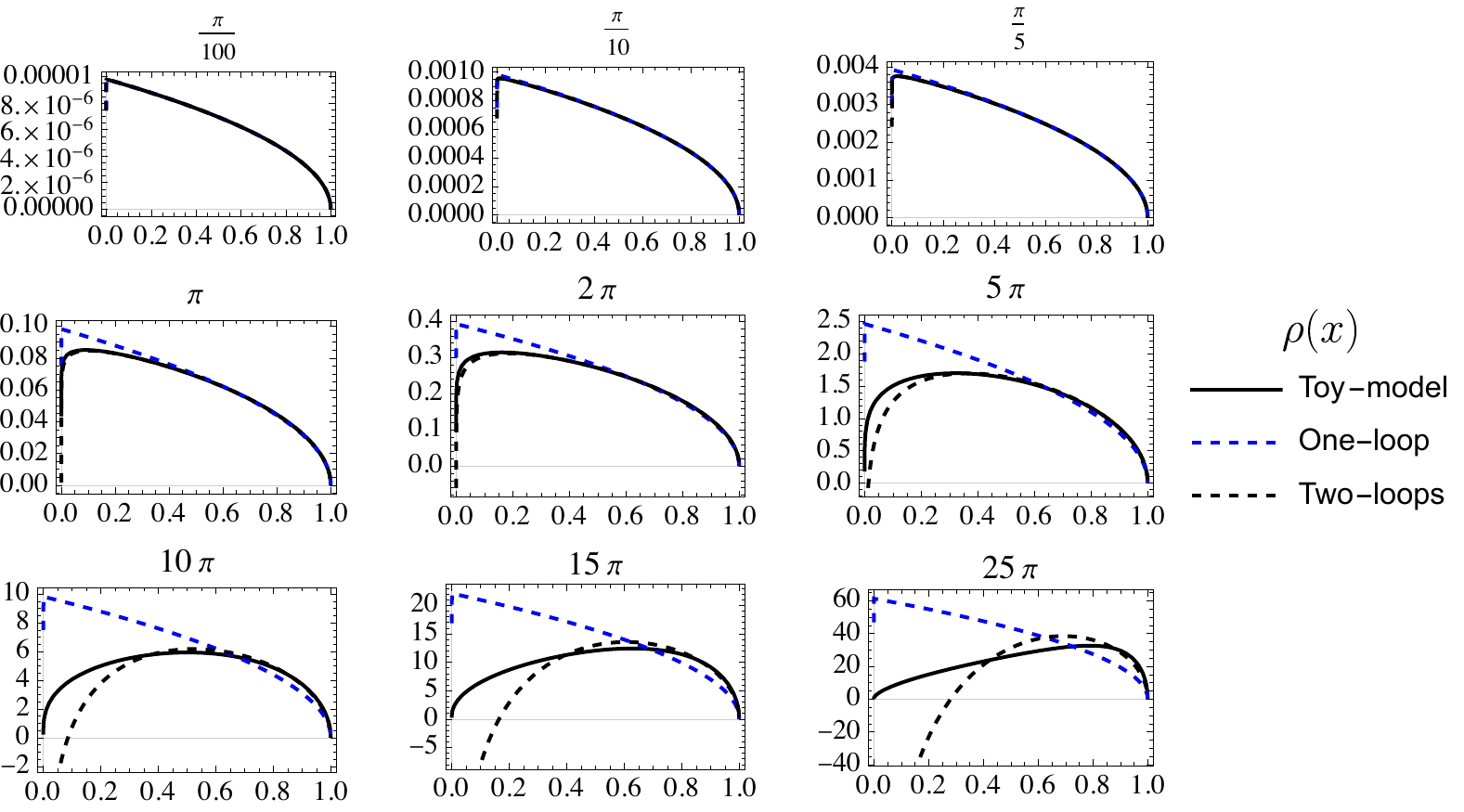}
    \caption{Comparison of the nonperturbative numerical solution for the toy-model in $d=4$ vs  perturbation theory: good at low energies, bad at high energies. $x$-axis is the variable $x=4m^2/s$, solid line is numerics and dashed is perturbation theory. This plot looks very similar to the one in $d=3$, except with Regge and two-particle threshold regions interchanged.%
    }
    \label{fig:4d-rho-2loops}
\end{figure}

\begin{figure}[htb!]
  \centering
  \includegraphics[scale=1.2]{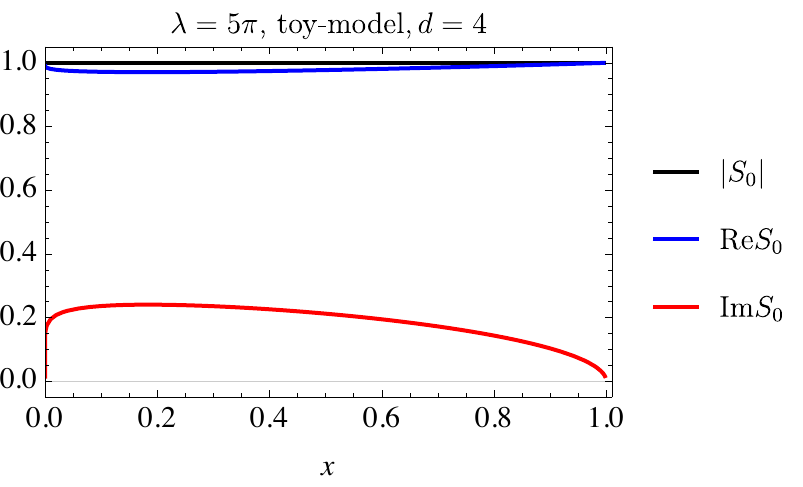}
  \caption{Spin zero partial wave $S_0$ in $d=4$ for the toy-model amplitude at $\lambda = 5 \pi$.}
  \label{fig:plS04dtoy5pi}
\end{figure}

\begin{figure}
    \centering
    \includegraphics{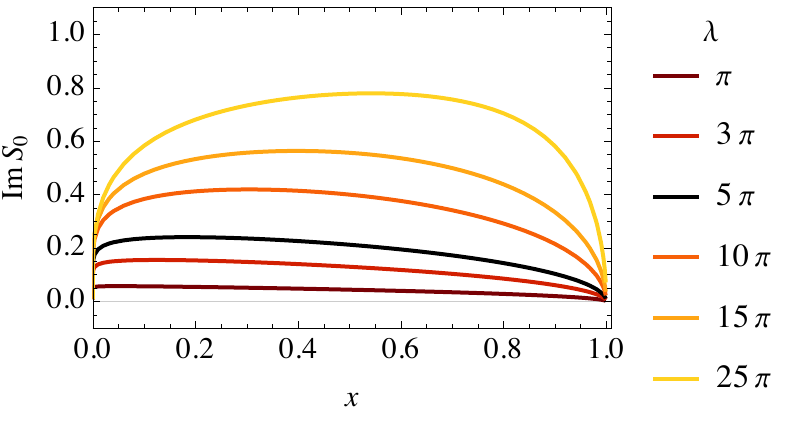}
    \includegraphics{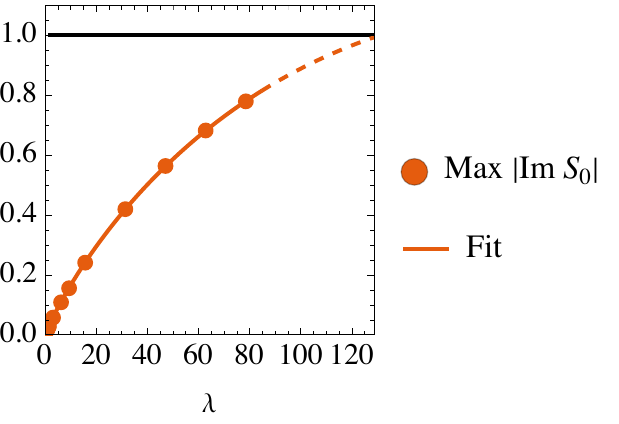}
    \caption{{\bf Left panel}: ${\rm Im} S_0$ for various $\lambda$ in $d=4$the. {\bf Right panel}: maximal value of ${\rm Im} S_0$ for various $\lambda$. Right: extrapolation of the maximum value of $\Im S_0$, which would seem to hit $1$ around $40\pi$. Consistently, our algorithm starts to struggle around $25-30\pi$.}
    \label{fig:plS04dtoy5pilambda}
\end{figure}

\begin{figure}[htb!]
  \centering
  \includegraphics[scale=1.]{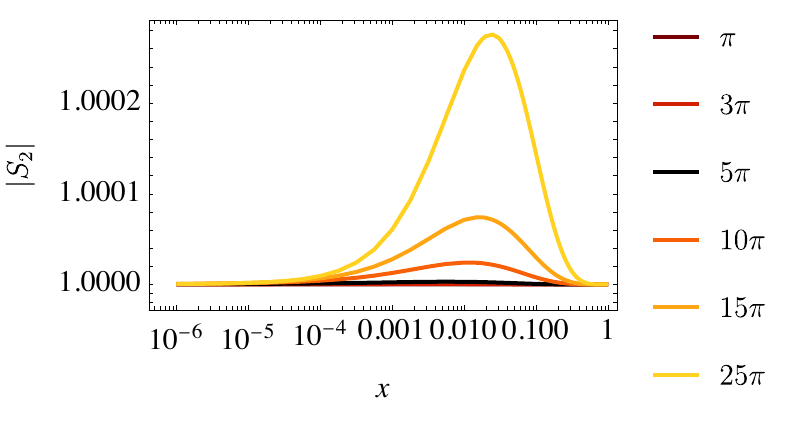}
  \caption{Violation of unitarity in the spin two partial wave $S_2$ in the $d=4$ toy model. Recall that unitarity implies that $|S_2|\leq 1$.}
  \label{fig:unit-viol-S2-4d}
\end{figure}

\begin{figure}[htb!]
  \centering
\includegraphics{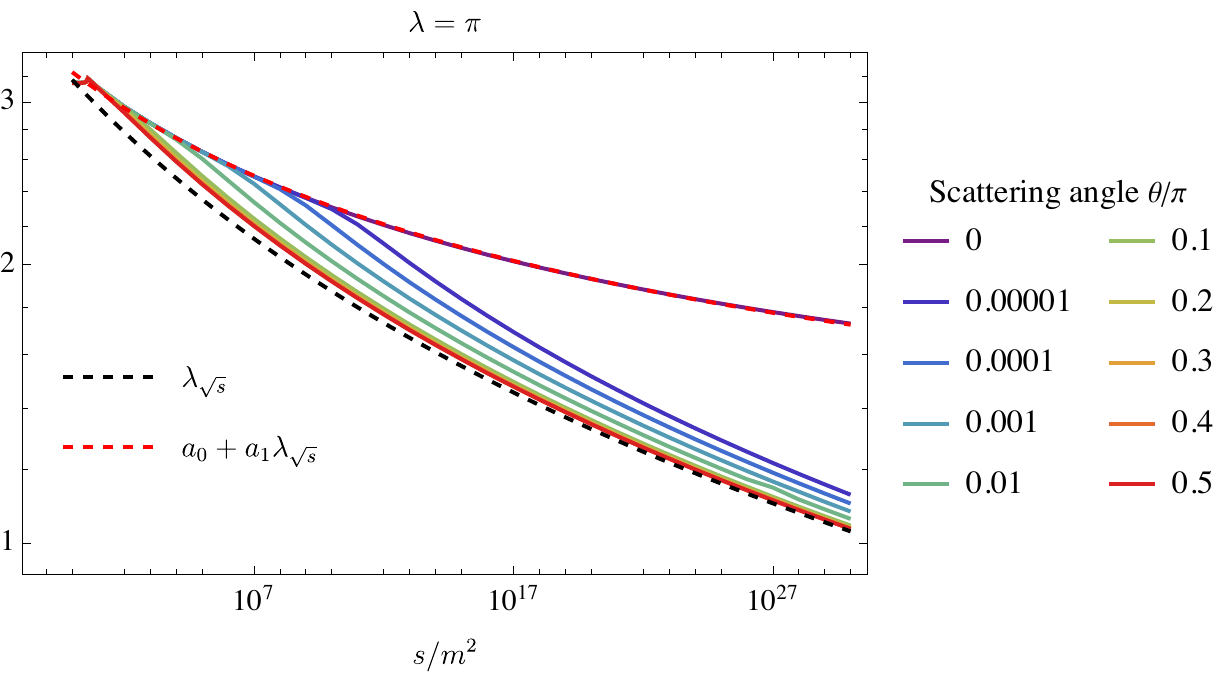}
  \caption{$| {\cal T}(s,\theta)|$ as a function of the scattering angle $\theta$ in $d=4$. We see that the amplitude decays as ${1 \over \log s}$ and transitions between two asymptotes: the Regge asymptote (red dashed curve) and the fixed-angle asymptote (black dashed curve), given by the beta function, and defined in \eqref{eq:running}. The transition occurs at energies ${s \over m^2} \sim \theta^{-2}$ (or impact parameters $b \sim {1 \over m}$, see footnote \ref{footnote_b_def}). We have not tried to compute $a_0$ and $a_1$ and obtained them by fitting.}
  \label{fig:allangles-rhos-4d}
\end{figure}

We present our results according to the following list:
\begin{itemize}
    \item In \figref{fig:rhos-all}, we show various $\rho(s)$ functions for different couplings, up to $\lambda=25\pi$.
    \item In \figref{fig:4d-rho-2loops}, we compare these $\rho(s)$ to perturbation theory up to one and two loops.
    \item In \figref{fig:plS04dtoy5pi}, we display the S-wave of the amplitude at $\lambda=5\pi$. We choose this intermediate coupling because it coincides with the maximum coupling we can reach with the full algorithm later. In addition, the back-reaction created by $\rho(s,t)$ is very small in $S_0$, thus the reader is invited to think of this partial wave as that of the amplitude coming from the full amplitude.
    \item In \figref{fig:plS04dtoy5pilambda}, we display the growth of $\Im S_0$ with $\lambda$, to illustrate that our algorithm reaches the end of its convergence range when $\Im S_0 \sim 1$. Since amplitudes with larger couplings are generated by other methods, \cite{Chen:2022nym,EliasMiro:2022xaa}, this means that our algorithm needs to be adapted in order to capture these amplitudes.
    \item In \figref{fig:unit-viol-S2-4d}, we display the unitarity violation observed in $S_2$. Again,in the toy-model we do not impose $|S_J|\leq1$ for $J>0$ so this is expected.
    \item In \figref{fig:allangles-rhos-4d}, we plot the toy-model resulting scattering amplitudes. At low energies, we simply have $T(s,t)\sim \lambda$. At high energies, at fixed angles ($s,t\to\infty$, $s/t$ fixed), our amplitude asymptotes to the effective running coupling $\lambda_p$, given in \eqref{eq:running} in the UV. In the Regge limit, ($s\to\infty$, $t$ fixed), the amplitude goes to a constant, which satisfies the sum rule given in \eqref{eq:sum-rule-rhos}. Increasing the scattering angle $\theta$ from 0 (Regge) to $O(1)$, we observe numerically that the amplitude remains of the Regge-type up to scattering energies $s$ of order $1/\theta^2$, and then transitions to the fixed-angle regime. In the impact parameter space this transition is associated to $b \sim m$.
\end{itemize}


\FloatBarrier

\subsection{Proof of existence of the amplitudes in $3<d<4$}

In the discussion above, we have constructed the desired amplitudes numerically. A natural question arises: what happens in the continuum limit? Our explorations of the grid support the idea that nothing dramatic happens and the amplitudes we have constructed exist in the space of continuous functions. This expectation can be made rigorous in $4>d>3$ following the argument of Atkinson \cite{Atkinson:1970zza}. The idea is to argue that iterations of unitarity converges to a unique fixed point in the space of H\"older continuous functions. By doing so, we will see that going to $d=3$ and $d=4$ requires an extension of the proof, which we currently do not have. 

Let us summarize the main steps of the proof leaving the technical details to \appref{app:atkinsonproofd34}.
We introduce $s=4 m^2/x$ so that $x \in [0,1]$ and consider a class of $\rho(x)$ such that
\be
\label{eq:bcrhos}
\rho(0) = \rho(1) = 0 .
\ee
We also consider $\rho(x)$ to be H\"older-continuous
\be
\label{eq:holderbound}
| \rho(x_1) -  \rho(x_2) | \leq \xi |x_1 - x_2|^{\mu}, ~~~ 0 < \mu < \min({{d-3 \over 2}}, {4-d \over 2}) ,
\ee
where $\mu < {d-3 \over 2}$ will arise from imposing unitarity close to the two-particle threshold, whereas $\mu < {4-d \over 2}$ comes from unitarity in the Regge limit. We introduce the following norm in this functional space
\be
|| \rho || = {\rm sup}_{0 \leq x_1 , x_2 \leq 1} {|  \rho(x_1) - \rho(x_2) | \over | x_1 - x_2|^{\mu}}. 
\ee
One can check that such functions form a complete, normed, linear space, i.e. the Banach space, see \cite{Atkinson:1970zza}.
We use S-wave unitarity to write down an iteration procedure
\be
\label{eq:unitarityAtkinson}
  \rho' = \Phi [ \rho] ,
\ee
see \eqref{eq:iterationtoyddim} for the explicit form.

\begin{figure}[thb!]
    \centering
    \includegraphics[scale=0.6]{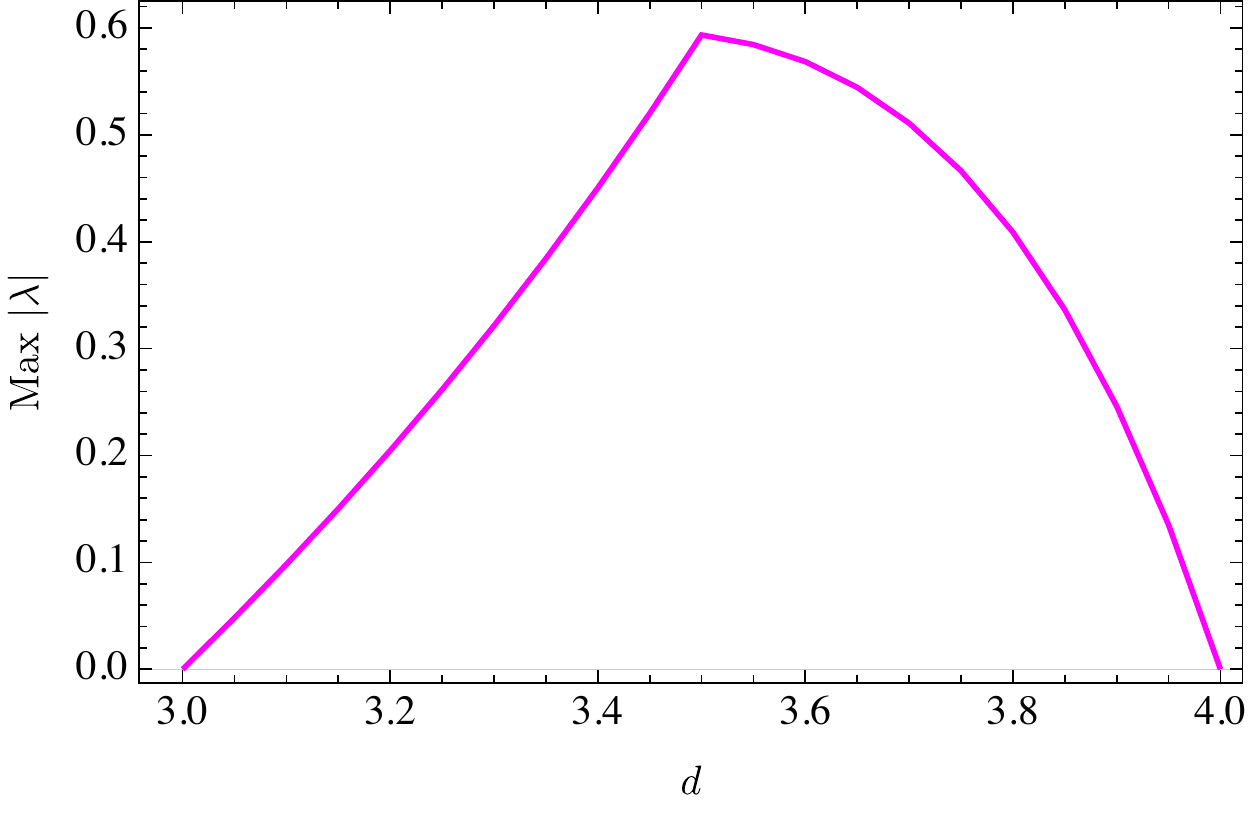}
    \caption{The maximal absolute value of the coupling $|\lambda|$ for which we can proof convergence of the iterations as a function of the number of spacetime dimensions $d$.}
    \label{fig:convergatk}
\end{figure}

Given a bound on the norm $||\rho|| \leq B$ we would like to show that $|| \rho' || \leq B'$ and impose $B' \leq B$ to make sure that the iterated spectral density stays in the same space of functions. We then would like to show that the map is actually contracting meaning that
\be
\label{eq:contractionindex}
|| \rho'_2 - \rho_1' || \leq k || \rho_2 - \rho_1 ||, ~~~ k<1.
\ee
Given \eqref{eq:contractionindex}, the fixed point of the unitarity equation \eqref{eq:unitarityAtkinson} $\rho_* = \Phi[\rho_*]$ is unique and iterations of unitarity converge exponentially fast
\be
||\rho_{*}-\rho_N||\leq 2 B {k^N \over 1-k}.
\ee
In appendix \ref{app:atkinsonproofd34} we derive $B'$ and $k$ explicitly in terms of $(\lambda,\mu, d, B)$. 
Here we simply plot the maximal value of $|\lambda|$ as a function of $d$ for which we can prove that the map to be contracting, see \figref{fig:convergatk}.

The characteristic feature of this plot is that the range of $\lambda$ for which the convergence is proven shrinks to zero as $d \to 3$ or $d \to 4$. How is it consistent with our explicit results described above? Intuitively, the reason is clear. The amplitudes there involved large amount of cancelations between various terms in the amplitude, see \eqref{eq:sumrule3dlambda} and \eqref{eq:sum-rule-rhos}. Such cancelations are not taken into account in the proof in \appref{app:atkinsonproofd34}, or in the original Atkinson proofs, where we bound each term separately.

Similarly, we observed both in $d=2$, see \cite{Tourkine:2021fqh}, and throughout the whole parameter space explored in this work that the iterations perform better than what is expected from the proofs.

There is another way to understand how we managed to find the fixed point in $d=3$ and $d=4$. Effectively, what our algorithm does is it modifies the unitarity equation (due to the cutoff at some maximal or minimal energy) as follows. 
Let us for concreteness discuss the case of $d=4$.\footnote{The case of $d=3$ is essentially identical, where the Regge limit is replaced by the near two-particle threshold region.} The basic idea is to modify the unitarity equation 
\be
 \rho'(x) &= {h(x) \over 32 \pi} (1-x)^{{1 \over 2}} \left( ( 16 \pi {\rm Re} f_0[\rho](x) )^2 +  \rho(x)^2 \right) ,
\ee
where we can for example choose (the precise form is not essential, as long as $h(0)=0$)
\be
h(x) = \theta(x-x_0) +{x \over x_0} \theta(x_0-x) .
\ee
The role of $h(x)$ is to effectively soften the behavior of $\rho'(x)$ in the Regge limit $x \to 0$. With this modification we can consider a class of Holder-continuous functions with $\mu = 1/2$ and run the argument of the previous section. All the bounds will now depend on $x_0$.

Physically, having such a cut-off is not desirable. Indeed, in a physical theory in the presence of S-wave particle production we have
\be
\rho'(x) &= {1 \over 32 \pi} (1-x)^{{1 \over 2}} \left( ( 16 \pi {\rm Re} f_0[\rho](x) )^2 +  \rho(x)^2 \right) + \rho_{\text{MP}}(x), ~~~ \rho_{\text{MP}}(x) \geq 0.
\ee
It is therefore curious to see how we effectively solved this problem in the numerical implementation of the algorithm. 

Effectively, when adopting the hotstart for a family of grids, we considered \emph{iteration-dependent} modification of the unitarity equation
\be
h_n(x) = \theta(x-x_{n}) +{x \over x_n} \theta(x_n-x) ,
\ee
where we chose $x_n$ such that $\lim_{n \to \infty} x_n = 0$. We observed convergence and smooth extrapolation of the solutions to the UV as we remove the cutoff $x_n$, where we could smoothly match to the analytic UV complete solution of the model.

\subsection{$d>4$: nonrenormalizability and unitarity}

Above we discussed the cases $d=3,4$ and $3<d<4$. 
Here we briefly comment what happens if we try to extend the same ideas to $d>4$, and we also make contact with the notion of renormalizability.%

The main difference comes from the form of the unitarity equation that takes the form
\be
2 {\rm Im}f_0(s) = {(s-4m^2)^{{d-3 \over 2}} \over \sqrt{s}} | f_0(s) |^2 , ~~~ s \geq 4 m^2 .
\ee
Compared to $d\leq4$, the phase space pre-factor ${(s-4m^2)^{{d-3 \over 2}} \over \sqrt{s}}$ now grows in the Regge limit as $s^{d-4 \over 2}$. This leads to the fact iterations of unitarity with the starting point $T^{(0)}(s,t) = \lambda$ lead to polynomially growing amplitudes with the power growing with the number of iterations.\footnote{At the first step, we get $f_0^{(1)}\sim s^{(d-4)/2}$, which then gives $f_0^{(2)} \sim s^{(d-4)/2} s^{d-4}$, and so on.} In particular, the dispersion relations with a finite number of subtractions will eventually break down. Of course, this is nothing but the fact that ${\lambda \over 4!} \phi^4$ is nonrenormalizable in $d>4$, and here we just see it as a simple consequence of unitarity. Therefore, simple iterations of unitarity starting from $T^{(0)}(s,t) = \lambda$ will not work in this case.

However, as hopefully became clear from the discussion of the $d=4$ case, this is not a problem per se. Indeed, even in $d=4$ to get a convergent amplitude we effectively had to ``UV complete'' the amplitude by supplementing it with the correct nonperturbative Regge limit (namely that $\rho(s) \to 0$), which then made the iterations convergent upon increasing the UV cutoff. One can imagine using the same strategy in $d>4$. For example, we can try to ``eikonalize'' the leading order amplitude, which should also be relevant for applying the methods discussed here to gravity. 
We do not explore the case of $\phi^4$ in  $d>4$, or other nonrenormalizable theories such as gravity here, and leave this interesting problem for the future.

Finally, note that a related problem occurs in 2d \cite{Tourkine:2021fqh}, where to describe a certain class of CDD amplitudes (with unequal number of zeros and poles), we had to specify boundary conditions, and use the Newton-Raphson method to force the iteration to preserve them.

\section{Quasi-elastic amplitudes in $d=3$}
\label{sec:QE3d}

In this section, we present the results of the numerical implementation of the iteration algorithm presented in \secref{sec:algorithm} in $d=3$ spacetime dimensions. The key difference compared to the previous section is that here $\rho(s,t) \neq 0$, and that we can now implement unitarity for partial waves with $J>0$.

We start with a presentation of the amplitudes with a reasonably large coupling, $\lambda=-3\pi$, which is close to the boundary of the range of convergence of our algorithm. Then, we move to describe the coupling dependence of certain elements of the amplitude, partial waves, etc. We also distinguish two iteration schemes for $\Big(\lambda, \eta_{\text{MP}}(s), \rho_{\text{MP}}(s,t) \Big)$ introduced earlier: 2QE given by $(\lambda, 0,0)$; 2PR given by $(\lambda, \eqref{eq:analyticityinspinJ0},0)$. We start the section with the 2QE case.

\subsection{2QE amplitude for $\lambda = -3 \pi$}

Below we present results for the 2QE amplitude in $d=3$ and $\lambda = - 3 \pi$. Recall that this algorithm is characterized by the fact that the $J=0$ partial wave is purely elastic at all energies.

\begin{figure}[t]
    \centering
    \includegraphics[scale=0.6]{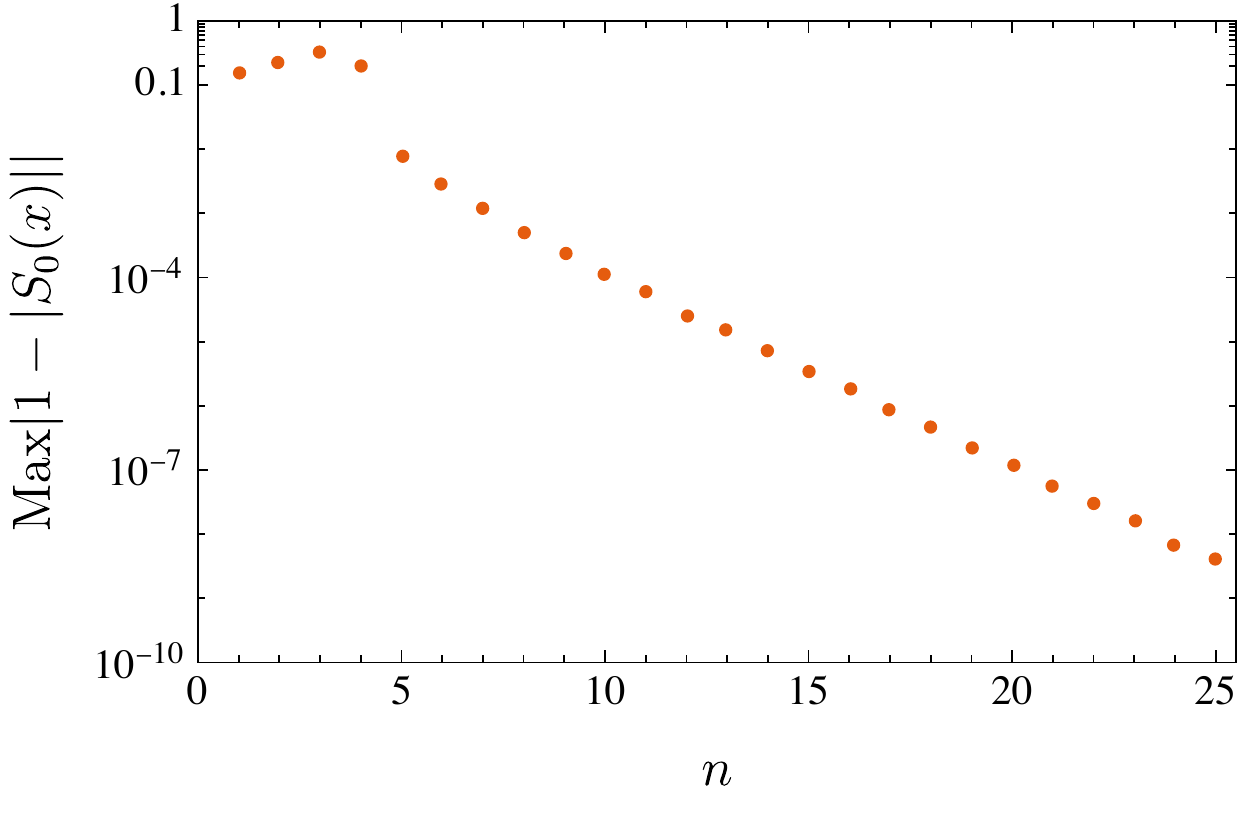}
    \caption{Maximal deviation of the absolute value of $J=0$ partial wave from $1$ as a function of number of iterations $n$ for $\lambda=-3\pi$ in $d=3$. We observe linear convergence, see \eqref{eq:convergencerate}.}
    \label{fig:3piconvergence}
\end{figure}

To illustrate the convergence of the algorithm, we first plot the maximal deviation of the absolute value of the $J=0$ partial wave from $1$ in \figref{fig:3piconvergence}. We observe linear convergence and reach the precision $\sim 10^{-9}$ after $\sim 25$ iterations. For smaller couplings, the convergence is much faster and usually only about 5 iterations are needed.

\subsubsection{Single and double spectral functions}

\begin{figure}[htb!]
  \centering
  \includegraphics{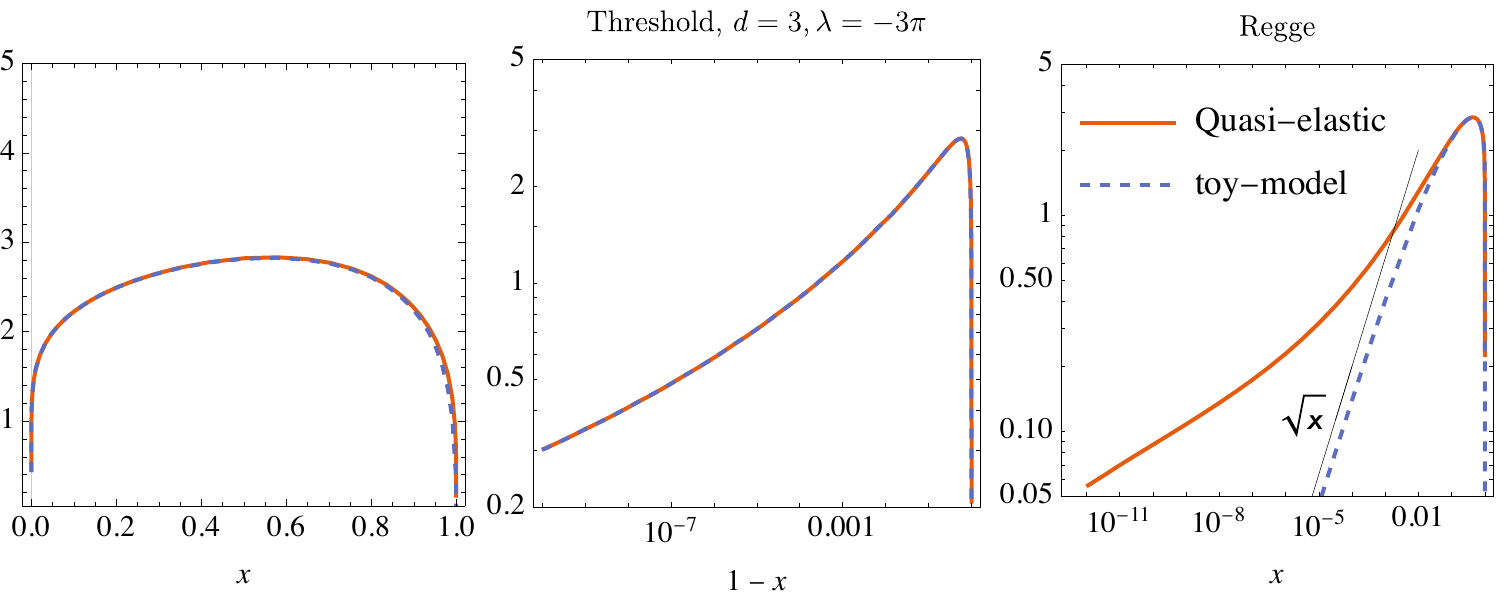}
    \caption{{\bf Left panel:} Single discontinuity $\rho(x)$ as a function of energy $x$. The results for $\rho(s,t) \neq 0$ obtained in this section are depicted in red, whereas the toy-model results of the previous section are shown in blue (dashed). {\bf Middle panel:} the two-particle threshold region $x\sim1$, we do not observed any visible difference compared to the toy-model result. {\bf Right panel:} the Regge region $x\sim0$. Here we see that $\rho(s)$ is significantly corrected compared to the toy model.}
    \label{fig:3pi3dsingledisc}
\end{figure}

Next we present various features of the resulting amplitude. The single spectral function $\rho(s)$ is depicted in \figref{fig:3pi3dsingledisc}. 
It exhibits logarithmic behavior close to the two-particle threshold $s \to 4 m^2$, and decays in the Regge limit $s\to \infty$ in a way which is distinctively different from $\sqrt{x}$, as observed from results of \secref{sec:3dnp}. 

\begin{figure}[h!]
    \centering
     \begin{subfigure}{0.5\textwidth}
    \centering
        \includegraphics[scale=0.55]{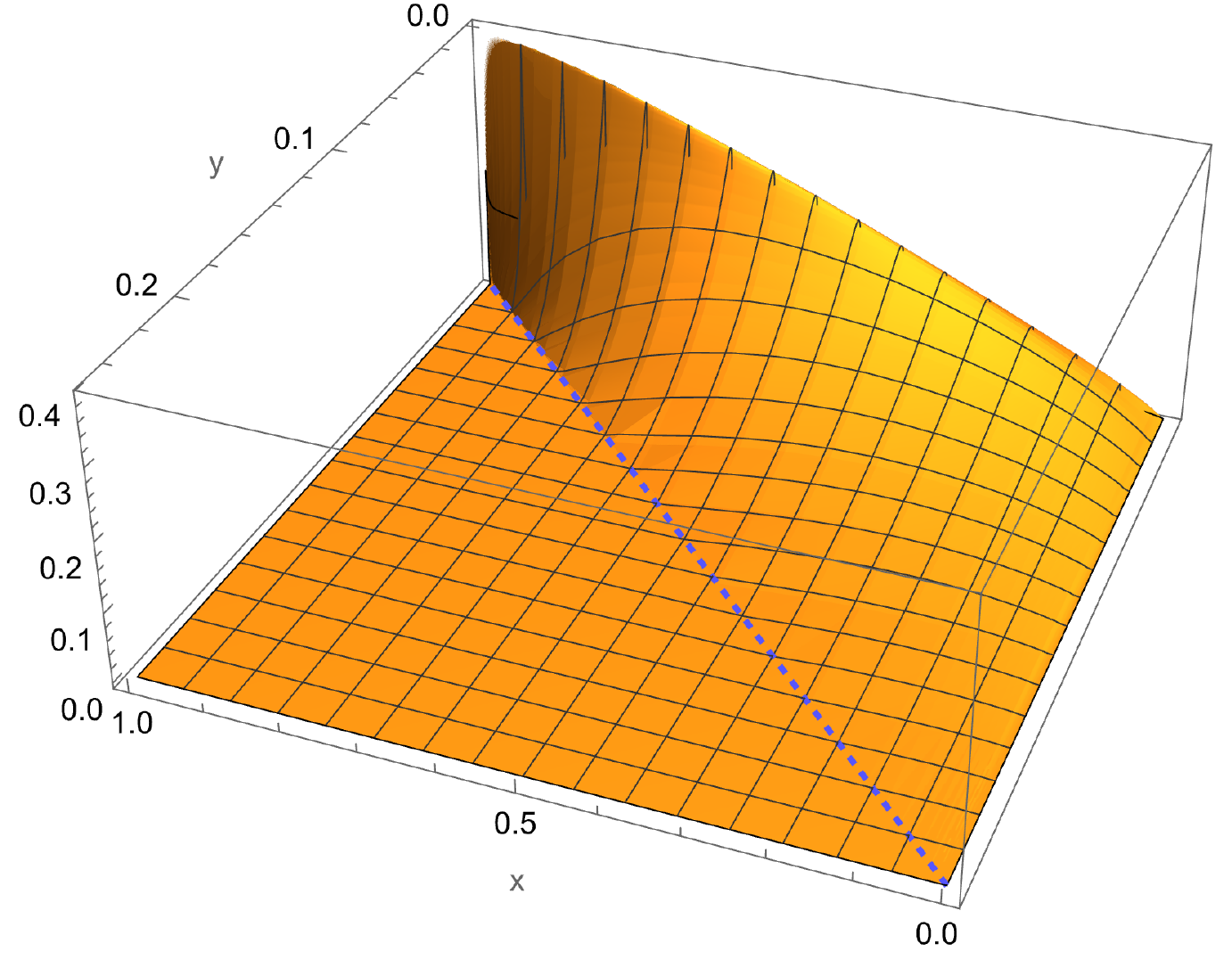}
  \end{subfigure}%
  \begin{subfigure}{0.5\textwidth}
    \centering
     \includegraphics[scale=0.45]{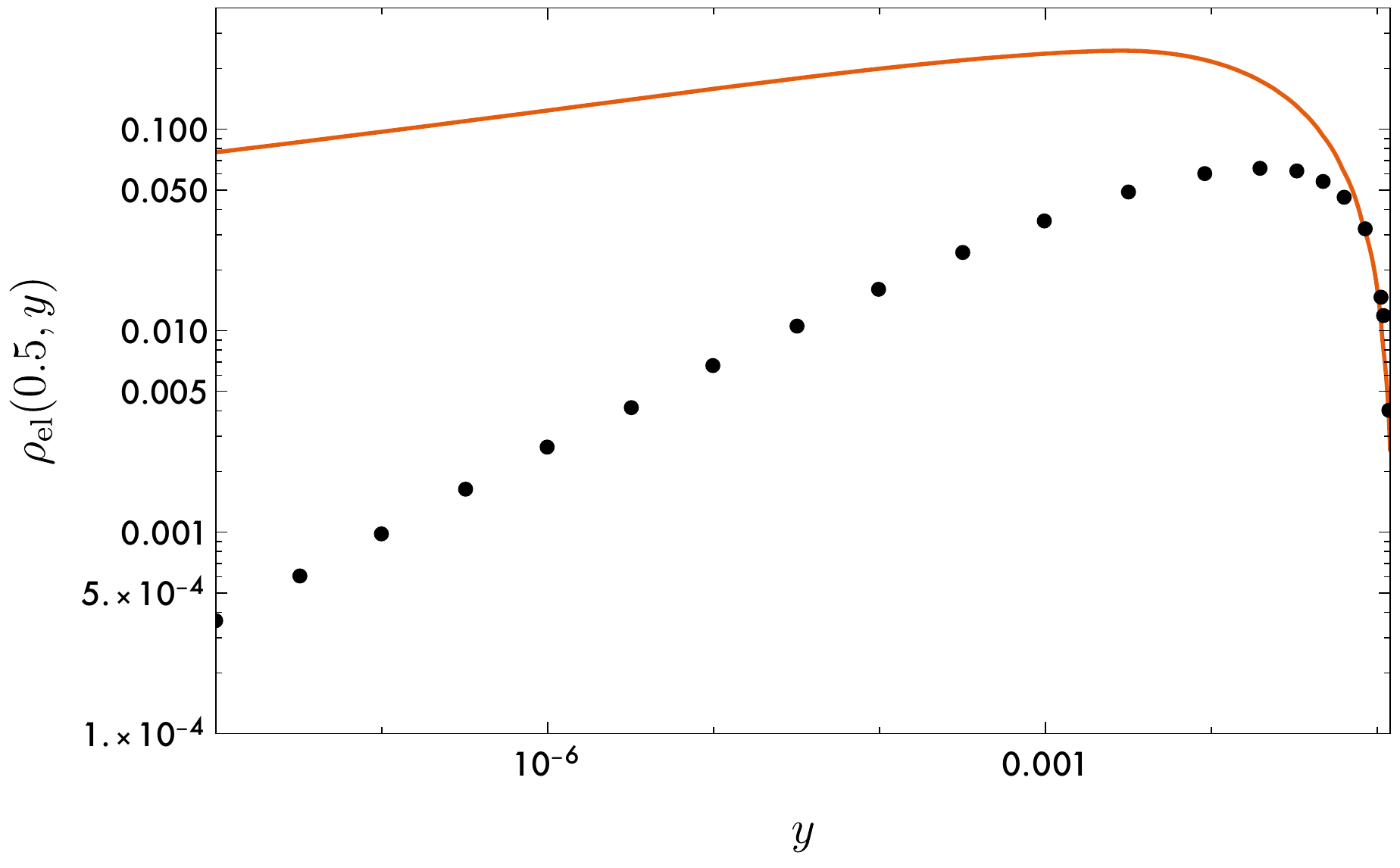}
  \end{subfigure}
    \caption{Double discontinuity $\rho_{\text{el}}(x,y)$ as a function of energies in $d=3$, $\lambda=-3 \pi$. {\bf Left panel:} the blue line represents the leading Landau curve to the left of which $\rho_{\text{el}}(x,y)=0$. {\bf Right panel:} we take the fixed energy slice $\rho_{\text{el}}(0.5,y)$ to exhibit a nontrivial emergent Regge limit as $y \to 0$. By the dashed line we exhibit the double discontinuity coming from the three-loop Aks graph,  see \figref{fig:threeloopgraphsforswaveinelasticity}.}
    \label{fig:3pidoubledisc}
\end{figure}

The double spectral density $\rho_{\text{el}}(s,t)$ is depicted in  \figref{fig:3pidoubledisc}. 
It peaks close to $x=1$, and decays in the Regge limit $x\to 0$. Recall that $\rho(x,y) = \rho_{\text{el}}(x,y)+\rho_{\text{el}}(y,x)$, where $\rho_{\text{el}}(x,y)$ satisfies the Mandelstam equation \eqref{eq:mandelstam} and has nonzero support only above the leading Landau curve, namely $y \leq {1-x \over 4}$.

In contrast to the toy-model consideration, we do not have an analytic solution to the Regge limit of the $\rho(x)$ and $\rho(x,y)$ for the 2QE amplitudes. Due to the Gribov's theorem \cite{Gribov:1961fm}, it cannot be a simple power, e.g. $x^{1/2}$. Indeed, this is what we observe in the right panel of \figref{fig:3pi3dsingledisc}.
It would be very interesting to try to solve the Regge limit of the 2QE model analytically.

\subsubsection{Partial waves, impact parameter.}

Next we analyze the partial waves of the amplitude. 
We find that the the spin-zero partial wave, the S-wave, is essentially identical to what
we got in the toy model, see \figref{fig:plS03dtoy}. In particular, the S-wave scattering decays both in the Regge limit $x\to 0$ and near the two-particle threshold $x \to 1$.

\begin{figure}[h]
    \centering
    \includegraphics[scale=1.0]{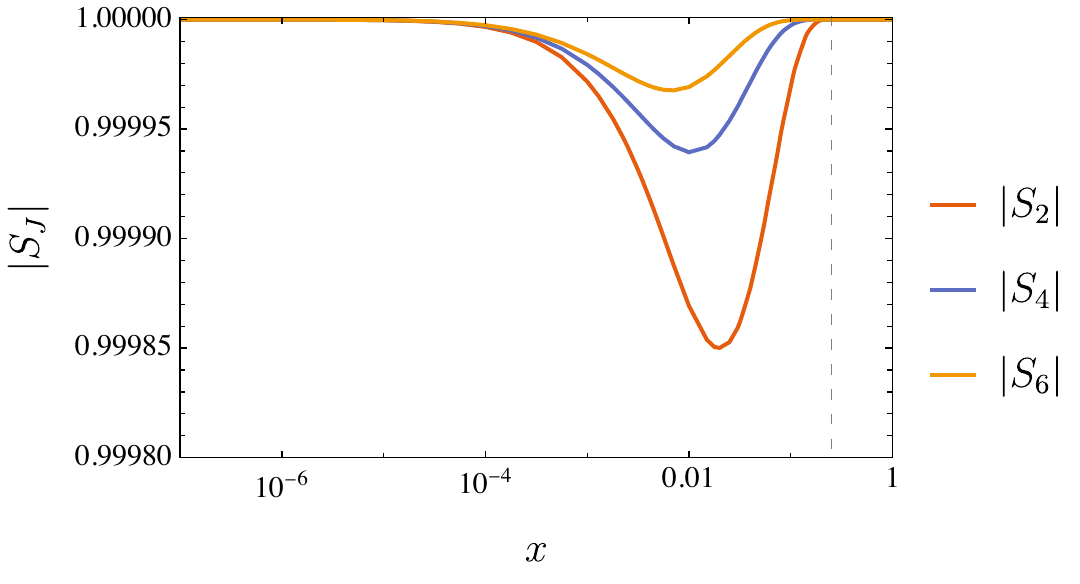}
    \caption{The absolute value of $|S_J|$ as a function of energy $x$ in $d=3$, $\lambda=-3 \pi$. We see that the amount of scattering in the higher spin partial waves quickly decreases as a function of spin $J$. Note that there is a dynamically emergent scale in the problem where each partial wave peaks at $x \sim 0.01$, which corresponds to ${s \over 4 m^2} \sim 10^2$. The dashed line is at $x = {1 \over 4}$, or, equivalently, $s=16 m^2$, which separates the elastic region $x \geq {1 \over 4}$ from the inelastic one $x < {1 \over 4}$. }
    \label{fig:3piSJ}
\end{figure}

In contrast the structure of the higher spin partial wave is quite different. 
We plot the absolute value of the higher spin partial waves $J>0$ in \figref{fig:3piSJ}. We see that this time they obey unitarity. Moreover, the departure from $1$ is the signal of particle production. The appearance of particle production is expected from the Aks theorem \cite{Aks:1965qga}, but here we see the quantitative amount of particle production necessary for elastic unitarity and crossing. As expected, the amplitude is \emph{quasi-elastic}, in other words, the amount of particle production is quite small. 
In the physical $\phi^4$ theory we expect particle production to be bigger, since in that case $\eta_{\text{MP}}(s)$ and $\rho_{\text{MP}}(s,t)$ are non-zero, essentially, due to multi-particle unitarity. We shall see below, when we analyze the coupling dependence, the magnitude of this production is of order $\lambda^4$ at small lambda, and deviates from it at larger couplings.

\begin{figure}[h]
    \centering
    \includegraphics[scale=1]{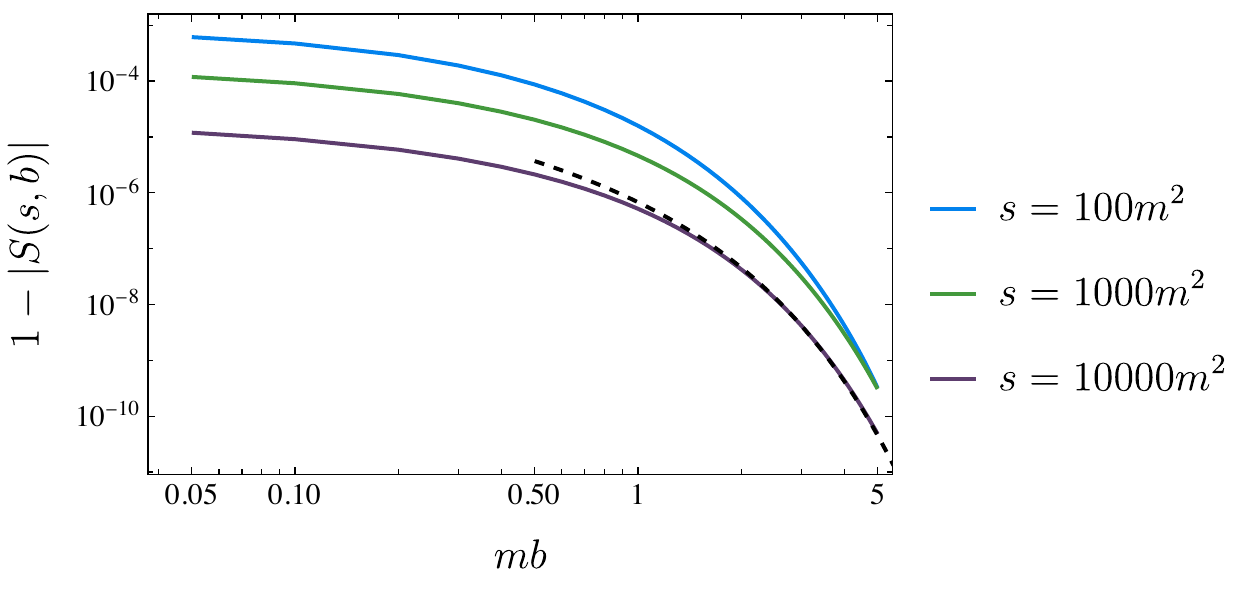}
    \caption{The absolute value of $1-|S(s,b)|$ as a function of impact parameter $b$ in $d=3$, $\lambda=-3 \pi$. As expected interactions quickly decay as a function of impact parameter, in a Yukawa-like, exponential fashion, see \appref{app:impactFG}. Here the dashed line stands for the fit $\sim{e^{- 2 b m} \over b m}$.}
    \label{fig:3piImpact}
\end{figure}

We can also consider scattering at fixed impact parameters $b \equiv {2 J \over \sqrt{s-4m^2}}$ where we recall that $p=\sqrt{s-4m^2}$ is the spatial momentum. As expected, the amplitude decreases exponentially fast as a function of impact parameters, see ~\figref{fig:3piImpact}.

\begin{figure}
    \centering
    \includegraphics[scale=1.1]{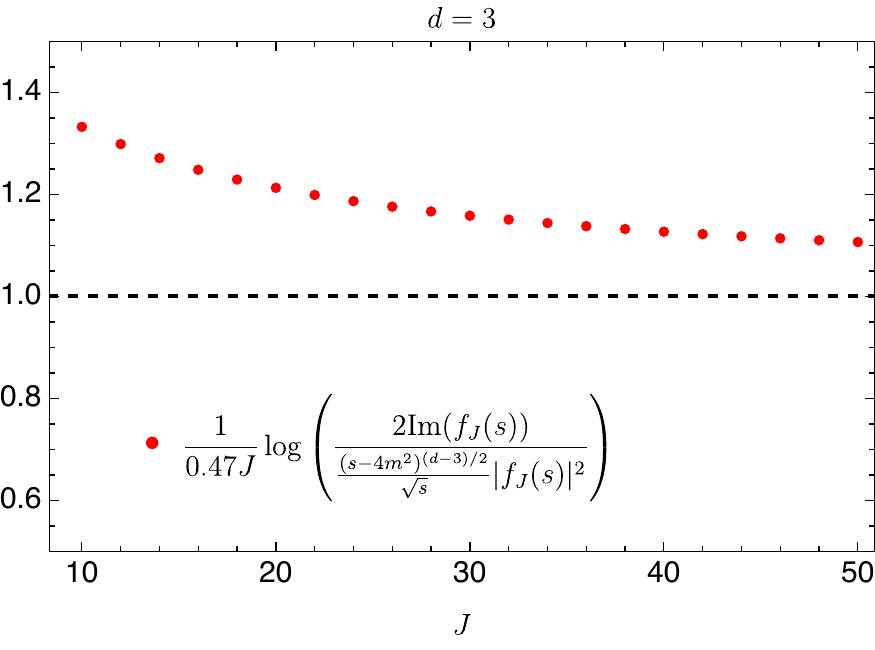}
    \caption{Higher-$J$ inelasticity in $d=3$ as a function of spin $J$. Elastic scattering corresponds to $0$ on this plot, meaning that at large spin scattering is weak but mostly inelastic as it goes to $1$. This property of partial waves is a direct consequence of elastic unitarity and crossing, see \cite{Correia:2020xtr} for the derivation of this fact.}
    \label{fig:higher-spin-3d}
\end{figure}

Finally, we can analyze unitarity in the partial waves at higher spins. In agreement with general arguments \cite{dragt1967amount,Correia:2020xtr} it becomes mostly inelastic, see \figref{fig:higher-spin-3d}.

\subsubsection{Amplitude} Finally, we can look at the amplitude itself. We plug our expressions for the single and double spectral function into the Mandelstam representation, and compute the resulting amplitude numerically. 

Since the double spectral function is numerically small compared to $\rho(s)$, by eye, the amplitude look indistinguishable from that obtain in the toy-model, so we do not represent them again here and refer the reader to \figref{fig:amp-toy-model-3d}. In particular, it also asymptotes the running coupling at high energies for non zero angles.

\begin{figure}[h]
    \centering
    \includegraphics[scale=0.8]{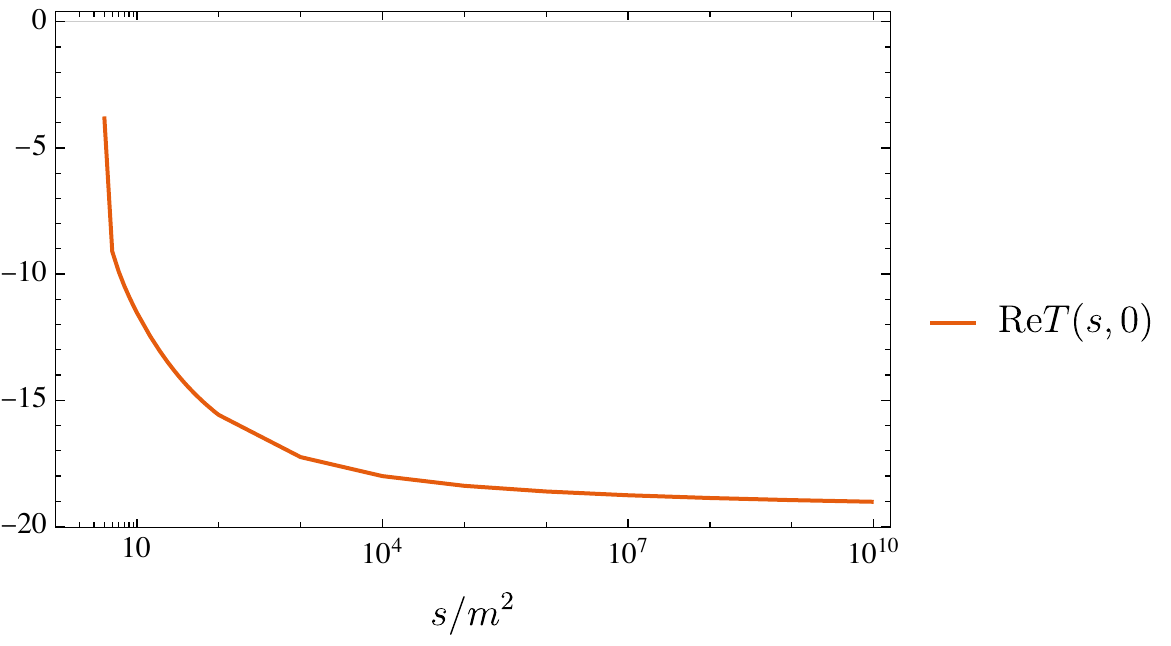}
    \caption{We plot the real part of the forward amplitude $T(s,0)$ for $s \geq 4 m^2$ in $d=3$, $\lambda=-3 \pi$. We see that it monotonically decreasing and it approaches a constant at infinite $s$.}
    \label{fig:3piforwardReal}
\end{figure}

\begin{figure}[h]
    \centering
    \includegraphics[scale=0.8]{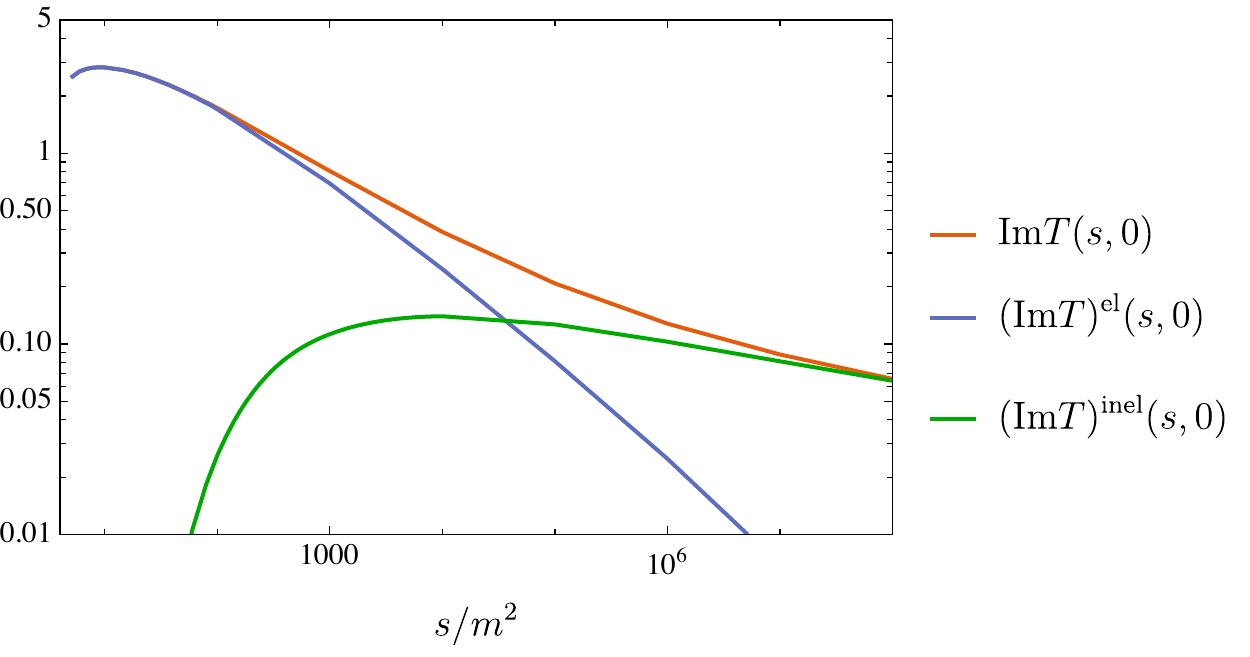}
    \caption{We plot the imaginary part of the forward amplitude $T(s,0)$ in $d=3$, $\lambda=-3 \pi$. Via the optical theorem it is trivially related to the scattering cross section. We plot three different cross sections: total, elastic, inelastic. Inelastic part of scattering in this model comes fully from the crossed terms $\rho_{\text{el}}(y,x)$ in the double spectral density. We see that at high energies scattering becomes more and more inelastic.}
    \label{fig:3piforwardIm}
\end{figure}

Next we plot the forward scattering by setting $t=0$. Via the optical theorem, this is related to the cross-section. For the real part we find \figref{fig:3piforwardReal}, and for the imaginary part \figref{fig:3piforwardIm}. Interestingly, we see that at high energies scattering becomes \emph{mostly inelastic} albeit weak.

\subsection{2PR amplitude}

\begin{figure}[h]
    \centering
    \includegraphics[scale=1.1]{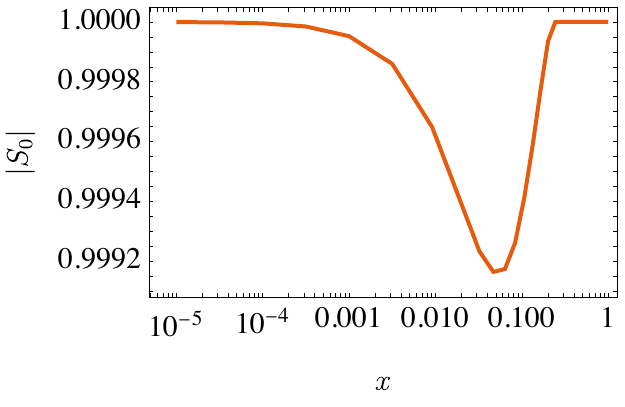}
    \includegraphics[scale=1.1]{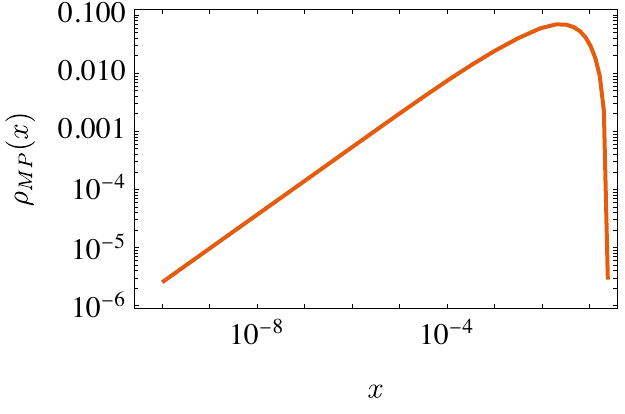}
    \caption{The spin zero partial wave in the 2PR scheme, $d=3$, $\lambda=-3\pi$. {\bf Left panel:} the absolute value which is strictly below $1$ signifies particle production. {\bf Right panel:} The multi-particle spectral density $\rho_{\text{MP}}(s)$ as given by \eqref{eq:2prdefinition} at the fixed point of the iteration algorithm.}
    \label{fig:3pi2pr3dS0}
\end{figure}

Next consider the 2PR iteration scheme. In practice, it means that instead of setting $\eta_{\text{MP}}(s)$ to zero, we iterate it using \eqref{eq:analyticityinspinJ0}. 
Essentially, the only visible difference for this scheme is that spin zero scattering is not purely elastic anymore, see \figref{fig:3pi2pr3dS0}. We also plot the corresponding ``multi-particle'' spectral density.

\subsection{Coupling dependence}

\begin{figure}
    \centering
    \includegraphics[scale=1]{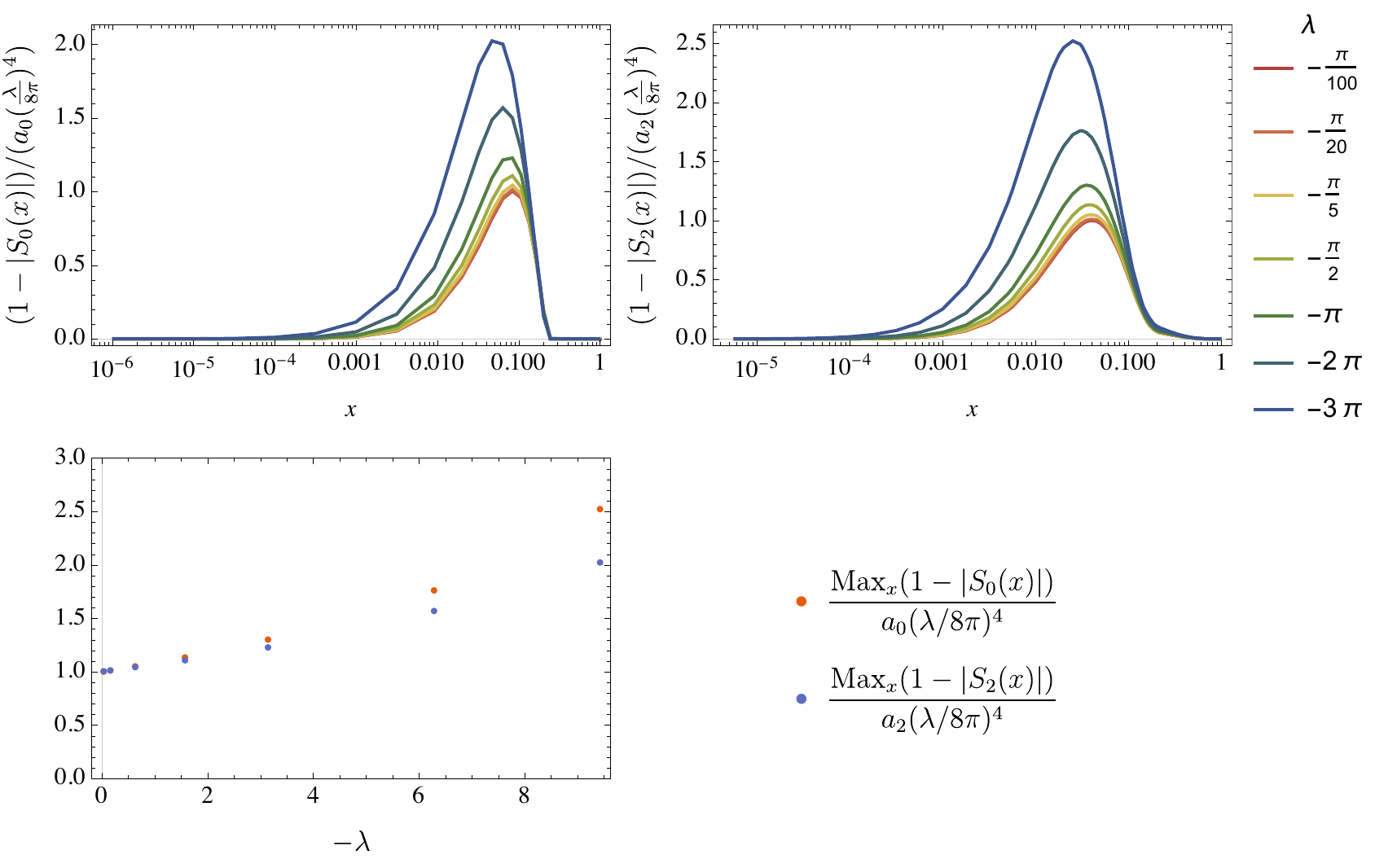}
    \caption{Inelasticities in $S_J$ for $J=0$ (2PR, top left) and $J=2$ (QE, top right). The partial waves are renormalized by a factor proportional to $\lambda^4$, which is the leading order effect for particle production. In these plots, $a_0=0.021$ and $a_2=0.00021$. The bottom row shows the maximum value of $1-|S_J|$ renormalized by $\lambda^4$: at small lambda, $a_0$ and $a_2$ are fitted to go to 1, at larger $\lambda$ the inelasticity deviates from pure $\lambda^4$ effects.}
    \label{fig:pl3dS0S2all}
\end{figure}

Here we present some elements of the coupling-dependence of the quasi-elastic amplitudes in $d=3$, in \figref{fig:pl3dS0S2all}. In $d=3$, we control all couplings from small to the largest we can reach, contrary to $d=4$ where we cannot control the small couplings, as explained below. The partial waves look smooth, and depend at leading order on $\lambda^4$, which corresponds to 3-loop Aks  graph, and receive corrections from higher orders.

\section{Quasi-elastic amplitudes in $d=4$}
\label{sec:QE4d}

In this section, we present results for the quasi-elastic amplitude in four dimensions for $\lambda = 5 \pi$. The result was obtained by hot-starting on the corresponding solution to the toy-model (that had zero double discontinuity). Why this procedure helps can be easily understood: it is because $\rho_{\text{el}}(s,t)$ is numerically small, and thus its back reaction on $\rho(s)$ is not big and thus it is worth coming close to the final answer first for $\rho(s)$ with the toy-model.

\subsection{2QE amplitude for $\lambda = 5 \pi$}

\begin{figure}[h]
    \centering
    \includegraphics[scale=0.4]{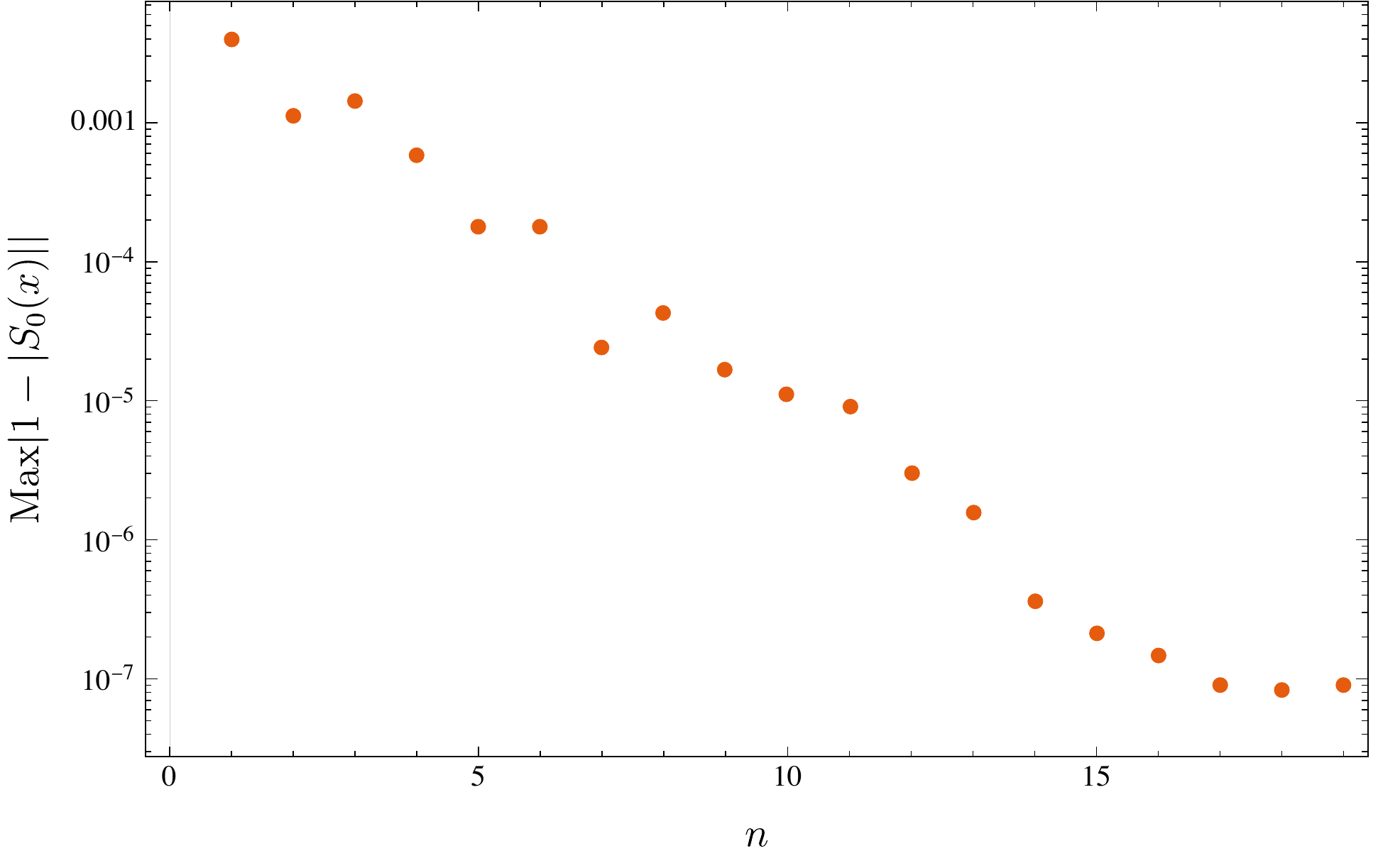}
    \caption{Maximal deviation of the absolute value of S-wave from $1$ in $d=4$ as a function of number of iterations $n$. We observe linear convergence and reach the precision $\sim 10^{-7}$ after $\sim 20$ iterations. }
    \label{fig:5piconvergence}
\end{figure}

To evaluate the convergence of the algorithm, we can measure how fast the absolute value of the S-wave converges to one. Recall that in the quasi-elastic case, we force the S-wave to be purely elastic at all energies. We have found that within 20 iterations we have reached the precision accuracy of our computation which for $|S_0(x)|$ is $\sim 10^{-7}$, see \figref{fig:5piconvergence}.

We now present various aspects of the amplitude after $20$ iterations. 

\subsubsection{Single and double spectral functions}

\begin{figure}[h]
    \centering
\includegraphics{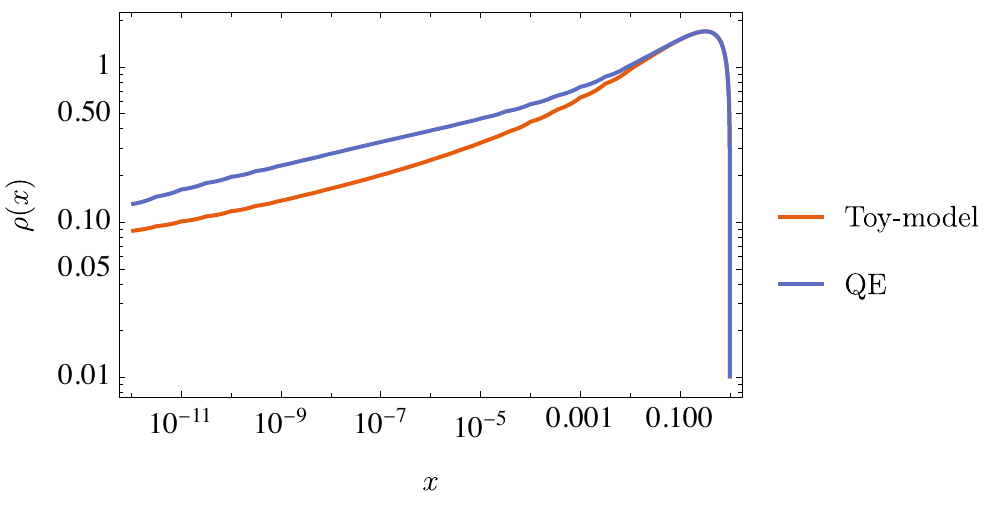}
    \caption{Single discontinuity $\rho(x)$ as a function of energy $x$ in $d=4$, $\lambda=5 \pi$. Here we focus on the Regge limit $x \to 0$ which reveals some emergent nontrivial behavior that differs from the toy-model result.
    }
    \label{fig:5pisingledisc}
\end{figure}

The single discontinuity is shown in \figref{fig:5pisingledisc}. As in 3d, we observe that the spectral density stays identical to the toy-model one at low energies, $x \to 1$, and deviates from it at high energies, $x \to 0$.

\begin{figure}[h]
    \centering
     \begin{subfigure}{0.5\textwidth}
    \centering
        \includegraphics[scale=0.55]{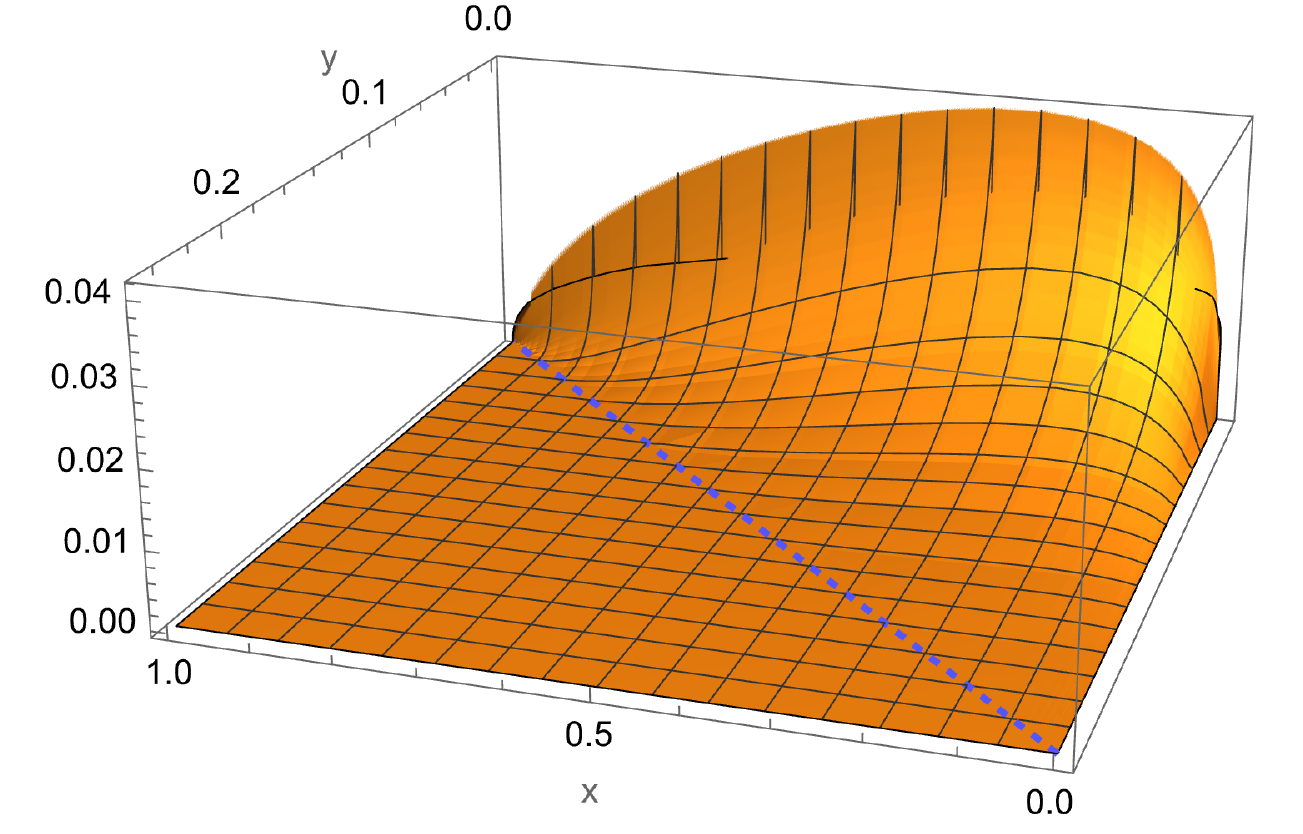}
  \end{subfigure}%
  \begin{subfigure}{0.5\textwidth}
    \centering
     \includegraphics[scale=0.6]{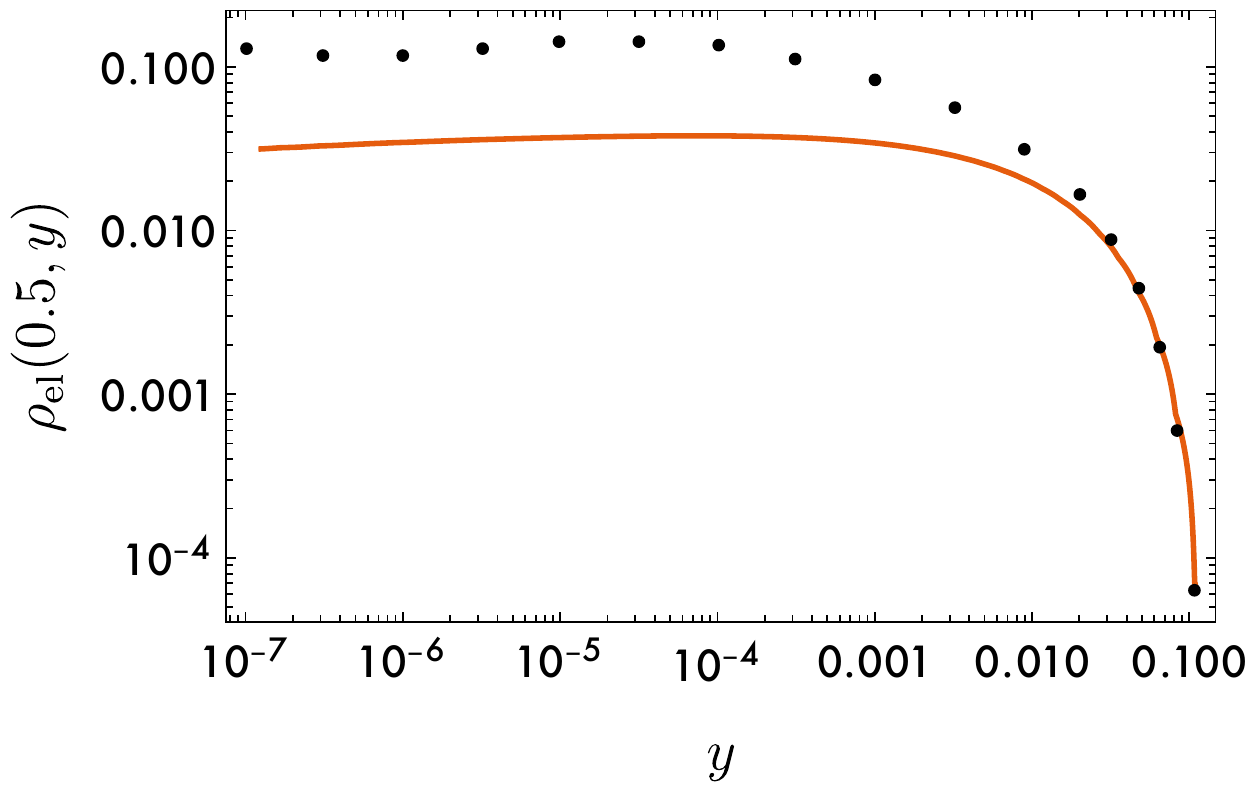}
  \end{subfigure}
    \caption{Double discontinuity $\rho_{\text{el}}(x,y)$ as a function of energies in $d=4$, $\lambda=5 \pi$. {\bf Left panel:} the blue line represents the leading Landau curve to the left of which $\rho_{\text{el}}(x,y)=0$. {\bf Right panel}: we take the fixed energy slice $\rho_{\text{el}}(0.5,y)$ to exhibit the nontrivial emergent Regge limit as $y \to 0$. The dashed line represents the result from the three-loop \emph{Aks} diagram, see \figref{fig:diagramspic} the third diagram in the third row. We see good agreement at low energies, $x \to 1$, and different behavior at high energies, $x \to 0$.}
    \label{fig:5pidoubledisc}
\end{figure}

Next we plot the double discontinuity in \figref{fig:5pidoubledisc}. Recall that $\rho(x,y) = \rho_{\text{el}}(x,y)+\rho_{\text{el}}(y,x)$, where $\rho_{\text{el}}(x,y)$ satisfies the Mandelstam equation \eqref{eq:mandelstam} and has nonzero support only above the leading Landau curve, namely $y \leq {1-x \over 4}$.

\subsubsection{Partial waves, impact parameter}

Consider next the structure of the partial waves as a function of energy $S_J(x)$, where $x={4 m^2 \over s}$. The scattering is dominated by the spin $J=0$ partial wave which is by indistinguishable from the corresponding toy-model result shown in \figref{fig:plS04dtoy5pi}.

\begin{figure}[h]
    \centering
    \includegraphics[scale=1.0]{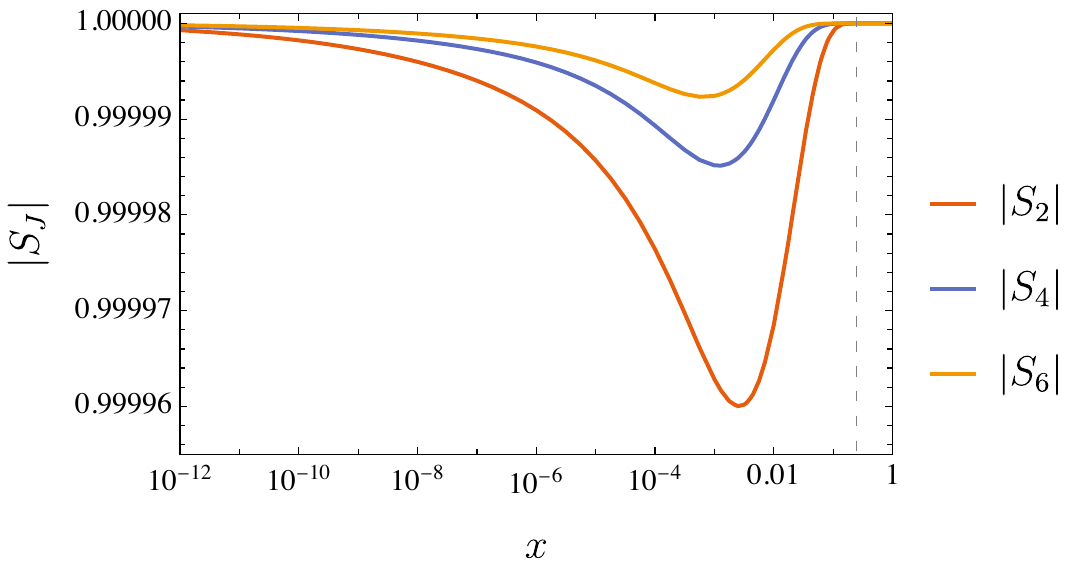}
    \caption{The absolute value of $|S_J|$ as a function of energy $x$ in $d=4$, $\lambda=5 \pi$. We see that the amount of scattering in the higher spin partial waves quickly decreases as a function of spin $J$. The dashed line is at $x = {1 \over 4}$, or, equivalently, $s=16 m^2$, which separates the elastic region $x \geq {1 \over 4}$ from the inelastic one $x < {1 \over 4}$. Note that there is a dynamically emergent scale in the problem where each partial wave peaks.}
    \label{fig:5piSJ}
\end{figure}

Scattering in the higher partial waves $S_{J>0}$ is much weaker, nevertheless it is nonzero in agreement with the expectation based on elastic unitarity and the Aks theorem \cite{Aks:1965qga}. We present a first few partial waves in \figref{fig:5piSJ}.

\begin{figure}[h]
    \centering
    \includegraphics[scale=1.0]{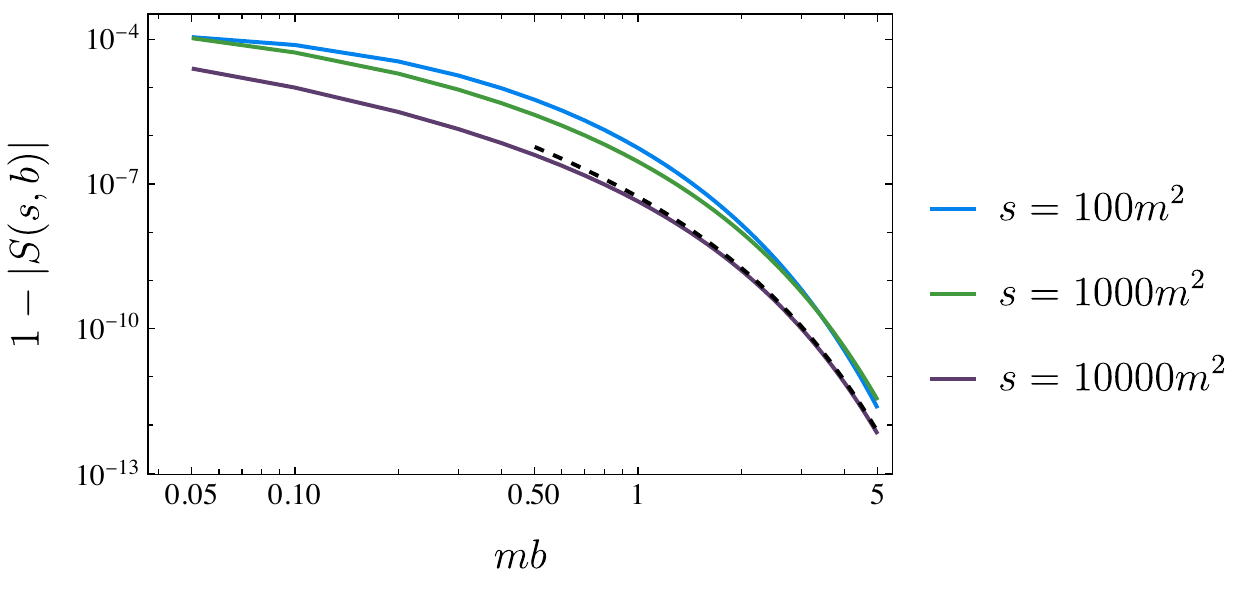}
    \caption{The absolute value of $1-|S(s,b)|$ as a function of impact parameter $b$ in $d=4$, $\lambda=5 \pi$.  As expected interactions quickly decay as a function of impact parameter, in a Yukawa-like, exponential fashion, see appendix \ref{app:impactFG}. Here the dashed line stands for the fit $\sim{e^{- 2 b m} \over (b m)^2}$.}
    \label{fig:5piImpact}
\end{figure}

We can also consider scattering at fixed impact parameters $b \equiv {2 J \over \sqrt{s-4m^2}}$. As expected the amplitude decreases quickly as a function of impact parameters, see figure \ref{fig:5piImpact}.

\subsubsection{Amplitude}

Next let us consider scattering at fixed angle scattering. We find that the fixed angle amplitude looks essentially the same as the corresponding plot for the toy model, see figure \figref{fig:allangles-rhos-4d}. 
We find that the amplitude acquires its maximum value at $s=4m^2$ and closely follow the one-loop running of the coupling $\lambda(s) = {\lambda \over 1 + {3 \lambda \over 16 \pi^2} \log {\sqrt{s} \over \sqrt{4/3}m}}$ in $\phi^4$ theory.
Therefore, the scattering amplitude that we constructed describes interactions that weaken in the UV, namely asymptotic freedom.

\begin{figure}[h]
    \centering
    \includegraphics[scale=1.0]{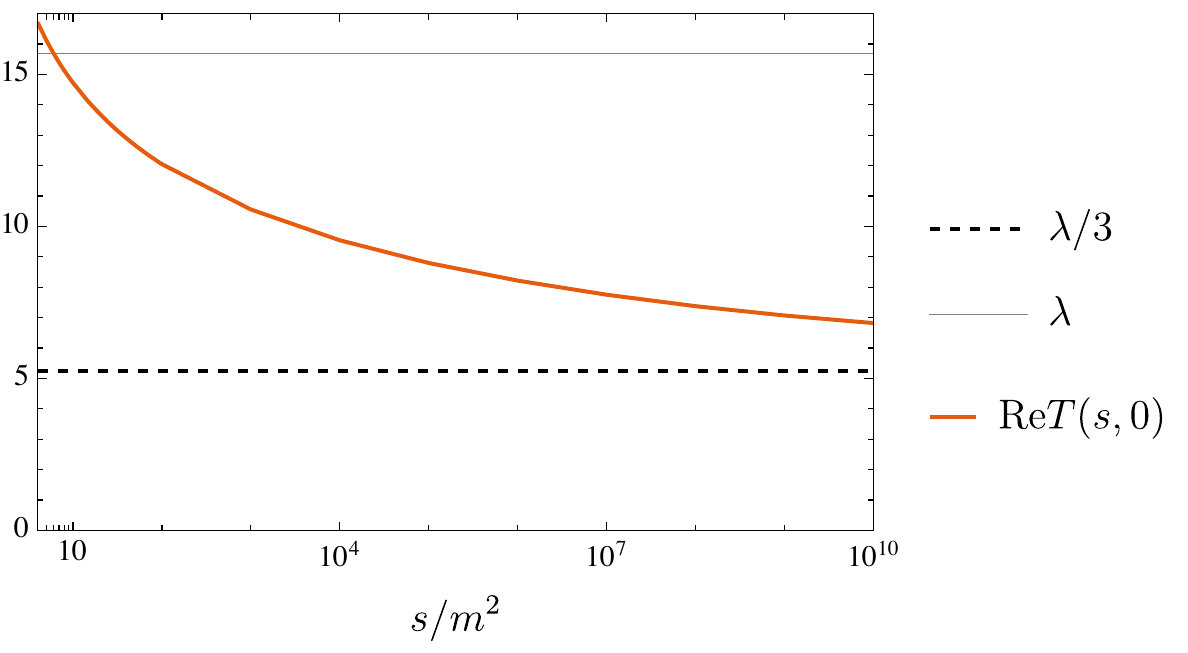}
    \caption{We plot the real part of the forward QE amplitude $T(s,0)$ in $d=4$. At low energies it start at the values around $\lambda$ and it goes to a constant at high energies. The constant is $\sim {\lambda \over 3}$ which is related to the sum rule \eqref{eq:sum-rule-rhos} in the toy model. Here we simply observe that the same behavior continues in the full back-reacted model.}
    \label{fig:5piforwardReal}
\end{figure}

\begin{figure}[h]
    \centering
    \includegraphics[scale=1.0]{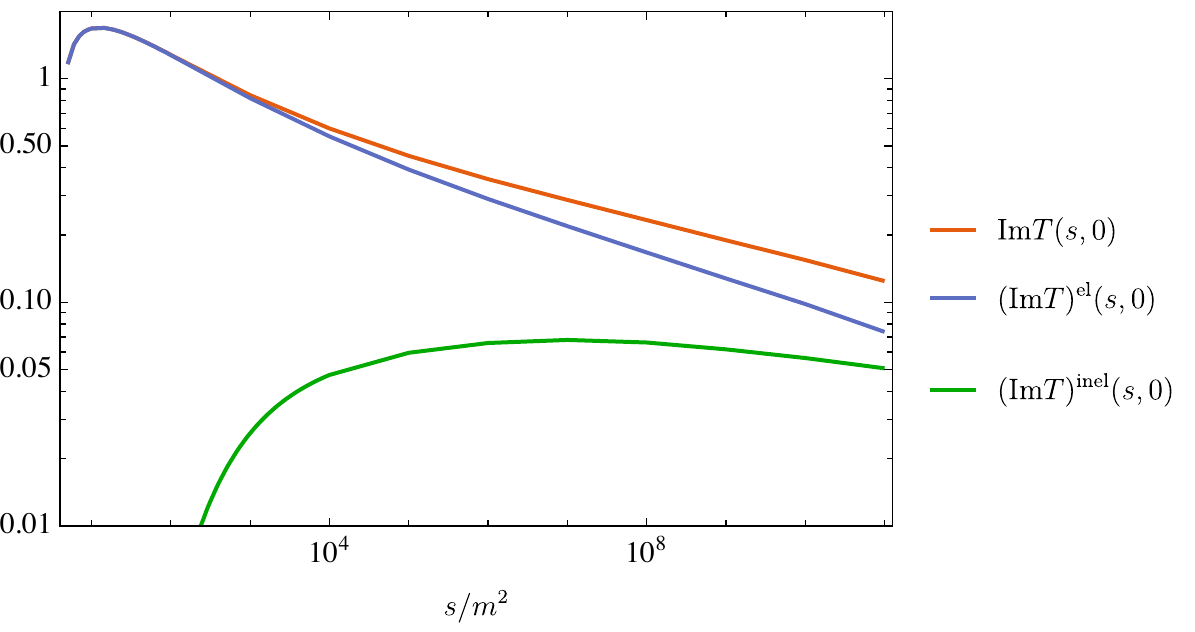}
    \caption{We plot the imaginary part of the forward amplitude, $\Im T(s,0)$. Via the optical theorem it is trivially related to the scattering cross section. We show three different cross sections: total, elastic, inelastic. The inelastic part of scattering in this model comes fully from the crossed terms $\rho_{\text{el}}(y,x)$ in the double spectral density. We see that at high energies scattering becomes more and more inelastic.}
    \label{fig:5piforwardIm}
\end{figure}

Finally, let us consider the forward scattering (or zero angle scattering). It is convenient to consider the real, \figref{fig:5piforwardReal}, and imaginary parts, \figref{fig:5piforwardIm},  separately.

Let us summarize the salient features of the nonperturbative scattering amplitude that we have constructed:

\begin{itemize}
    \item it obeys elastic and inelastic unitarity (in addition to crossing and analyticity). Let us clarify what we mean by this exactly. We impose elastic unitarity numerically on a finite set of grid points. For example, for $|S_0|$ we observed that violations of elastic unitarity are $\sim 10^{-7}$. For inelastic unitarity we have checked both a few low spin partial waves, as well as scattering at various impact parameters and energies. 
    \item it is elastic at any energies for the S-wave projection. This is what we called quasi-elastic amplitude in this paper.
    \item it has non-zero particle production for partial waves with $J>0$. This is related to the fact that the double spectral density of the amplitude has a correct support above the leading Landau curves. In fact, we see that as energy grows scattering becomes more and more inelastic.
    \item it exhibits asymptotic freedom. By this we mean that the scattering amplitude at fixed angles decays logarithmically as a function of energy.
    \item it has a nontrivial Regge limit. For example, for $t=0$ we have found that the real part of the amplitude goes to a constant, whereas the imaginary part decays logarithmically in energy. The nontrivial Regge limit of the amplitude is forced upon us by elastic unitarity \cite{Gribov:1961fm}.
\end{itemize}

Finally, we believe that the plots presented in this section will not change as one takes the continuous limit of the amplitude. Our belief is based on experimenting with various grids and observing stability of the results presented here, but of course an actual proof using the fixed point methods in the space of continuous functions is highly desirable.

\subsection{2PR amplitudes}

Next we consider the same amplitudes but in the 2PR scheme which means that scattering in the $J=0$ partial wave is not purely elastic. Structurally and for the couplings $\lambda$ analyzed in this paper, the 2PR amplitudes look very similar to the 2QE amplitudes. The main difference is that now $S_0$ has nonzero inelasticity \figref{fig:5pi2pr4dS0}:

\begin{figure}[h]
    \centering
    \includegraphics[scale=1]{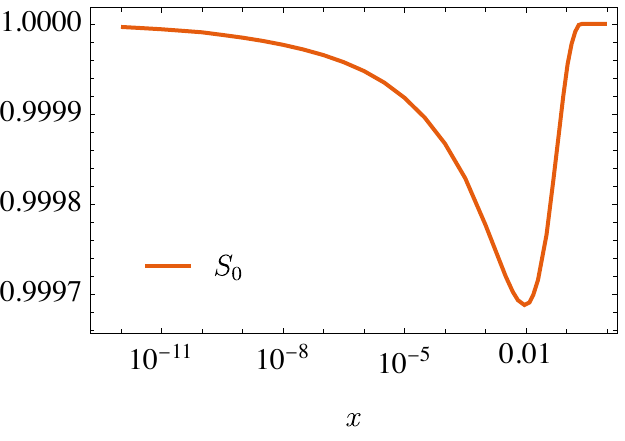}
        \includegraphics[scale=0.75]{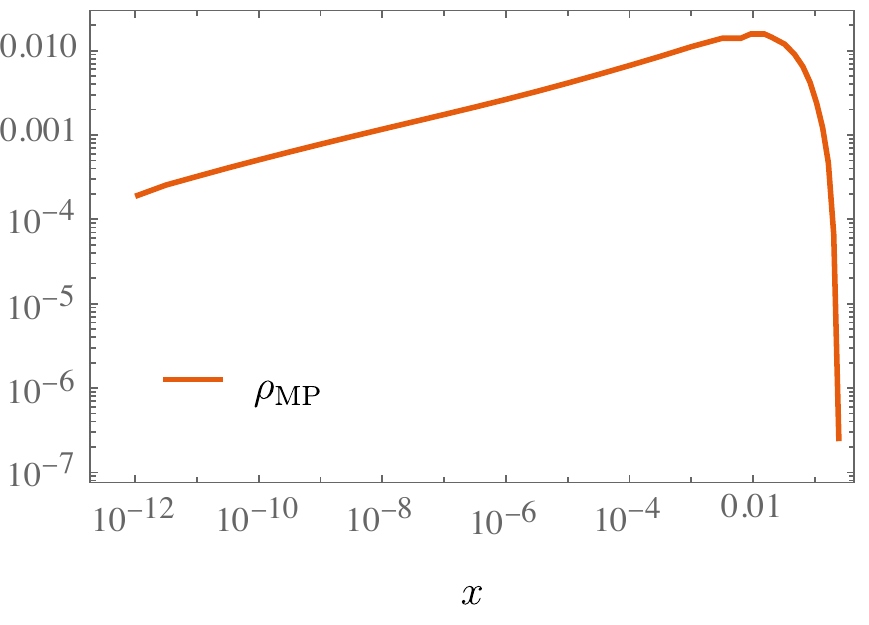}
    \caption{Spin zero partial wave $S_0(x)$, and $\rho_{\text{MP}}(x)$ in the 2PR scheme, $d=4$, $\lambda = 5 \pi$, as a function of energy. We see that the spin zero partial wave takes the same characteristic shape, as the higher spin partial waves depicted in \figref{fig:5piSJ}.}
    \label{fig:5pi2pr4dS0}
\end{figure}

\subsection{Coupling-dependence}
\label{sec:coupling-dependence}

\begin{figure}
    \centering
    \includegraphics{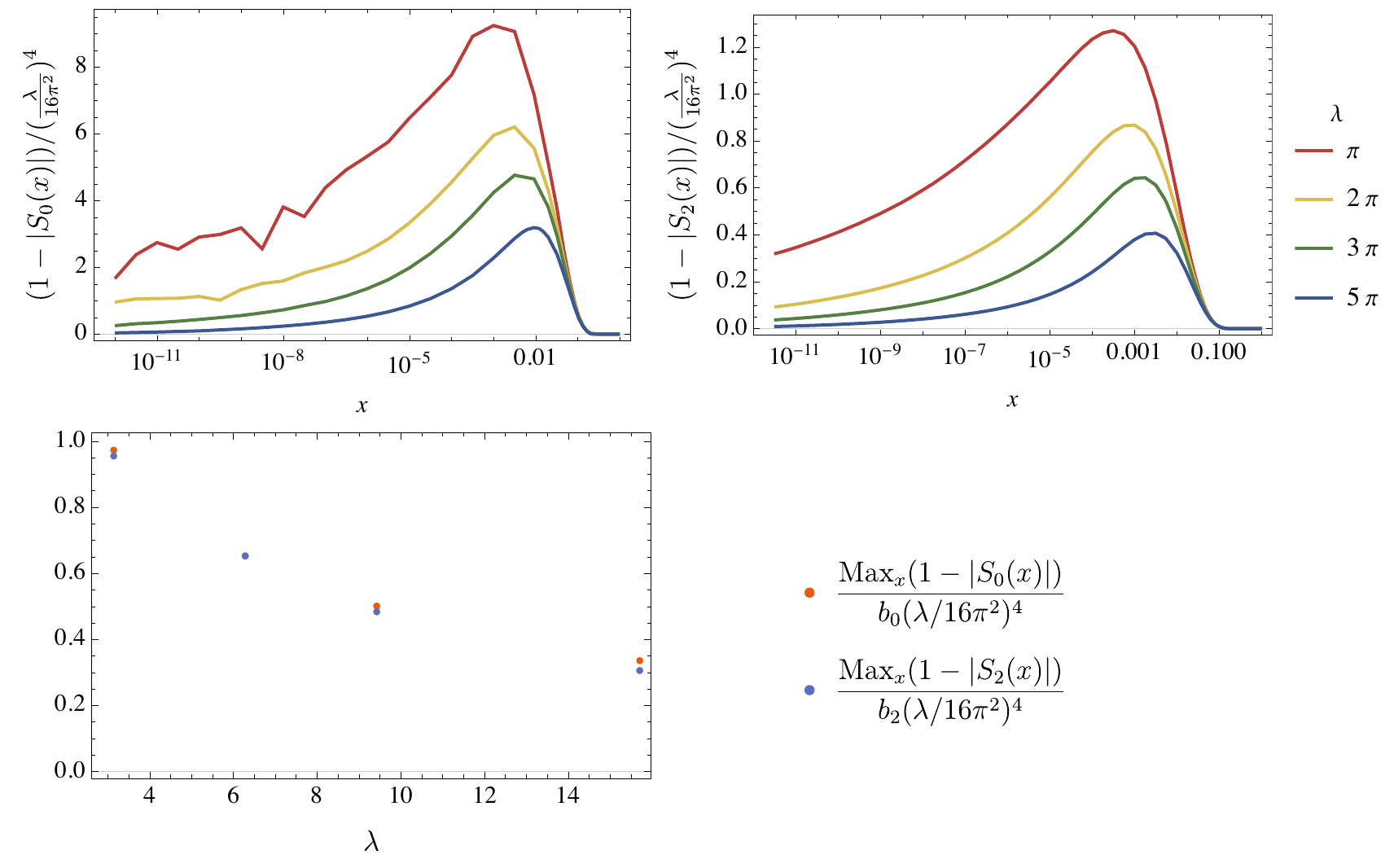}
    \caption{All the inelasticities, renormalized by leading order $(\lambda/(16\pi^2))^4$ Aks graph effect, in $d=4$ for $J=0$ (top left, 2PR models) and $J=2$ (top right, 2QE models) for couplings $\lambda\in[\pi;5\pi]$. The higher waves $J>0$ are visually indistinguishable in 2PR and 2QE models so we display a mixture of the two models. The rescaling by $\lambda^4$ of $1-|S_J|$ allows to simultaneously show various couplings on the same plot, for otherwise only the $5\pi$ curve would be visible and the other would appear completely flat, due to the approximate $\lambda^4$ dependence.
    Bottom: maximum value of $(1-|S_J|)$ renormalized by $\lambda^4$, and a an extra coefficient $b_J$ such that at $\lambda=\pi$, the curves go to one. This choice is arbitrary and simply allow to show that, in $d=4$, contrary to $d=3$, the coupling dependence is approximately identical in $J=0$ and $J=2$. We used empirical fitting to find $b_0=9.5$ and $b_2=1.33$. Finally, importantly, we did not display smaller couplings $\lambda$ because the numerics appeared not trustable, possibly due to a grid/cut-off effect. We show the full set of data in \appref{app:extraSj4d}.}
    \label{fig:pl4dS0S2all}
\end{figure}

We find again that inelasticity in $S_0$ and $S_2$ is driven by $\lambda^4$ at low energies and deviates from it at larger energies, see fig.~\ref{fig:pl4dS0S2all}.

\section{Multi-particle double discontinuity: a case of acnode in $d=4$}
\label{sec:acnode}

In the previous sections we worked with the approximation where the multi-particle double spectral density was set to zero, $\rho_{\text{MP}}(s,t)=0$. While this approximation might be a good leading order calculation, e.g. for $\phi^4$ theory in $d=3$, it is not exactly true in physical theories. Moreover, in confining gauge theories, for example in QCD, we expect $\rho_{\text{MP}}(s,t)$
to be sizable and important \cite{Kupsch:1982aa}. Therefore it is important to understand better the role of $\rho_{\text{MP}}(s,t)$ and its effect on the physical amplitude.

\begin{figure}[h]
    \centering
    \includegraphics[scale=0.24]{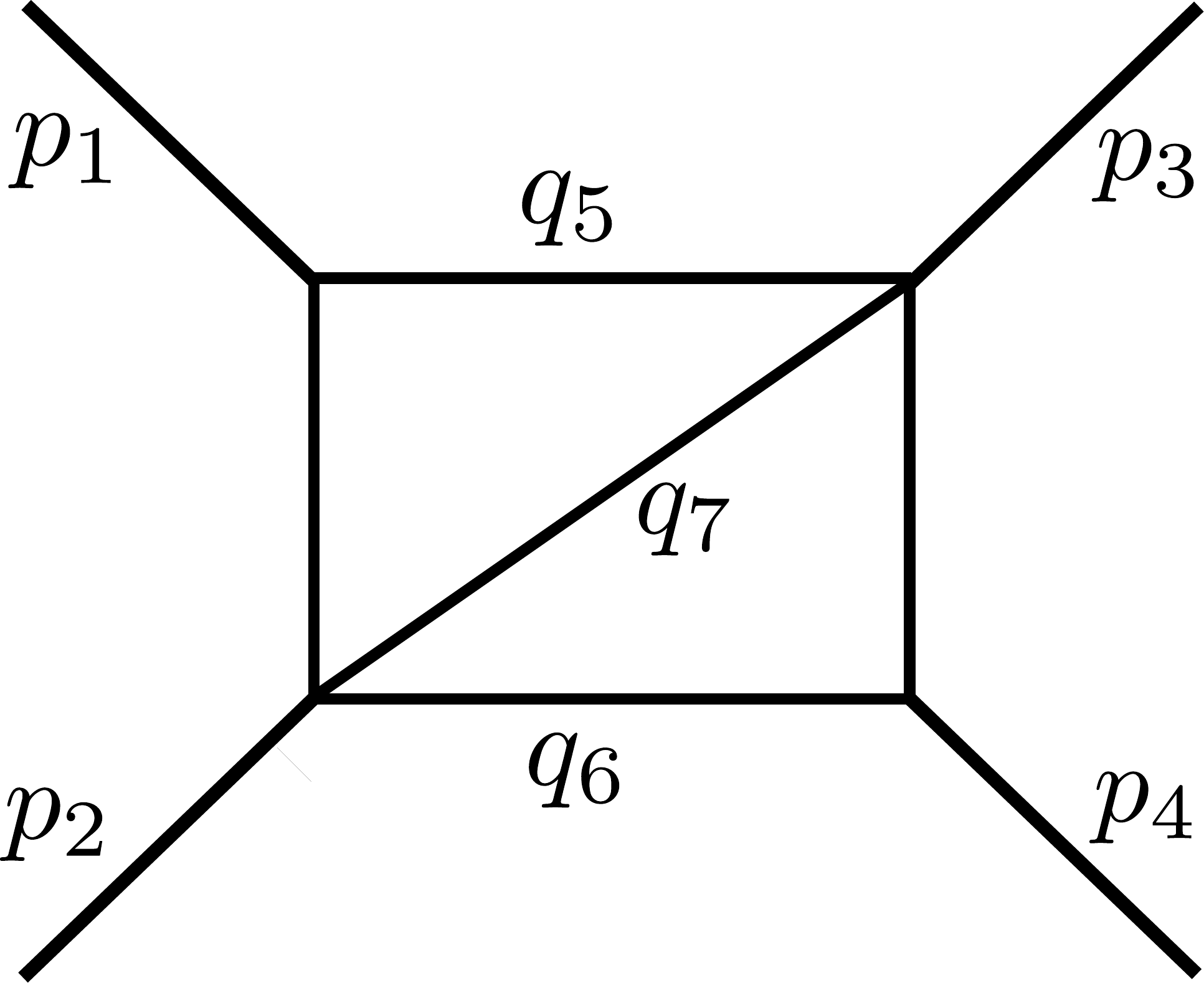}
    \caption{The acnode graph produces the leading contribution to $\rho_{\text{MP}}$ in a theory with both the $\phi^3$ and $\phi^4$-type vertices.}
    \label{fig:acnode}
\end{figure}

Here we consider the simplest graph which has three-particle cut both in the $s$ and $t$ channel, namely the acnode graph, see \figref{fig:acnode}.\footnote{This graph is also known as the four-point kite graph, see e.g. \cite{Lairez:2022zkj,Doran:2023yzu} for the recent analysis of this graph using geometric methods.} If we are to run the iteration process in the presence of bound states it would be the simplest graph that develops $\rho_{\text{MP}}(s,t)$. For the $\phi^4$-type theory (or $\mathbb{Z}_2$ symmetric scattering) the first analogous graph is the open envelope, see \figref{fig:diagramspic}.

We would like to compute $\rho_{\text{MP}}(s,t)$ for the acnode. This problem was analyzed by Gribov and Dyatlov in a short, insightful, but not fully explicit paper \cite{gribov1962contribution}. Moreover, their results have not been checked to the best of our knowledge. Here we fill these gaps, by writing down an explicit formula for the double discontinuity and checking that it correctly reproduces the single discontinuity upon doing a dispersive integral.

Let us describe the derivation that goes through several steps. The starting point is the three-body unitarity integral for the discontinuity of the acnode that takes the form
\be
\label{eq:threebodyunitarity}
2 T_s = {1 \over 3!} \int \prod_{i=5}^7 {d^3 \vec q_i \over (2 \pi)^3 (2 E_{\vec q_i})}  (2 \pi)^4 \delta(p_1+p_2-\sum_{i=5}^7 q_i) T(p_1,p_2; q_5,q_6, q_7) T^*(q_5,q_6,q_7; p_3, p_4) ,
\ee
where $s = (p_1 + p_2)^2$ and $t=(p_1-p_3)^2$.  The explicit formulas for the amplitudes that enter into the unitarity integral take the form
\be
T(p_1,p_2; q_5,q_6, q_7) &=- {1 \over t_{15} - m^2}, \\
T^*(q_5,q_6,q_7; p_3, p_4) &=-  {1 \over t_{46} - m^2},
\ee
where $t_{15}^2 = (q_5 - p_1)^2$ and similarly for $t_{46}$.

The unitarity integral can be simplified to the following form, see \appref{app:acnodeapp} for details and the definitions of various parameters,
\be
\label{eq:singlediscacnode}
2 T_s &= {1 \over 32 s} {1 \over 3!} {1 \over (2 \pi)^3}  \int^{(\sqrt{s}-m)^2} {d s_{57} d s_{67} \theta(1- (z_{56}^0)^2)  \over 4 | \vec p_1 |  | \vec p_4 | | \vec q_5| | \vec q_6| } \int_{-1}^{1} d z_{26} {1\over \sqrt{K(z_{15}^0,z_{56}^0, -z_{26})  K(z, z_{26}, z_{46}^0)}}.
\ee
This agrees with the expression in \cite{gribov1962contribution} (up to $1/3!$ due to the identical particles that is not imposed in that paper). The expression above is amenable for numerical evaluation. It also admits the following dispersive representation
\be
\label{eq:dispersivesingle}
T_s(s,t) = \int_{t_{LC}(s)}^\infty {dt' \over \pi} {\rho_{\text{ac}}(s,t') \over t'-t} ,
\ee
where $t_{LC}(s)$ is the position of the acnode Landau curve. A convenient representation of the acnode Landau curve is given by the following formula \cite{eden1961acnodes}
\be
\label{eq:acnodeLC}
s(\phi) &= 5 + 4 \cos \phi + 2 ({3 \over 2} + \cos \theta + \cos \phi) {\sin \phi \over \sin \theta} , \\
t(\phi) &= 5 + 4 \cos \phi + 2 ({3 \over 2} + \cos \theta + \cos \phi) {\sin \theta \over \sin \phi} ,   
\ee
where $\theta+\phi ={\pi \over 3}$ and $0 \leq \phi \leq {\pi \over 3}$.

Our task below will be to derive a formula for $\rho_{\text{ac}}(s,t')$ in \eqref{eq:dispersivesingle} starting from the unitarity integral \eqref{eq:singlediscacnode}. For simplicity below we set $m=1$.

\subsection{Computing the $s_{67}$ integral}

 A convenient strategy is to start with the $s_{67}$ integral that can be easily done explicitly, see e.g. \cite{islam1965analytic}.
Indeed, one can check that the integral takes the following form %
\begin{equation}
    \int_{s_{67}^{(1)}}^{s_{67}^{(2)}} {d s_{67} \over \sqrt{(s_{67} - s_{67}^+) (s_{67} - s_{67}^-) }}\,.
\end{equation}

From now on we switch $z_{26} \to t_{26}$ via the linear map
\be
t_{26} = \frac{1}{2} \left(z_{26} \sqrt{\frac{s-4}{s}} \sqrt{(s-1)^2+s_{57}^2-2 (s+1)
   s_{57}} -s+s_{57}+3\right) .
\ee

To present the result for the single discontinuity $T_s$ it is convenient to introduce the following variables
\be
\tilde t_{26}^{\pm} &= (\sqrt s_{57} \pm 1)^2 , \nn \\
t_{26}^{(1,2)} &= 1 + {s_{57} \over 2} \mp {\sqrt{3} \over 2} \sqrt{s_{57} (4-s_{57})} , \nn \\
t^{\mp}_{26} &= \frac{\pm \frac{2 \sqrt{s^2-\left(s_{57}+2\right)
   s+\left(s_{57}-1\right){}^2} \sqrt{t}
   \sqrt{s+t-4}}{\sqrt{s}}+\left(-s+s_{57}+1\right) t}{4-s}+1 , \nn \\
   t' &= \frac{1}{2} \left(-s+s_{57}-\sqrt{\frac{s-4}{s}}
   \sqrt{(s-1)^2+s_{57}^2-2 (s+1) s_{57}}+3\right), \nn \\
t'' &= \frac{1}{2} \left(-s+s_{57}-\sqrt{\frac{s-4}{s}}
   \sqrt{(s-1)^2+s_{57}^2-2 (s+1) s_{57}}+3\right) .
\ee
Note that they are functions of $s,t,s_{57}$ only.

In terms of these variables the discontinuity of the amplitude takes the following compact form
\be
\label{eq:singledisc}
2 T_s(s,t) &={1 \over 768 \pi^3 \sqrt{s(s-4)}} \int_{4}^{(\sqrt{s}-1)^2} d s_{57} \int_{t'}^{t''} d t_{26} {1 \over \sqrt{(t_{26} - t_{26}^{-})(t_{26} - t_{26}^{+})}} \nn \\
& {\log { t_{26}-s_{57}+1 + \sqrt{(1-4/s_{57})(t_{26} - \tilde t_{26}^+)(t_{26} - \tilde t_{26}^{-})} \over t_{26} - s_{57} +1  - \sqrt{(1-4/s_{57})(t_{26} - \tilde t_{26}^+)(t_{26} - \tilde t_{26}^{-})}} \over  \sqrt{(t_{26} - \tilde t_{26}^+)(t_{26} - \tilde t_{26}^{-})}} .
\ee

\subsection{Analytic continuation in $t$}

Next we would like to analytically continue the expression above from negative to positive $t$. We note that the last two integrals in \eqref{eq:singledisc} look too hard to be computed explicitly, but it is possible to understand qualitatively how the integration contour deforms during the continuation.%

Let us introduce the function
\be
F(s,t,s_{57}) = \int_{t'}^{t''} d t_{26} {1 \over \sqrt{(t_{26} - t_{26}^{-})(t_{26} - t_{26}^{+})}}  {\log { t_{26}-s_{57}+1 + \sqrt{(1-4/s_{57})(t_{26} - \tilde t_{26}^+)(t_{26} - \tilde t_{26}^{-})} \over t_{26} - s_{57} +1  - \sqrt{(1-4/s_{57})(t_{26} - \tilde t_{26}^+)(t_{26} - \tilde t_{26}^{-})}} \over  \sqrt{(t_{26} - \tilde t_{26}^+)(t_{26} - \tilde t_{26}^{-})}} . 
\ee
In terms of $F$, we have that
\be
\label{eq:lastintegral}
2 T_s(s,t) &={1 \over 768 \pi^3 \sqrt{s(s-4)}} \int_{4}^{(\sqrt{s}-1)^2} d s_{57}  F(s,t, s_{57}) . 
\ee

\begin{figure}[h]
  \centering
   \includegraphics[scale=0.5]{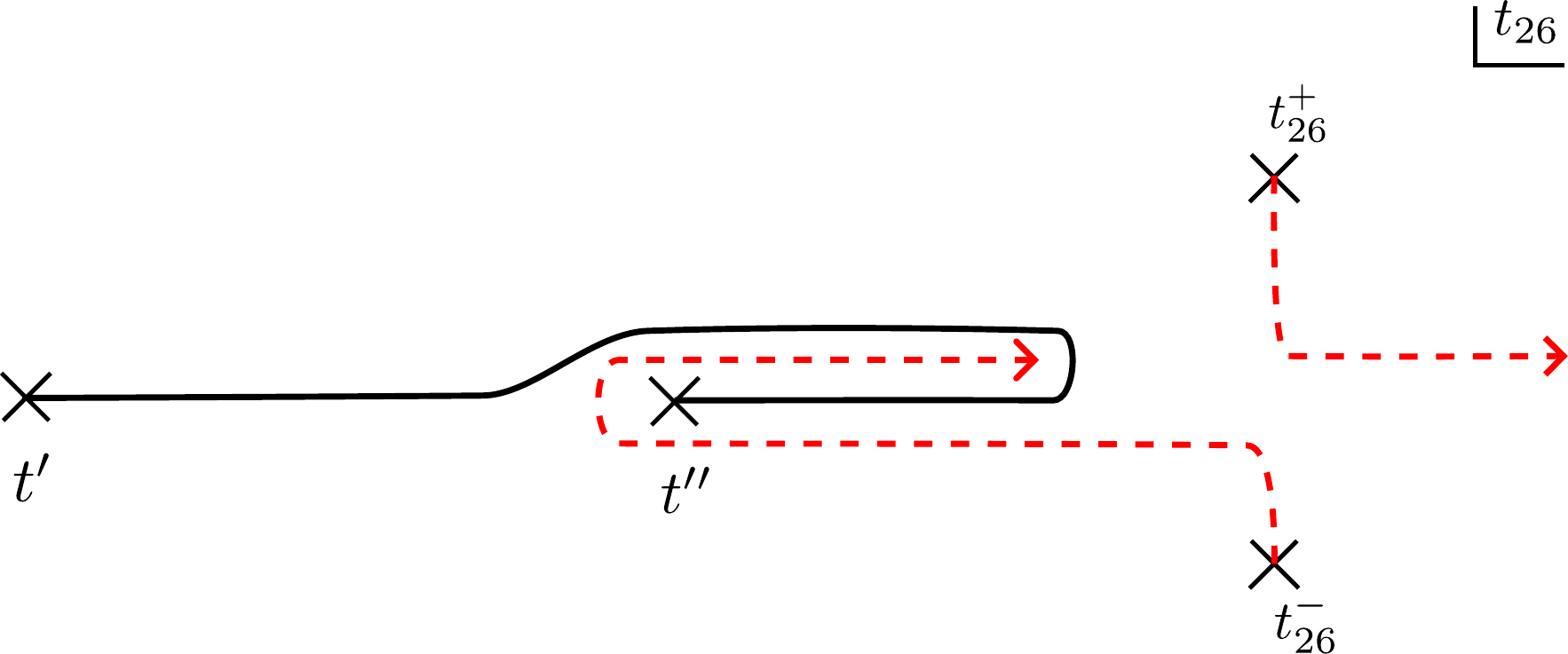}
  \caption{Deformation of the $t_{26}$ contour relevant for the double spectral density computation. The original contour goes along the real axis $t'' \geq t_{26} \geq t'$. As we increase $t$ the singularity of the integrand $t_{26}^-$ enters the integration contour and drags it away from the original contour. When $t_{26}^-$ collides with other singularities after that the integral develops a singularity, due to the pinching of the contour.}
\label{fig:t26contour}
\end{figure}

We now would like to understand the singularities of $F(s,t, s_{57})$ as we increase $t$. One can trace the motion of the singularities of the integrand and observe that as we increase $t$, $t_{26}^{(-)}$ enters the integration contour and drags it, see figure \ref{fig:t26contour}.

The singularities of $F(s,t,s_{57}) $ arise when the deformed contour gets pinched, which happens for $s_{57}^{\pm}(s,t)$ such that
\be
s_{57}^{\pm}(s,t):~~~ t_{26}^{-}(s,t,s_{57}) = t_{26}^{(2)}(s,t,s_{57}).
\ee%
Let us denote these singularities as $s_{57}^{\pm}(s,t)$ (they can be written explicitly). Another relevant for us singularity is given by
\be
s_{{\rm max}}(s,t):~~~ t_{26}^{-}(s,t,s_{57}) = \tilde t_{26}^{+}(s_{57}) . 
\ee

\begin{figure}[h]
  \centering
   \includegraphics[scale=0.5]{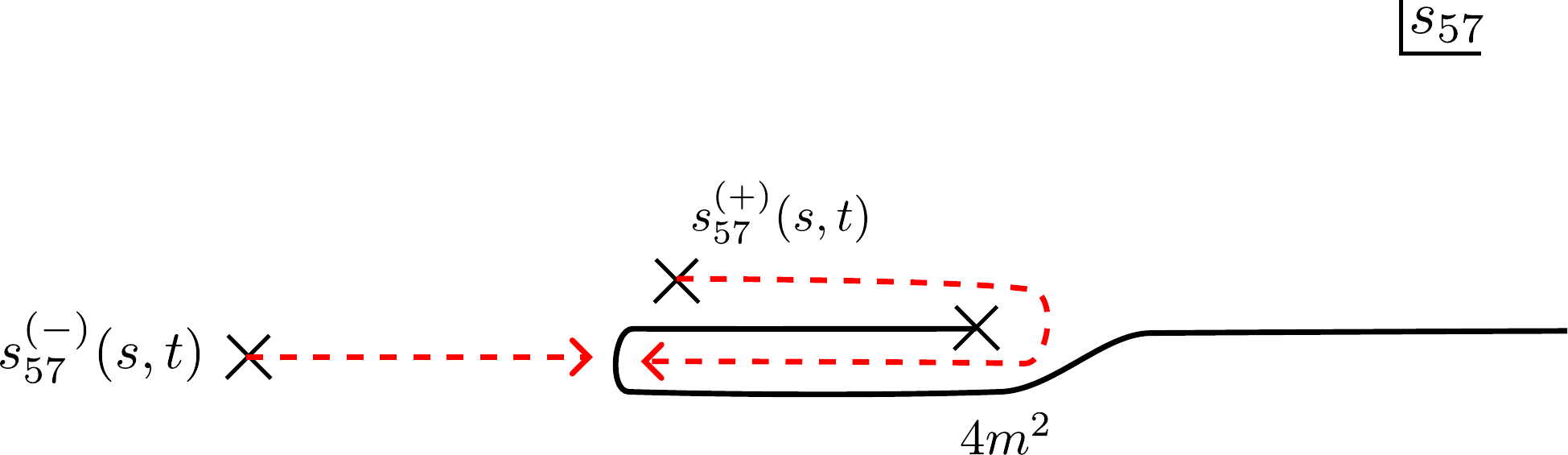}
  \caption{Deformation of the $s_{57}$ contour relevant for the double spectral density computation. The original integration integral goes along the real axis with $s_{57} \geq 4 m^2$ which is the condition that the invariant mass of a pair of on-shell particles should be greater than $4m^2$. As we increase $t$ to the positive values, the singularity of the integrand $s_{57}^+(s,t)$ enters the integration contour and drags it in the region $s_{57} < 4 m^2$. The Landau curve develops when the integration contour is pinched between the two singularities of the integrand $s_{57}^+(s,t)$ and $s_{57}^-(s,t)$. As we further increase $t$ these singularities move into the complex plane, leading eventually to the contour depicted in \figref{fig:contouracnode}. }
 \label{fig:s57contour}
\end{figure}

Next we analyze the last integral \eqref{eq:lastintegral} as we increase $t$. We can follow the motion of singularities $s_{57}^{\pm}(s,t)$ which is depicted in \figref{fig:s57contour}. 
The pinch occurs on the real axis when 
\be
\text{Landau curve}: ~~~ s_{57}^+(s,t) = s_{57}^-(s,t) .
\ee
One can check that this is indeed the correct location of the Landau curve for the acnode, namely it coincides with \eqref{eq:acnodeLC}. Note that in terms of the original unitarity integral it involves a pinch in $s_{57}$ away from the physical values of this invariant $s_{57} \geq 4 m^2$. In fact from the point of view of $T_{2 \to 4} \sim T_{2 \to 2} \times T_{1 \to 2}$  it involves analytic continuation to the second sheet! 

To compute $\rho(s,t)$ we need to understand what happens as we continue further. The singularities $s_{57}^{\pm}$ then recede in the complex plane. In the $t_{26}$-plane we take the discontinuity across $t_{26}^{-}(s,t,s_{57}) = t_{26}^{(2)}(s,t,s_{57})$
which is just logarithmic. As a result the relevant integral takes the form
\be
{\cal I}(s,t,s_{57}) &=  \int_{t_{26}^{(2)}}^{t_{26}^{-}} d t_{26}  {1 \over  \sqrt{(t_{26} - \tilde t_{26}^+)(t_{26} - \tilde t_{26}^{-})}}  {1 \over \sqrt{(t_{26} - t_{26}^{+})(t_{26} - t_{26}^{-})}} \nn \\
&=- 2 {({\cal F}( \arcsin \chi_1^{1/2} | \chi_2) - {\bf K}(\chi_2)) \over \sqrt{(\tilde t_{26}^+ - t_{26}^- ) (t_{26}^+ - \tilde t_{26}^{-})} } \ , \\
\chi_1 &= {(t_{26}^{(2)} - \tilde t_{26}^-) (t_{26}^{-} - \tilde t_{26}^+) \over (t_{26}^{(2)} - \tilde t_{26}^+) (t_{26}^{-} - \tilde t_{26}^-)} \ , ~~~\chi_2 = {(t_{26}^{-} - \tilde t_{26}^-) (t_{26}^{+} - \tilde t_{26}^+) \over (t_{26}^{-} - \tilde t_{26}^+) (t_{26}^{+} - \tilde t_{26}^-)} \ . 
\ee
In the expression above ${\cal F}(\phi|\chi)$ is the elliptic integral of the first kind, given by \texttt{EllipticF}$[\phi,\chi]$ in Mathematica, and ${\bf K}(\chi)$ is the complete elliptic integral of the first kind, given by \texttt{EllipticK}$[\chi]$ in Mathematica.

\begin{figure}[h]
  \centering
   \includegraphics[scale=0.85]{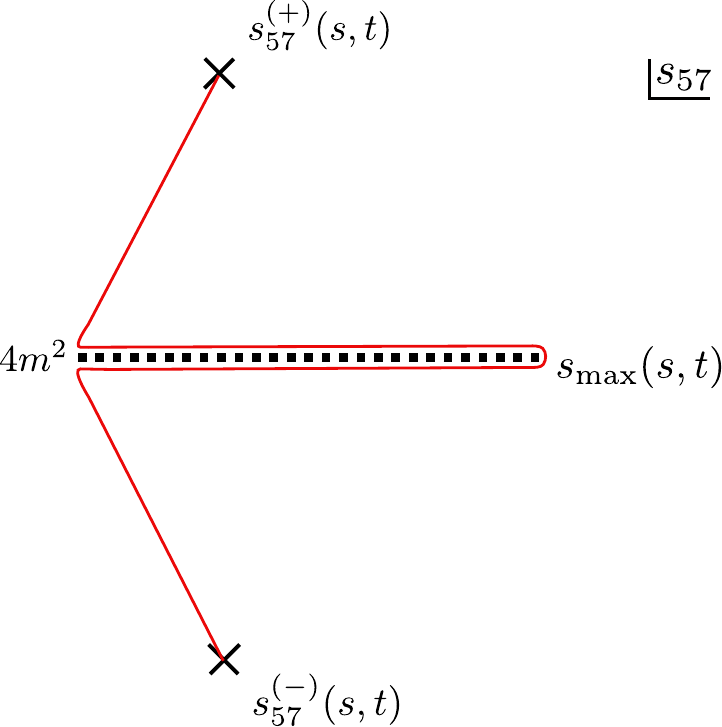}
  \caption{The final integration contour for ${\cal I}(s,t,s_{57})$ in the $s_{57}$ complex plane which computes the double spectral density $\rho(s,t)$. The horizontal part is given by ${\rm Disc}_{s_{57}} {\cal I}(s,t,s_{57})$.}
  \label{fig:contouracnode}
\end{figure}

We then need to integrate ${\cal I}(s,t,s_{57})$ across the contour in the $s_{57}$ plane which is given in \figref{fig:contouracnode}, and we obtain finally
\be
\label{eq:doublediscacnode}
\rho_{\text{acnode}}(s,t) = {1 \over 3 (8 \pi)^3} {2 \pi \over \sqrt{s(s-4)}}  \left( \int_{s_{57}^{(-)}}^{s_{57}^{(+)}} {d s_{57} \over i} {\cal I}(s,t,s_{57}) - 2 \int_{4m^2}^{s_{\text{max}}(s,t)} d s_{57} {\rm Disc}_{s_{57}} {\cal I}(s,t,s_{57}) \right),
\ee
where ${\rm Disc}_{s_{57}} {\cal I}(s,t,s_{57}) = 2 {{\bf K}({1 \over 1- \chi_2}) \over \sqrt{1-\chi_2} \sqrt{(\tilde t_{26}^+ - t_{26}^- ) (t_{26}^+ - \tilde t_{26}^{-})} }$.
We attach a Mathematica notebook together with the arXiv submission that numerically computes $\rho_{\text{acnode}}(s,t)$ using the formula above.

\begin{figure}[h]
  \centering
   \includegraphics[scale=0.7]{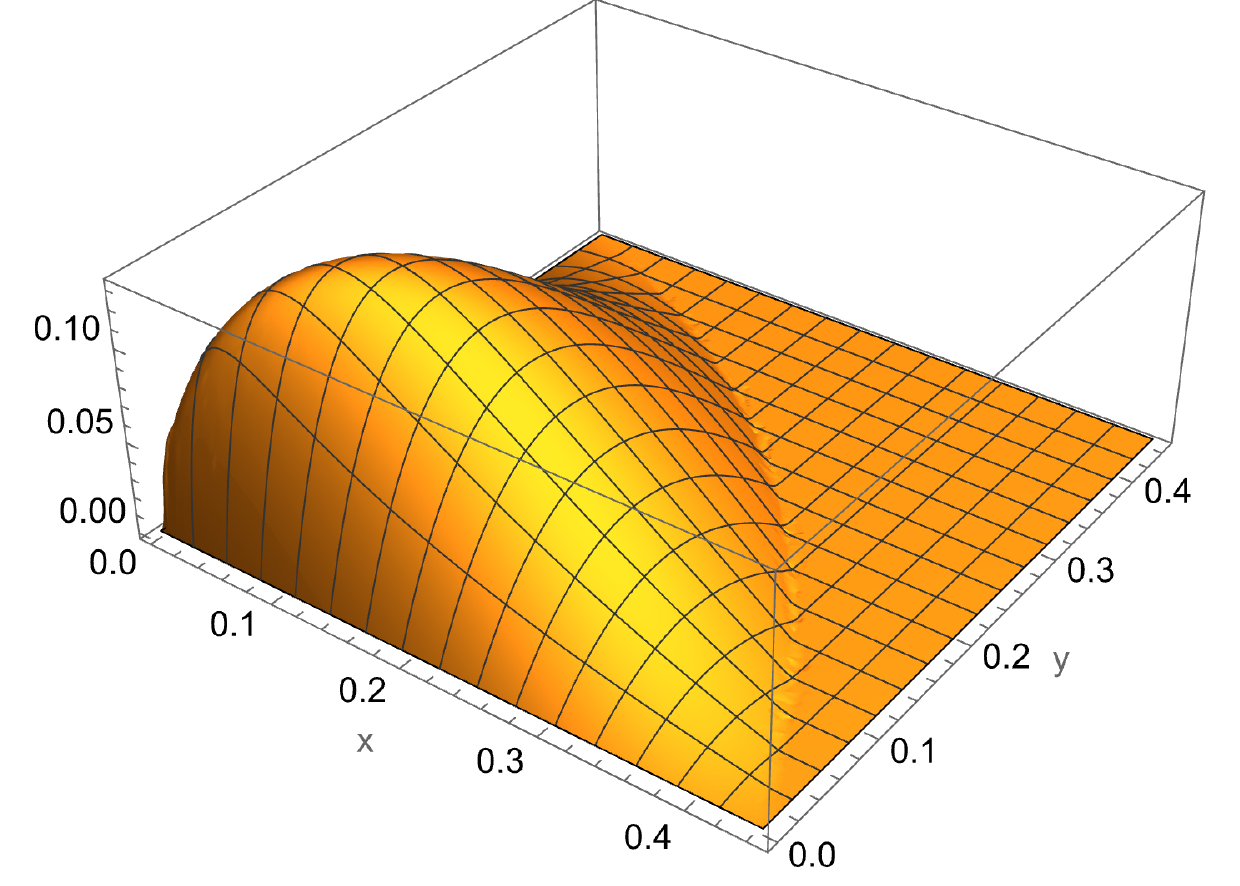}
  \caption{Acnode double spectral density given in $768 \pi^3 \times$\eqref{eq:doublediscacnode} evaluated for $(x,y)=({4 m^2 \over s}, {4 m^2 \over t})$, such that $0 \leq x, y \leq 1$. The nontrivial support of the double spectral density is controlled by the Landau curve \eqref{eq:acnodeLC} and is localized in the region $s,t> 9m^2$, or, equivalently, $0 \leq x,y, \leq {4 \over 9}$.}
 \label{fig:acnodePlot}
\end{figure}

There are several nontrivial tests for the result \eqref{eq:doublediscacnode}. First of all, it should be crossing-symmetric
\be
\rho_{\text{acnode}}(s,t) = \rho_{\text{acnode}}(t,s) .
\ee
This is highly non-obvious from \eqref{eq:doublediscacnode}, but we have checked numerically that it is indeed the case. Second, it should reproduce the single discontinuity \eqref{eq:singlediscacnode} upon doing the dispersive integral \eqref{eq:dispersivesingle}, which we again have tested numerically and found perfect agreement. We plot the double spectral density in  \figref{fig:acnodePlot}. 

\subsection{Nonperturbative lesson}

What does the computation of the double discontinuity of the acnode graph teach us about the nonperturbative scattering amplitudes?
The starting point in this case is the nonperturbative relation, which includes the three-particle amplitude for $s>9m^2$ as in \eqref{eq:threebodyunitarity}. We can then analytically continue the unitarity relation in $t$ to find the double discontinuity in terms of the analytically continued $T_{2 \to 3}$ amplitudes, as reviewed for example in \cite{Correia:2021etg}. In the case of the acnode the relevant singularity was due to one the one-particle poles ${1 \over t_{15}-m^2}$ and ${1 \over t_{16}-m^2}$. In the theory with the cubic coupling, these will still be present nonperturbatively. The residue of the pole however, instead of being just a constant will involve a nonperturbative $T_{2 \to 2}$ amplitude. Therefore, some of the formulas that rely on the detailed form $T_{2 \to 2}$ will change, however the most striking feature of the acnode double discontinuity calculation is that it probes  $T_{2 \to 2}$ on the second sheet, see \figref{fig:s57contour}! This fact has implications for closing the system of equations in the iteration scheme.

The nonperturbative version of the equation \eqref{eq:doublediscacnode} schematically will presumably take the form 
\be
\label{eq:nonperturbativeacnode}
\rho_{\text{MP}}(s,t) \stackrel{?}{=} \int d s_{57} d s_{67} K(s,t,s_{57},s_{67}) T^{(2)}_{2 \to 2}(s_{57}) (T^{(2)}_{2 \to 2}(s_{67}))^* ,
\ee
where $T^{(2)}_{2 \to 2}$ stands for the $2 \to 2$ scattering amplitude analytically continued to the second sheet through the elastic unitarity cut. This analytic continuation and the structure of the amplitude on the second sheet was recently discussed in \cite{Mizera:2022dko}. Note that finding the amplitude on the second sheet continued through elastic unitarity cut might also be possible via an iteration process, see \cite{Mizera:2022dko}, therefore equations of the type \eqref{eq:nonperturbativeacnode} do provide some interesting multi-particle extension of the current $2 \to 2$ S-matrix bootstrap program.

\subsection{Atkinson scattering-from-acnode}

To explore the effect of $\rho_{\text{MP}}(s,t)$ on the amplitudes constructed in the previous sections, we next construct amplitude functions for which, we set
\be
\label{eq:acnodeinitialize}
\rho_{\text{MP}}(s,t) = c_{\text{a}} 768 \pi^3 \rho_{\text{acnode}} \Big({16 s\over 9},{16 t\over 9} \Big),
\ee
where $c_{\text{a}}$ is a constant, $\rho_{\text{acnode}}(s,t)$ is given by \eqref{eq:doublediscacnode}, and we rescaled the arguments to make the acnode Landau curve asymptote to $16 m^2$.

\begin{figure}[hbt!]
    \centering
    \includegraphics[scale=1]{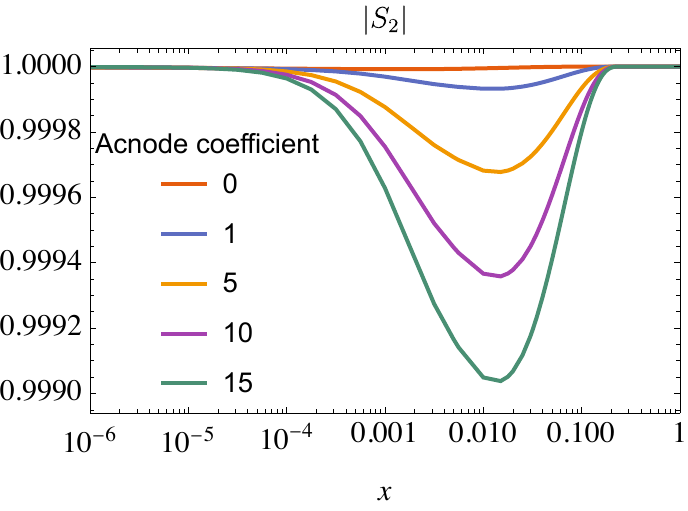}
    \includegraphics[scale=1]{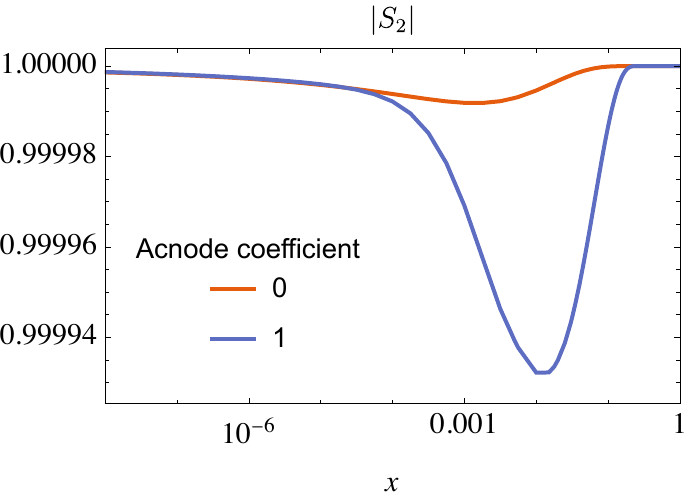}
    \caption{$S_2$ absolute value for the 2QE+acnode $\rho_{\text{MP}}(s,t)$ scheme in $d=4$, $\lambda = 3 \pi$. {\bf Left panel:} we see the expected growth of inelasticity as we increase $c_{\text{a}}$, given in \eqref{eq:acnodeinitialize}.  {\bf Right panel:} the same plot, zoomed in to compare the 2QE iteration scheme with $\rho_{\text{MP}}(s,t)=0$, versus $c_{\text{a}}=1$.}
    \label{fig:acnode-example-QE-4d-S2}
\end{figure}

We therefore initialize our iteration algorithm with the following input $(\lambda,0,\eqref{eq:acnodeinitialize})$ which generates an amplitude parameterized by $(\lambda, c_{\text{a}})$. Let us present the results for such amplitudes in $d=4$. In essence, the effect is to introduce more inelasticity to our amplitudes, see \figref{fig:acnode-example-QE-4d-S2}.

\begin{figure}[hbt!]
    \centering
    \includegraphics[scale=1]{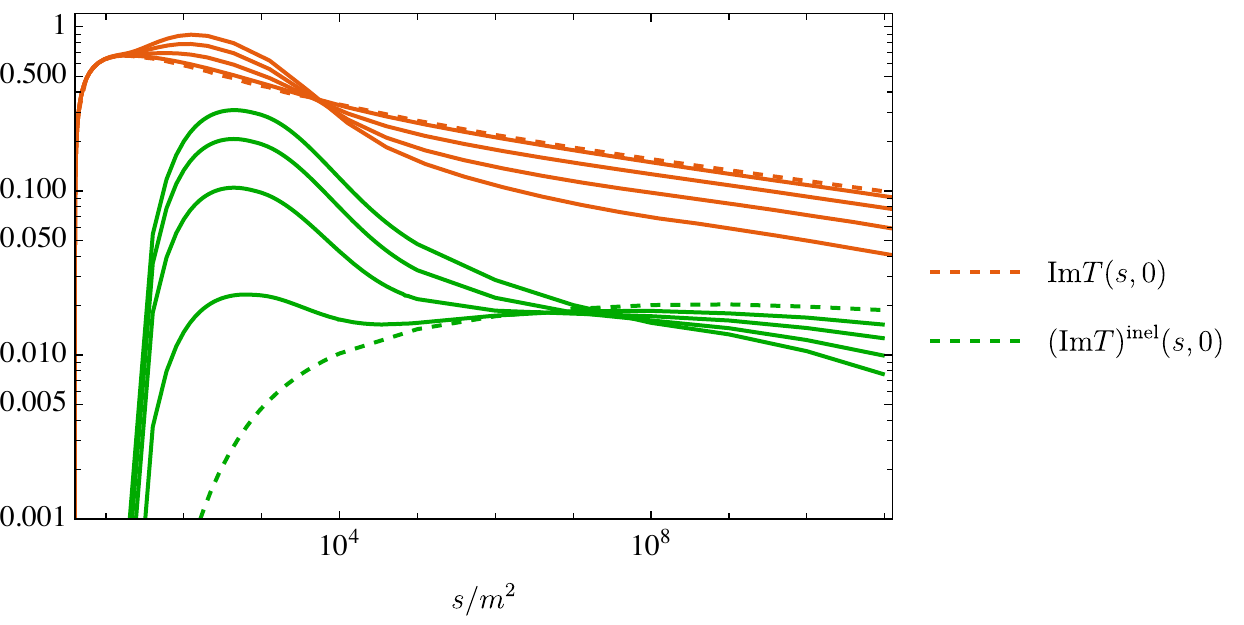}
    \caption{We plot the imaginary part of the amplitude (red), proportional to the total cross section, versus the inelastic cross section (green). The dashed curves correspond to $\lambda = 3 \pi$, $d=4$, 2QE amplitude. Other four curves are given by introducing the acnode $\rho_{\text{MP}}(s,t)$, as in \eqref{eq:acnodeinitialize}, with $c_{\text{a}}=1,5,10,15$.}
    \label{fig:acnode-example-totalcrosssection}
\end{figure}

We use the results for these amplitudes below in section \ref{sec:lowenergy} for the low energy observables and navigate the space of allowed theories.

\section{Low energy observables}
\label{sec:lowenergy}

In this section, we look at the comparison of our approach to the current main approach, initiated in \cite{Paulos:2016fap,Paulos:2016but,Paulos:2017fhb}, and we compare with results in with the recent works \cite{Chen:2022nym,EliasMiro:2022xaa}. In these works, the authors investigated in particular some bootstrap bounds on some low energy observables, akin to nonperturbative QFT couplings.

These approaches are extremely powerful to derive bounds, but do not provide actual S-matrices whose various properties like the presence of LC  or fine details of unitarity like elastic unitarity or Aks production can be tested.

Below, we consider the same low energy couplings and are able to make very precise statements about theories that live extremely close to the boundary. Following \cite{Chen:2022nym} and \cite{EliasMiro:2022xaa}, we thus look at two low-energy pairs of observables, $(\lambda,c_2)$ and $(\tau_{0,0},\tau_{1,0})$. They are defined as:
\begin{equation}
\begin{aligned}
c_2 &\equiv {1 \over 4} \partial_s^2 T(s,t) |_{s=t={4 m^2 \over 3}} \ , \cr %
\tau_{0,0} &=2^{d-1} m^{d-4} f_0(2m^2) \ , \cr %
\tau_{1,0} &=2^{d-1} m^{d-2} \partial_s f_0(2m^2) .
\end{aligned}
\end{equation}
Both $(\lambda,c_2)$ and $(\tau_{0,0},\tau_{1,0})$ probe the amplitude at low energies and very similar almond-shape, two-sided bounds on the values of these parameters were derived in the aforementioned works, see \cite[fig.~3]{Chen:2022nym} and \cite[fig.~2]{EliasMiro:2022xaa}.

While our methods 
allow us to resolve the fine structure near the boundary, they do not allow us to explore the whole parameter space of allowed theories discussed in these papers. Our amplitudes are located near the origin $(0,0)$ on both plots (on the scale set by the size of the almond/butterfly/leaf shape of allowed couplings). More precisely, we find that our amplitudes are located very close to the lower-bound on $c_2$ for given $\lambda$, or $\tau_{1,0}$ for given $\tau_{0,0}$. 

We are thus interested in the following question: which amplitudes among the ones that we have constructed are more extremal? Here, by extremal we imagine amplitudes that minimize $c_2$ for given $\lambda$, or $\tau_{1,0}$ for given $\tau_{0,0}$. We address this question in two steps. First, we consider 2QE and 2PR amplitudes for which $\rho_{\text{MP}}(s,t)=0$. We find that QE amplitudes are more extremal. Second, we turn on $\rho_{\text{MP}}(s,t)$ by taking as a proxy the double spectral density of the acnode graph. We find, that adding $\rho_{\text{acnode}}(s,t)$ moves the amplitude \emph{inside} the allowed region, or away from extremality. 

Looking at the plots and observing that the iterated amplitudes end up close to the boundary of the allowed region, it is tempting to ask: how close to the boundary are they? This question requires higher precision calculations, and thus a detailed and systematic analysis of our algorithm as we take the continuum limit. Leaving this for future work, we have chosen a particular 2QE amplitude for which the primal bootstrap data is available from \cite{EliasMiro:2022xaa}, namely  $\lambda = 2.012434211 \pi$, $d=4$
\be
{c_{2,\text{min}}^{\text{primal}} \over 32 \pi} \simeq 0.00000414818 .
\ee
It would be also interesting to explore how close to this value the dual bounds on this observable \cite{Guerrieri:2021tak,He:2021eqn,EliasMiro:2022xaa} can be brought.

Constructing the corresponding 2QE amplitude for the same value of $\lambda$ we have gotten that
\be
{c_{2}^{\text{2QE}} \over 32 \pi} \simeq 0.0000041484 - 0.00000002 = 0.0000041482 ,
\ee
where we indicated the contribution to $c_2$ coming from the single and double discontinuity of the amplitude. In particular, the contribution of particle production to $c_2$ is $\sim 1 \times 10^{-9}$.

To get this value we have explored how our results are affected by changing various grids and cutoffs. Therefore with the current precision we cannot definitively answer the question whether the 2QE amplitude is truly extremal. 

Also let us compare the results above with the two-loop result in $\phi^4$, see \appref{app:perturbativephi4},
\be
{c_{2}^{\text{two-loop}} \over 32 \pi} \simeq 0.000041475 \ .
\ee
It would be interesting to do the three-loop computation of $c_2$ since it is expected to be even closer to the numbers quoted above.

More conceptually, to understand if $T_{\text{2QE}}(s,t)$ is extremal, we would need to study ${\delta c_2 \over \delta \eta_{\text{MP}}(s)}$ and ${\delta c_2  \over \delta \rho_{\text{MP}}(s,t)}$ for infinitesimal deformations in the functional space of multi-particle data. More generally, we can write down two extra equations for fixed $\lambda$
\begin{equation}
\label{eq:extremal}
\text{Extremal } c_2:~~~{\delta c_2 \over \delta \eta_{\text{MP}}(s)} = 0, ~~~ {\delta c_2 \over \delta \rho_{\text{MP}}(s,t)} = 0 \ ,
\end{equation}
and we try to use them to solve for $( \eta_{\text{MP}}(s),  \rho_{\text{MP}}(s,t))$. In general we observed that turning on inelasticity tends to move the amplitude inside the allowed region. This motivates the expectation that setting $\eta_{MP}=0$ is extremal. We do not have a similar argument for $\rho_{\text{MP}}(s,t)$. Recall that particle production in the spin $J>0$ sector $\eta_J(s) \equiv 1-|S_J|^2$ is given by the Froissart-Gribov formula
\be
&\eta_{J}(s) = {(s-4m^2)^{{d-3 \over 2}} \over \sqrt{s}} {4 {\cal N}_d \over \pi} \int_{z_1}^\infty d z (z^2 - 1)^{{d-4 \over 2}} Q_J^{(d)}(z) \left( \rho_{\text{el}}(t(z),s) +  \rho_{\text{MP}}(s,t(z))   \right) \ .
\label{eq:analyticityinspinJgen}
\ee
Minimizing production in $J>0$ sector is not the same as setting $\rho_{\text{MP}}(s,t)=0$.
In particular, to make contact with amplitudes constructed in \cite{Chen:2022nym,EliasMiro:2022xaa}, it would be interesting to try iterating $\rho_{\text{MP}}(s,t)$ in a way that cancels as much particle production as possible in some finite energy (and spin) range. We do not explore this possibility in the present paper. A fixed point of such a procedure would be an extremal amplitude.
We leave exploration of this possibility for future work.

\subsection{$(\tau_{0,0},\tau_{0,1})$ plots in $d=3$, $d=4$}

\begin{figure}[h]
    \centering
    \includegraphics[scale=1]{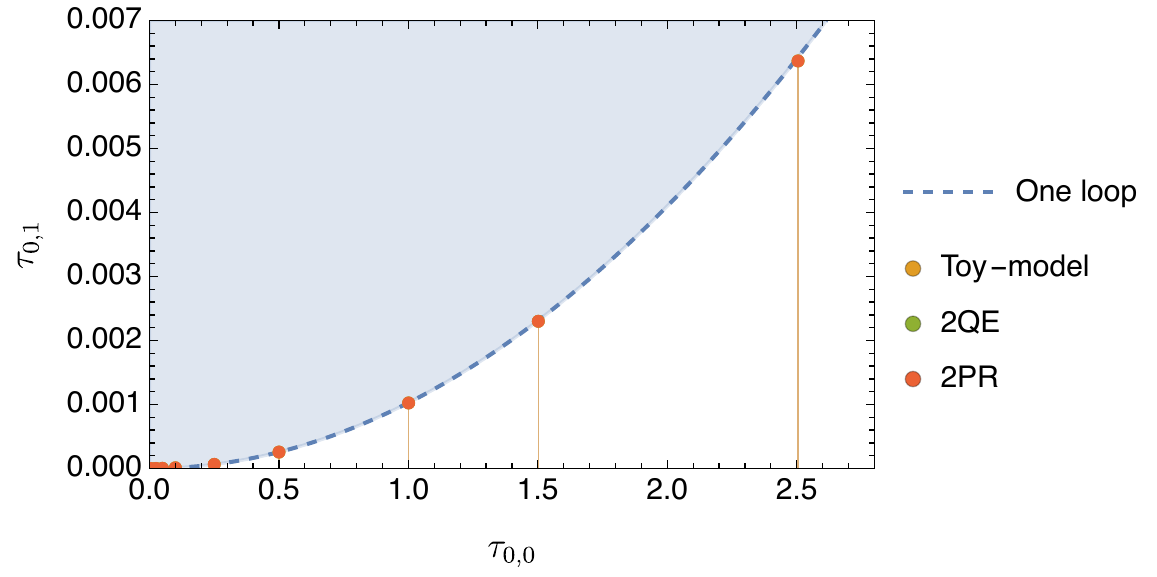}
    \caption{We plot the $(\tau_{0,0},\tau_{1,0})$ plane in $d=4$. It was observed in \cite{Chen:2022nym} that the space of allowed values is very well approximated by the one-loop perturbative result. We see that on this plot all our amplitudes: toy-model amplitudes with zero double discontinuity, 2QE and 2PR all lie on the boundary of the allowed space and are not resolvable.}
    \label{fig:pl4dtau0itot}
\end{figure}

\begin{figure}[h]
    \centering
    \includegraphics[scale=1]{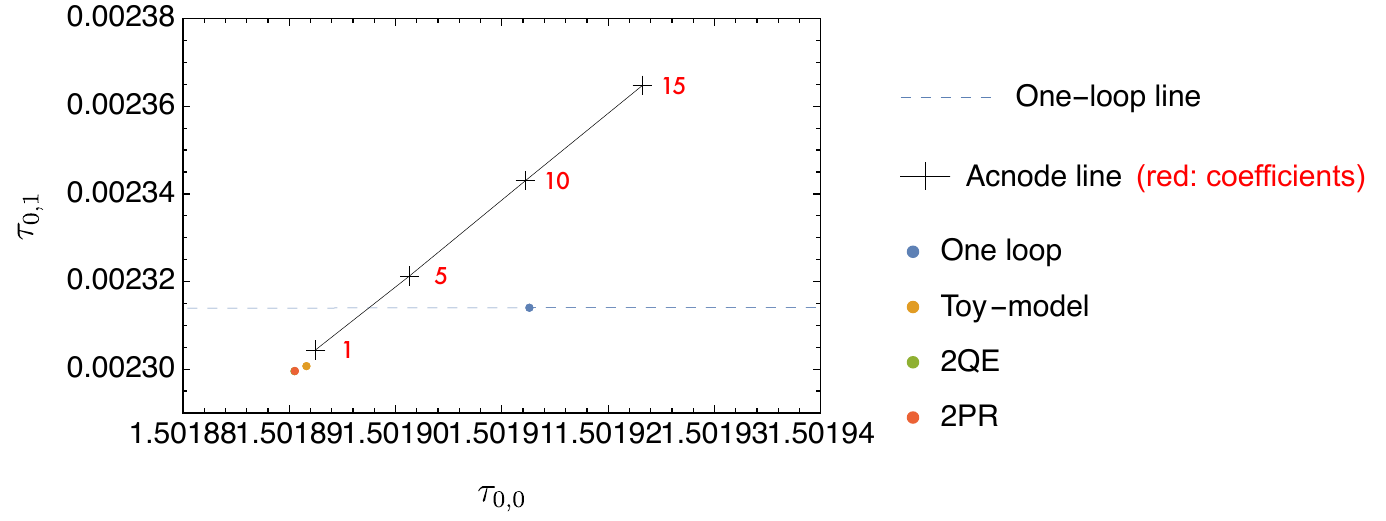}
    \caption{The same plot as above still in $d=4$, only now we zoomed on the point close to $\tau_{0,0} \simeq 1.5$. On this plot we still cannot resolve the 2QE and 2PR amplitudes, but we see that the toy model amplitude, as well as the amplitudes with $\rho_{\text{acnode}}(s,t)$ turned on have higher values of $\tau_{1,0}$, and are thus less extremal.}
    \label{fig:4dzoomed1}
\end{figure}

\begin{figure}[h]
    \centering
    \includegraphics[scale=1]{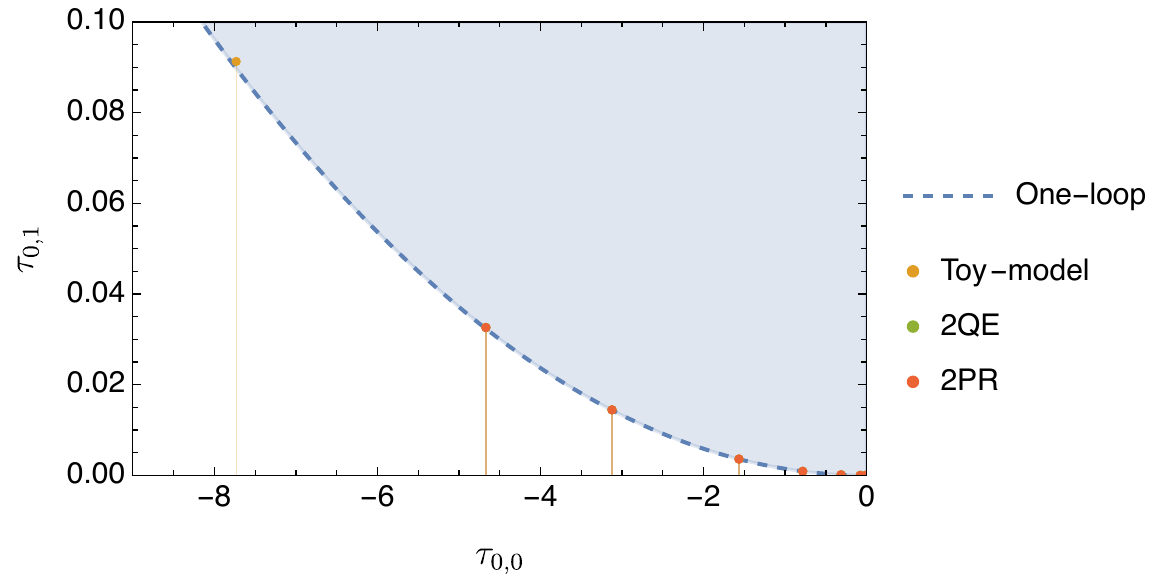}
    \caption{We plot the $(\tau_{0,0},\tau_{1,0})$ plane in $d=3$. It was observed in \cite{Chen:2022nym} that the space of allowed values is well-approximated by the one-loop result. We see that on this plot all our amplitudes: toy-model amplitudes with zero double discontinuity, 2QE and 2PR all lie on the boundary of the allowed space and are not resolvable.}
    \label{fig:pl3dtautot}
\end{figure}

\begin{figure}[h]
    \centering
    \includegraphics[scale=1]{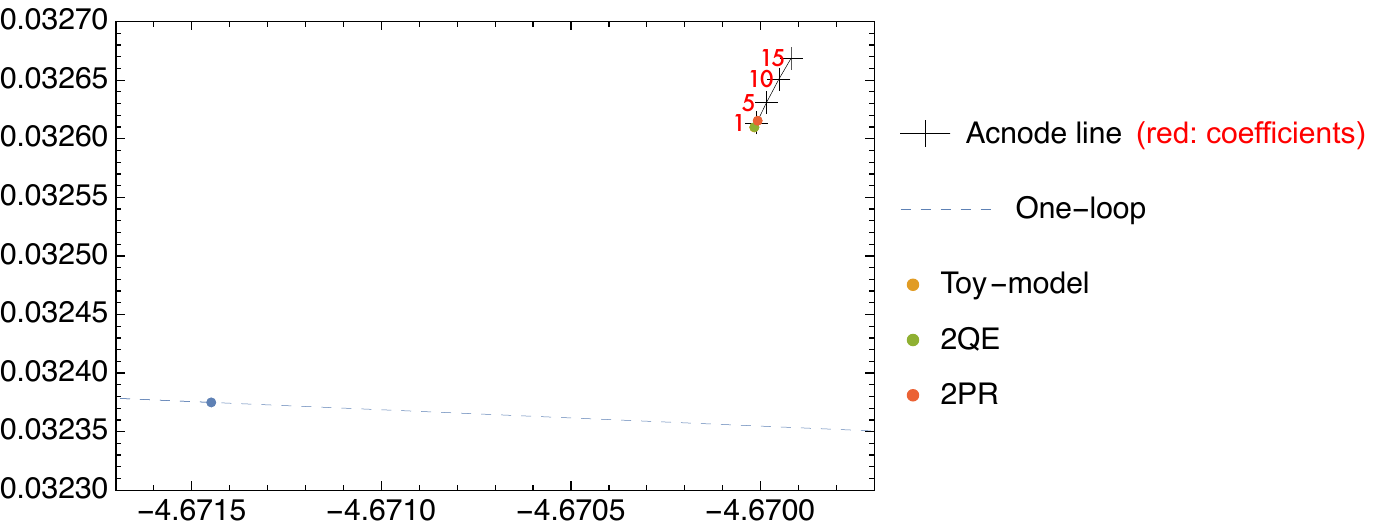}
    \caption{The same plot as above in $d=3$, only now we zoomed on the point close to $\tau_{0,0} \simeq -4.67$. On this plot we still cannot resolve the 2QE and 2PR amplitudes very well, but we see that the toy model amplitude, as well as the amplitudes with $\rho_{\text{acnode}}(s,t)$ turned on have higher values of $\tau_{1,0}$, and are thus less extremal.}
    \label{fig:pltau3dzoomed}
\end{figure}

Here we provide the plots for the $(\tau_{0,0},\tau_{0,1})$ observables in $d=4$, see \figref{fig:pl4dtau0itot} and \figref{fig:4dzoomed1}, and in $d=3$, see \figref{fig:pl3dtautot} and \figref{fig:pltau3dzoomed}.

\FloatBarrier

\subsection{$c_2$ plots}

\begin{figure}[h]
    \centering
    \includegraphics{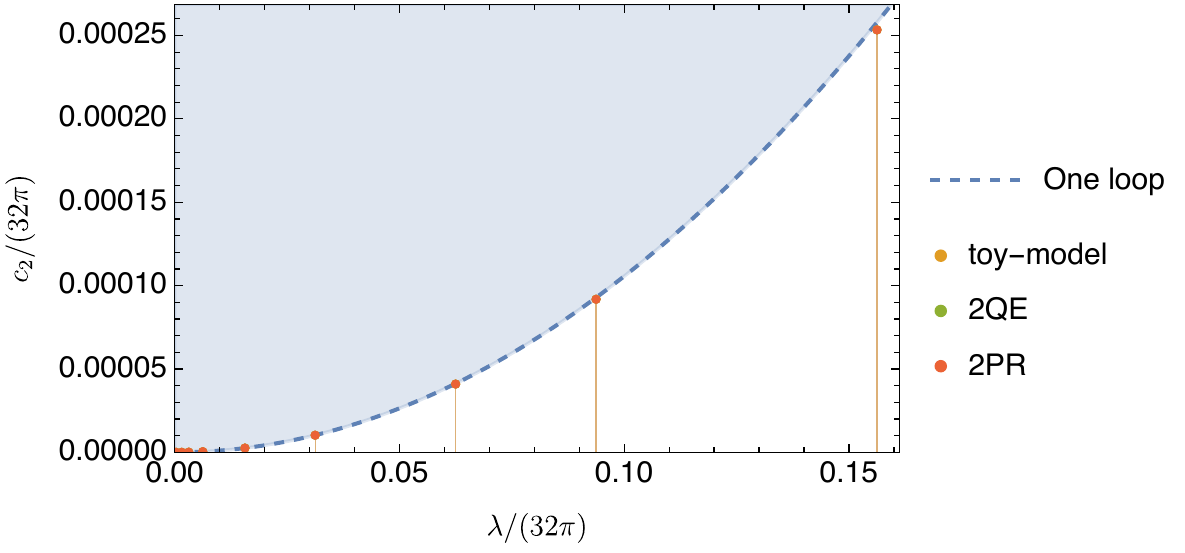}
    \caption{$({\lambda \over 32 \pi},{c_2 \over 32 \pi})$ plane in $d=4$. The space of allowed amplitudes, see \cite{Chen:2022nym,EliasMiro:2022xaa}, (to a good approximation) lies above the one-loop result in $\phi^4$ theory (dashed line).  Various amplitudes constructed in this paper all lie very close to the one-loop result. In particular, on this scale the toy-model, 2QE and 2PR amplitudes are indistinguishable.}
    \label{fig:c24d}
\end{figure}

\begin{figure}[h]
    \centering
    \includegraphics{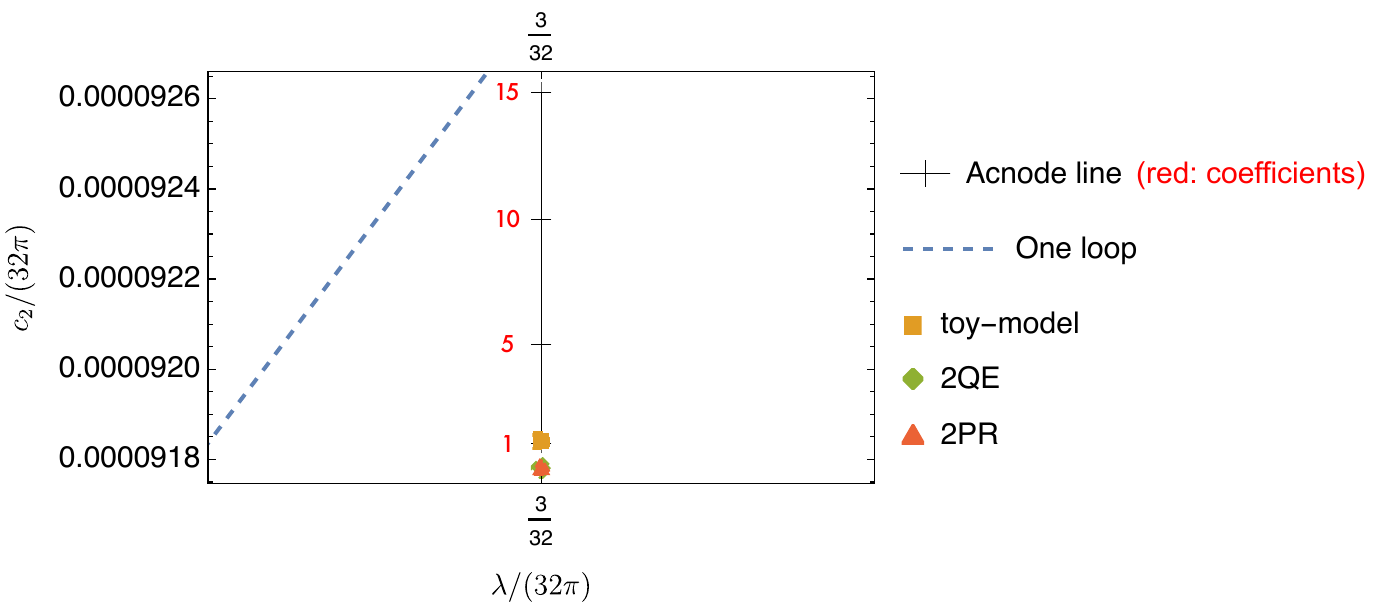}
    \caption{$({\lambda \over 32 \pi},{c_2 \over 32 \pi})$ plane in $d=4$, zoomed around $\lambda=3\pi$. In addition to the amplitudes on the previous plot we add the data points generated by turning on $\rho_{\text{MP}}$ proportional to the double spectral density of the acnode graph (rescaled in a way to have the Landau curve asymptotes to $16m^2$). We see that among the amplitudes constructed in this paper the 2QE amplitude has the minimal $c_2$ for a given $\lambda$. }
    \label{fig:c24dzoom1}
\end{figure}

\begin{figure}[h]
    \centering
    \includegraphics{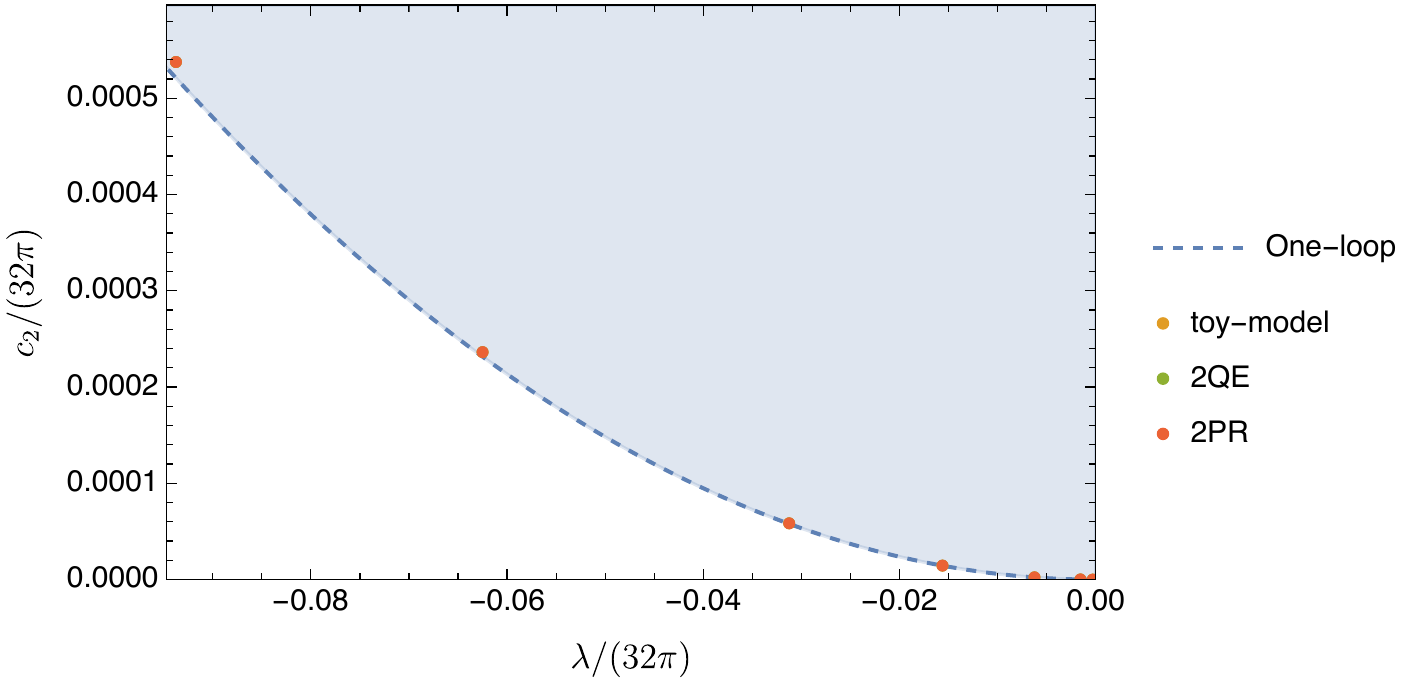}
    \caption{$(\frac{\lambda}{32\pi},\frac{c_2}{32\pi})$ plane in $d=3$. The space of allowed amplitudes, see \cite{Chen:2022nym}, (to a good approximation) lies above the one-loop result in $\phi^4$ theory (dashed line). Various amplitudes constructed in this paper all lie very close to the one-loop result. In particular, on this scale the toy-model, 2QE and 2PR amplitudes are indistinguishable.}
    \label{fig:c23d}
\end{figure}

\begin{figure}[h]
    \centering
    \includegraphics[scale=0.9]{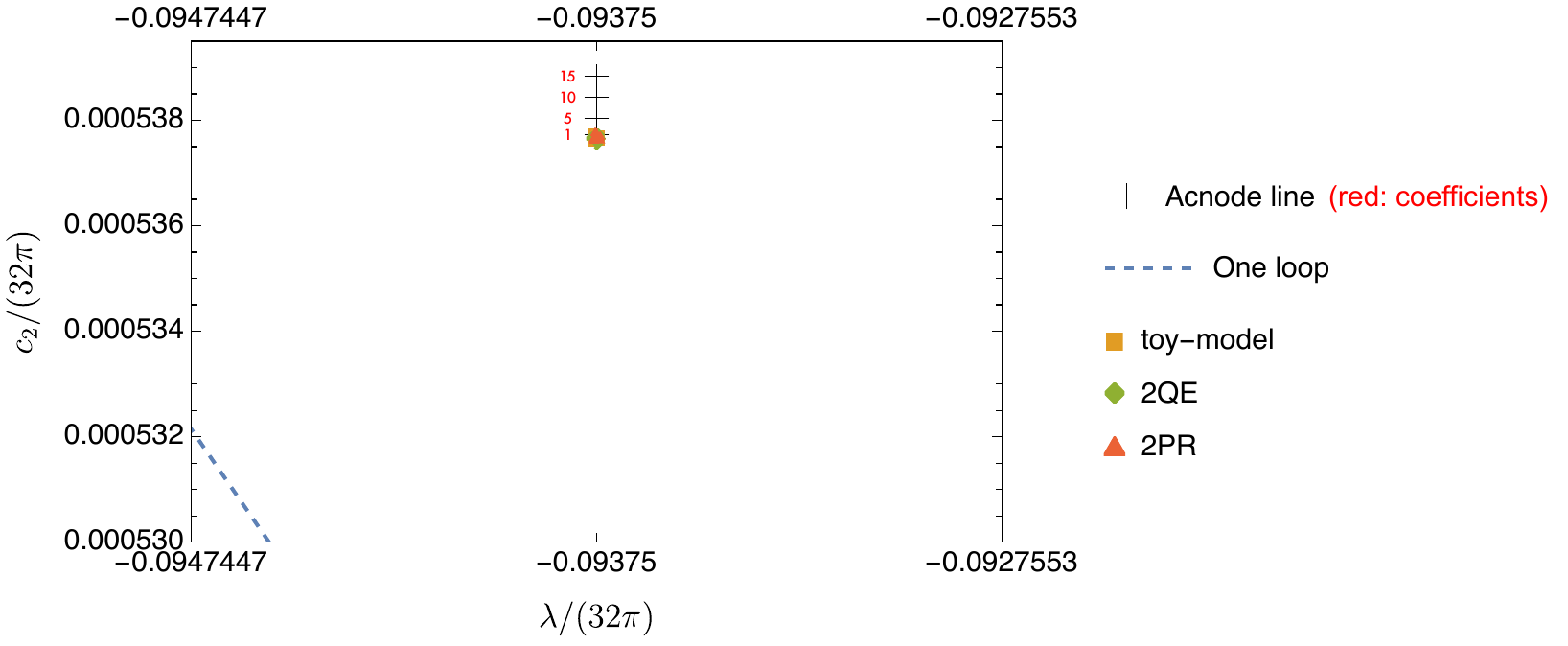}
    \caption{$(\frac{\lambda}{32\pi},\frac{c_2}{32\pi})$ plane in $d=3$, zoomed around $\lambda=-3\pi$.}
    \label{fig:c23dzoom1}
\end{figure}

Here we provide the plots for the $c_2$ observables in $d=4$, \figref{fig:c24d}, \ref{fig:c24dzoom1} and in $d=3$ in \figref{fig:c23d}, \ref{fig:c23dzoom1}.

\FloatBarrier

\subsection{Grid dependence}

One could ask that since our result seem very precise, what is the convergence with the number of points and can one add error bars? In this paper, as was discussed above, the main error we try to quantify is to what precision unitarity is satisfied. 

With the perspective of producing an actual model for the pion-S-matrix, an actual study of the errors needs to be performed. Here, we display the result of varying the $\rho(s,t)$-grid. We investigate the effect of two different cutoffs, $10^{-10}$ and $10^{-12}$, and for these we vary the number of points.

\begin{figure}[h]
    \centering
    \includegraphics[scale=1]{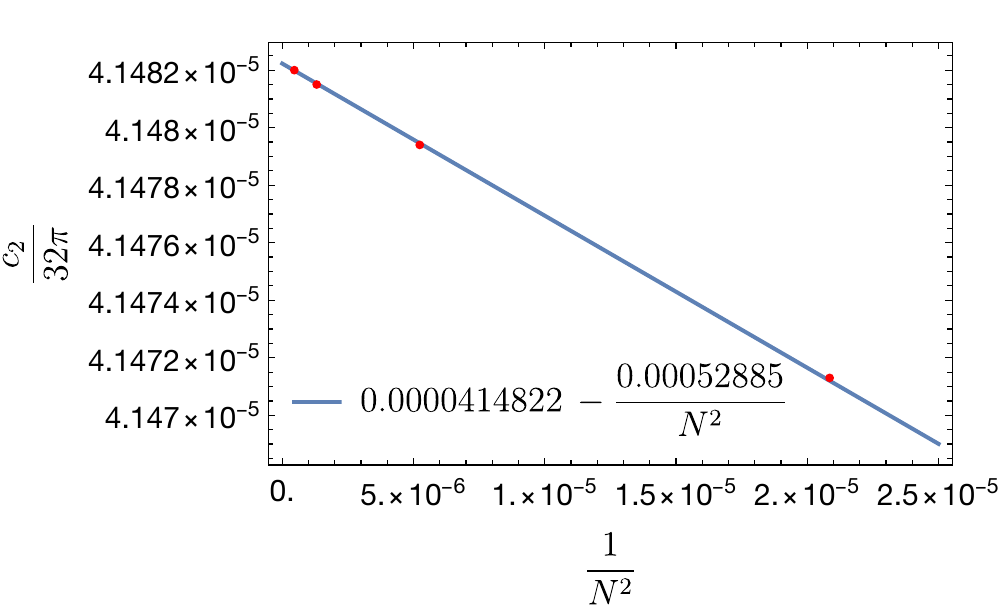}
    \caption{Here we plot the dependence of ${c_2 \over 32 \pi}$ for $\lambda = 2.012434211 \pi $ in $d=4$ on the number of points $N$ in the grid used for the single discontinuity $\rho(s)$. We see that a simple power-like fit agrees with the data very well. The power of $N$ agrees with the expectation value for the error made when approximating an integral with the trapezoidal rule (which is what linear interpolation used in the paper effectively does). As we varied the grid for the single discontinuity we  kept the grid for the double discontinuity fixed.}
    \label{fig:plerrorbar4dc2sgrid}
\end{figure}

\begin{figure}
    \centering
    \includegraphics{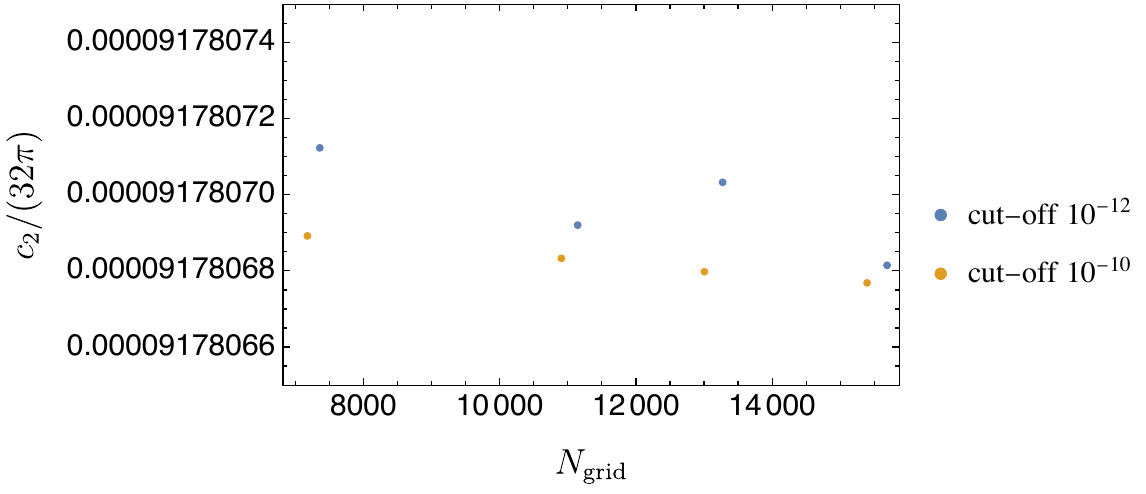}
    \caption{Dependence of $c_2/(32\pi)$ on the number of points $N_{\text{grid}}$ in the double spectral density $\rho(s,t)$ two-dimensional grid. The number of points has been generated by our meshing procedure taking into account some density profile function, see \appref{app:grids}. These grids correspond to $\mathtt{nx}=70,90,100,110$. The two different cutoffs in the Regge limit $10^{-12}$ and $10^{-10}$ are the points closest to zero after which we connect to $0$ with a linear segment, which means that we force a $1/s$ or $1/t$ fall-off at infinity. As discussed in the text, this does not pose a problem in the numerical integrals because they are dispersive, hence have a $1/(s'-s)$ kernel that renders the difference between a linear fall-off or a true logarithmic fall-off very small for such high cut-offs.}
    \label{fig:plc2errorsplotrhost}
\end{figure}

\section{Mandelstam representation and quantum gravity}
\label{sec:mandelstamQG}

It is interesting to ask if methods developed in this paper can be applied to quantum gravity. In this section, we will argue that one important element for our approach, the Mandelstam representation, should not hold in quantum gravity.

The starting point of our analysis is the Mandelstam representation which is based on two a priori independent assumptions about the amplitude:
\begin{itemize}
    \item maximal analyticity
    \item polynomial boundedness in $s$ for any $t$ (and vice versa)
\end{itemize}
Both assumptions are highly non-obvious and have not been proven neither in the context of nonperturbative QFT, nor for gravitational theories. It is interesting to ask a simpler question: is the Mandelstam representation consistent with other properties expected from gravitational amplitudes? Note that the Mandelstam representation trivially does not hold in perturbative string theory \cite{Veneziano:1968yb,Amati:1990xe,Eberhardt:2023xck}, where the polynomial boundedness assumption breaks down: for any $N$, there exists $t>0$ such that the Regge growth $s^{\alpha' t} > s^N$.

In this section we review the old argument of Cerulus and Martin \cite{Cerulus:1964cjb} that relates polynomial boundedness of the amplitude for unphysical values of $t$ to certain properties of the amplitudes in the Regge limit and high energy scattering at fixed angles.\footnote{Recently, this argument was generalized to rely on the axiomatic QFT analyticity only \cite{Epstein:2019zdn}. The resulting bound however is very weak.} While the original argument was done for gapped theories, it is a straightforward exercise to relax this assumption. 

Consider the scattering amplitude as a function of fixed angle $z \equiv \cos \theta$, namely ${\cal T}(s,z) \equiv T \Big(s, t=-{s \over 2}(1-z) \Big)$.
We fix $s$ to be real and positive,\footnote{We approach the real axis from above as usual.} and we consider ${\cal T}(s,z)$ in the complex $z$-plane.
\emph{Maximal analyticity} implies that ${\cal T}(s,z)$ is analytic, modulo the two cuts $z \in (- \infty,-1] \cup [1,\infty)$ which correspond to scattering in the $u$- and $t$-channel correspondingly.

We take three real $z$'s such that $0<z_1<z_2<z_3<1$ and we map the $z$-plane to the $\tau$-plane. We first transform
\be
w(z) = {1 \over z} (1 - \sqrt{1-z^2}) ,
\ee
which maps the cut $z$-plane inside the unit circle in the $w$-plane. We then consider the following map
\be
\tau(w) = {1 \over w(z_1)} ( w+\sqrt{w^2 - w(z_1)^2} ) .
\ee
This mapping maps the region $-z_1 \leq z \leq z_1$ in the $z$-plane to the unit circle $|\tau | = 1$. 

\begin{figure}[h]
    \centering
     \begin{subfigure}{0.5\textwidth}
    \centering
        \includegraphics[scale=0.8]{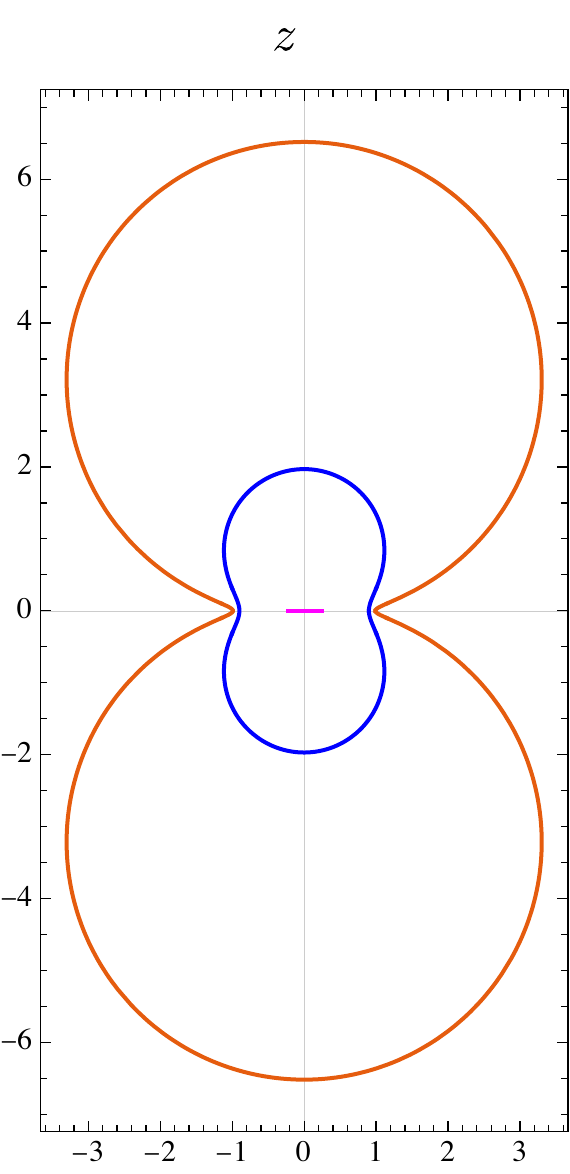}
  \end{subfigure}%
  \begin{subfigure}{0.5\textwidth}
    \centering
     \includegraphics[scale=0.85]{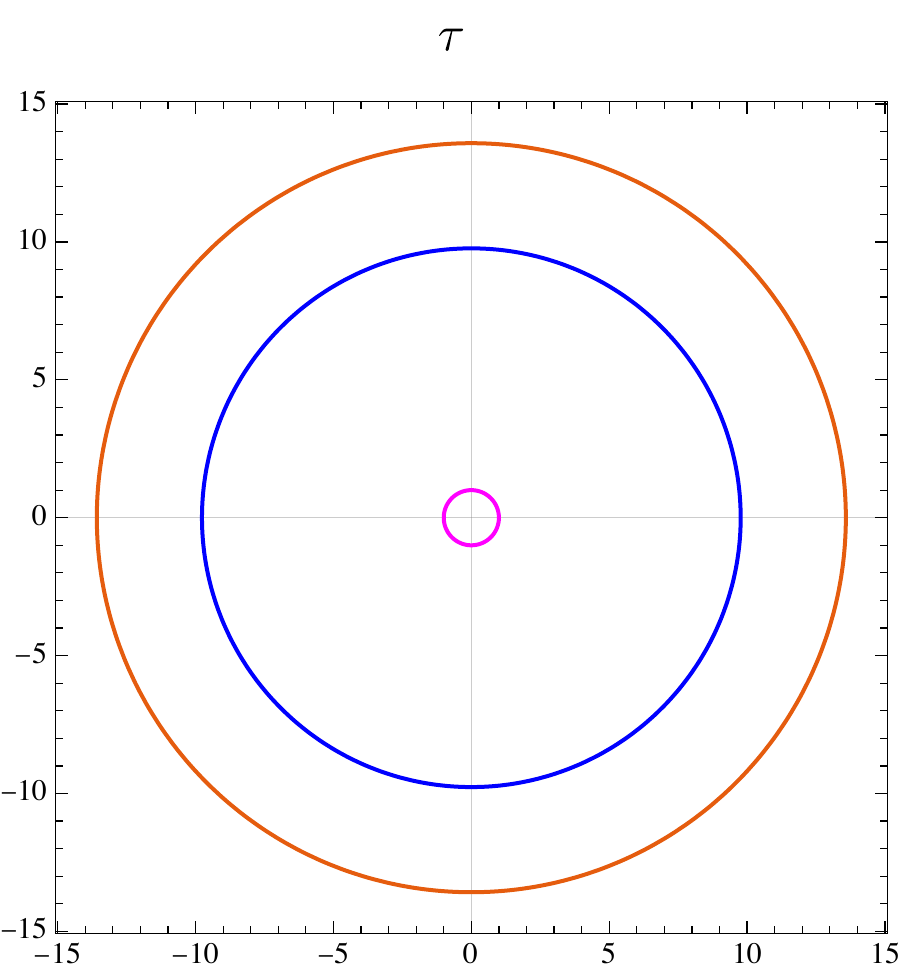}
  \end{subfigure}
    \caption{Here we depict regions in the $z$-plane which map to the concentric circles in the $\tau$-plane. Here we consider $z_1 = {1 \over 4}$ so that the region (magenta) $-{1 \over 4} \leq z \leq {1 \over 4}$ is mapped to the unit circle in the $\tau$-plane. The blue region corresponds to $z_2=0.9$ and the red region comes from $z_3 = 0.99$ in the argument.}
    \label{fig:cermartinmandelstam}
\end{figure}

Consider now another pair of circles in the $\tau$-plane of radii $r_2 = \tau(w(z_2))$ and $r_3 = \tau(w(z_3))$. In the original $z$-plane the circles map in the oval shape region, see \figref{fig:cermartinmandelstam}.

By assumption, maximal analyticity implies that the scattering amplitude is analytic in the annulus $1 \leq |\tau| \leq r_3$.
Let us introduce the maximal value of the amplitude on a given circle
\be
M_r \equiv {\rm max}_{|\tau| = r} | {\cal T}(s, z) | .
\ee
We then apply the Hadamard three-circle theorem, see e.g. chapter 23 in \cite{ullrich2008complex}, that states that for $1 < r_2 < r_3$ we have the following inequality 
\be
\label{eq:inequalitythreecircles}
M_{r_2} \leq M_1^{1 - {\log r_2 \over \log r_3}} M_{r_3}^{{\log r_2 \over \log r_3}}.
\ee
This constraint becomes particularly interesting if we choose $z_{2,3} = 1+{2 t_{2,3} \over s}$, where $t_2<t_3<0$ are fixed momenta and we take $s \to \infty$.

Let us introduce, following Cerulus and Martin, the upper bound on the fixed angle scattering amplitude at high energies
\be
|{\cal T}(s,z)| \leq e^{- \phi(s)}, ~~~ - a \leq z \leq a < 1.
\ee

We can then rewrite \eqref{eq:inequalitythreecircles} as follows
\be
| T(s,t_2) | \leq e^{-C(a) \phi(s) {\sqrt{-t_2} - \sqrt{-t_3} \over \sqrt{s}} } {\rm max}_{|\tau| = r(s,t_3)} | {\cal T}(s, z) |, ~~~ s \to \infty ,
\ee
where $C(a)>0$ and we used that $M_{r_2} \geq | T(s,t_2) | $. This equation bounds the Regge limit of the amplitude, the LHS, in terms of the fixed angle scattering, and the value of the amplitude in certain sub-domain of the $z$-plane, the RHS.

The validity of the Mandelstam representation implies that there should exist an integer $N$, such that
\be
\label{eq:polymand}
{\rm max}_{|\tau| = r(s,t_3)} | {\cal T}(s, z) | \leq |s|^N , ~~~ |s| \to \infty .
\ee
Assuming that $| T(s,t_2) |$ is polynomially bounded from below (it cannot decay too fast), we then get the Cerulus-Martin bound on the fixed angle scattering for amplitudes that admit Mandelstam representation
\be
\label{eq:cermartin}
\phi(s) \leq c_0 \sqrt{s} \log s , ~~~ s \to \infty .
\ee
In other words, maximal analyticity implies the following schematic Cerulus-Martin relation
\be
\text{Regge} \leq \text{Fixed angle} \times \text{Mandelstam}.
\ee

The Regge limit for gravitational amplitudes is controlled by the large impact parameter scattering and its absolute value behaves polynomially in $s$, see e.g. \cite{Muzinich:1987in,Amati:1990xe,Caron-Huot:2022ugt,Haring:2022cyf}. The fixed angle scattering, on the other hand, is believed to be entropically suppressed, see e.g. \cite{Amati:1990xe,Arkani-Hamed:2007ryv,Giddings:2007qq,Bah:2022uyz},
\be
\phi_{\text{QG}} \sim (\sqrt{s})^{1+{1 \over d-3}} ,
\ee
which clearly violates \eqref{eq:cermartin}.

Our conclusion is that the \emph{expected} properties of the scattering amplitude in gravitational theories for physical $t$ (polynomial behavior in the Regge limit and exponentially faster than $\sqrt{s}$ decay for fixed angle scattering) are \emph{not} compatible with polynomial boundedness needed for the Mandelstam representation.

This does not mean that a version of our approach cannot be used to the question of unitarizing gravity or string theory, but simply that the basic iteration presented in the present paper will not work. Another method should be found, or another regime studied, for instance the eikonal one. We believe that the general idea initiated in this paper should be eventually applicable to gravity as well.

Let us also mention, that in confining gauge theories, e.g. in QCD, we get a power-law decay at high energies for fixed angle scattering \cite{Lepage:1980fj,Polchinski:2001tt}, therefore we expect that it should be possible to use the Mandelstam representation to accurately model pion scattering.

\section{Conclusions and open directions}
\label{sec:conclusions}
We now briefly summarize the results of the paper, emphasizing some of the physics we learned on the way. We then move to listing a few open directions.

\subsection{Summary}
\label{sec:summary}

In this paper we have explored models of nonperturbative scattering constructed by iterations of the Mandelstam representation and unitarity. The scattering amplitudes are generated from some input data $(\lambda, \eta_{\text{MP}}(s), \rho_{\text{MP}}(s,t))$: 
\begin{itemize}
    \item  the subtraction constant in the Mandelstam representation \eqref{eq:mandelstamRep1}, $\lambda \equiv T \Big( {4 m^2 \over 3} , {4 m^2 \over 3} \Big)$;
    \item the S-wave ($J=0$) inelasticity $1 \geq \eta_{\text{MP}}(s) \geq 0$ defined in \eqref{eq:partialwavezero}; 
    \item the multi-particle double spectral density $\rho_{\text{MP}}(s,t)$ defined in \eqref{eq:iterationdoublespectral}.
\end{itemize}
As an output we get  amplitude functions
that satisfy: maximal analyticity, crossing, elastic and inelastic unitarity (with the level of precision controlled by the numerical algorithm that we used). The amplitude functions that we get are also UV-complete in the sense that we have control over their behavior at all energies and scattering angles.

The amplitudes that we obtained exhibit interesting physical properties:
Landau curves, particle production, nontrivial Regge and fixed angle behavior, nonperturbative scaling of various quantities with the ``coupling'' $\lambda$. In the present paper we focused mostly on the two classes of amplitudes:
the two-particle quasi-elastic scattering amplitudes (2QE) initialized by $(\lambda, 0, 0)$; and the two-particle reducible amplitudes (2PR) given by $(\lambda,\eqref{eq:analyticityinspinJ0},0)$ for which inelasticity is analytic in spin all the way to $J=0$. The amplitudes obtained in this way are close to saturating certain minimal coupling bounds derived by other methods \cite{Chen:2022nym,EliasMiro:2022xaa}. 

We now list a few notable facts that we learned from studying these amplitudes.

\paragraph{Dispersiveness and sign of $\phi^4$ coupling.}

Interestingly, we have found that for the amplitudes constructed in the paper the subtraction constant $\lambda$ is actually \textit{dispersive}. We call a quantity dispersive when it is given by a dispersion relation integral of a discontinuity of the amplitude with some kernel.
For the quartic coupling $\lambda$, this comes out as a surprise because a naive Cauchy argument $\lambda = \oint {d s' \over 2 \pi i} {T(s',{4 m^2 \over 3}) \over s'-{4 m^2 \over 3}}$
fails. Indeed, as we deform the contour to express $\lambda$ through the discontinuity of the amplitude, we find that we cannot drop the contour at infinity because $T(s,{4 m^2 \over 3}) \to \text{const}$.
The way $\lambda$ turned out to be dispersive is more subtle, and comes through S-wave unitarity, where $\lambda$ contributes to $\Re f_0(s)$.

In three dimensions, elastic unitarity together with the asymptotic $\Im f_0(4m^2)=0$ of our amplitudes imply that $\Re f_0(4m^2)=0$.\footnote{This was argued to be true in the full $d=3$ $\phi^4$ theory in \cite{Chadan:1998qm,Bros:1998tt}.} In four dimensions, our amplitudes satisfy $\Im f_0(\infty)=0$, which through inelastic unitarity implies that $\Re f_0(\infty)=0$. Both $\Re f_0(4m^2)=0$ in $d=3$, and $\Re f_0(\infty)=0$ in $d=4$ provide an extra equation which expresses $\lambda$ through the discontinuity of the amplitude. In the toy model, these were the sum rules \eqref{eq:sumrule3dlambda} and \eqref{eq:sum-rule-rhos} correspondingly. 

In the context of the toy models, for which $\rho(s,t)=0$ and we have from unitarity $\rho(s) \geq 0$, these sum rules immediately fix the sign of $\lambda$ to the one used in the paper ($\lambda <0$ in $d=3$, $\lambda>0$ in $d=4$). For the opposite signs of $\lambda$ we thus conclude that the double discontinuity contribution to the sum rules (and to the amplitude itself) should be essential. 

Note also that the 4d amplitudes constructed in the present paper correspond to the ``wrong'' sign of the potential $V(\phi) = - {\lambda \over 4!} \phi^4$ for which we do not expect to have a stable vacuum. Of course, the construction of the present paper concerns itself only with a particular observable, namely the $ 2 \to 2$ scattering amplitude, and it does not in any sense define a theory. Moreover, we can imagine a theory where negative $\phi^4$ potential emerges as a leading low-energy approximation of a healthy UV complete theory.\footnote{We thank Joan Elias Miro for discussions on this point.} Still, if we are to insist on having ``pure'' $\phi^4$ theory with an unbounded potential we do not expect to have a consistent unitarity S-matrix. At the level of the amplitudes constructed in the paper we did not see any inconsistencies. It would be however very interesting to understand how vacuum instability can be detected in the S-matrix language.\footnote{We thank Nima Arkani-Hamed for discussions on this point.} 

\paragraph{Non-trivial Regge behavior}
The amplitudes constructed in the paper exhibit a nontrivial Regge behavior. For toy models we could solve for the Regge limit analytically at a price of setting $\rho(s,t)=0$. We then observed that the Regge limit gets modified as we consider the full back-reacted solution with $\rho(s,t) \neq 0$ for which we do not have an analytic control. It would be interesting to develop better understanding of the Regge behavior of the amplitudes constructed in the present paper.

\paragraph{Transition between Regge / fixed-angle regimes.} Continuing with the high-energy limits of the amplitudes, we also observed a very clear transition from Regge to the fixed-angle regime. We recall that the Regge limit corresponds to fixed $t$, large $s$, while the fixed-angle is large $s$ and large $-t$, with $s/t$ fixed. Since the scattering angle is defined by $\cos(\theta) = 1-\frac{2t}{s-4m^2}$, these two regions correspond to $\theta\ll1$ and $\theta\sim O(1)$ correspondingly, and in theory as $|t|$ is increased, or equivalently $\theta$, scattering amplitudes should transition from Regge to fixed-angle behavior. 

We observed a very clean transition, both in $d=3$ ad $d=4$ for 
\begin{equation}
    \frac{s}{m^2}\sim \theta^{-2}
\end{equation}
which is illustrated in \figref{fig:amp-toy-model-3d}, \figref{fig:allangles-rhos-4d}. This regime is reached when $t/m^2 \sim O(1)$, which in impact parameter space corresponds to $b\sim 1/m$.\footnote{This relation holds for our amplitudes because they have a simple Regge limit, $T(s,t)\to \text{const}$ which leaves the saddle-point of the impact-parameter Fourier transform unchanged, $T(s,b) \sim \int d^2 b e^{-i\vec q \dot \vec b} T(s,-q^2) $. For reggeized string-like particles for which $T(s,t)\sim s^{-\alpha' t}$, $t\sim m^2$ corresponds to $b\sim \log(s)$ which is the famous logarithmic transverse growth of strings at high-energies.\label{footnote_b_def}}

\subsection{Open directions}
\label{eq:open}

There are many future important directions to explore,  we list some of them below:

\paragraph{Are the two-particle quasi-elastic (2QE) amplitudes physical?} In this paper we have argued that there exist a natural one-parameter family of amplitudes defined by the triad $(\lambda,0,0)$, with no bound states below the first threshold. We have also observed that these amplitudes are located very close to the bounds on couplings derived by other methods~\cite{Chen:2022nym,EliasMiro:2022xaa}. Apart from the question of whether the 2QE amplitudes are truly extremal on which we comment below, one obvious question is: are they clearly unphysical? 
Recall that they are defined by the requirement $\eta_{MP}(s)=0$, which means that in their graphic, $\phi^4$-like expansion, the $J=0$ part of any graph with a multiparticle-cut is projected out (the simplest of which being the Aks graph and the frog graph). Naively this implies a certain level of non-locality to be present in the theory. 
However, at the level of the constraints imposed in this paper there is no problem in doing so. In other words, this non-locality is compatible with the constraints imposed on the scattering amplitude in the present paper.  We expect that the 2QE amplitudes could be excluded using multi-particle unitarity and it would be very interesting to show this explicitly. 

An obvious extension of the present work is to consider 2QE amplitudes in the presence of bound states. For example, it would be interesting to see if the 2QE amplitudes with a bound state at the two-particle threshold, as first observed in \cite{Paulos:2017fhb}, are close to saturating the bound on the maximal coupling found in \cite{Chen:2022nym,EliasMiro:2022xaa}. A preliminary attempt of ours to implement the fixed-point algorithm in this case appears to diverge. We observed some similar phenomena in $d=2$ before \cite{Tourkine:2021fqh}, and there we restored convergence by implementing the Newton-Raphson method. It would be interesting to try this here as well.

Turning to gravity, it would be very interesting to construct quasi-elastic amplitudes in this case as well. These tentative amplitudes would be both crossing-symmetric and eikonalize correctly.

\paragraph{Nonperturbative unitarization via two-particle reducible resummation?} Another interesting case that we considered is given by the following choice of the iteration data $ (\lambda,\eqref{eq:analyticityinspinJ0},0) $. In this scheme,  $J=0$ inelasticity is fixed using the Froissart-Gribov formula.

This corresponds to re-summation of the two-particle reducible graphs in $\phi^4$ theory which are generated by iterations of two-particle unitarity and crossing, see graphs in black in \figref{fig:diagramspic}. These graphs are the only graphs in $\phi^4$ theory at one and two loops, but starting from three loops there are two-particle irreducible graphs, namely the frog and the open envelope which have $\eta_{\text{MP}}(s) \neq 0$ and $\rho_{\text{MP}}(s,t) \neq 0$. As we proceed to higher loops, we generate a web of planar diagrams made of bubbles, triangles, and various connected ladders of those. 

Since the planar sector of QFTs is usually expected to be summable and for $\phi^4$ is even known to have a finite 
radius of convergence \cite{Brezin:1977sv}, one can already suspect that a perturbative scheme could work in our set-up. 
As a matter of fact, a simple diagrammatic counting given in \secref{sec:algorithm} and with explicit examples up to $n=10$ in \tabref{table:graphs}, indicates that the number of such planar, 
two-particle reducible graphs grows indeed like a power, which, by standard reasoning suggests a finite radius of convergence.
Indeed, this is what we also find numerically, since we observe convergence of that scheme for finitely many couplings, summarized in~\tabref{tab:conv}.

An interesting result of our analysis is that such a truncation of the Feynman series is self-consistent at the level of the two-to-two scattering amplitude in the following sense: it generates the amplitude function $T(s,t)$ that satisfies both elastic and inelastic unitarity, as well as crossing. It would be very interesting to study this unitarization scheme in other models. It would be also interesting to understand if similar iteration techniques can be applied to re-sum families of nPR planar graphs, e.g. melonic graphs of which the frog diagram is the first example.

\paragraph{Are extremal amplitudes physical?} An interesting problem is to find amplitudes that satisfy the basic constraints studied in the paper and minimize/maximize certain couplings. In \secref{sec:lowenergy}, we have found that all the amplitudes constructed in this paper lie close to saturating the bounds on the low-energy observables derived in \cite{Chen:2022nym,EliasMiro:2022xaa}. 
The most extremal amplitudes among the ones constructed in this paper are the 2QE amplitudes initialized with zero multi-particle input $\eta_{\text{MP}}(s)=\rho_{\text{MP}}(s,t)=0$. Understanding if the QE amplitudes are truly extremal would require a search in the space of multi-particle inelasticities encoded by $(\eta_{\text{MP}}(s),\rho_{\text{MP}}(s,t))$. We see the following possibility to perform this search. We can start with one of the nontrivial solutions found in this paper, and then explore infinitesimal variations $\delta \eta_{\text{MP}}(s)$ and $\delta \rho_{\text{MP}}(s,t)$ while choosing a direction in which the coupling decreases (if there is one). At the extremum, we expect to find \eqref{eq:extremal}.

Finally, the works \cite{Paulos:2017fhb,Chen:2022nym,EliasMiro:2022xaa} observed  inelasticity very different from ours. In that case  it seems to be generically pushed to the UV. For us, inelasticity comes at low/medium energy due to graphs with simple multi-particle cuts. It would be interesting to understand if our approach can also push inelasticity to the UV and see what constraints this poses on $ \rho_{\text{MP}}(s,t)$.
Conversely, we can also imagine using the results of the present paper as an input to the primal S-matrix bootstrap, see e.g. \cite{Antunes:2023irg} for a related discussion.

\paragraph{Is inclusion of multi-particle unitarity feasible?} The existing S-matrix bootstrap methods focus on the $ 2 \to 2$ scattering. In the same spirit, in the present paper, the multi-particle information, contained in $\eta_{\text{MP}}(s)$ and $ \rho_{\text{MP}}(s,t)$, was used as an input. It is however interesting to notice that our approach has a natural extension in which some of the multi-particle information enters the iteration algorithm. The simplest one concerns the nontrivial support of $\rho_{\text{MP}}(s,t)$ which is very easy to impose in our approach, that is to say the choice of the Landau-curve above which $\rho_{\text{MP}}(s,t)$ is non-vanishing, see \cite{Correia:2021etg}.

For more complicated constraints, consider, for example, the open envelope diagram in \figref{fig:threeloopgraphsforswaveinelasticity}. As reviewed for example in \cite{Correia:2021etg}, in the nonperturbative theory this diagram leads to an equation which schematically takes the following form
\be
\label{eq:openenvelopenonperturbative}
\rho_{\text{MP}}(s,t) \stackrel{?}{=} \int K_4 (T_{2 \to 2})^2 (T_{2 \to 2}^*)^2 , 
\ee
which holds for the full nonperturbative amplitude in some finite region of the Mandelstam plane. It would be important to try implementing \eqref{eq:openenvelopenonperturbative} and see how computationally costly it is.

In this paper we analyzed the simplest example of this type of relation in \secref{sec:acnode}, where we derived the double spectral density of the acnode graph. An interesting outcome of this analysis is that the two-to-two scattering amplitudes that enter into the relation analogous to \eqref{eq:openenvelopenonperturbative} turned out to be evaluated on the second sheet reached through the elastic unitarity cut. As recently discussed in \cite{Mizera:2022dko}, the amplitude there in principle can also be found by iterations very similar to the one considered in the present paper. It would be very interesting to derive the nonperturbative relation of the type  \eqref{eq:openenvelopenonperturbative} explicitly (as well as the one for the acnode graph) and to implement them numerically. As for $\eta_{\text{MP}}(s)$, there are similar relations which require further exploration. For example cutting the frog graph, we get $T_{2 \to 4}$ diagram which contains, schematically, a pole ${T_{2 \to 2}^2 \over s_{123}-m^2}$, where $s_{123}$ is an invariant mass of the three produced particles. We can imagine adding this contribution to our iteration scheme for $\eta_{\text{MP}}(s)$. 

Lastly, for $\eta_{\text{MP}}(s)$, we can imagine taking experimental data as an input given that it has a very simple physical meaning, the situation is more complicated for $\rho_{\text{MP}}(s,t)$ whose moments are related to particle production in partial waves with spin via the Froissart-Gribov formula. It would be also interesting to explore if semi-classical methods for multi-particle production, see e.g. \cite{Libanov:1994ug,Son:1995wz,Demidov:2022ljh}, could shed some light on multi-particle input of the iterative algorithm.

\paragraph{Iteration methods and various physical models.} Given that in our algorithm physics is naturally encoded in the multi-particle data, it would be very helpful to understand how various familiar physical models are encoded in the $(\lambda, \eta_{\text{MP}}(s), \rho_{\text{MP}}(s,t))$: amplitudes that saturate the Froissart bound \cite{Kupsch:1982aa,Kupsch:2008hq}, pion scattering \cite{Martin:1969wv}, scattering in Chern-Simons matter theories \cite{Mehta:2022lgq}, stringy amplitudes \cite{Eberhardt:2023xck}, gravitational amplitudes.\footnote{For the case of gravitational amplitudes, an interesting question is: do quantum gravity amplitudes satisfy the Mandelstam representation? Our conclusion in \secref{sec:mandelstamQG} based on the lore about the high-energy fixed angle behavior of scattering amplitude in gravity is that the answer to this question is ``No''.} It would be then interesting to generalize the methods applied in this paper to these theories. 

\paragraph{Taking the continuum limit.} The algorithm we presented in this paper is based on discretizing the space of Mandelstam variables $4m^2\leq  s,t <\infty$
and then interpolating between the grid points. To make sure that the amplitudes of interest exist we would like to argue that the amplitudes that
we have constructed numerically admit a continuum limit. We have considered two approaches to this problem, which we have explained in the text. First, we have increased the grid resolution and checked
that the results do not change in any significant way. Relatedly, we increased the grid cutoffs and checked that the amplitudes that we interpolate smoothly to the IR in $d=3$ and to the UV in $d=4$. Secondly, we looked into convergence of the iteration algorithm in the space of continuous functions directly. In the latter case we showed convergence for the toy model discussed in \secref{sec:toymodel} in $3<d<4$. Extending the existing functional analysis proofs to $d=3$ and $d=4$ requires developing more sophisticated methods which go beyond what has been done in the literature or in the present paper.

\paragraph{Increasing the convergence range.} 
One feature of the amplitudes presented in this paper is that both particle production and the coupling constant, as measured by $\lambda$, are relatively small. The amplitudes that we obtained are also weakly coupled at high energies in the sense that partial waves $S_J(s) \to 1$ as $s \to \infty$ (but $T(s,t) \to \text{const}$ for fixed $t$ and $s\to \infty$).  It would be very interesting to extend the method of the present paper to larger couplings and inelasticities.

To approach this class of amplitudes, we need to generalize the simple fixed point method developed in the present paper. 
In general, there should be nothing fundamental about the \emph{iterative solution} to these functional equations. Any method numerical functional analysis that can work is equally good.

For instance, in $d=2$~\cite{Tourkine:2021fqh}, we observed that gradient methods such as Newton-Raphson allowed us to extend the range of convergence of the iterative algorithm, and to describe new-topologies of amplitudes (with extra zeros or poles), sensitive to the starting point of the algorithm. We therefore expect that such methods should also improve the performance of our algorithm in $d=3,4$. One immediate bottleneck relates to the discretization procedure of our numerical integrals. One-dimensional piecewise-linear integrals are not really integrals, they can be performed analytically, because on each segment, they amount to integrating a combination of a constant and a linear function times an integration kernel: all these integrals can be done beforehand and the whole integration of an interpolating function $\rho(x)$, $\int_0^1 K(x',x)\rho(x') dx' $ becomes the action of a matrix on the vector $\rho(0),\rho(x_1),\dots,\rho(x_{n-1}),\rho(1)$ (for more details, see~\cite{Paulos:2016but,Tourkine:2021fqh}). For two-dimensional integrals, our grids intersect the integration domains in some non-trivial way and a similar procedure is not immediately applicable. This problem should be solved, in order for a gradient descent or Newton-Raphson method to be applied in the present case of $d=3,4$.
Lastly, it would be very interesting to explore the possibility of using modern machine learning methods in the present context~\cite{Schwartz:2021ftp}.

\begin{acknowledgments}
  We thank Nima Arkani-Hamed, David Atkinson, 
  Pierre Aubert, Lucia Cordova, Miguel Correia, Joan Elias Miro, Giulia Isabella, Hofie Hannesdottir, Kelian H\"{a}ring, Enrico Herrmann, Andreas Juettner, Denis Karateev, Madalena Lemos, Andrew McLeod, 
  Sebastian Mizera, Julio Parra-Martinez, Joao Penedones, Jiaxin Qiao, Balt van Rees, %
  Matthew Schwartz, Amit Sever, Jaroslav Trnka, Pierre Vanhove, Pedro Vieira, Matt Walters, Xiang Zhang for useful discussions. We thank the authors of \cite{Chen:2022nym,EliasMiro:2022xaa} for sharing and explaining their detailed results to us. We thank Dmitrii Vorobev and Dmitri Vasiliev for exploring the $d$-dimensional formulae as a part of the Physics Practicum 2022. This project has received
  funding from the European Research Council (ERC) under the European Union's Horizon 2020 research and innovation programme (grant
  agreement number 949077) and from Agence Nationale de la Recherche (ANR), project ANR-22-CE31-0017. This research was supported in part by the National Science Foundation under Grant No. NSF PHY-1748958.
  % For the purpose of open access, the authors have applied a CC-BY public copyright licence to any Author Accepted Manuscript (AAM) version arising from this submission.
\end{acknowledgments}

\newpage 
\appendix

\section{Definitions}
\label{app:definitions}

\paragraph{Partial wave expansion}
Here we summarize our conventions for the partial wave expansion following \cite{Correia:2020xtr}:
\be
\label{eq:nJd}
T(s,t)&=\sum_{J=0}^{\infty} n_J^{(d)}f_J(s)P_J^{(d)}\(\cos\theta\)\ , \\
P^{(d)}_J(z) &= {}_2F_1\(-J,J+d-3,\frac{d-2}{2},\frac{1-z}{2}\)\ . \nn 
\ee
The partial wave coefficients can be found from the amplitude using the orthogonality relation
\be
{1\over2}\int\limits_{-1}^1 d z\, (1-z^2)^{{d-4 \over 2}} P^{(d)}_J(z) P^{(d)}_{\tilde J}(z) ={\delta_{J \tilde J}\over {\cal N}_d\, n_J^{(d)}}\ .
\ee
The normalization coefficients are taken to be
\beq
\label{eq:nJd2}
{\cal N}_d={(16\pi)^{2-d\over2}\over\Gamma\({d-2\over2}\)}\ ,\qquad n_J^{(d)}=\frac{(4\pi)^{d\over2}(d+2J-3) \Gamma (d+J-3)}{\pi\,\Gamma \({d-2\over2}\) \Gamma (J+1)}\ .
\eeq
Using these formulas we get for scattering of identical particles
\beq
\label{eq:pwprojection}
f_J(s)={{\cal N}_d\over2} \int\limits_{-1}^1dz\,(1-z^2)^{d-4\over2}P_J^{(d)}(z)\,T\(s,t(z)\)\ , ~~~ t(z) = -{s-4m^2 \over 2}(1-z) \ .
\eeq

The $Q$-functions that enter the Froissart-Gribov formula
\beq\label{eq:FGformula}
 f_J(s) ={2 \, {\cal N}_d\over\pi}\int\limits_{z_1}^\infty d z \,(z^2-1)^{{d-4 \over 2}}Q^{(d)}_J(z) T_t(s,t(z))\  , ~~~ z_1 = 1+{4 m^2 \over s-4m^2}, 
\eeq
take the following form
\beq
Q^{(d)}_J(z)=\frac{\sqrt{\pi}\Gamma(J+1)\Gamma(\frac{d-2}{2})}{2^{J+1}\Gamma(J+\frac{d-1}{2})} {1\over z^{J+d-3}} {}_2F_1\(\frac{J+d-3}{2},\frac{J+d-2}{2},J+\frac{d-1}{2}, {1 \over z^2}\)\ .
\eeq

\paragraph{Spin zero partial wave in $d$ dimensions}
\label{sec:s0-projection}

The $J=0$ partial wave for an amplitude given by a once-subtracted Mandelstam representation \eqref{eq:mandelstamRep1}, in general $d$, can be computed to be:
\be
\label{eq:spinzeroAgenerald}
f_0(s) &={(16 \pi)^{{3-d \over 2}} \over 8 \Gamma({d-1 \over 2})} \left( \lambda + \int_{4 m^2}^\infty {d s' \over \pi} { (s-s_0)\rho(s') \over (s'-s)(s'-s_0)} \right. \nn \\
 &+\left. 2  \int_{4 m^2}^\infty {d t' \over \pi} \rho(t') [ {1 \over t_0 - t'} + {1 \over t'} \ _2 F_1 (1, {d-2 \over 2}, d-2, {s-4m^2 \over t'})\right. \nn \\
&+ \left. 2 \int_{4 m^2}^\infty {ds' d t' \over \pi^2} {(s-s_0) \rho(s',t')  \over (s'-s)(s'-s_0)} [ {1 \over t_0 - t'} + {1 \over t'} \ _2 F_1 (1, {d-2 \over 2}, d-2, {s-4m^2 \over t'})]  \right. \nn \\
&+ \left. \int_{4 m^2}^\infty {dt' d u' \over  \pi^2} {\rho(t',u') \over (t'-t_0)(u'-u_0)}  [1 - {2 A_2^{(d)}(s,t',u') \over s+t'+u' - 4 m^2 }] \right) , \nn \\
 A_2^{(d)}(s,t',u')&=\frac{\left({t_0}- t'\right) \left({s_0+t_0} - (s+t')\right) \, _2F_1\left(\frac{1}{2},1;\frac{d-1}{2};\frac{\left(s-4 m^2\right)^2}{\left(-4
   m^2+s+2 t'\right)^2}\right)}{-4 m^2+s+2 t'} \nn \\
   &+\frac{\left({u_0} - u'\right) \left({u_0+s_0}- (s+u')\right) \,
   _2F_1\left(\frac{1}{2},1;\frac{d-1}{2};\frac{\left(s-4 m^2\right)^2}{\left(-4 m^2+s+2 u'\right)^2}\right)}{-4 m^2+s+2 u'} \ ,
\ee
where recall that in the text we use $s_0=t_0=u_0={4 m^2 \over 3}$.

\section{Grid generation}
\label{app:grids}

In this section, we explain the Mathematica routines that we used to generate our meshes. They rely on the use of the Finite Element Methods package, that can be called with \texttt{<<NDSolve`FEM`}. Mathematica documentation is provided at \url{https://reference.wolfram.com/language/FEMDocumentation/tutorial/ElementMeshCreation.html}. 

This package is very handy and in particular it provides interpolating routines via the command \texttt{ElementMeshInterpolation}.

\paragraph{1d meshes}

The one-dimensional meshes are straightforward to generate. A grid with \texttt{Nelem} elements and Gradient Ratio \texttt{r} is generated as follows:

\begin{mmaCell}[morefunctionlocal={x}]{Input}
<<NDSolve`FEM`
\end{mmaCell}
\begin{mmaCell}{Input}
meshx=ToGradedMesh[Line[\{\{0\},\{1\}\}],<|"Alignment"->"BothEnds","ElementCount"->Nelem,"GradingRatio"->r|>,"MeshOrder"->1];
\end{mmaCell}
The grading ratio $r$, is quantity that enforces some exponential density in the grid, near the extremities (through the \texttt{"Alignment"$->$"BothEnds"} option, which can also be set to \texttt{Right} or \texttt{Left}, see documentation). When $r=1$ the grid is equally spaced, when $r$ increases, more points are set to the extremities.

Note that we encountered interpolation troubles with non-linear interpolation. Mathematica  struggled with the singular behavior near $0$ in the $x,y$ variables, and we had to resort to enforcing linear interpolation everywhere, through the \texttt{"MeshOrder"$->$1} command.

\paragraph{2d, $\rho(s,t)$ mesh}

The most delicate grid to generate in our work is the 2d grid for $\rho(s,t)$. This grid should have an exponential density of points near the Regge region, but as few points as possible in the bulk, where not much physics happens. In particular, we empirically observed that the double spectral densities generated by the iteration algorithm vary slowly at fixed $y$ in the bulk, $x\in[0.1,0.9]$. One could have thought that many points are needed close to the Landau curves too. However due to elastic unitarity the spectral densities decay close to the leading Landau curve in a simple known fashion, see \cite{Correia:2020xtr}, and correspondingly we have observed in our analysis that this region does not require any special care.

The \texttt{FEM} package has a useful command \texttt{ProductMesh} that takes two meshes and produce a tensorial product of them. The elements of this tensor-product are quadrilateral elements which are rectangles. 
However, we are not interested in a mesh that covers the whole range $[0;1]\times [0;1]$, we only need to cover the wedge below the Landau curve $y=(1-x)/4$ for the double spectral density $\rho_{el}(x,y)$, see \secref{sec:num-impl}. 

The construction starts from a product of two meshes covering $[0;1/4]$ and $[0;1]$. Then one  removes points above the Landau curve, add the boundary Landau curve. At this point we have a grid of points, from which Mathematica can in theory generate a new mesh.

This procedure turns out to not work well, because given a set of points, the automatic meshing occurs through ``most neighboring points''. Since the density of points in the $x$ and $y$ directions can be very different in some regions, this sometimes gives rise to  triangle mesh elements, especially near the Landau curve, which are very unnatural.

Therefore the re-meshing has to be done by hand. One keeps all the quadrilateral elements which are not cut by the Landau curve in the original tensor product, and constructs the triangle elements that connect these quadrilaterals to the Landau curve. This gives rise to a mesh such as the one showed in \figref{fig:mesh}, with a mixture or rectangles and triangles, which now looks uniform.

Overall,  we have the following steps to generate the two-dimensional grid for the double spectral density
\begin{enumerate}
    \item Generate x-grid \texttt{meshX},
    \item Generate y-grid \texttt{meshY},
    \item Perform tensor product,
    \item Remove points above the Landau curve and add points on the Landau curve boundary.
\end{enumerate}

In our model, there are a few parameters which determine the grids. Let us enter the details of each grids. 

\paragraph{x-grid}
This grid describes the interval $[0;1]$. It is mostly adapted to probe the Regge limit in $d=4$. It is parametrized by 3 parameters, \texttt{nx,mdx,maxD} which are, respectively, the number of points, the Regge cutoff, and the maximal spacing in the bulk. The grid glues three meshes: one with exponential density of points to the cutoff that spans 0 to some number determined by \texttt{nx}, another linear one for the bulk, and another rigidly fixed grid with 30 points that ends on $1-10^{-12}$. We work with fixed $\texttt{maxD}=1/15$ and $\texttt{mdx}=10^{-12}$ in general.

\begin{mmaCell}{Input}
xmesh[nx_,mdx_,maxD_]:=Module[\{r0,imax\},
(*From the number of points nx and the minimal distance mdx, determine a grading ratio r0 so as to glue the first exponential grid to the second uniform grid at xmax:*)
r0=FindRoot[(1/2)*(r^(1/nx) - 1)/(r-1)-mdx==0, 
\{r,1.0000000000001,10*1/mdx\}, Method -> "Brent",WorkingPrecision->100][[1, 2]];
imax=Floor[FindRoot[(r0^((i+1)/nx) - r0^(i/nx) )/(r0-1)==maxD,\{i,1,nx\}][[1,2]]];
(*Solve xmax*)
xmax=(r0^((imax)/nx) - 1 )/(r0-1);
ToGradedMesh[\{\{Line[\{\{0\},\{xmax\}\}],<|"Alignment"->function,"ElementCount"->nx,"MinimalDistance"->mdx|>\},\{Line[\{\{xmax\},\{80/100\}\}],<|"Alignment"->"Uniform","MinimalDistance"->maxD|>\},\{Line[\{\{80/100\},\{1-10^-12\}\}],<|"Alignment"->functionReverse,"ElementCount"->30,"MinimalDistance"->10^-5|>\}\},"MeshOrder"->1]]
\end{mmaCell}

where the density functions had to be defined by hand as:
\begin{mmaCell}{Input}
Options[function]=\{"ElementCount"->Automatic, "MinimalDistance" -> Automatic\};
function[\{start_,end_\}, OptionsPattern[]]:=
Module[
\{n,len,r, base, minD\},
len=end-start;
n=OptionValue["ElementCount"];
minD=OptionValue["MinimalDistance"];
r =Abs@FindRoot[len*(r^(1/n) - 1)/(r-1)-minD==0, 
\{r,1.0000000000001,10*1/minD\}, Method -> "Brent",WorkingPrecision->100][[1, 2]];
base=Table[(r^(i/n) - 1)/(r - 1),\{i, 0, n\}];
base=len*base + start;
base
]
\end{mmaCell}
 and 
\begin{mmaCell}{Input}
Options[functionReverse]=\{"ElementCount"->Automatic, "MinimalDistance" -> Automatic\};
functionReverse[\{start_,end_\}, OptionsPattern[]]:=
Module[
\{n,len,r, base, minD\},
len=end-start;
n=OptionValue["ElementCount"];
minD=OptionValue["MinimalDistance"];
r =Abs@FindRoot[len*(r^(1/n) - 1)/(r-1)-minD==0, 
\{r,1.0000000000001,10*1/minD\}, Method -> "Brent",WorkingPrecision->100,MaxIterations->10000][[1, 2]];
base=Table[(r^(i/n) - 1)/(r - 1),\{i, 0, n\}];
base=end-len*base;
base//Reverse
\end{mmaCell}
 
\paragraph{y-grid}
The y-grid has 2 parameters, \texttt{ny,mdy} is more straightforward. It just has \texttt{ny} points with an exponential density of points to the cutoff given by \texttt{mdy}, which in general we set to $10^{-12}$, to be consistent with the $x$ cut-off.

\begin{mmaCell}{Input}
ymesh[ny_,mdy_]:=ToGradedMesh[Line[\{\{0\},\{1/4\}\}],<|"Alignment"->function,"ElementCount"->ny,"MinimalDistance"->mdy|>,"MeshOrder"->1];
\end{mmaCell}

\paragraph{Landau-curve points}
Finally, we add points on the Landau curve which are located at the position of the points on the x-grid, i.e. they are located at $(0,1/4),(x_1,lc(x_1)),\dots,(1,0)$ where $lc(x)=(1-x)/4$, is, again, the Landau curve that defines the domain in which $\rho_{el}(x,y)$ is non-vanishing.

\section{Elastic amplitudes in $d=2$: product of CDD zeros}
\label{app:CDDs}

This appendix reviews some material presented in \cite{Tourkine:2021fqh}.

In two spacetime dimensions there is no momentum transfer, $t=0$, and we only have $s\leftrightarrow u$ crossing symmetry. In this case, eq.~\eqref{eq:unitarity0spin} describes a consistent scattering amplitude, but without particle production. Such S-matrices are well known in two spacetime dimensions and they are given by the product of the so-called CDD factors, see e.g. \cite{Paulos:2016but}. These S-matrices can correspond either to poles, or zeros. Given our assumption that there are no bound states, the relevant S-matrices are given by a product of the CDD zeros
\be
S(s) &= \prod_{i=1}^{N_{\text{tot}}} S_{\text{CDD}}^{\text{zero}}(s,m_i) , \nn \\
S_{\text{CDD}}^{\text{zero}}(s,m_z) &={\sqrt{s(4m^2-s)}-\sqrt{m_z^2(4m^2-m_z^2)} \over \sqrt{s(4m^2-s)}+\sqrt{m_z^2(4m^2-m_z^2)}} .
\ee
Following the conventions of \cite{Tourkine:2021fqh}, we can define the connected scattering amplitude $T(s)$ as follows
\be
S(s) = 1+i {T(s) \over \sqrt{s(s-4m^2)}}. 
\ee
The connected amplitude $T(s)$ satisfies the unitarity condition
\be
2 {\rm Im} T(s) = {1 \over \sqrt{s(s-4m^2)}} | T(s) |^2 ,
\ee
as well as crossing
\be
T(s) = T(4m^2 - s).
\ee
At large energies $|s| \to \infty$, adding the correct $i\epsilon$ prescription, one sees easily that $T(s)$ goes to a constant given by
\be
c_\infty = - 2 \sum_{i = 1}^{N_{\text{tot}}} \sqrt{m_{i}^2 (4 m^2 - m_i^2)} . 
\ee
In this way we can write a single subtracted dispersion relation
\be
\label{eq:dispersive2d}
T(s) = c_\infty + \int_{4 m^2}^\infty {d s' \over \pi} {\rm Im} T(s') \Big( {1 \over s'-s} + {1 \over s' - (4 m^2 - s)} \Big), 
\ee
where at large energies $ {\rm Im} T(s') \sim {1 \over s'}$ so that the dispersive integral in \eqref{eq:dispersive2d} converges. The representation \eqref{eq:dispersive2d} is clearly very similar to \eqref{eq:mandelstamone} which we will be using in the higher dimensions.

It is also interesting to consider the behavior of the amplitude close to the two-particle threshold
\be
T(s) &= 4 i m \sqrt{s-4 m^2} + O(s-4), ~~~ N_{\text{tot}}~\text{odd}, \\
T(s) &= i c_0 (s-4 m^2)^{3/2} + O(s-4), ~~~ N_{\text{tot}}~\text{even}. 
\ee

One can ask if we can construct the S-matrices above using iterations of unitarity and dispersion relations. In $d=2$, we analyzed this question in our previous work \cite{Tourkine:2021fqh}.
One simple scenario is to fix $c_\infty$ and try to iterate unitarity, starting from $T^{(0)}(s) = c_\infty$. Curiously, for the problem above, this will lead to divergent results. Therefore, while the 2d was reasonably easy to understand, it happens so that the class of problem which we are interested in this paper (with no bound states) is not easily amenable to dispersive iterations in $d=2$. It would be interesting to understand how to generalize our methods to this case and if this generalization can be useful in higher dimensions. Consequently, our approach in this previous work was to solve the problem \textit{with} bound-states.

The situation is better in higher dimensions ($d=3,4$) where the simple iterations of unitarity and dispersion relations leads to convergent results shown in the text.

\section{Analytic Regge limit}
\label{sec:analyticregge}

To derive the analytic Regge solution in $d=4$ given by \eqref{eq:ansatz-rhos-4d}, we need to evaluate the large $s$ expansion of the following integrals
\be
I_1(\alpha, {s \over m^2}) &= 2 \int_{4m^2}^\infty {d s' \over \pi} {1 \over (\log {s' \over m^2})^\alpha} {\log {s'+s-4m^2 \over s'} \over s-4 m^2} \  , \nn \\
I_2(\alpha, {s \over m^2}) &=  \int_{4 m^2}^\infty {d s' \over \pi} {1 \over (\log {s' \over m^2})^\alpha} \Big[ {1 \over s'-s} \Big]_{P.V.}
\ee
This can be done for example by introducing the Mellin representation for the integration kernels, then performing the integral over $s'$, and finally deforming the Mellin contour
to generate the large $s$ expansion. The result for the first few terms take the following form
\be
I_1(\alpha, {s \over m^2}) &= {2 \over \pi} {(\log s/m^2)^{1-\alpha} \over \alpha - 1} \Big( 1 + {\alpha - 1 \over \log s/m^2} + \Big. \nn \\
\Big. &+ (1+{\pi^2 \over 6}){\alpha (\alpha-1) \over (\log s/m^2)^2}+(1+{\pi^2 \over 6}){(\alpha+1) \alpha (\alpha-1) \over (\log s/m^2)^3} + ... \Big) + ... \  ,
\ee
\be
I_2(\alpha, {s \over m^2}) &= {1 \over \pi} {(\log s/m^2)^{1-\alpha} \over \alpha - 1} \Big( 1 - {\pi^2 \over 3} {\alpha (\alpha-1) \over (\log s/m^2)^2} - {\pi^2 \over 45} {(\alpha+2)(\alpha+1) \alpha (\alpha-1) \over (\log s/m^2)^4} + ... \Big) + ...  \ .
\ee
Similarly, by taking the derivatives with respect to $\alpha$ we can generate the integrands of the type ${(\log \log {s' \over m^2})^k \over (\log {s' \over m^2})^\alpha}$. Writing down the expansion
\eqref{eq:ansatz-rhos-4d}, plugging it into the unitarity relation \eqref{eq:unitarity4d}, and using the formulas from this appendix one can check our claim.

\section{Proof of convergence \`a la Atkinson for toy-model  in $3<d<4$}
\label{app:atkinsonproofd34}

In this appendix we prove the contracting properties of the unitarity map for the toy-model amplitude in $3<d<4$. Many of the estimates of various integrals closely follow the $d=4$ case analyzed in \cite{Atkinson:1970zza}.

Consider the following amplitude 
\be
T(s,t) = \lambda + \int_{4m^2}^\infty {d s' \over \pi } {\rho(s') \over s' - {4 m^2 \over 3}} \left( {s-{4 \over 3} m^2 \over s'-s}+  {t - {4 m^2 \over 3} \over s' - t}+  {u - {4 m^2 \over 3} \over s' - u}\right) ,~~~ s+t+u=4m^2 . 
\ee
We introduce $s=4m^2/x$ so that $x \in [0,1]$ and consider class of $\rho(x)$ such that
\be
\rho(0) = \rho(1) = 0 .
\ee
We also consider $\rho(x)$ to be H\"{o}lder-continuous
\be
\label{eq:holderbound2}
| \rho(x_1) -  \rho(x_2) | \leq \xi |x_1 - x_2|^{\mu}, ~~~ 0 < \mu < \min({{d-3 \over 2}}, {4-d \over 2}) ,
\ee
where $\mu < {d-3 \over 2}$ will arise from imposing unitarity close to the two-particle threshold, whereas $\mu < {4-d \over 2}$ comes from unitarity in the Regge limit. H\"{o}lder-continuity is natural because it is preserved under iterations of unitarity and dispersion relations.

Following Atkinson, introduce the following norm in this functional space
\be
|| \rho || = {\rm sup}_{0 \leq x_1 , x_2 \leq 1} {|  \rho(x_1) - \rho(x_2) | \over | x_1 - x_2|^{\mu}}. 
\ee

We re-write unitarity $|S_0|=1$ in terms of a map $\Phi$ as:
\be
  \rho = \Phi [ \rho] .
\ee
which we solve iteratively. If $\rho$ is given, let us define
\be
  \rho' := \Phi [ \rho] .
\ee

Given a bound on the norm $||\rho|| \leq B$, in this appendix, we would like to show that $|| \rho' || \leq B$ to make sure that the iterated spectral density stays in the same space defined by $||\rho|| \leq B$. Via continuity of the map, this will guarantee that some solutions exist. We then would like to show that the map is actually contracting, which will prove uniqueness.

Let us write down the iteration equation and thus the map $\Phi$ explicitly
\be
\label{eq:iterationtoyddim}
 \rho'(x) &= {\pi^{{3-d \over 2}} \over 2^{d+2}\Gamma({d-1 \over 2})} x^{{4-d \over 2}} (1-x)^{{d-3 \over 2}} \left( ( c_d {\rm Re} f_0[\rho](x) )^2 +  \rho(x)^2 \right) \ , \nn \\
c_d {\rm Re} f_0[\rho](x) &= \lambda + \int_0^1 {d x' \over \pi} \rho(x') \left( {\rm PV} {3 - x \over 3 - x'}  {1 \over x - x'} + K_d(x,x')\right) , \nn \\
K_d(x,x') &= {2 \over x'} \left( \ _2 F_1 \left(1,{d-2 \over 2}, d-2, - {(1-x)x' \over x} \right) - {3 \over 3-x'} \right) , 
\ee
where $c_d=2^{2 d-3} \pi ^{\frac{d-3}{2}} \Gamma\left(\frac{d-1}{2}\right)$. Note that $K_d(x,x')$ is regular when $x'=0$. The universal prefactor $x^{{4-d \over 2}} (1-x)^{{d-3 \over 2}}$ in the unitarity relation is the source of the bounds on $\mu$ in \eqref{eq:holderbound}.

\subsection{Bounding the norm $||\rho'||$}

Our first task is to bound
\be
&| \rho'(x_1) - \rho'(x_2) | \Big( {\pi^{{3-d \over 2}} \over 2^{d+2}\Gamma({d-1 \over 2})} \Big)^{-1}    \nn \\
&=| x_1^{{4-d \over 2}} (1-x_1)^{{d-3 \over 2}} \left( ( c_d {\rm Re} f_0[\rho](x_1) )^2 +  \rho(x_1)^2 \right) - x_2^{{4-d \over 2}} (1-x_2)^{{d-3 \over 2}} \left( ( c_d {\rm Re} f_0[\rho](x_2) )^2 +  \rho(x_2)^2 \right) | \nn \\
&\leq | x_1^{{4-d \over 2}} (1-x_1)^{{d-3 \over 2}} - x_2^{{4-d \over 2}} (1-x_2)^{{d-3 \over 2}}| (( c_d {\rm Re} f_0[\rho](x_1) )^2 +  \rho(x_1)^2) \nn \\
&+x_2^{{4-d \over 2}} (1-x_2)^{{d-3 \over 2}} \left( |(c_d {\rm Re} f_0[\rho](x_1) )^2 -  ( c_d {\rm Re} f_0[\rho](x_2))^2 | + | \rho(x_1)^2 - \rho(x_2)^2 | \right).
\ee

Let us now bound each of the terms. First of all, we notice that
\be
| x_1^{{4-d \over 2}} (1-x_1)^{{d-3 \over 2}} - x_2^{{4-d \over 2}} (1-x_2)^{{d-3 \over 2}}| \leq |x_1 - x_2|^{\mu}, ~~~ 0 < \mu \leq  \min({{d-3 \over 2}}, {4-d \over 2}) .
\ee
Second, using the fact that $\rho(0)=0$ we get
\be
\label{eq:boundonrhoapp}
{| \rho(x) | \over |x|^{\mu} } \leq || \rho || \leq B , 
\ee
which in particular implies
\be
| \rho(x) | \leq B .
\ee
For the difference we also have $|\rho(x_1) - \rho(x_2) | \leq B |x_1 - x_2|^\mu$. Finally, we need to bound the terms that involve the real part ${\rm Re} f_0$. We do it in two parts by first bounding the principal value integral, and then the part that involves $K_d(x,x')$.

\subsubsection{Bounding the principal value integral}

Here we use Atkinson tricks. We first extend the principal value integral
\be
\int_{-2}^{2} {d x' \over \pi} \rho(x') {\rm PV} {3 - x \over 3 - x'}  {1 \over x - x'} ,
\ee
where we simply continuously set $\rho(x')=0$ for $x \notin (0,1)$. This extension preserves Holder continuity
\be
| \rho(x_1) - \rho(x_2) | \leq B |x_1 - x_2|^\mu, ~~~ -2 \leq x_1, x_2 \leq 2 \ . 
\ee
The only non-trivial case to check is when one of the variables, for example, $x_1 \in [0,1]$ and $x_2 \notin [0,1]$.

This allows us to use a simple trick
\be
&| \int_{-2}^{2} {d x' \over \pi} \rho(x') {\rm PV} {3 - x \over 3 - x'}  {1 \over x - x'} | = | \int_{-2}^{2} {d x' \over \pi}  {3 - x \over 3 - x'}  {\rho(x')- \rho(x)  \over x - x'} + {\rho(x) \over \pi} \log {2+x \over 10- 5 x} | \nn \\
&\leq {B \over \pi} ( {2^\mu ( \ _2 F_1 (1, \mu, 1+ \mu, -2/3)+ \ _2 F_1 (1, \mu, 1+ \mu, 2/3)) \over \mu} + \log 5) .
\ee
Note that the bound becomes weak as $\mu \to 0$.

Similarly, we can bound
\be
| \int_0^1 {d x' \over \pi} \rho(x') K_d(x,x') | \leq {B \over \pi} \left( {2 \over \mu} + {2 \over 3} {\ _2 F_1 (1, 1+\mu, 2+ \mu, 1/3) \over \mu + 1} \right).
\ee

In this way we get the following bound
\be
&|(c_d {\rm Re} f_0[\rho](x_1) | \leq  | \lambda | + {B \over \pi}  C_{{\rm Re} f_0} \nn \\
&C_{{\rm Re} f_0}={2^\mu ( \ _2 F_1 (1, \mu, 1+ \mu, -2/3)+ \ _2 F_1 (1, \mu, 1+ \mu, 2/3)) \over \mu} + \log 5 +  {2 \over \mu} + {2 \over 3} {\ _2 F_1 (1, 1+\mu, 2+ \mu, 1/3) \over \mu + 1} . 
\ee
It is clear that this bound could be highly sub-optimal given the fact that it does not allow for cancellations.

Finally, we need to bound $|(c_d {\rm Re} f_0[\rho](x_1) - (c_d {\rm Re} f_0[\rho](x_2) |$. Consider first the principal value part
\be
&| \int_{-2}^{2} {d x' \over \pi} \rho(x') {\rm PV} \left( {3 - x_1 \over 3 - x'}  {1 \over x_1 - x'} - {3 - x_2 \over 3 - x'}  {1 \over x_2 - x'} \right) | \leq B_1 + B_2 , \nn \\
B_1&= \Big| \int_{-2}^{2} {d x' \over \pi}  {3 - x_1 \over 3 - x'}  {\rho(x')- \rho(x_1)  \over x_1 - x'} - \int_{-2}^{2} {d x' \over \pi}  {3 - x_2 \over 3 - x'}  {\rho(x')- \rho(x_2)  \over x_2 - x'} \Big| \nn \\
B_2 &=|  {\rho(x_1) \over \pi} \log {2+x_1 \over 10- 5 x_1}  -  {\rho(x_2) \over \pi} \log {2+x_2 \over 10- 5 x_2} | \ . 
\ee

To bound $B_1$ let us assume that $x_2 \geq x_1$ (the case $x_1 \geq x_2$ leads to the same bound) and split the integration into two regions: $\Omega: |x'-x_1| \leq 2 (x_2 - x_1)$, and $\bar \Omega$ which is the rest. Both points $x'=x_1$ and $x'=x_2$ lie inside $\Omega$. We also have  $| x_1 \pm  2 (x_2 - x_1) | \leq 2$ so that $\Omega$ is inside the integration interval for any $x_1, x_2 \in [0,1]$.

In this way we get
\be
B_1 &\leq B_{11} + B_{12} + B_{13} , \nn \\
B_{11} &= \int_{\Omega} {d x' \over \pi}  {3 - x_1 \over 3 - x'} \Big| {\rho(x')- \rho(x_1)  \over x_1 - x'} \Big| + \int_{\Omega} {d x' \over \pi}  {3 - x_2 \over 3 - x'}  \Big| {\rho(x')- \rho(x_2)  \over x_2 - x'} \Big| , \nn \\
B_{12} &= \Big| \int_{\bar \Omega} {d x' \over \pi} (\rho(x') - \rho(x_1) ) \left( {3 - x_1 \over 3 - x'} {1  \over x_1 - x'} - {3 - x_2 \over 3 - x'} {1  \over x_2 - x'} \right) \Big|  \ , \nn \\
B_{13} &= \Big| \int_{\bar \Omega} {d x' \over \pi} (\rho(x_1) - \rho(x_2) ) {3 - x_2 \over 3 - x'} {1  \over x_2 - x'} \Big| \ .
\ee
For $B_{11}$ we directly use the Holder condition to get
\be
B_{11} \leq {B \over \pi \mu} \left( 3 \times 2^{1+\mu} + 2 (1+ 3^\mu) \right) |x_1 -x_2|^\mu. 
\ee
For $B_{12}$ we get 
\be
B_{12} &\leq B \int_{\bar \Omega} {d x' \over \pi} |x'-x_1|^{\mu} {|x_1-x_2| \over |x'-x_1||x'-x_2|} = {B |x_1 -x_2| \over \pi} \int_{\bar \Omega} d x' |x'-x_1|^{\mu-1} |x'-x_2|^{-1} \nn \\
&\leq {2^\mu B |x_1 -x_2|^\mu \over (1-\mu)\pi}.
\ee
For $B_{13}$  the bound takes the form
\be
B_{13} &\leq {B |x_1-x_2|^\mu \over \pi}  | \int_{\bar \Omega} dx' {3 - x_2 \over 3 - x'} {1  \over x_2 - x'} | \leq \log 15 {B |x_1-x_2|^\mu \over \pi} .
\ee
Next we bound $B_2$ as follows
\be
B_2 &\leq  {|\rho(x_1)-\rho(x_2)| \over \pi} |\log {2+x_1 \over 10- 5 x_1}| + |{\rho(x_2) \over \pi}||\log {2+x_1 \over 10- 5 x_1}-\log {2+x_2 \over 10- 5 x_2}| \nn \\
&\leq \log 5 {B \over \pi} |x_1-x_2|^\mu +{B \over \pi} \log (1 + {4(x_2-x_1) \over (2+x_1)(2-x_2)}).
\ee
Using
\be
\log (1 + A) \leq A , ~~~ A \geq 0 ,
\ee
we finally get
\be
B_2 \leq {B \over \pi} (2 + \log 5) |x_1-x_2|^\mu . 
\ee
Our next step is to bound the difference term that involves $K_d(x,x')$. We get that
\be
\left| \int_0^1 {d x' \over \pi} \rho(x') \left(K_d(x_1,x')-K_d(x_2,x')\right) \right| \leq {B |x_1-x_2|^\mu \over \pi} \frac{\sqrt{\pi } 2^{d-2} \Gamma \left(\frac{d-1}{2}\right) \csc (\pi  \mu ) \Gamma \left(\frac{d}{2}-\mu -1\right)}{\Gamma (d-\mu -2)} \ .
\ee
To derive this formula we used \eqref{eq:boundonrhoapp}, after which the integral can be taken exactly.

To summarize, in this part of the computation we have derived the following bound
\be
&|| (c_d {\rm Re} f_0[\rho] || \leq {B \over \pi} C_{\text{norm}} , \\
C_{\text{norm}} &=\frac{\sqrt{\pi } 2^{d-2} \Gamma \left(\frac{d-1}{2}\right) \csc (\pi  \mu ) \Gamma \left(\frac{d}{2}-\mu
   -1\right)}{\Gamma (d-\mu -2)}+\frac{2 \left(3\ 2^{\mu }+3^{\mu }+1\right)}{\mu }+\frac{2^{\mu }}{1-\mu }+2+\log
   75 .
\ee

Putting everything together we get the desired bound on the norm of the spectral density after the unitarity iteration
\be
|| \rho' || &\leq B' , \\
B' &={\pi^{{3-d \over 2}} \over 2^{d+2}\Gamma({d-1 \over 2})} \left( (|\lambda|+{B \over \pi}  C_{{\rm Re} f_0})^2+ 2 B^2 {C_{\text{norm}} C_{{\rm Re} f_0} \over \pi^2} + 3 B^2 \right) ,
\ee
where $C_{{\rm Re} f_0}$ and $ C_{\text{norm}}$  are only functions of $\mu$ and $d$. One can easily check using the explicit formula above that for small enough $\lambda$ and $3<d<4$ we have $B'\leq B$. 

\subsection{Proving the contracting property of the map}

As a next step we would like to bound the norm of the images in terms of the norm of the pre-images. In other words, we would like to derive a bound of the type
\be
\label{eq:contractionindex2}
|| \rho'_2 - \rho_1' || \leq k || \rho_2 - \rho_1 ||.
\ee
To derive such a bound we start by noticing that the following bound on the real part holds
\be
&|c_d {\rm Re} f_0[\rho_1](x) - c_d {\rm Re} f_0[\rho_2](x) |=|\int_0^1 {d x' \over \pi} (\rho_1(x')-\rho_2(x')) \left( {\rm PV} {3 - x \over 3 - x'}  {1 \over x - x'} + K_d(x,x')\right)|\nn \\
&\leq C_{{\rm Re} f_0} {|| \rho_2 - \rho_1 || \over \pi} \nn \\
\ee
where we trivially used the results of the previous section.

Similarly, we have
\be
&||c_d {\rm Re} f_0[\rho_1] - c_d {\rm Re} f_0[\rho_2] || \leq C_{\text{norm}} {|| \rho_2 - \rho_1 || \over \pi} \nn \\
\ee

We are now ready to derive the desired bound
\be
& \left( \rho'_1(x_1) - \rho'_2(x_1) -(\rho'_1(x_2) - \rho'_2(x_2)) \right) \Big( {\pi^{{3-d \over 2}} \over 2^{d+2}\Gamma({d-1 \over 2})} \Big)^{-1} \nn \\
&=x_1^{{4-d \over 2}} (1-x_1)^{{d-3 \over 2}} \left( ( c_d {\rm Re} f_0[\rho_1](x_1) )^2 +  \rho_1(x_1)^2  -  ( c_d {\rm Re} f_0[\rho_2](x_1) )^2 -  \rho_2(x_1)^2 \right)  \nn \\
&-x_2^{{4-d \over 2}} (1-x_2)^{{d-3 \over 2}} \left( ( c_d {\rm Re} f_0[\rho_1](x_2) )^2 +  \rho_1(x_2)^2  - ( c_d {\rm Re} f_0[\rho_2](x_2) )^2 -  \rho_2(x_2)^2 \right) 
\ee
As before we split the RHS in three parts
\be
I_1 &=(x_1^{{4-d \over 2}} (1-x_1)^{{d-3 \over 2}} - x_2^{{4-d \over 2}} (1-x_2)^{{d-3 \over 2}}) \Big( ( c_d {\rm Re} f_0[\rho_1](x_1) )^2 +  \rho_1(x_1)^2 -  ( c_d {\rm Re} f_0[\rho_2](x_1) )^2 -  \rho_2(x_1)^2 \Big) \nn \\
I_2&=x_2^{{4-d \over 2}} (1-x_2)^{{d-3 \over 2}} \left( (c_d {\rm Re} f_0[\rho_1](x_1) )^2 - (c_d {\rm Re} f_0[\rho_2](x_1) )^2 -  ( c_d {\rm Re} f_0[\rho_1](x_2))^2  +  (c_d {\rm Re} f_0[\rho_2](x_2) )^2\right) \nn \\
I_3 &=x_2^{{4-d \over 2}} (1-x_2)^{{d-3 \over 2}} \left( \rho_1(x_1)^2 - \rho_2(x_1)^2 -  \rho_1(x_2)^2  +  \rho_2(x_2)^2 \right) .
\ee
and bound each part separately.

Using the results that we have already derived we get
\be
| I_1 | \leq |x_1 - x_2|^\mu  || \rho_2 - \rho_1 || \left( 2{C_{{\rm Re} f_0} \over \pi} ( | \lambda | + {B \over \pi}  C_{{\rm Re} f_0}) + 2 B  \right) .
\ee
Next we proceed to $I_3$ where we add and subtract $(\rho_1(x_2)+\rho_2(x_2))(\rho_1(x_1)-\rho_2(x_1))$ to get
\be
|I_3|\leq  4 B |x_1 - x_2|^\mu || \rho_2 - \rho_1 ||.
\ee
Let us present this step in a bit more detail, we write
\be
I_3 &=(\rho_1(x_1)-\rho_1(x_2))(\rho_1(x_1)-\rho_2(x_1))+(\rho_2(x_1)-\rho_2(x_2))(\rho_1(x_1)-\rho_2(x_1)) \nn \\
&+(\rho_1(x_2)+\rho_2(x_2))(\rho_1(x_1)-\rho_2(x_1)-\rho_1(x_2)+\rho_2(x_2)),
\ee
and then bound each of the three terms separately.

And finally we get using the same trick for the real part of the partial wave
\be
|I_2|\leq |x_1 - x_2|^\mu || \rho_2 - \rho_1 || {2 \over \pi}   \left( {B \over \pi} C_{\text{norm}} C_{{\rm Re} f_0} + (| \lambda | + {B \over \pi}  C_{{\rm Re} f_0}) C_{\text{norm}}  \right) .
\ee

In this way we get for $k$ in \eqref{eq:contractionindex}
\be
k = {\pi^{{3-d \over 2}} \over 2^{d+2}\Gamma({d-1 \over 2})} \left(|\lambda| {2 (C_{{\rm Re} f_0} +C_{\text{norm}} ) \over \pi}  +{B \over \pi^2}(4 C_{{\rm Re} f_0} C_{\text{norm}}+2 C_{{\rm Re} f_0}^2 )  + 6 B \right).  
\ee

We can now find $d, \mu, B, \lambda$ such that
\be
B' \leq B, ~~~ k<1 .
\ee 
Given such parameters we can show that iterations converge to a unique fixed point. First, let us recall that the space of H\"{o}lder continuous function satisfying $||\rho||\leq B$ is complete. In other words, if we find a Cauchy sequence of functions $||\rho_n|| \leq B$, such that for any $\eps>0$ there exists $N$ for which
\be
\label{eq:convergence}
||\rho_{N+p}-\rho_N|| < \eps, ~~~ p \in \mathbb{N},
\ee
then $\rho_n \to \rho_*$ and $||\rho_*|| \leq B$.

To demonstrate \eqref{eq:convergence} we proceed as follows
\be
||\rho_{N+p}-\rho_N|| \leq \sum_{m=N}^{N+p-1} || \rho_{m+1} - \rho_m ||  \leq  || \rho_{1} - \rho_0 || \sum_{m=N}^{N+p-1} k^m \leq || \rho_{1} - \rho_0 || {k^N \over 1-k} \leq 2 B {k^N \over 1-k}.
\ee
Therefore, given $\eps$ we choose $N$ so that $2 B {k^N \over 1-k} < \eps$ is satisfied. This is always possible since $k<1$. By setting $p=\infty$ in the formula above we converge to the limiting solution as follows
\be
||\rho_{*}-\rho_N||\leq 2 B {k^N \over 1-k}.
\ee

\subsection{Generalization to $d=3$ and $d=4$}

Let us discuss possible generalizations of the argument above to $d=3$ and $d=4$. An immediate problem is that from \eqref{eq:holderbound} it follows that $\mu=0$, and many of the bounds of the previous section blow up for $\mu=0$. Here we discuss a method to fix this problem inspired both by the success of our numerical analysis of these cases, as well as the trick used by Atkinson in \cite{Atkinson:1969eh}.

Let us for concreteness discuss the case of $d=4$.\footnote{The case of $d=3$ is essentially identical, where the Regge limit is replaced by the near two-particle threshold region.} The basic idea is to modify the unitarity equation 
\be
 \rho'(x) &= {h(x) \over 32 \pi} (1-x)^{{1 \over 2}} \left( ( 16 \pi {\rm Re} f_0[\rho](x) )^2 +  \rho(x)^2 \right) ,
\ee
where we choose
\be
h(x) = \theta(x-x_0) +{x \over x_0} \theta(x_0-x) .
\ee
The role of $h(x)$ is to effectively soften the behavior of $\rho'(x)$ in the Regge limit $x \to 0$. With this modification we can consider a class of H\"{o}lder-continuous functions with $\mu = 1/2$ and run the argument of the previous section. All the bounds will now depend on $x_0$.

In a physical theory we would like to set $h(x)=1$ or $x_0=0$. Naively, as explained above, the algorithm will stop converging (or at least we cannot argue for convergence). It is therefore interesting to see how we effectively solved this problem in the numerical implementation of the algorithm in $d=4$.

Effectively, we considered \emph{iteration-dependent} modification of the unitarity equation
\be
h_n(x) = \theta(x-x_{n}) +{x \over x_n} \theta(x_n-x) ,
\ee
where we chose $x_n$ such that $\lim_{n \to \infty} x_n = 0$. We observed convergence and smooth extrapolation of the solutions to the UV as we remove the cutoff $x_n$, where we could smoothly match to the analytic UV complete solution. 

The fact that such an algorithm works is based on cancellation between various terms as we approach the solution better and better, see for example the sum rule \eqref{eq:sum-rule-rhos}, and therefore it clearly goes beyond the proof of the previous subsection based on a brute-force bounding of each term separately. 

An alternative approach in $d=4$ that does not require introduction of $h(x)$ and subtleties discussed here is to consider the Mandelstam equation with no subtractions (so that only the double spectral density is present), and construct amplitudes starting from a given $\rho_{\text{MP}}(s,t)$, as shown by Atkinson in \cite{Atkinson:1968exe}, in this case one can again argue that the iterations of unitarity lead to a unique solution. 

In $d=3$ the modified unitarity equation takes the form
\be
\rho'(x) &=  {\sqrt{x} h(x) \over 32} \left( (8 {\rm Re} f_0)^2 + \rho(x)^2 \right) ,
\ee
and in this case the regulator is needed close to the two-particle threshold
\be
h(x) = \theta(x_0-x) +{1-x \over 1-x_0} \theta(x-x_0) .
\ee

\section{Perturbative $\phi^4$}
\label{app:perturbativephi4}

In this appendix we explicitly perform the first two unitarity iterations for the 2PR amplitude.
It is completely equivalent to the two-loop computation in ${\lambda \over 4!} \phi^4$ theory. 

\subsection{$d=4$}
We expand the amplitude in the powers of $\lambda$
\be
T(s,t) = \lambda T^{(0)}(s,t)  + \lambda^2 T^{(1)}(s,t) + \lambda^3 T^{(2)}(s,t)+ ... \ . 
\ee
We would like to generate the usual perturbative expansion using dispersive iterations of unitarity with one subtraction described in the main text of the paper.

We initialize the iteration by setting 
\be
T^{(0)} =1 . 
\ee
From this we get using \eqref{eq:iterationrho0}
\be
\label{eq:rhosq4d}
\rho^{(1)}(s) = {1 \over 32 \pi} \sqrt{1 - {4 m^2 \over s}} ,
\ee
where we used the fact that $f^{(0)}_0(s)= {1 \over 16 \pi}$.

Plugging $\rho^{(1)}(s)$ into the Mandelstam representation \eqref{eq:mandelstamRep1} and using the fact that
\be
&\int_{4 m^2}^\infty {d s' \over \pi} { (s-s_0) \over (s'-s)(s'-s_0)} {1 \over 32 \pi} \sqrt{1 - {4 m^2 \over s'}} =- {1 \over 16 \pi^2} \left( I(s) - I(s_0) \right) , \\
&I \Big({4 m^2 \over 1+y^2} \Big) \equiv \int_0^1 d x {1 \over 1 + {x^2 \over y^2}} = {i y \over 2} \log {y -i \over y+ i} ,
\ee
we get that
\beq
\label{eq:secondorderamp4d}
T^{(1)}(s,t)  ={3 \over 16 \pi^2} I(s_0) - {1 \over 16 \pi^2} \left( I(s) + I(u) + I(t)   \right)
\eeq

Next we compute $\rho^{(2)}(s)$ which takes the form
\be
\rho^{(2)}(s) &=8 \pi  {(s-4 m^2)^{1/2} \over \sqrt{s}} \Big( f^{(0)}_0(s) (f^{(1)}_0(s))^* + (f^{(0)}_0(s))^* f^{(1)}_0(s) \Big) ={(s-4 m^2)^{1/2} \over \sqrt{s}} {\rm Re} f^{(1)}_0(s) .
\ee
We can use \eqref{eq:secondorderamp4d} to get
\be
&{\rm Re} f^{(1)}_0(s) = {1 \over 32 \pi} \int_{-1}^{1} d z {\rm Re}  T^{(1)}(s,t(z))  \nn \\ 
&= {1 \over 16 \pi} \Big( {3 \over 16 \pi^2} I(s_0) - {1 \over 16 \pi^2} {\rm Re} I(s)  - {1 \over 16 \pi^2} \int_{-1}^{1} d z \ I \Big(-{(s-4 m^2)(1-z) \over 2} \Big) \Big) \nn \\
&=- {1 \over 256 \pi^3} \Big(3I(s_0) - {\rm Re} I(s) + {m^2 \over s - 4m^2} \log^2
   \left(\frac{1-\sqrt{1-\frac{4 m^2}{s}}}{1+\sqrt{1-\frac{4 m^2}{s}}}\right)-{1 \over \sqrt{1-\frac{4 m^2}{s}}} \log
   \left(\frac{1-\sqrt{1-\frac{4 m^2}{s}}}{1+\sqrt{1-\frac{4 m^2}{s}}}\right)- 1 \Big).
\ee

Finally, we do the dispersive integrals \eqref{eq:mandelstamRep1} to compute $T^{(2)}(s,t)$. The result takes the following form
\be
T^{(2)}(s,t) = {3 \over 8 \pi^2} I(s_0) T^{(1)}(s,t)  + {1 \over 256 \pi^4} \left( I(s)^2 + I(u)^2 + I(t)^2  \right)  - {1 \over 128 \pi^4} \left( I_3(s) + I_3(u) +I_3(t) \right) -c_0^{(2)},
\ee
where
\be
I_3({4 m^2 \over 1+y^2}) 
   &\equiv \frac{I({4 m^2 \over 1+y^2})^2}{y^2}+\left(\frac{\pi ^2 \left(y^2+1\right)}{12 y^2}+1\right) I({4 m^2 \over 1+y^2})-\frac{\left(y^2+1\right)
   I({4 m^2 \over 1+y^2})^3}{3 y^4},
\ee
and the constant $c_0^{(2)}$ is fixed by the condition $T^{(2)}({4 m^2 \over 3},{4 m^2 \over 3})=0$. We see therefore that to the first two non-trivial orders dispersive iterations of unitarity and the usual Feynman perturbation theory coincide. 

 It is also interesting to examine the leading Regge behavior of the amplitude at each order computed so far. We get
\be
\label{eq:amplphi4regge}
\lim_{s \to \infty} T(s,t) = \lambda \left( 1 - {\lambda \over 16 \pi^2} \log s + {3 \over 2} \left( {\lambda \over 16 \pi^2} \log s \right)^2 + ...  \right) .
\ee
The behavior of the amplitude is correlated with the sign of the $\beta$-function. In our convention $\lambda<0$ corresponds to the Landau pole in the UV and stable vacuum, whereas $\lambda>0$ corresponds to asymptotic freedom, with the potential not bounded from below. We see from \eqref{eq:amplphi4regge} that for negative $\lambda$ all the terms add up and the amplitude becomes big at high energies, whereas for $\lambda>0$ we have cancelations.

At order $\lambda^4$ two new features appear. First of all, using $\rho^{(2)}(s)$ inside the Mandelstam equation \eqref{eq:mandelstam}, we get a nontrivial $\rho_{\text{el}}^{(4)}(s,t)$
\be
\rho_{\text{el}}^{(4)}(s,t) =  {(s - 4 m^2)^{1 \over2} \over 16 \pi^2 \sqrt{s} } {1 \over (32 \pi)^2}\int\limits_{z_1}^{\infty} d \eta'  \int\limits_{z_1}^{\infty} d \eta''\, \sqrt{1 - {4 m^2 \over t(\eta')}} \sqrt{1 - {4 m^2 \over t(\eta'')}} {\theta(z- \eta_+) \over \sqrt{(z - \eta_-)(z - \eta_+)}} ,
\ee
which can be straightforwardly evaluated numerically. This corresponds to the ``Aks'' diagram \figref{fig:threeloopgraphsforswaveinelasticity}.

Another novelty is that both the S-wave production amplitude $\eta^{(4)}_{\text{MP}}(s)$, as well as the multi-particle double spectral density are non-zero at this order $\rho^{(4)}_{\text{MP}}(s,t)$. These are due to the frog and open envelope graphs, see \figref{fig:diagramspic}. Therefore, for our iteration method to continue agreeing with the Feynman graph computation at order $\lambda^4$,  $\rho^{(4)}_{\text{MP}}(s,t)$ and $\eta^{(4)}_{\text{MP}}(s)$ should be independently computed and provided as an input.

\subsection{$d=3$}
\label{sec:d3phi4twoloops}

We now repeat the same exercise in $d=3$. We initialize the iteration by setting 
\be
T^{(0)} =1 . 
\ee
From this we get using \eqref{eq:iterationrho0}
\be
\label{eq:rhosq3d}
\rho^{(1)}(s) = {1 \over 16 \sqrt{s}},
\ee
where we used the fact that $f^{(0)}_0(s)= {1 \over 8}$.

Plugging $\rho^{(1)}(s)$ into the Mandelstam representation \eqref{eq:mandelstamRep1} and using the fact that
\be
&\int_{4 m^2}^\infty {d s' \over \pi} { (s-s_0) \over (s'-s)(s'-s_0)} {1 \over 16  \sqrt{s}} ={1 \over 16 \pi} \left( \tilde I(s) - \tilde I(s_0) \right) , \\
&\tilde I (s) \equiv \int_{4 m^2}^\infty {d s' \over s'-s} {1 \over \sqrt{s'}} = {\log {1 + {\sqrt s \over 2 m} \over 1 - {\sqrt s \over 2 m}} \over \sqrt{s}},
\ee
we get that
\beq
\label{eq:secondorderamp3d}
T^{(1)}(s,t)  =-{3 \over 16 \pi} \tilde I(s_0) + {1 \over 16 \pi}\left( \tilde I(s) + \tilde I(u) + \tilde I(t)   \right) . 
\eeq
Next we compute $\rho^{(2)}(s)$ which takes the form
\be
\rho^{(2)}(s) &= {4 \over \sqrt{s}} \Big( f^{(0)}_0(s) (f^{(1)}_0(s))^* + (f^{(0)}_0(s))^* f^{(1)}_0(s) \Big) ={1 \over \sqrt{s}} {\rm Re} f^{(1)}_0(s) .
\ee
We can use \eqref{eq:secondorderamp3d} to get
\be
{\rm Re} f^{(1)}_0(s) &= {1 \over 16 \pi} \int_{0}^{2 \pi} d \theta {\rm Re}  T^{(1)}(s,t(\theta)) \nn \\
&={1 \over 128 \pi} \left( {\rm Re} \tilde I(s) - 3 \tilde I(s_0) \right) + {1 \over 32 \pi} {\log {\sqrt s \over 2 m} + \sqrt{{s \over 4 m^2} - 1} \over \sqrt{s-4 m^2}},
\ee
where ${\rm Re} \tilde I(s) = {\log {1 + {\sqrt s \over 2 m} \over {\sqrt s \over 2 m}-1} \over \sqrt{s}}$ for $s>4 m^2$.

Finally, we do the dispersive integrals \eqref{eq:mandelstamRep1} to compute $T^{(2)}(s,t)$. The result takes the following form
\be
T^{(2)}(s,t) = - {3 \over 8 \pi} \tilde I_1(s_0) T^{(1)}(s,t) + {\tilde I_2(s)+ \tilde I_2(t)+\tilde I_2(u) \over 128 \pi^2}+{\tilde I_3(s)+ \tilde I_3(t)+\tilde I_3(u) \over 64 \pi^2}  - \tilde c_0^{(2)},
\ee
where
\be
s \tilde I_2(s) &= \text{Li}_2\left(\frac{1}{2}-\frac{\sqrt{s}}{4 m}\right)+\text{Li}_2\left(\frac{1}{2}+\frac{\sqrt{s}}{4 m}\right)+ \frac{1}{4} \log ^2\left(\frac{1}{4}-\frac{s}{16 m^2}\right) \nn \\
&+\frac{1}{4} \log ^2\left(\frac{2 m+\sqrt{s}}{2 m-\sqrt{s}}\right) - {\pi^2 \over 6} \ , 
\ee
and
\be
\tilde I_3(s) &=\frac{\text{Li}_2\left(\frac{-2 m^2+s+\sqrt{s} \sqrt{s-4 m^2}}{2 m^2}\right)-\text{Li}_2\left(\frac{2 m^2}{-2 m^2+s+\sqrt{s} \sqrt{s-4
   m^2}}\right)}{\sqrt{s} \sqrt{s-4 m^2}} ,
\ee
and the constant $\tilde c_0^{(2)}$ is fixed by the condition $T^{(2)}({4 m^2 \over 3},{4 m^2 \over 3})=0$. 

Expanding the amplitude at large $s$ we get
\be
T(s,0) \simeq \lambda - 0.04817 \lambda^2 + 0.0026 \lambda^3 , \nn \\
{\cal T}(s, \theta) \simeq \lambda - 0.06807 \lambda^2 + 0.0038 \lambda^3 ,
\ee
where in the first line we consider the forward amplitude $t=0$, and in the second line we consider scattering at fixed angles $\theta \neq 0, \pi$. Setting $\lambda = - \pi$, we get good agreement with \figref{fig:amp-toy-model-3d}.

\section{Aks graph inelasticity}

In this appendix we first compute the simplest possible $\eta_{\text{MP}}(s)$ (multi-particle production) in $\phi^4$ in $d=3$ and $d=4$. 
Let us start by recalling the definition of $\eta_{\text{MP}}(s)$ given in \eqref{eq:partialwavezero}. Using the fact that $S_J(s) = 1 + i { (s - 4m^2)^{{d-3 \over 2}} \over \sqrt{s}} f_J(s)$, we can rewrite it as follows
\be
2 {\rm Im} f_J(s) = {(s - 4 m^2)^{{d-3 \over 2}} \over \sqrt{s}} | f_J(s)|^2 + {\sqrt{s} \over (s - 4 m^2)^{{d-3 \over 2}} } \eta_{\text{MP}}(s) . 
\ee

\begin{figure}
    \centering
    \includegraphics[scale=0.7]{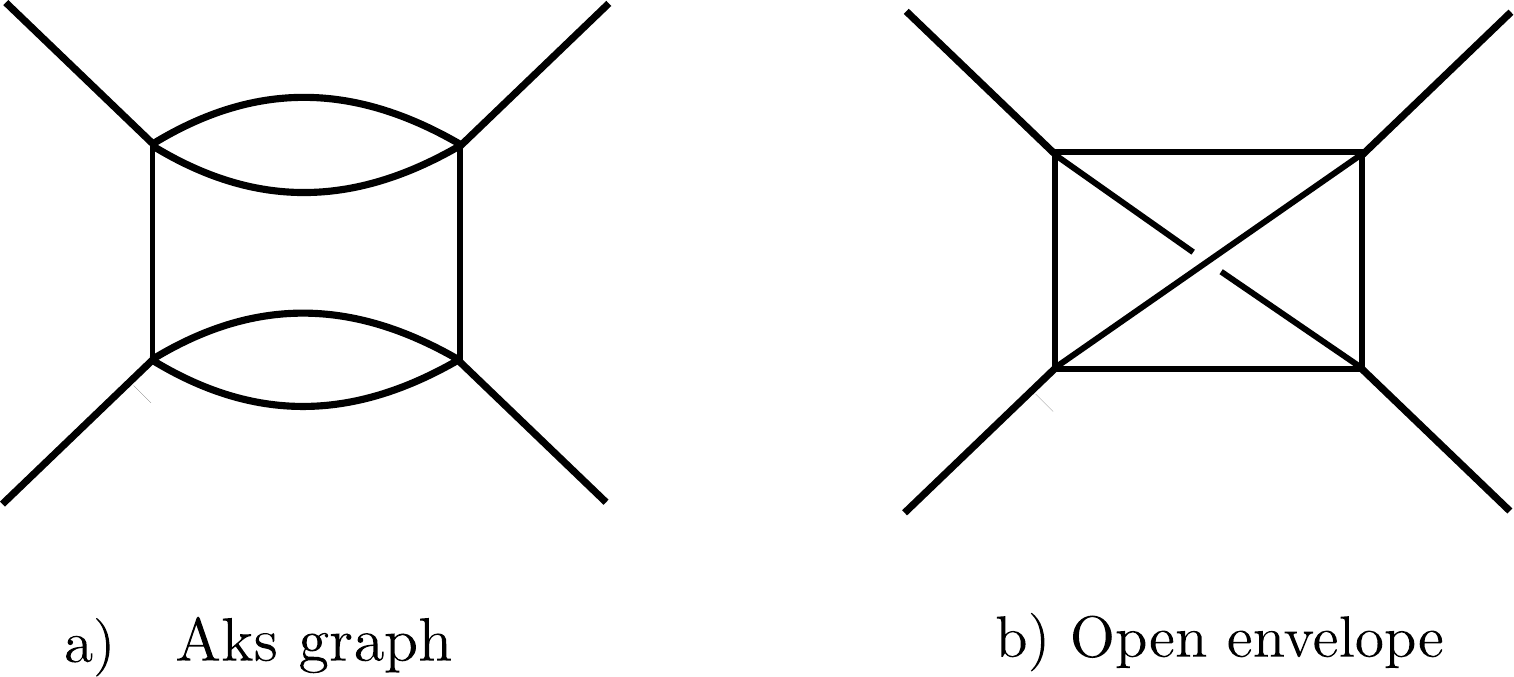}
    \caption{Three-loop diagrams in $\phi^4$ that contribute to the multi-particle production. Double discontinuity of the Aks diagram is captured by our iteration process. The open envelope diagram is the leading contribution to $\rho_{\text{MP}}(s,t)$, which we do not compute in the present paper. Both diagrams contribute to $\eta_{\text{MP}}(s)$.}
    \label{fig:threeloopgraphsforswaveinelasticity}
\end{figure}

The leading order contribution to $\eta_{\text{MP}}(s)$ enters at order $\lambda^4$ and is given by two graphs, see figure \eqref{fig:threeloopgraphsforswaveinelasticity}, which we can call \emph{Aks} and \emph{open envelope} graphs.
The complete $2 \to 4$ amplitude that is relevant for our computation is
\be
T(p_1,p_2; q_5,q_6, q_7,q_8)  &= - \lambda^2 \Big( {1 \over (p_1 - q_{56})^2-m^2} + {1 \over (p_1 - q_{57})^2-m^2} + {1 \over (p_1 - q_{58})^2-m^2} +  (p_1 \leftrightarrow p_2) \Big) ,\\
T^*(q_5,q_6,q_7,q_8; p_3, p_4) &=- \lambda^2 \Big( {1 \over (p_3 - q_{56})^2-m^2} + {1 \over (p_3 - q_{57})^2-m^2} + {1 \over (p_3 - q_{58})^2-m^2}  +  (p_3 \leftrightarrow p_4) \Big) ,
\ee
where $q_{56}^\mu = q_5^\mu+q_6^\mu$, etc. Doing unitarity gluing with these amplitudes leads to 
\be
\eta_{\text{MP}}(s) = 2 \times 6 \eta_{\text{Aks}}(s)  + 2 \times 12 \eta_{\text{oe}}(s) ,
\ee
where $\text{Aks}$ stands for the \emph{Aks graph}, and $\text{oe}$ for the \emph{open envelope} graph, see figure \ref{fig:threeloopgraphsforswaveinelasticity}.\footnote{We call it the Aks graph because it represents inelasticity that enters into the proof of the Aks theorem \cite{Aks:1965qga}.} To keep formulas shorter we do not write below the $\lambda$ factor.

\subsection{$d=3$}

The four-body discontinuity takes the form
\be
2 T_s = {1 \over 4!} \int {d^2 \vec q_5 \over (2 \pi)^2 (2 E_{\vec q_5})} {d^2 \vec q_6 \over (2 \pi)^2 (2 E_{\vec q_6}) } {d^2 \vec q_7 \over (2 \pi)^2 (2 E_{\vec q_7})}  {d^2 \vec q_8 \over (2 \pi)^2 (2 E_{\vec q_8})} \nn \\ 
(2 \pi)^3 \delta(p_1+p_2- q_5 - q_6 - q_7-q_8) T(p_1,p_2; q_5,q_6, q_7,q_8) T^*(q_5,q_6,q_7,q_8; p_3, p_4) ,
\ee
where $s = (p_1 + p_2)^2$ and $t=(p_1-p_3)^2$. We will be interested in the amplitudes
\be
\label{eq:dbamp}
T(p_1,p_2; q_5,q_6, q_7,q_8) &=- {1 \over (q_5+q_6-p_1)^2 - m^2}\nn \\
T^*(q_5,q_6,q_7,q_8; p_3, p_4) &=- {1 \over (q_7+q_8-p_4)^2 - m^2} .
\ee

Let us first transform the measure using the fact that the amplitudes \eqref{eq:dbamp} only depend on the combinations of momenta $q_5^\mu+q_6^\mu$ and $q_7^\mu+q_8^\mu$. We define the effective invariant masses of these pairs of particles $(q_5+q_6)^2=m_{56}^2$, $(q_7+q_8)^2=m_{78}^2$ and introduce integration over the corresponding 3-momenta $q_{56}^\mu \equiv q_5^\mu+q_6^\mu$, $q_{78}^\mu \equiv q_7^\mu+q_8^\mu$
\be
2 T_s&={1  \over 4!} \int_{4 m^2}^{(\sqrt{s}-2m)^2} d m_{56}^2 d m_{78}^2 \int d^3 q_{56} d^3 q_{78} \delta^+(q_{56}^2 - m_{56}^2) \delta^+(q_{78}^2 - m_{78}^2) T(p_1,q_{56}) T^*(p_4,q_{78}) \nn \\
 &\times (2 \pi)^3 \delta(p_1+p_2- q_{56} - q_{78}) {\cal I} ,
\ee
where the maximum value of $m_{56}^2$ is achieved by setting $\vec q_7 = \vec q_8 =0$, and
\be
&{\cal I} \equiv {1 \over (2 \pi)^{8}} \Big( \int d^3 q_5 \delta^+(q_5^2 - m^2) \int d^3 q_5 \delta^+(q_6^2 - m^2) \delta^{(3)}(q_{56} - q_5 - q_6) \Big) \nn \\
& \Big( \int d^3 q_7 \delta^+(q_7^2 - m^2) \int d^3 q_8 \delta^+(q_8^2 - m^2) \delta^{(3)}(q_{78} - q_7 - q_8) \Big) .
\ee

It is convenient to evaluate integrals in the brackets in the center-of-mass frame, for example $q_{78}= (m_{78}, \vec 0)$ to get
\be
 &\int d^3 q_7 \delta^+(q_7^2 - m^2) \int d^3 q_8 \delta^+(q_8^2 - m^2) \delta^{(3)}(q_{78} - q_7 - q_8) =  \int {d^2 \vec q_7 \over 4 E_{\vec q_7}^2}  \delta(m_{78} - 2 E_{\vec q_7}) \nn \\
 &={\pi \over 2 m_{78}} , ~~~ m_{78} \geq 2 m . 
\ee
In this way we get ${\cal I} = {1 \over 16} {1 \over (2 \pi)^6} {1 \over m_{56} m_{78}}$ and we are left with the following unitarity integral
\be
2 T_s &={(2 \pi)^{-3} \over 4!}  {1 \over 16}  \int_{4 m^2}^{(\sqrt{s}-2m)^2} {d m_{56}^2 \over m_{56}} {d m_{78}^2  \over m_{78} } \nn \\
&\int d^3 q_{56} d^3 q_{78} \delta^+(q_{56}^2 - m_{56}^2) \delta^+(q_{78}^2 - m_{78}^2) T(p_1,q_{56}) T^*(p_4,q_{78}) \delta(p_1+p_2- q_{56} - q_{78}).
\ee

We proceed by evaluating the integral above in the center-of-mass frame $\vec p_1 + \vec p_2 = \vec q_{56}+\vec q_{78}= 0$
\be
&\int d^3 q_{56} d^3 q_{78} \delta^+(q_{56}^2 - m_{56}^2) \delta^+(q_{78}^2 - m_{78}^2) T(p_1,q_{56}) T^*(p_4,q_{78}) \delta(p_1+p_2- q_{56} - q_{78}) \nn \\
&=\int {d^2 \vec q_{56} \over 4 E_{\vec q_{56},m_{56}} E_{\vec q_{56},m_{78}} } \delta(\sqrt{s} - E_{\vec q_{56},m_{56}} - E_{-\vec q_{56},m_{78}} )T(p_1,q_{56}) T^*(p_4,q_{78})
\ee
For the amplitudes we have
\be
T(p_1,q_{56}) &=- {1 \over (q_{56}-p_1)^2 - m^2} =- {1 \over m_{56}^2 - 2 q_{56} \cdot p_1}\nn \\
T^*(p_4,q_{78}) &=- {1 \over (q_{78}-p_4)^2 - m^2}=- {1 \over m_{78}^2 - 2 q_{78} \cdot p_4} .
\ee
Note that due to the fact that $\vec q_{78} = - \vec q_{56}$ we have $z_{78,4} = - z_{56,4} = z_{56,3}$, where $z_{i,j} \equiv {\vec q_i \cdot \vec q_j \over |\vec q_i| |\vec q_j| }$.

As a result the unitarity integral takes the form 
\be
2 T_s &={(2 \pi)^{-3} \over 4!}  {1 \over 16}  \int_{4 m^2}^{(\sqrt{s}-2m)^2} {d m_{56}^2 \over m_{56}} {d m_{78}^2  \over m_{78} } {\theta(\tilde K) \over \tilde K} {\sqrt{s} \over s-4m^2} \int_{0}^{2 \pi} d \phi_{56} {1 \over z^0 - z_{56,1} } {1 \over z^0 - z_{56,3} }. 
\ee
where we have introduced
\be
\tilde K &= (m_{56}^2 - m_{78}^2)^2 - 2 (m_{56}^2 + m_{78}^2) s + s^2 , \nn \\
z^0 &={\sqrt{s} \over \sqrt{s-4m^2}} {s- m_{56}^2 - m_{78}^2 \over \sqrt{\tilde K} } >1 .
\ee
The constraint $\theta(K)$ arises from the fact that $| \vec q_{56}| \geq 0$.

The integral over angles gives
\be
 \int_{0}^{2 \pi} d \phi_{56} {1 \over z^0 - z_{56,1} } {1 \over z^0 - z_{56,3} } = {4 \pi z^0 \over \sqrt{(z^0)^2 - 1}} {1 \over 2 (z^0)^2 - 1 - z} , 
\ee
where we have used that $z \equiv z_{1,3}$.

\begin{figure}
    \centering
    \includegraphics[scale=0.7]{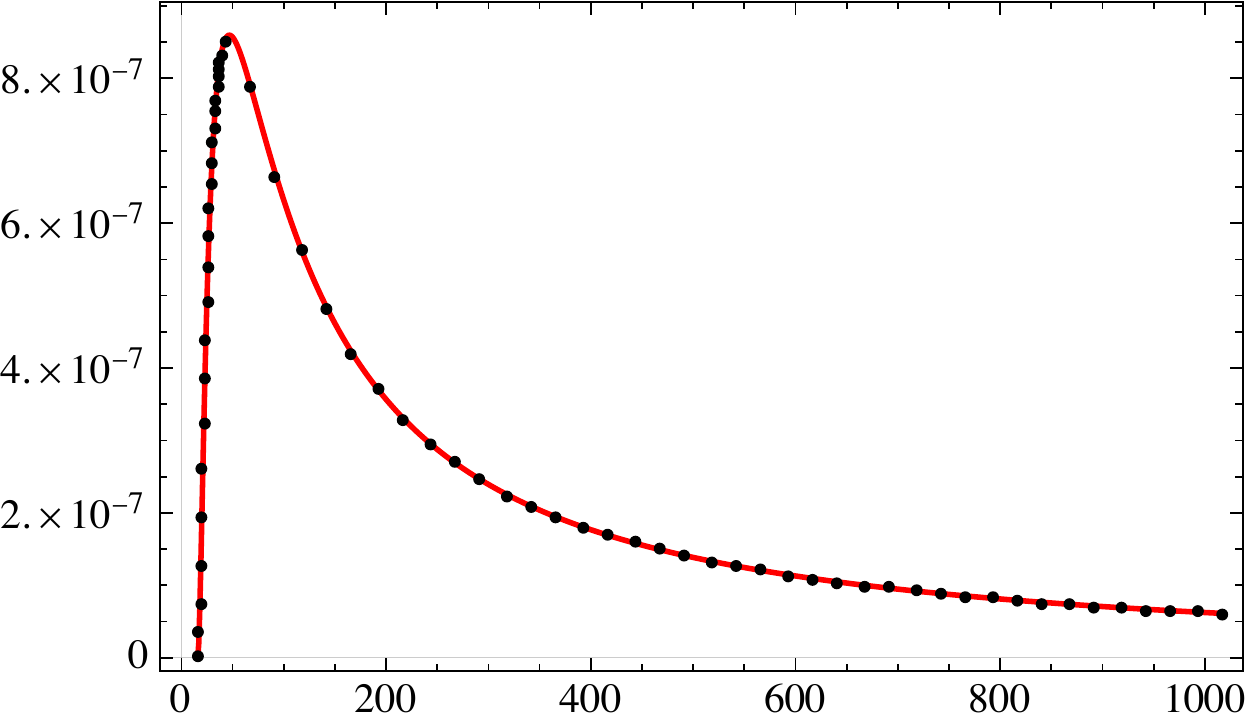}
    \caption{We plot S-wave inelasticity $\eta_{\text{Aks}}(s)$ in $d=3$ as a function of ${s \over m^2}$ for $\lambda=-\pi$ using \eqref{eq:3ddoublebubbleetares} (red). At large $s$, we have $\eta_{\text{Aks}}(s) \sim {1 \over s^{3/2}}$. With black dots we denote the result of initializing the iteration algorithm starting from $(\lambda,0,0)$ and applying the Froissart-Gribov formula to the first nontrivial result for $\rho_{\text{el}}(s,t)$. This serves us a nontrivial consistency check of many formulas used in the draft.}
    \label{fig:threeloopgraphsforswaveinelasticity3d}
\end{figure}

Restoring the overall normalization factor $\lambda^4$ we thus get
\be
T_s &={\lambda^4 \over 1536 \pi^2} \int_{4 m^2}^{(\sqrt{s}-2m)^2} {d m_{56}^2 \over m_{56}} {d m_{78}^2  \over m_{78} } {\theta(\tilde K) \over \tilde K} {\sqrt{s} \over s-4m^2}{z^0 \over \sqrt{(z^0)^2 - 1}} {1 \over 2 (z^0)^2 - 1 - z}
\ee
We can now write the formula for $\eta_{\text{Aks}}(s)$, which is simply the projection of the discontinuity to the spin zero partial wave
\be
{\sqrt{s} \over 2} \eta_{\text{Aks}}(s) = {1 \over 8 \pi} \int_{-1}^{1} d z {1 \over \sqrt{1-z^2}}T_s(s,z) . 
\ee
The relevant integral takes the form
\be
{1 \over 8 \pi} \int_{-1}^{1} d z  {1 \over \sqrt{1-z^2}}  {1 \over 2 (z^0)^2 - 1 - z} = {1 \over 16 } {1 \over z^0 \sqrt{(z^0)^2-1}}.
\ee
Putting everything together we get

\be
\eta_{\text{Aks}}(s) =\left( {\lambda \over 16} \right)^4 {4  \over 3 \pi^2} \int_{4 m^2}^{(\sqrt{s}-2m)^2} {d m_{56}^2 \over m_{56}} {d m_{78}^2 \over m_{78}} {\theta \Big( (m_{56}^2 - m_{78}^2)^2 - 2 (m_{56}^2 + m_{78}^2) s + s^2 \Big) \over m^2 s^2+\left(m_{56}^2 m_{78}^2-2 m^2 \left(m_{56}^2+m_{78}^2\right)\right) s+m^2
   \left(m_{56}^2-m_{78}^2\right){}^2} .
   \label{eq:3ddoublebubbleetares}
\ee
We plot $\eta_{\text{Aks}}(s)$ in \figref{fig:threeloopgraphsforswaveinelasticity3d}. At large $s$ it decays as ${1 \over s^2}$.

\subsection{$d=4$}

The four-body discontinuity takes the form
\be
2 T_s = {1 \over 4!} \int {d^3 \vec q_5 \over (2 \pi)^3 (2 E_{\vec q_5})} {d^3 \vec q_6 \over (2 \pi)^3 (2 E_{\vec q_6}) } {d^3 \vec q_7 \over (2 \pi)^3 (2 E_{\vec q_7})}  {d^3 \vec q_8 \over (2 \pi)^3 (2 E_{\vec q_8})} \nn \\ 
(2 \pi)^4 \delta(p_1+p_2- q_5 - q_6 - q_7-q_8) T(p_1,p_2; q_5,q_6, q_7,q_8) T^*(q_5,q_6,q_7,q_8; p_3, p_4) ,
\ee
where $s = (p_1 + p_2)^2$ and $t=(p_1-p_3)^2$. We will be interested in the amplitudes
\be
T(p_1,p_2; q_5,q_6, q_7,q_8) &=- {1 \over (q_5+q_6-p_1)^2 - m^2}\nn \\
T^*(q_5,q_6,q_7,q_8; p_3, p_4) &=- {1 \over (q_7+q_8-p_4)^2 - m^2} .
\ee

Let us first transform the measure. We define $(q_5+q_6)^2=m_{56}^2$, $(q_7+q_8)^2=m_{78}^2$
\be
2 T_s&={1  \over 4!} \int_{2 m^2}^{(\sqrt{s}-2m)^2} d m_{56}^2 d m_{78}^2 \int d^4 q_{56} d^4 q_{78} \delta^+(q_{56}^2 - m_{56}^2) \delta^+(q_{78}^2 - m_{78}^2) T(p_1,q_{56}) T^*(p_4,q_{78}) \nn \\
 &\times (2 \pi)^4 \delta(p_1+p_2- q_{56} - q_{78}) {\cal I} ,
\ee
where
\be
&{\cal I} \equiv {1 \over (2 \pi)^{12}} \int d^4 q_5 \delta^+(q_5^2 - m^2) \int d^4 q_5 \delta^+(q_6^2 - m^2) \delta^{(4)}(q_{56} - q_5 - q_6) \nn \\
& \int d^4 q_7 \delta^+(q_7^2 - m^2) \int d^4 q_8 \delta^+(q_8^2 - m^2) \delta^{(4)}(q_{78} - q_7 - q_8) .
\ee

Let's go to the COM frame $q_{78}= (m_{78}, \vec 0)$ to get
\be
 &\int d^4 q_7 \delta^+(q_7^2 - m^2) \int d^4 q_8 \delta^+(q_8^2 - m^2) \delta^{(4)}(q_{78} - q_7 - q_8) \nn \\
 &=  \int {d^3 \vec q_7 \over 4 E_{\vec q_7}^2}  \delta(m_{78} - 2 E_{\vec q_7}) ={\pi \over 2}  \sqrt{1 - {4 m^2 \over m_{78}^2}}, ~~~ m_{78}^2 \geq 4 m^2 . 
\ee

So we are left with the following equation
\be
2 T_s &={(2 \pi)^{-8} \over 4!}  {\pi^2 \over 4}  \int_{4 m^2}^{(\sqrt{s}-2m)^2} d m_{56}^2 d m_{78}^2  \sqrt{1 - {4 m^2 \over m_{56}^2}}  \sqrt{1 - {4 m^2 \over m_{78}^2}} \nn \\
&\int d^4 q_{56} d^4 q_{78} \delta^+(q_{56}^2 - m_{56}^2) \delta^+(q_{78}^2 - m_{78}^2) T(p_1,q_{56}) T^*(p_4,q_{78}) \delta(p_1+p_2- q_{56} - q_{78}).
\ee

Going to the COM frame we get
\be
&\int d^4 q_{56} d^4 q_{78} \delta^+(q_{56}^2 - m_{56}^2) \delta^+(q_{78}^2 - m_{78}^2) T(p_1,q_{56}) T^*(p_4,q_{78}) \delta(p_1+p_2- q_{56} - q_{78}) \nn \\
&=\int {d^3 \vec q_{56} \over 4 E_{\vec q_{56},m_{56}} E_{\vec q_{56},m_{78}} } \delta(\sqrt{s} - E_{\vec q_{56},m_{56}} - E_{-\vec q_{56},m_{78}} )T(p_1,q_{56}) T^*(p_4,q_{78})
\ee
We write
\be
T(p_1,q_{56}) &=- {1 \over (q_{56}-p_1)^2 - m^2} =- {1 \over m_{56}^2 - 2 q_{56} \cdot p_1}\nn \\
T^*(p_4,q_{78}) &=- {1 \over (q_{78}-p_4)^2 - m^2}=- {1 \over m_{78}^2 - 2 q_{78} \cdot p_4} .
\ee
Note that due to the fact that $\vec q_{78} = - \vec q_{56}$ we have $z_{78,4} = - z_{56,4} = z_{56,3}$.

For the angular measure we introduce
\be
d \Omega_{56} &=2 {d z_{56,1} d z_{56,3} \theta\Big( -K(z_{56,1}, z_{56,3}, z) \Big) \over \sqrt{-K(z_{56,1}, z_{56,3}, z)}} .
\ee

As a result the integral takes the form
\be
2 T_s &={(2 \pi)^{-8} \over 4!}  {\pi^2 \over 4}  \int_{4 m^2}^{(\sqrt{s}-2m)^2} d m_{56}^2 d m_{78}^2  \sqrt{1 - {4 m^2 \over m_{56}^2}}  \sqrt{1 - {4 m^2 \over m_{78}^2}} {\theta(\tilde K)  \over \sqrt{\tilde K} (s-4 m^2) } \nn \\
&\int_{-1}^1 d z_{56,1} d z_{56,3} {\theta\Big( -K(-z,  z_{56,1} , z_{56,3}) \Big) \over \sqrt{-K(-z,  z_{56,1} , z_{56,3}) }} {1 \over z^0 - z_{56,1} } {1 \over z^0 - z_{56,3} }. 
\ee
where
\be
\tilde K &= (m_{56}^2 - m_{78}^2)^2 - 2 (m_{56}^2 + m_{78}^2) s + s^2 , \nn \\
z^0 &={\sqrt{s} \over \sqrt{s-4m^2}} {s- m_{56}^2 - m_{78}^2 \over \sqrt{\tilde K} } >1 .
\ee
As before we can take one of the integrals easily using 
\be
\int_{-1}^1 d z_{56,1}  {\theta\Big( -K(z,  z_{56,1} , z_{56,3}) \Big) \over \sqrt{-K(z,  z_{56,1} , z_{56,3}) }} {1 \over z^0 - z_{56,1} }  = {\pi \over \sqrt{K(z,  z^0 , z_{56,3})}} .
\ee
Finally we define the last integral 
\be
{\cal I}(z,z^0) &\equiv \int_{-1}^1  d z_{56,3}  {1 \over \sqrt{K(z,  z^0 , z_{56,3})}} {1 \over z^0 - z_{56,3} } \nn \\
&=\frac{\tanh ^{-1}\left(\frac{2 \sqrt{(1-z) \left(2 (z^0)^2-1-z\right)} \left((z^0)^2-z\right)}{2 z^2-4 z (z^0)^2+(z^0)^4+2 (z^0)^2-1}\right)}{\sqrt{(1-z) \left(2
   (z^0)^2-1-z\right)}} . 
\ee

In this way we get (where we also have restored the $\lambda$ factor)

\be
T_s &={4 \pi^3 \over 3} \Big( {\lambda \over 16 \pi^2} \Big)^4 \int_{4 m^2}^{(\sqrt{s}-2m)^2} d m_{56}^2 d m_{78}^2  \sqrt{1 - {4 m^2 \over m_{56}^2}}  \sqrt{1 - {4 m^2 \over m_{78}^2}} {\theta(\tilde K)  \over \sqrt{\tilde K} (s-4m^2) } {\cal I}(z,z^0) .
\ee
We can now write the formula for $\eta_{\text{Aks}}(s)$
\be
{1 \over 2} {\sqrt{s} \over \sqrt{s-4m^2}}\eta_{\text{Aks}}(s) = {1 \over 32 \pi} \int_{-1}^{1} d z T_s(s,z) . 
\ee
We get that
\be
{1 \over 32 \pi} \int_{-1}^{1} d z {\cal I}(z,z^0) = {1 \over 32 \pi} \Big(\log {z^0 -1 \over z^0 + 1} \Big)^2.
\ee

\begin{figure}[h]
    \centering
    \includegraphics[scale=0.8]{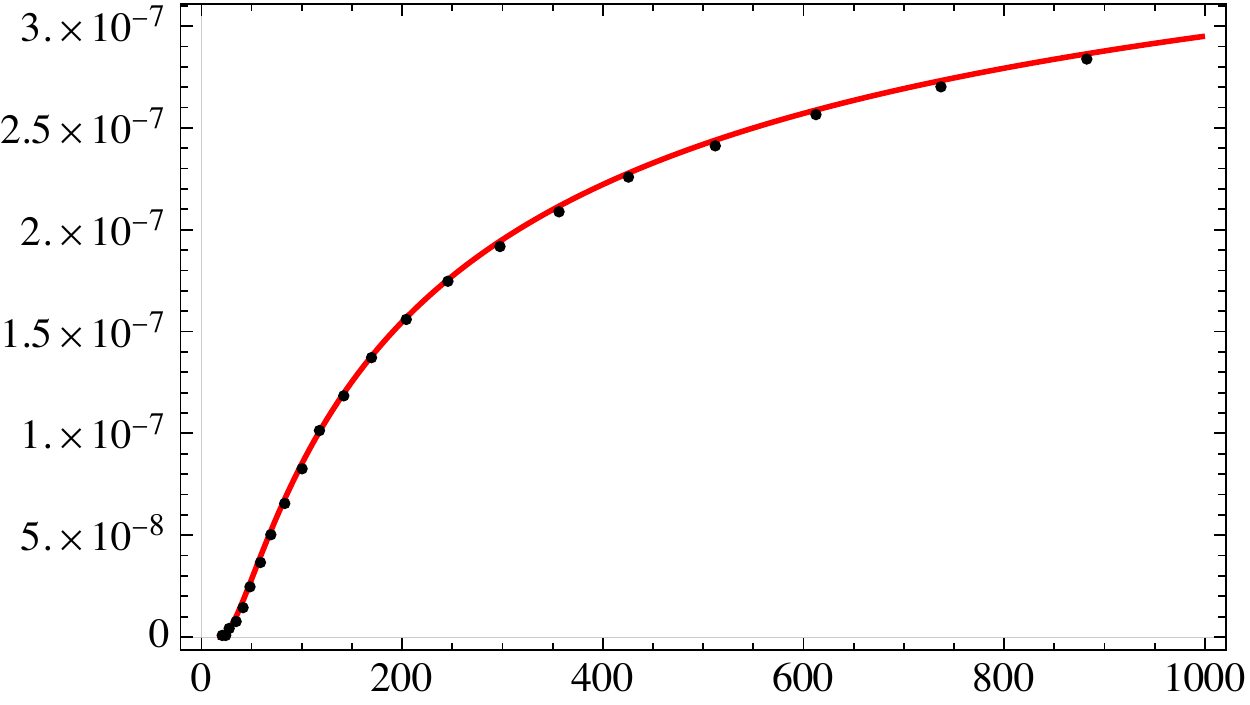}
    \caption{We plot S-wave inelasticity $\eta_{\text{Aks}}(s)$ in $d=4$ as a function of ${s \over m^2}$ for $\lambda=\pi$ using \eqref{eq:doublebubbleeta4d} (red). At large $s$, we have $\eta_{\text{Aks}}(s) \sim 1$. The black dots were produced by our numerical algorithm at first nontrivial iteration for double spectral density.}
    \label{fig:threeloopgraphsforswaveinelasticity4d}
\end{figure}

In this way we get
\be
\eta_{\text{Aks}}(s) ={\pi^2 \over 12}  \Big( {\lambda \over 16 \pi^2} \Big)^4 {1 \over \sqrt{s} \sqrt{s-4m^2}} \int_{4 m^2}^{(\sqrt{s}-2m)^2} d m_{56}^2 d m_{78}^2  \sqrt{1 - {4 m^2 \over m_{56}^2}}  \sqrt{1 - {4 m^2 \over m_{78}^2}} {\theta(\tilde K)  \over \sqrt{\tilde K}}  \Big( \log {z^0 -1 \over z^0 + 1} \Big)^2
\label{eq:doublebubbleeta4d}
\ee
We plot $\eta_{\text{Aks}}(s)$ in \figref{fig:threeloopgraphsforswaveinelasticity4d}, where we also compare it to our numerics as a consistency check.

\section{Full coupling dependence of $d=4$ 2QE and 2PR partial waves}
\label{app:extraSj4d}

\begin{figure}
    \centering
    \includegraphics{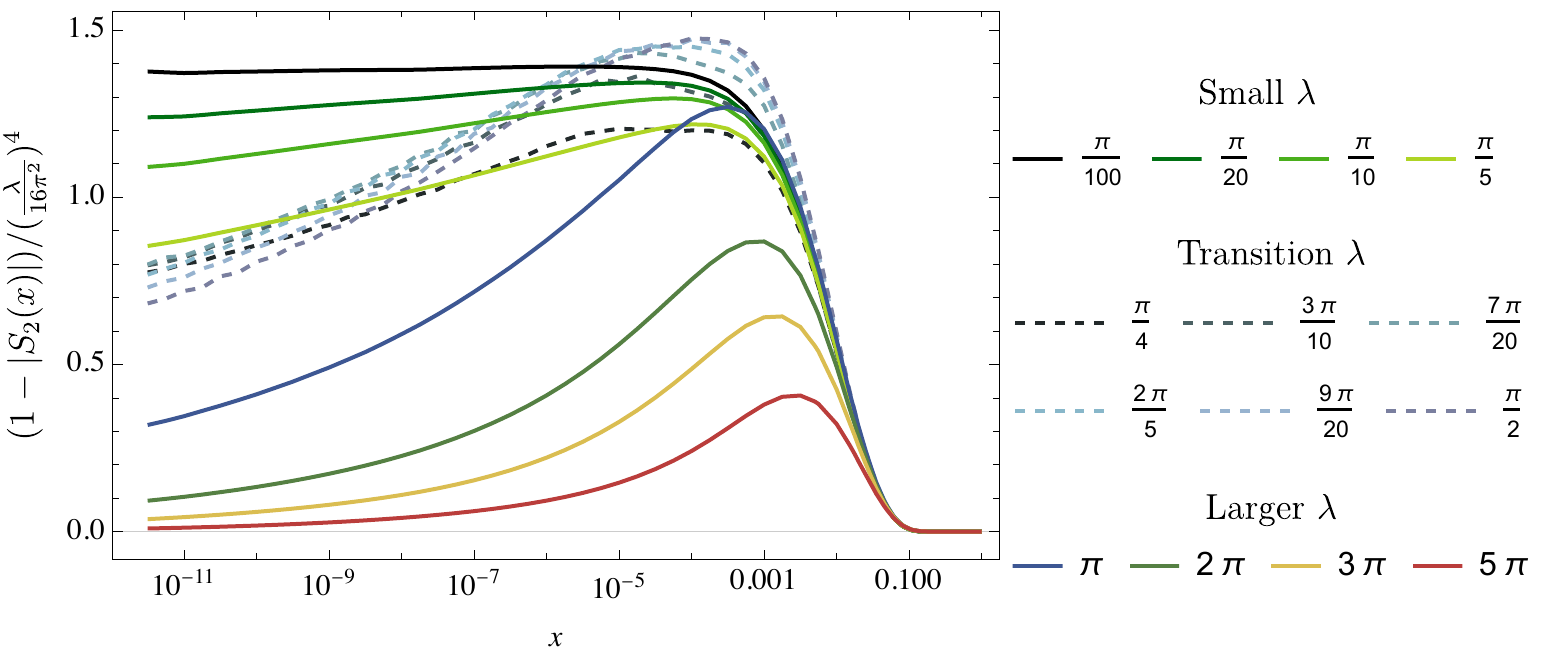}
    \caption{Inelasticities in the $J=0$ sector for $d=4$ 2QE amplitudes for different coupling constants $\lambda$. We identify three regions: small couplings, transition zone, and larger couplings. The later region is already plotted in \figref{fig:pl4dS0S2all} but we reproduce it here for convenience. The small couplings clearly seem to be dominated by some very large scale, inaccessible to our calculation. An irregular transition occurs between small and medium lambda, which would be interesting to understand. %
    }
    \label{fig:pl4dallS2}
\end{figure}

\begin{figure}
    \centering \includegraphics{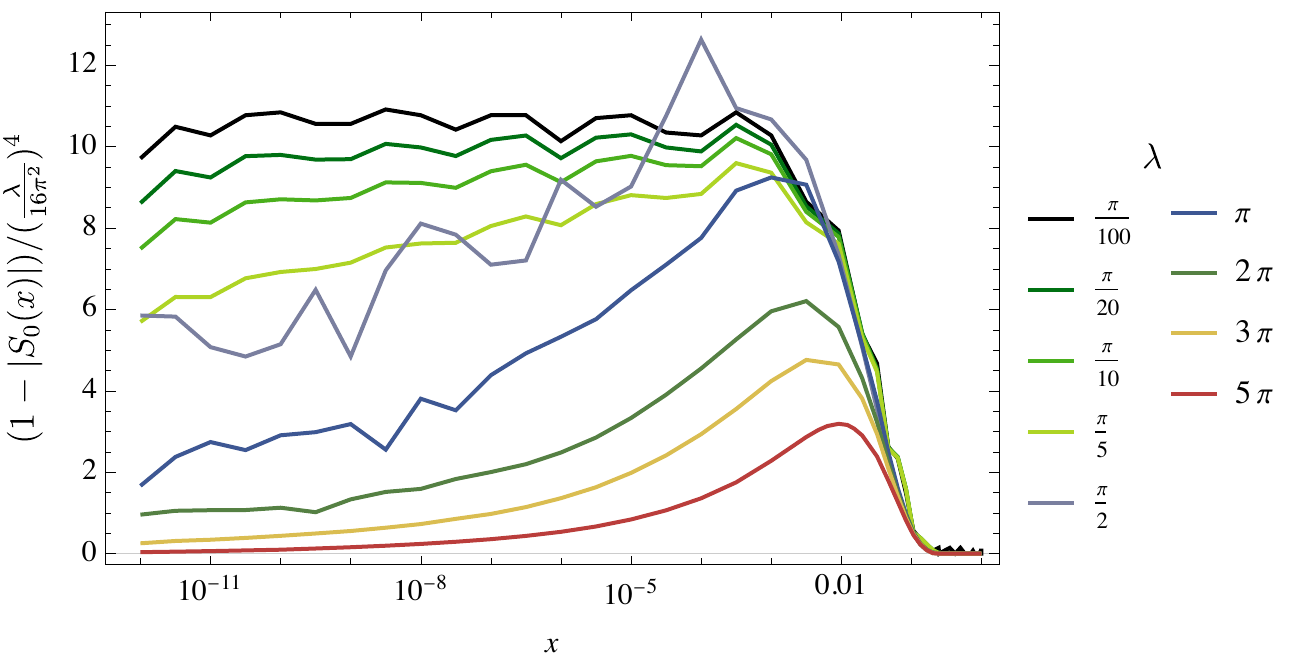}
    \caption{Inelasticities in the $J=0$ sector for $d=4$ 2PR amplitudes for different coupling constants $\lambda$. We did not generate data to probe accurately the transition range for the 2PR amplitudes, but the data for $\lambda=\pi/2$ shows a consistent behavior with the transition range in the $J=2$ case for the 2QE dataset in \figref{fig:pl4dallS2}.}
    \label{fig:pl4d2PRRallS0normalisedsqLargeLambdaFull}
\end{figure}

\begin{figure}
    \centering
    \includegraphics{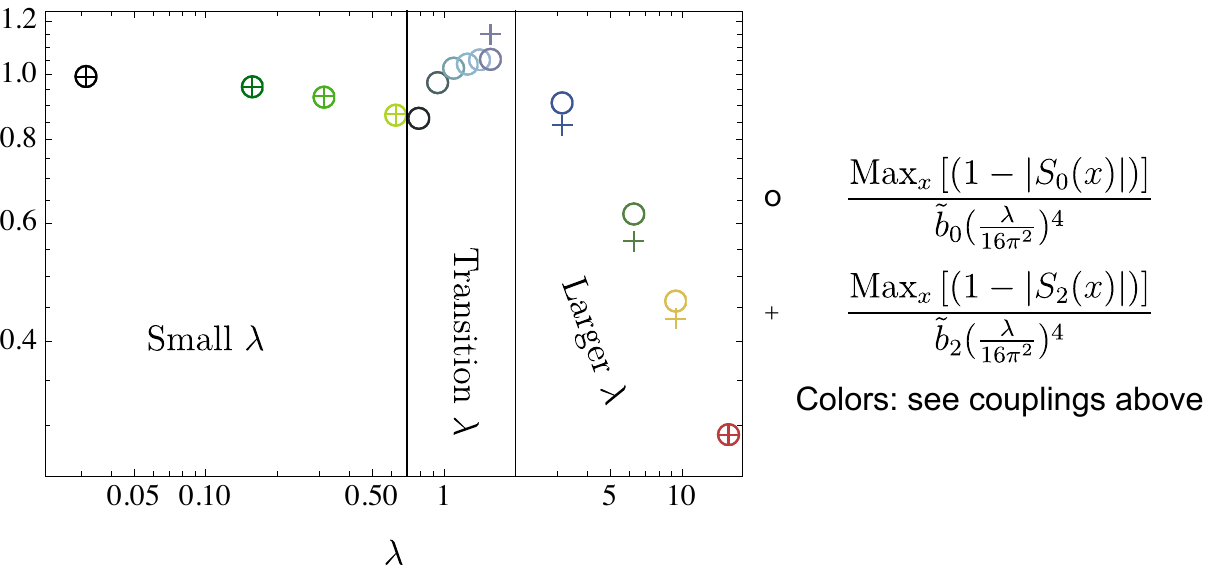}
    \caption{Maximum value of rescaled inelasticity in  the partial waves of the 2PR amplitude in $d=4$ as a function of $\lambda$, same color code as Fig.~\ref{fig:pl4dallS2}. The fitting parameters are given by $\tilde b_0 = 11$ and $\tilde b_2 = 1.4$. }
    \label{fig:plmaxS2QE4dflatFull}
\end{figure}

In this appendix we present the complete set of data for partial waves of the 2QE and 2PR amplitudes in $d=4$ for different values of $\lambda$. We identified three regions (small couplings, transition zone, and larger couplings), where the amplitude behaved in a slightly different way, see \figref{fig:pl4dallS2}, \figref{fig:pl4d2PRRallS0normalisedsqLargeLambdaFull}, \figref{fig:plmaxS2QE4dflatFull}. 

The most immediate explanation of the jaggedness would be that is related to sensitivity of the amplitude in this region to the high-energy cut-off and, possibly, to the finiteness of the grid. We leave a more detailed exploration of these amplitudes for future work.

\section{Impact parameter space and the Froissart-Gribov formula}
\label{app:impactFG}

In the main text we have analyzed the behavior of the partial waves in the impact parameter representation $S(s,b)$, see \eqref{eq:impactunitarity}. 
Here we derive a formula for the phase shift $\delta(s,b)$ defined via
\be
S(s,b) = e^{2 i \delta(s,b)} 
\ee
using the Froissart-Gribov formula.

Recall that we consider the limit $J \to \infty$, $s \to \infty$ with the ratio
\be
b \equiv {2 J \over \sqrt{s-4 m^2}}
\ee
held fixed. 

We would like to take the corresponding limit in the Froissart-Gribov representation which we write as follows
\beq\label{eq:FGform2}
 f_J(s) ={2 \, {\cal N}_d\over\pi}{2 \over s - 4 m^2} \int\limits_{4 m^2}^\infty d t \,(z^2-1)^{{d-4 \over 2}}Q^{(d)}_J(1+{2 t \over s- 4 m^2}) T_t(s,t) \ ,
\eeq
where we assume that as we take $s \to \infty$ the leading Regge intercept does not grow faster than $\sqrt{s}$ so that we have $J>J_0(s)$. This is manifestly true in the amplitudes that obey the Mandelstam representation since then $J_0(s)$ is bounded by some fixed number for any $s$.

The relevant limit of the $Q$-function takes the form
\be
\lim_{J \to \infty} Q_J^{(d)}(1+{t b^2 \over 2 J^2}) = 2^{{d-4 \over 2}} \Gamma({d-2 \over 2}) (\sqrt{t} b)^{2 -{d \over 2}} K_{{d-4 \over 2}}( \sqrt{t} b ) .
\ee

The relationship between the phase shift and the partial wave takes the form
\be
e^{2 i \delta(s,b)} = 1+i {(s- 4 m^2)^{{d-3 \over 2}} \over \sqrt{s}}f_J(s).
\ee

So that we get
\be
 i (1 - e^{2 i \delta(s,b)}) =
 {1 \over (2 \pi)^{d/2}}{1 \over  b^{d-4}} \int\limits_{4 m^2}^\infty {d t \over s} (\sqrt{t} b)^{{d-4 \over 2}} K_{{d-4 \over 2}}( \sqrt{t} b )  T_t(s,t) + ... \ ,
\ee
where $...$ stand for corrections which are subleading in ${1 \over \sqrt{s} b} \sim {1 \over J}$.

In the context of our paper high-energy phase shift is small. Therefore we can expand the LHS of the formula above to get
\be
2 \delta(s,b) =  {1 \over (2 \pi)^{d/2}}{1 \over  b^{d-4}} \int\limits_{4 m^2}^\infty {d t \over s} (\sqrt{t} b)^{{d-4 \over 2}} K_{{d-4 \over 2}}( \sqrt{t} b )  T_t(s,t) .
\ee
Notice that $K_{{d-4 \over 2}}( \sqrt{t} b ) \sim e^{- \sqrt{t} b}$ and therefore the integral localizes close to its minimal value $t = 4 m^2$ which produces the expected Yukawa potential $e^{- 2 b m}$. 

In the main text we were interested in the behavior of $1 - |S(s,b)|$ which is controlled by the leading behavior of the imaginary part of the phase shift
\be
1 - |S(s,b)| \simeq 2 {\rm Im} \delta(s,b) = {1 \over (2 \pi)^{d/2}}{1 \over  b^{d-4}} \int\limits_{4m^2}^\infty {d t \over s} (\sqrt{t} b)^{{d-4 \over 2}} K_{{d-4 \over 2}}( \sqrt{t} b )  \rho(s,t).
\ee
This time the integral is localized close to the leading Landau curve in the crossed channel ${4 m^2 s \over s - 16 m^2}$ which for $s \to \infty$ starts at $t=4m^2$. 

To get some feeling how the phase shift depends on the near-threshold behavior of the integrand consider the following ansatz $T_t(s,t) \sim c_0 ({t \over 4 m^2}-1)^\alpha$, we then get
\be
\delta(s,b) \sim {c_0  \over s} {e^{- 2 b m} \over (b m)^{{d-1 \over 2}+\alpha}}  \ .
\ee

\section{Acnode evaluation}
\label{app:acnodeapp}

To derive \eqref{eq:singlediscacnode} starting from \eqref{eq:threebodyunitarity}, we first eliminate $p_7$ completely using the energy-momentum conservation and restore the on-shell conditions to get for the measure
\be
{1 \over 3!}  (2 \pi) \int {d^4 q_5 \over (2 \pi)^3} {d^4 q_6 \over (2 \pi)^3}  \delta^+(q_5^2-m^2) \delta^+(q_6^2-m^2)  \delta^+((p_1+p_2 - q_5 - q_6)^2 - m^2) .
\ee
We next eliminate $| \vec q_i|$ integrals via
\be
&\int d^4 q_i \delta^+(q_i^2 - m^2)=\int \theta(q_i^0) d q_i^0 d |\vec q_i| |\vec q_i|^2 d \Omega_i \delta( (q_i^0)^2 - m^2 - |\vec q_i|^2) \nn \\
&=\int_{m}^\infty d q_i^0 {\sqrt{(q_i^0)^2 - m^2} \over 2} d \Omega_i .
\ee
In this way we get for the measure
\be
{1 \over 4} {1 \over 3!} {1 \over (2 \pi)^5} \int_{m}^\infty d q_5^0 \sqrt{(q_5^0)^2 - m^2} d \Omega_5 \int_{m}^\infty d q_6^0 \sqrt{(q_6^0)^2 - m^2} d \Omega_6 \delta^+((p_1+p_2 - q_5 - q_6)^2 - m^2) .
\ee
As a next step we write the angular measure in a more convenient way. First, we introduce the relative angles $z_{i j} \equiv \cos \vec n_i \cdot \vec n_j$, where $\vec n_i = {\vec q_i \over | \vec q_i|}$. Second, we switch for the measure
\be
{1 \over 4} d \Omega_5 d \Omega_6 &= {d z_{15} d z_{56} \theta\Big( -K(z_{15}, z_{56},z_{16}) \Big) \over \sqrt{-K(z_{15}, z_{56}, z_{16})}} {d z_{26} d z_{46} \theta\Big( -K(z_{24}, z_{26}, z_{46}) \Big) \over \sqrt{-K(z_{24}, z_{26}, z_{46})}} , \\
-K(z, z', z'') &= 1 - z^2 - z'^2 - z''^2 + 2 z z' z'' .
\ee
This formula holds if we use it for the integrand which only depends on the relative angles that appear in the RHS. This will be the case for us.

It is very convenient then to switch to the COM frame where $\vec p_2 = - \vec p_1$. In this frame we get
\be
(p_1+p_2 - q_5 - q_6)^2 - m^2 &= (\sqrt{s} - q_5^0 - q_6^0)^2 - (\vec q_5 -  \vec q_6)^2 - m^2 \nn \\
&= 2 \sqrt{(q_5^0)^2 - m^2} \sqrt{(q_6^0)^2 - m^2} (z_{56} - z_{56}^0), \nn \\
z_{56}^0 &={ s + m^2 + 2 q_{5}^0 q_6^0 - 2 \sqrt{s} (q_{5}^0 + q_6^0)  \over 2 \sqrt{(q_5^0)^2 - m^2} \sqrt{(q_6^0)^2 - m^2}} .
\ee
In this way we get for the phase space integral measure
\be
&{1 \over 2} {1 \over 3!} {1 \over (2 \pi)^5} \int_{m}^\infty d q_5^0 \int_{m}^\infty d q_6^0 \theta(\sqrt{s} - m - q_5^0 - q_6^0) \nn \\
&\int_{-1}^1 {d z_{15} d z_{56} \theta\Big( -K(z_{15}, z_{56},-z_{26}) \Big) \over \sqrt{-K(z_{15}, z_{56}, -z_{26})}} {d z_{26} d z_{46} \theta\Big( -K(z_{24}, z_{26}, z_{46}) \Big) \over \sqrt{-K(z_{24}, z_{26}, z_{46})}} \delta(z_{56} - z_{56}^0),
\ee
where we used that in the COM frame $z_{16} = - z_{26}$. 
Next we introduce the invariant momenta of the pairs of particles $s_{57} = (q_5 + q_7)^2$ and $s_{67} = (q_6 + q_7)^2$ so that
\be
q_{5}^0 = {s + m^2 - s_{67} \over 2 \sqrt{s}} , ~~~ q_{6}^0 = {s + m^2 - s_{57} \over 2 \sqrt{s}}  . 
\ee
We get
\be
\int_{m}^\infty d q_5^0 \int_{m}^\infty d q_6^0 \theta(\sqrt{s} - q_5^0 - q_6^0-m) ={1 \over 4 s} \int_{4m^2}^{(\sqrt{s}-m)^2} d s_{57} d s_{67} \theta(s_{57} + s_{67} - 2 m^2 -2 m \sqrt{s} ) .
\ee
If we set $\sqrt{s} = 3m$ the integration region shrinks to zero. Therefore we get the expected $\theta(s-9m^2)$.

The limits of integrations could be further simplified and we finally get the following formula
\be
2 T_s &= {1 \over 2} {1 \over 3!} {1 \over (2 \pi)^5} {1 \over 4 s} \int_{4m^2}^{(\sqrt{s}-m)^2} d s_{57} d s_{67} \theta(1 - (z_{56}^{0})^2) \nn \\
&\int_{-1}^1  {d z_{15} \theta\Big( -K(z_{15}, z_{56}^0,-z_{26}) \Big) \over \sqrt{-K(z_{15}, z_{56}^0, -z_{26})}} {d z_{26} d z_{46} \theta\Big( -K(z_{24}, z_{26}, z_{46}) \Big) \over \sqrt{-K(z_{24}, z_{26}, z_{46})}} T_{1,2\to 5,6,7} T^*_{5,6,7 \to 3,4}
\ee

We next write the explicit expressions for the amplitudes that appear in the acnode graph
\be
T(p_1,p_2; q_5,q_6, q_7) &=- {1 \over t_{15} - m^2} =- {1 \over 2 | \vec p_1 | | \vec q_5| } {1 \over z_{15} - z_{15}^0}, \\
T^*(q_5,q_6,q_7; p_3, p_4) &=-  {1 \over t_{46} - m^2} =  -{1 \over 2 | \vec p_4 | | \vec q_6| } {1 \over z_{46} - z_{46}^0},
\ee
where
\be
z_{15}^0 = {2 p_1^0 q_5^0 - m^2 \over 2 | \vec p_1 | | \vec q_5| }, ~~~ z_{46}^0 = {2 p_4^0 q_6^0 - m^2 \over 2 | \vec p_4 | | \vec q_6| } .
\ee
It is easy to check using the on-shell condition that $z_{15}^0, z_{46}^0 > 1$. 

Recall that 
\be
z_{56}^0 =\frac{s
   \left(s_{57}+s_{67}\right) - s^2 +(s_{57}-m^2)(s_{67}- m^2)}{\sqrt{m^4-2 m^2
   \left(s+s_{57}\right)+\left(s-s_{57}\right){}^2} \sqrt{m^4-2 m^2
   \left(s+s_{67}\right)+\left(s-s_{67}\right){}^2}} . 
\ee
We then get for the acnode
\be
2 T_s &= {1 \over 8 s} {1 \over 3!} {1 \over (2 \pi)^5}   \int_{4 m^2}^{(\sqrt{s}-m)^2} {d s_{57} d s_{67} \over 4 | \vec p_1 | | \vec p_5| | \vec p_4 | | \vec p_6| }d z_{15} d z_{26} d z_{46} \theta(1 - (z_{56}^{0})^2)  \nn \\
&\times { \theta\Big( -K(z_{15}, z_{56}^0, -z_{26}) \Big) \theta\Big( -K(z, z_{26}, z_{46}) \Big)  \over \sqrt{K(z_{15}, z_{56}^0, -z_{26}) K(z, z_{26}, z_{46})}} {1 \over z_{15} - z_{15}^0} {1 \over z_{46} - z_{46}^0} .
\ee
Next step is to do integrals over $z_{15}$ and $z_{46}$. This is done using the following useful formula
\be
\int_{-1}^1 d z_{15} {\theta(-K(z_{15}, z_{56}^0, -z_{26})) \over \sqrt{-K(z_{15}, z_{56}^0, -z_{26})}} {1 \over  z_{15}^0 - z_{15}}  = {\pi \over \sqrt{K(z_{15}^0,z_{56}^0, -z_{26})}} . 
\ee

In this way we arrive at the final formula for the unitarity cut of the acnode \eqref{eq:singlediscacnode}.

\newpage

\bibliographystyle{JHEP}
\bibliography{papers} 

\end{document}